\documentclass[a4paper,11pt,times,custombib,index]{PhDThesisPSnPDF}

\usepackage[grey]{quotchap}
\makeatletter
\renewcommand\chapter{%
	\if@openright\cleardoublepage\else\clearpage\fi
	\thispagestyle{plain}%
	\global\@topnum\z@
	\null\hfill\@printcites\par
	\@afterindentfalse
	\secdef\@chapter\@schapter
}
\renewcommand{\@makechapterhead}[1]{%
	\chapterheadstartvskip%
	{\size@chapter{\sectfont\raggedright
			{\chapnumfont
				\ifnum \c@secnumdepth >\m@ne%
				\if@mainmatter\thechapter%
				\fi\fi
				\par\nobreak}%
			{\raggedright\advance\leftmargin10em\interlinepenalty\@M #1\par}}
		\nobreak\chapterheadendvskip}}
\makeatother
\input{preamble}

\newcommand{\aap}{    {\it A \& A}}
\newcommand{\aaps}{   {\it A \& A Suppl.}}

\newcommand{\aj}{     {\it AJ}}
\newcommand{\apj}{    {\it ApJ}}
\newcommand{\apjl}{   {\it ApJL}}
\newcommand{\apjs}{   {\it ApJ Suppl.}}

\newcommand{\mnras}{  {\it MNRAS}}
\newcommand{\nat}{    {\it Nature}}
\newcommand{\pasp}{   {\it Pub. Astron. Soc. Pac.}}

\newcommand{\araa}{   {\it Anu Rev. in Astronomy and Astrophysics.}}
\newcommand{\na}{     {\it Nature Astronomy}}
\newcommand{\pasa}{   {\it Publications of the Astronomical Society of Australia}}

\usepackage{adjustbox}
\usepackage{pdflscape}
\usepackage{rotating}
\usepackage{graphicx}
\usepackage{epstopdf}
\usepackage{epsfig}
\usepackage{wrapfig}

\usepackage{threeparttable}
\usepackage{lipsum}
\usepackage{mathtools}
\usepackage{cuted}
\usepackage{rotating}



\def \be{\begin{equation}}
	\def \ee{\end{equation}}
\newcommand       \ba           {\begin{eqnarray}}
	\newcommand       \ea           {\end{eqnarray}}
\def \bea{\begin{eqnarray}}
	\def \eea{\end{eqnarray}}

\newcommand{\comments}[1]{}

\newcommand   \la          {\langle}

\usepackage{multicol}
\DeclareMathOperator{\sech}{sech}
\newcommand{\kms}{\mbox{$\>{\rm km\, s^{-1}}$}}
\newcommand{\pc}{\>{\rm pc}}
\newcommand{\kpc}{\mbox{$\>{\rm kpc}$}} 

\newcommand{\Gyr}{\mbox{$\>{\rm Gyr}$}}
\newcommand{\Myr}{\mbox{$\>{\rm Myr}$}}

\newcommand\degrees{^\circ}
\newcommand\gaia{{\it Gaia}}
\newcommand{\avg}[1]{\mbox{$\left<{#1}\right>$}}

\newcommand{\vb}{\mbox{$V_{\rm breath}$}}

\usepackage{mathrsfs}
\allowdisplaybreaks
\definecolor{dimgray}{rgb}{0.41, 0.41, 0.41}

\definecolor{webgreen}{rgb}{0,.5,0}
\definecolor{webbrown}{rgb}{.6,0,0}

\hypersetup{%
	colorlinks=true,hyperfootnotes=false,%
	breaklinks=true,%
	plainpages=false, bookmarksnumbered, bookmarksopen=true,
	bookmarksopenlevel=1,%
	urlcolor=blue, linkcolor=blue, citecolor=blue,
}

\begin{document}
	\title{A study of the evolution of the bulges and disks of spiral galaxies in interacting and isolated environments}
	\submitdate{November 2022}
	\degree{Doctor of Philosophy}
	\dept{Joint Astronomy Programme 
	\\ Department of Physics}
	\faculty{Faculty of Science}
	\author{Ankit Kumar}
	
	\begin{titlepage}
		\maketitle
	\end{titlepage}
	\null
	\thispagestyle{empty} 
	\newpage
	
	\vspace*{\fill}
	\begin{center}
		\large\bf \textcopyright \ Ankit Kumar\\
		\large\bf November 2022\\
		\large\bf All rights reserved
	\end{center}
	\vspace*{\fill}
	\thispagestyle{empty}
	%
	
	\setcounter{secnumdepth}{3}
	\setcounter{tocdepth}{3}
	
	\pagenumbering{roman}
	
	\begin{declaration}
I hereby declare that the work reported in this doctoral thesis titled ``A study of the evolution of the bulges and disks of spiral galaxies in interacting and isolated environments'' is entirely original and is the result of investigations carried out by me in the Department of Physics, Indian Institute of Science, Bangalore, under the supervision of Prof. Mousumi Das at the Indian Institute of Astrophysics, Bangalore and Dr. Nirupam Roy at the Indian Institute of Science, Bangalore.

I further declare that this work has not formed the basis for the award of any degree, diploma, fellowship, associateship or similar title of any University or Institution.

\end{declaration}


\begin{dedication} 
\centering
\Large{To}

\Huge{\it My Family\\}

\end{dedication}

	\begin{acknowledgements}
	
\vspace{-3pt}
To reach where I am today, there is the contribution of each and every individual. Here, in this thesis, I am extending my gratitude to all of them.

Firstly, I am very obliged to have my supervisor, Prof. Mousumi Das. Her active encouragement, fruitful scientific discussions, constant support, and freedom to work helped me a lot to dig up deep insight into my reseach topic during this journey. Her care, like my mother, always makes me feel being with family. I want to thank her for never stopping me from pursuing my hobbies throughout my Ph.D.

I thank my nominal guide, Prof. Nirupam Roy, for his help and support in my academic and professional work at the Indian Institute of Science. I am grateful for his kind nature, friendly behaviour, keeping me up to date about academics, and his instant help whenever needed.

I would like to thank my senior, Dr Sandeep Kataria, for his active discussions and insightful suggestions related to research. I am grateful to have him beside me like a brother. I am thankful to him for encouraging me to participate actively in non-academic activities as well.

I am fortunate to work with Dr Soumavo Ghosh. His beneficial suggestions and comments helped me learn to frame research professionally and work in collaboration. I thank Prof. Victor Debattista for his constructive feedback and insightful comments during one of my research projects.

I am thankful to Prof. Annapurni Subramaniam for letting me study in her group. I really appreciate her kind nature, supportive behaviour, and giving her precious time to discuss various topics in stellar astrophysics.

My sincere thank to Prof. Stacy McGaugh for his suggestions, feedback and valuable time for discussion. I sincerely thank Prof. Volker Springel and Prof. Chien Peng for their help and advice regarding codes over email communications. I express my gratitude to my coursework guide, Prof. Firoza Sutaria, for helping me learn IRAF, fv-tool, JUDE and other tools for observational science. 

I extend my gratitude to faculties from various institutions in Bangalore (in alphabetical order): Prof. Banibrata Mukhopadhyay, Prof. Biman Nath, Prof. Chanda Jog, Prof. GC Anupama, Prof. Gajendra Pandey, Prof. M Sampoorna, Prof. Maheswar Gopinathan, Prof. Panditi Vemareddy, Prof. Piyali Chatterjee, Prof. Prateek Sharma, Prof. Pravabati Chingangbam, Prof. Rajeev K Jain, Prof. Ravi Joshi, Prof. Ravinder K banyal, Prof. Sanved Kolekar, Prof. SP Rajaguru, Prof. Smitha Subramanian, Prof. Sudhanshu Barway, Prof. Tarun Souradeep, Prof. Thirupathi Sivarani, Prof. Vivek M, Prof. Wageesh Mishra.

I want to thank the director, dean and BGS at IIA for giving me a chance to work in this institute and providing all the premises that are useful for my work. I like to thank all the academic staff, the administrative officers, and the store staff for their active help. I thank all the librarians for their service in getting valuable books and access to the journals. I am thankful to all the computer center staff, Anish, Fayaz, and Ashok, for their help with computers, supercomputers and the internet. I also want to thank all the Physics department office (IISc) staff, Sumithra, Meena, and Maryappa, for their help and support.

I am thankful to all my JNU professors, BSc Professors and school teachers for being part of my journey at various stages. I appreciate their help, support and advice.

I express my gratitude to my friends, Suman, Sahel, Prerana, Manju, and Sudeb, for their help. I am very fortunate to have a close friend, Pallavi, who is always there to support and motivate me for work. 

I am blessed to have amazing colleagues (in alphabetical order): Abhinaya, Akhil, Amrutha, Anirban, Anohita, Aratrika, Aritra, Athira, Avinash, Bibhuti, Deepak, Deepthi, Ekta, Fazlu, Gurwinder, Harsh, Hema, Indrani, Joby, Jyoti, Khushbu, Manika, Manoj, Masroor, Megha, Partha, Parvathy, Pavan, Payel, Prasanta, Prerana, Priya, Priyanka, Raghubar, Ramya, Ravi, Renu, Rishabh, Ritesh, Rubinur, Saili, Samriddhi, Satabdwa, Sharmila, Shashank, Sioree, Sipra, Sonith, Soumya, Soumyaranjan, Subham, Sudheer, Swastik, Tridib, Vikrant, Vishnu. Thank you, guys, for being part of my journey.

Many thanks to my parents, brothers and sister for being there and providing me with every possible facility. Whatever I could achieve in my life is because of you wonderful people. Your love, affection, and care are next level. You always motivate me to keep moving.

I express my heartfelt gratitude to all the doctors, especially during covid times, for all their help. I salute the army and navy forces of INDIA for their service at the border. Because of their sacrifices, we are living a peaceful life.

{\bf HPC:} I thank the high-performance computing facility ‘NOVA’ at the Indian Institute of Astrophysics, India, where I executed all the simulations. 

{\bf Travel Support:} I thank International Astronomical Union (IAU) Symposium-353 and IIA for supporting my travel to China during 2019. 

{\bf IllustrisTNG:} The IllustrisTNG simulations were undertaken with compute time awarded by the Gauss Centre for Supercomputing (GCS) under GCS Large-Scale Projects GCS-ILLU and GCS-DWAR on the GCS share of the supercomputer Hazel Hen at the High Performance Computing Center Stuttgart (HLRS), as well as on the machines of the Max Planck Computing and Data Facility (MPCDF) in Garching, Germany.

{\bf SDSS:} Funding for the SDSS and SDSS-II has been provided by the Alfred P. Sloan Foundation, the Participating Institutions, the National Science Foundation, the U.S. Department of Energy, the National Aeronautics and Space Administration, the Japanese Monbukagakusho, the Max Planck Society, and the Higher Education Funding Council for England. The SDSS Web Site is http://www.sdss.org/.

The SDSS is managed by the Astrophysical Research Consortium for the Participating Institutions. The Participating Institutions are the American Museum of Natural History, Astrophysical Institute Potsdam, University of Basel, University of Cambridge, Case Western Reserve University, University of Chicago, Drexel University, Fermilab, the Institute for Advanced Study, the Japan Participation Group, Johns Hopkins University, the Joint Institute for Nuclear Astrophysics, the Kavli Institute for Particle Astrophysics and Cosmology, the Korean Scientist Group, the Chinese Academy of Sciences (LAMOST), Los Alamos National Laboratory, the Max-Planck-Institute for Astronomy (MPIA), the Max-Planck-Institute for Astrophysics (MPA), New Mexico State University, Ohio State University, University of Pittsburgh, University of Portsmouth, Princeton University, the United States Naval Observatory, and the University of Washington.

{\bf S$^{4}$G:} This research has made use of the NASA/IPAC Infrared Science Archive, which is funded by the National Aeronautics and Space Administration and operated by the California Institute of Technology.

{\bf VizieR:} This research has made use of the VizieR catalogue access tool, CDS, Strasbourg, France (DOI:10.26093/cds/vizier). The original description of the VizieR service was published in A\&AS 143, 23.

Last but not least, I would like to thank those who have helped during this beautiful journey, and I forgot to mention it here. 

\end{acknowledgements}

	\begin{abstract}

Galaxies are vast collections of stars, gas, dust, and invisible dark matter. They are usually found in groups and clusters, where they interact can with each other gravitationally. These interactions affect the morphology and kinematics of galaxies over the course of their evolution. In this thesis, we have studied the effect of flyby interactions and dark matter distributions on the evolution of bulges and disks of spiral galaxies using N-body simulations. All of the galactic components in our simulations are treated live. We have also studied the cosmic evolution of bulges since z=0.1 using SDSS DR7 data. Additionally, the statistical properties of bulgeless galaxies in a state-of-art cosmological simulation, IllustrisTNG, have been compared with observed bulgeless galaxies in order to test our understanding of galaxy formation. In the following paragraphs, we describe our findings.

To investigate the effect of flyby interactions on the bulges, disks, and spiral arms of Milky Way mass galaxies, we modelled disk galaxies with two types of bulges: classical bulges and boxy/peanut pseudo-bulges. We have performed N-body simulations of galaxy flybys of 10:1 and 5:1 mass ratios with varying pericenter distances. Using photometric and kinematic bulge-disk decompositions of the major galaxy at regular time steps, we found that the disks become shorter and thicker during flyby interactions. Flyby induced spiral arms are transient. They form just after pericenter passage, quickly reach maximum strength, and then slowly decay after reaching maximum strength. There is no effect of flyby interactions on the classical bulges. However, pseudo-bulges in the host galaxy become dynamically hotter at the cost of the disk. We also found no effect of flyby interactions on the strength and the formation time of bar buckling.

We found that flyby induced spiral arms are density waves in nature in contrast to material arms. These spiral arms show two winding phases: the initial rapid winding phase and the subsequent slow winding phase. We confirmed that the spiral arms are the main drivers of the observed wave-like vertical breathing motion in the Milky Way, and the effect of tidal interactions does not directly induces breathing motion. The strongest spiral arms produce the largest breathing motion. For a given pericenter distance, the co-planar prograde orbit produces the strongest spiral arms and the largest breathing motion. On the other hand, co-planar retrograde orbits produce no spiral arms and hence no breathing motion.

In another work, we have simulated the effect of dark matter halo shape on the formation of boxy/peanut/x-shape bulges via bar buckling. We found that the presence of oblate dark matter halos delay bar formation and so bar buckling is also delayed. All the models show two buckling events but the most extreme prolate halo exhibits three distinct buckling events. As a result of multiple buckling events, the boxy/peanut structures in prolate halos show the maximum thickness. Since ongoing buckling events are rarely observed, our study suggests that the most barred galaxies may have oblate or spherical halos rather than prolate halos. 

We have also studied the cosmic evolution of bulges since z=0.1 using SDSS data. We compared two well know bulge classification schemes based on the S\'ersic index and Kormendy relation. We have shown that both schemes have their own drawbacks and suggest an alternative scheme which combines these two schemes. We have also found that disk-like pseudo-bulges are growing in number as the Universe is getting older. The pseudo-bulges appear optically diffuse compared to classical bulges and are commonly found in low mass galaxies. In the local volume, pseudo-bulges overcome the classical bulges even in bulge dominated galaxies, and so more than 75\% of the local volume is rotation dominated. 

Finally, we have tested galaxy formation models of the state-of-art cosmological simulation, Illustris TNG, using bulgeless galaxies. We selected Illustris TNG50 galaxies having mass greater than 10$^{9}$M$_\odot$ and performed photometric decomposition to find bulgeless galaxies. We found that the bulgeless galaxies in TNG50 are metal-poor and have high specific angular momentum as compared to the galaxies with bulges and fall at the lower end of the baryonic to dark matter mass ratio. We also found that the fraction of bulgeless galaxies in TNG50 is equivalent to that determined from observations of galaxies. Thus the TNG galaxy formation model is capable of producing observed characteristics of bulgeless galaxies in the low redshift Universe.

\end{abstract}

	\begin{publication}
\begin{enumerate}

\item \textbf{Kumar, A.}, Ghosh, S., Kataria, S. K., Das, M., and Debattista, V. P., 2022, \textit{Excitation of vertical breathing motion in disc galaxies by tidally-induced spirals in fly-by interactions},\\
\href{https://ui.adsabs.harvard.edu/abs/2022MNRAS.516.1114K/abstract}{Monthly Notices of the Royal Astronomical Society, 516, 1, 1114-1126}.

\item \textbf{Kumar, A.}, and Kataria, S. K., 2022, \textit{Growth of disc-like pseudo-bulges in SDSS DR7 since z = 0.1},\\
\href{https://ui.adsabs.harvard.edu/abs/2022MNRAS.514.2497K/abstract}{Monthly Notices of the Royal Astronomical Society, 514, 2, 2497-2512}.

\item \textbf{Kumar, A.}, Das, M., and Kataria, S. K., 2022, \textit{The effect of dark matter halo shape on bar buckling and boxy/peanut bulges},\\
\href{https://ui.adsabs.harvard.edu/abs/2022MNRAS.509.1262K/abstract}{Monthly Notices of the Royal Astronomical Society, 509, 1, 1262-1268}.

\item \textbf{Kumar, A.}, Das, M., and Kataria, S. K., 2021, \textit{Galaxy flybys: evolution of the bulge, disc, and spiral arms},\\
\href{https://ui.adsabs.harvard.edu/abs/2021MNRAS.506...98K/abstract}{Monthly Notices of the Royal Astronomical Society, 506, 1, 98-114}.

\item \textbf{Kumar, A.}, Das, M., and Kataria, S. K., 2020, \textit{The evolution of bulges of galaxies in minor fly-by interactions},\\
\href{https://ui.adsabs.harvard.edu/abs/2020IAUS..353..166K/abstract}{Proceedings of the International Astronomical Union, 353, 166-167}.

\item \textbf{Kumar, A.}, and Das, M., 2022, \textit{Bulgeless galaxies in the Illustris TNG50 simulations: A test for angular momentum problem},\\
(In preparation).

\end{enumerate}
\end{publication}


	\tableofcontents
	\listoffigures
	
	\listoftables
	
	
	
	
	\mainmatter
	\setcounter{page}{1}
	\begin{savequote}[100mm]
``In all chaos there is a cosmos, in all disorder a secret order.''
\qauthor{\textbf{$-$ Carl Gustav Jung}}
\end{savequote}

\chapter{Introduction}
\label{chapter1}

Not all the point-like sources we observe in the night sky are stars. Several of them are actually groups of stars known as stellar systems. Their distances from us are too large to identify without zooming into the vastness of the Universe with high resolution telescopes. Galaxies are the stellar systems comprising a huge number of stars ranging from 10$^{5}$ to 10$^{12}$. Apart from a copious number of stars, galaxies also consist of gas, dust, and enormous amount of dark matter. The self-gravitational force of the galaxy holds the stars together. Galaxies can be considered as the fundamental constituents of the Universe because they are the main visible sources that contribute to the observable Universe \citep{Gott2005}. There are trillions of galaxies in our observable Universe \citep{Conselice2016}. The hazy patch of light seen stretching from north to south in the
night sky is our home galaxy, the Milky Way or simply the Galaxy (with capital ‘G’).

\section{A brief history of the discovery of our Galaxy}
\label{sec:brief_history}
Until 1925, it was not clear whether the Milky Way was the only galaxy or was instead one of several million such galaxies in the Universe. This question gave rise to a debate in astronomy known as ``The Great Debate'' \citep{Hetherington.1970}. Two astronomers, named Harlow Shapley and Heber Doust Curtis, argued their views about the Milky Way before the National Academy of Science (NAS) in April 1920. Curtis had studied spiral nebulae and concluded that they are comparable to the Galaxy. On the other hand, using period-luminosity relation of Cepheid variable stars, Shapley had determined the size of the Milky Way and concluded that spiral nebulae are the part of the Galaxy. This is what brought them before the National Academy of Science.

Shapley assumed that the globular clusters (GCs) were part of the Galaxy and considered the maximum extent of the globular clusters to be the size of the Galaxy. To estimate distance of the globular clusters, he assumed that the stars in the other part of the Galaxy and in globular clusters are similar to the nearby stars. Shapley's assumptions led him to conclude that the spiral nebulae had to be at unconvincingly large distances for being as big as the Galaxy \citep{Shapley.Curtis.1921}. Although Shapley provided several arguments to support his estimate for the size of the Galaxy, Curtis was not convinced and raised objections against the assumptions Shapley made. Curtis did not give up his belief that the spiral nebulae were comparable to the Galaxy \citep{Hetherington.1970}. 

According to Shapley, the Galaxy was big and the Sun was away from the center of the Galaxy. On the other side, in Curtis' model, the Galaxy was small and the Sun was near the center of the Galaxy. This issue was resolved in 1925, when Edwin Hubble showed that the distance to the M31 spiral nebula was greater than the size of the Galaxy that Shapley estimated. Hubble used the 2.5m Hooker Telescope (which was the biggest telescope of that time) installed at Mount Wilson Observatory to identify Cepheid variable stars and measured the distance to M31. He confirmed that M31 was indeed like the Galaxy \citep{Hubble.1925}. His findings concluded that Shapley was right in estimating the size of our Galaxy and the location of the Sun. In contrast, Curtis correctly identified spiral nebulae as galaxies similar to ours. Nowadays these spiral nebulae are known as galaxies.


\section{Classification of galaxies}
\label{sec:classification_of_galaxis}
As the number of observed galaxies increased, it became essential to separate them in different groups for their systematic studies. \cite{Hubble.1926} presented the first detailed classification of the galaxies based on their structures on the photographic plate. It was latter updated to include other morphological types \citep{de_Vaucouleurs.1959}.

\subsection{Hubble's classification}
\label{subsec:hubble_classification}
\begin{figure*}[ht!]
    \centering
	\includegraphics[width=\textwidth]{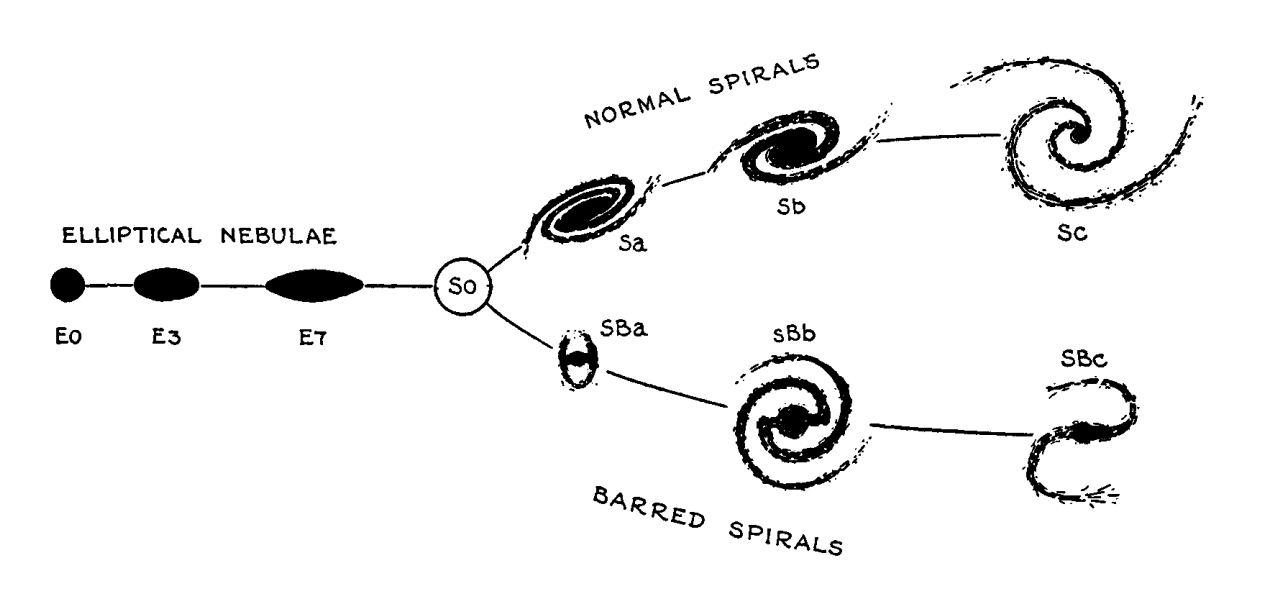}
    \caption{Hubble's classification of galaxies taken from `The Realm of the Nebulae'. Ellipticals are arranged along the stem of the fork, and the two sub-classes of spirals are positioned along the prongs of the fork. Furthermore, lenticular galaxies are placed at the base of the fork. Picture courtesy: \citep{Hubble.1936.book}.}
    \label{fig:hubble_classification}
\end{figure*}

Edwin Hubble collected a large sample of galaxies and divided them into two main classes, regular and irregular, primarily based on the morphology of the galaxies on the photographic plates \cite{Hubble.1926}. The galaxies that show rotation symmetry about their centers were assigned to the regular class. Depending on the flatness of the galaxy and the presence of spiral structure, regular galaxies are divided into two sub-classes: Ellipticals (E) and Spirals. Spiral galaxies are further separated into normal spirals (S) and barred spirals (SB). On the other hand, all the remaining galaxies lacking rotation symmetry are classified as irregular (Irr) galaxies. See Fig.~\ref{fig:hubble_classification} for Hubble's classification of the galaxies \citep{Hubble.1936.book} in the form of a tuning-fork. It is commonly known as Hubble's tuning-fork diagram.

Elliptical galaxies are featureless and show smoothly fading luminosity from center to the outer region. Their projected shapes vary from circular to elongated. The shape of the elliptical is quantified with its ellipticity defined as $\epsilon = 1 - \frac{b}{a}$, where $a$ and $b$ are the semi-major and semi-minor axis. Ellipticity of the elliptical galaxies falls between 0 (for circular) and 0.7 (for the most elongated). Depending on the ellipticity, an elliptical galaxy is represented by the letter `E$n$', where $n=10\epsilon$. Therefore, circular ellipticals are denoted by E0 and the most elongated is designated as E7. We do not have any elliptical galaxy designated as E8, E9 or E10, because they cannot be dynamically stable.

Spiral galaxies are flat and usually show distinct features such as spiral arms and a central bar. Spirals without the central bar are called normal spirals (S). In contrast, spirals having central bar are known as barred spirals (SB). Spiral galaxies show different level of winding in their spiral arms. Spirals with closely wound and unresolved spiral arms are called early-type galaxies (Sa or SBa), whereas spirals with clearly open and resolved spiral arms are known as late-type spirals (Sc or SBc). In between, there lie the intermediate-type spirals which show relatively open and slightly resolved spiral arms (Sb or SBb). Moving from early-type to late-type galaxies, we see the following smooth transitions: spiral arms start opening, luminosity of the galaxies begins to decrease, gas fraction in the galaxies increases, the relative importance of centrally bright component decreases. Edwin Hubble placed lenticular galaxies (S0) at the transition point from ellipticals to spirals. These galaxies look like the lens, so they are called lenticular galaxies.

\subsection{de Vaucouleurs' classification}
\label{subsec:de_vaucouleurs_classification}
\cite{de_Vaucouleurs.1959} took one step ahead of Hubble and updated Hubble's classification to include more refined features seen in spiral galaxies. The following are the changes de Vaucouleurs made in Hubble's classification scheme,

\begin{figure*}[ht!]
    \centering
	\includegraphics[width=0.8\textwidth]{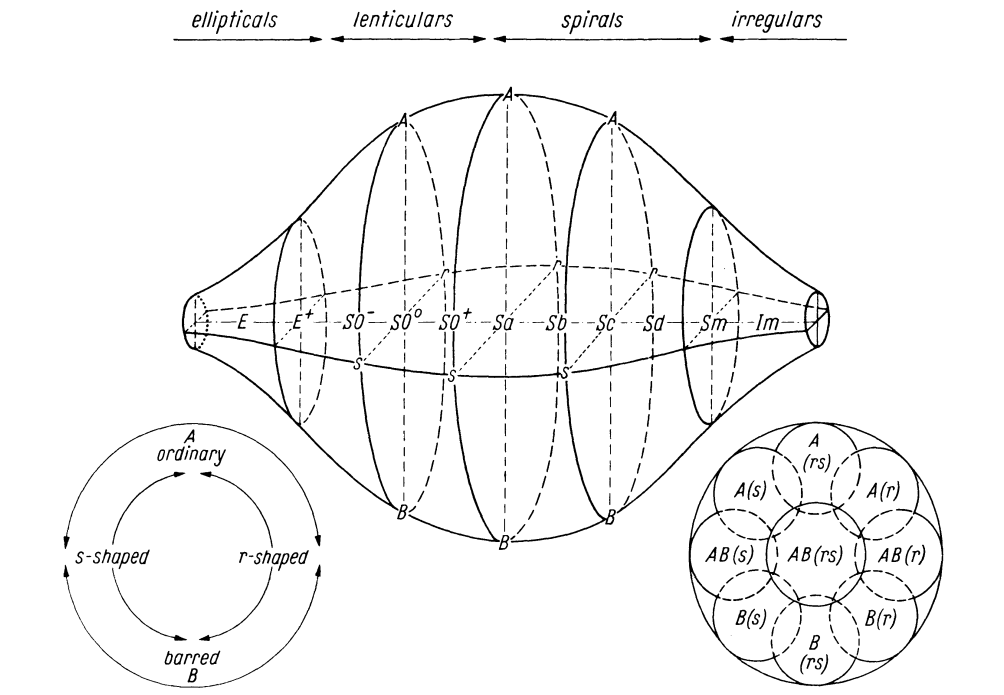}
    \caption{de Vaucouleurs' classification of galaxies taken from `Classification and Morphology of External Galaxies'. This 3-dimensional representation shows E to I galaxies from left to right, normal spirals at the top, barred spirals at the bottom, ring-shaped (r-shaped) on the far side, and spiral-shaped (s-shaped) on the near side. Two cross-sections at the bottom show the classification system at any location along the axis. Picture courtesy: \citep{de_Vaucouleurs.1959}.}
    \label{fig:de_Vaucouleurs_classification1}
\end{figure*}

\begin{enumerate}
    \item de Vaucouleurs used Hubble's designation of ellipticals (E) and barred spirals (SB) as is and renamed normal spirals as SA to include a new class SAB for the galaxies with mixed characteristics (e.g. weakly barred galaxies). 
    \item To recognize the nuclear structure of the galaxies, he used `(r)' for spirals emerging from ring surrounding the nucleus, `(s)' for spirals emerging from the nucleus, and `(rs)' for the mixed type nucleus.
    \item Along with Hubble's early-type (`a'), intermediate-type (`b') and late-type (`c') stages of spiral galaxies, he introduced a very late-type stage (`d') and spiral to irregular transition stage (`m'). In between these main stages, he denoted intermediate stages by `ab', `bc', `cd', etc.
    \item He preferred to used `$-$' and `$+$' to represent respectively early and late stages of non-spiral galaxies. For example, E$^+$ shows the late stage of E.
    \item Magellanic type irregular galaxies are designated as I(m).
    \item To include the information of outer ring like structure seen in spiral galaxies symbol `(R)' is prepended in the main class of the galaxy.
\end{enumerate}

\begin{figure*}[ht!]
    \centering
	\includegraphics[width=0.8\textwidth]{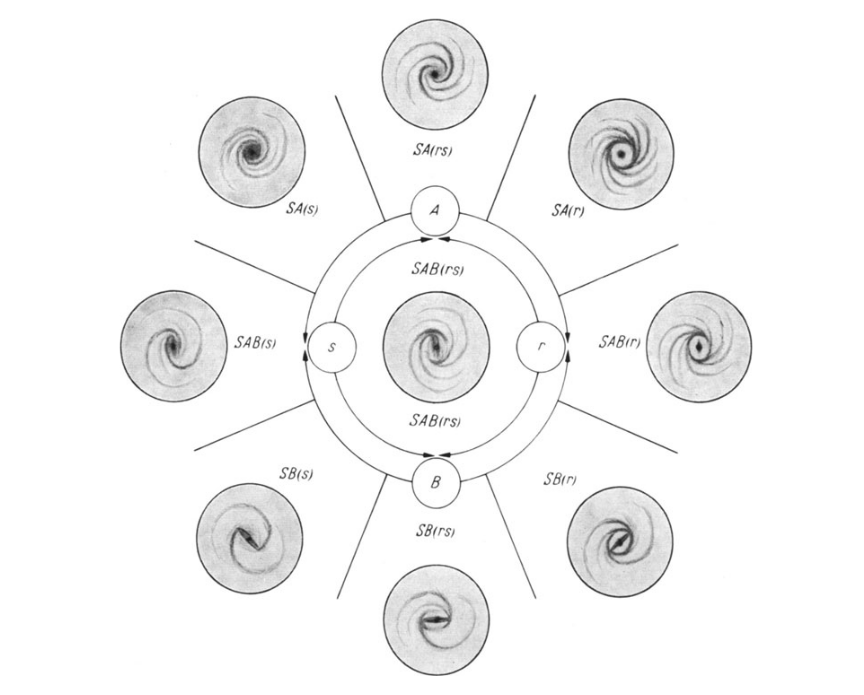}
    \caption{A visual illustration of de Vaucouleurs' classification using a cross-section near Sb stage of the galaxies. See Section~\ref{subsec:de_vaucouleurs_classification} for the nomenclature scheme shown in this figure. Picture courtesy \citep{de_Vaucouleurs.1959}.}
    \label{fig:de_Vaucouleurs_classification2}
\end{figure*}

The detailed de Vaucouleurs' classification for galaxies has been shown in Fig.~\ref{fig:de_Vaucouleurs_classification1}. It shows E to I galaxies from left to right, normal spirals at the top, barred spirals at the bottom, ring-shaped (r-shaped) on the far side, and spiral-shaped (s-shaped) on the near side. Two cross-sections at the bottom of figure show the classification system at any position along the axis. Fig.~\ref{fig:de_Vaucouleurs_classification2} illustrates the classification scheme using a cross-section near Sb stage of the galaxies.

\section{Evolution of galaxies}
\label{evolution_of_galaxies}
\begin{figure*}[ht!]
    \centering
	\includegraphics[width=0.6\textwidth]{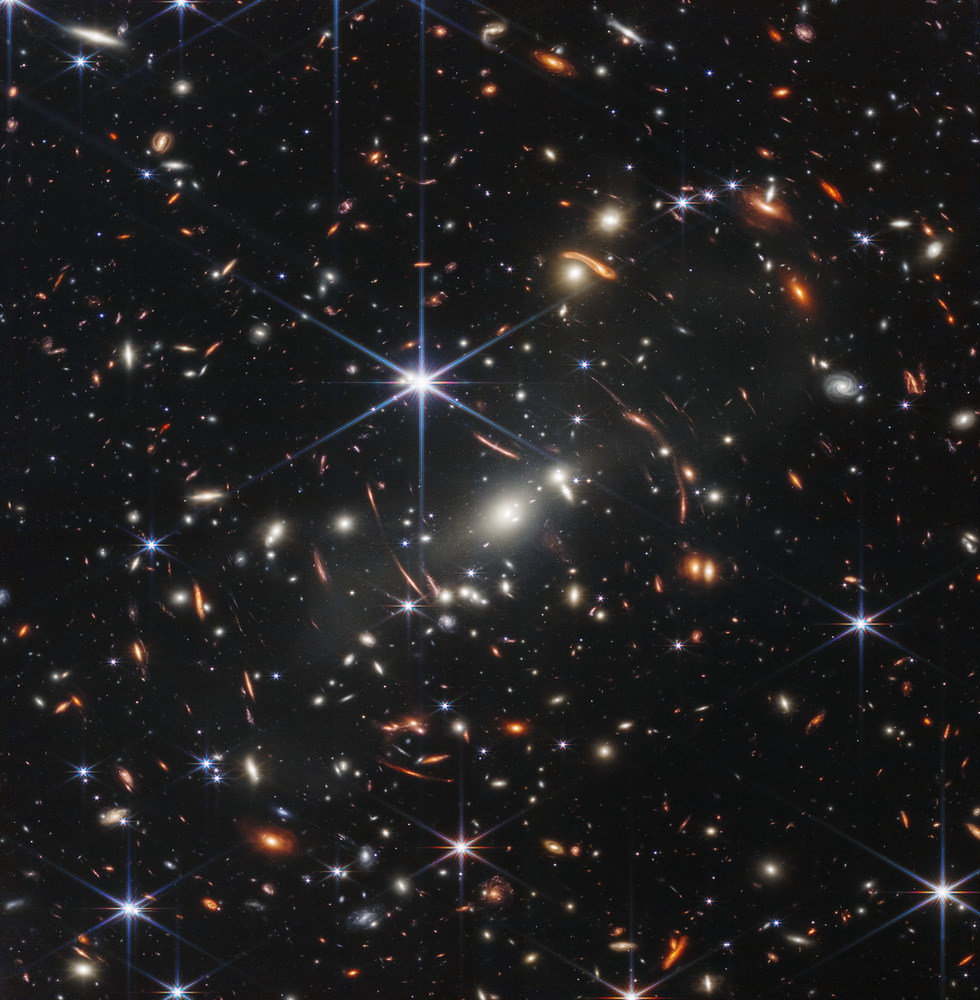}
    \caption{A cluster of galaxies, SMACS 0723, as seen from Near-Infrared Camera (NIRCam) instrument onboard NASA's James Webb Space Telescope (JWST). It shows thousands of galaxies in a tiny part of the sky represented by a sand grain on arm-length. The arc-shaped structures are more distant galaxies behind the cluster which are magnified due to enormous gravity of the cluster. Objects with sharp spikes are foreground stars. Picture courtesy NASA, ESA, CSA, and STScI.}
    \label{fig:jwst_deep_field}
\end{figure*}

Galaxies are normally found in groups and clusters where they interact with each other gravitationally. Fig.~\ref{fig:jwst_deep_field} shows a cluster of galaxies named SMACS 0723 imaged from Near-Infrared Camera (NIRCam) installed on NASA's James Webb Space Telescope (JWST). It contains thousands of galaxies bound by the huge gravity of the cluster. Gravitational interactions between galaxies plays a crucial role in their formation and evolution over cosmic time. They impact the morphology and kinematics of the galaxies. Interaction of galaxies are categorized in two types: mergers and fly-by interactions.

\begin{itemize}
    \item {\bf Mergers:} Mergers are events where two or more galaxies come closer as result of their tremendous gravitational attraction and fall into each other. Mergers are also called collisions of galaxies. These interactions are very violent and affect galaxies very significantly. Depending on the mass ratio of two merging galaxies, mergers can be classified into major mergers and minor mergers. Suppose, M$_{\rm p}$ is the mass of primary (high-mass) galaxy and M$_{\rm s}$ is the mass of secondary (low-mass) galaxy in a merger event. In general, mergers with M$_{\rm p}$:M$_{\rm s} \leq$ 3:1 are called major mergers, and the mergers with M$_{\rm p}$:M$_{\rm s} >$ 3:1 are termed minor mergers.
    
    \item {\bf Fly-bys:} In contrast to the mergers, fly-bys are events where two or more galaxies come closer, exerting enormous gravitation pull on each other, and then separate without merging. Fly-bys are commonly referred to as tidal encounters or interactions. Similar to mergers, fly-bys can also be categorized into major flybys and minor flybys depending on the mass ratio of interacting galaxies. For example, M$_{\rm p}$ is the mass of the primary (high-mass) galaxy and M$_{\rm s}$ is the mass of the secondary (low-mass) galaxy in a flyby event. Then the fly-bys with M$_{\rm p}$:M$_{\rm s} \leq$ 3:1 are called major fly-bys, and the fly-bys with M$_{\rm p}$:M$_{\rm s} >$ 3:1 are termed minor fly-bys.
\end{itemize}

Two interacting galaxies can also go through several fly-by events prior to the final merger event. It mainly depends on the initial orbits of the interacting galaxies. If both the galaxies approach radially towards each other, they will directly collide and merge. On the other hand, if both the galaxies approach on elliptical orbits, they will keep flying past each other until their orbits decay significantly (due to dynamical friction).

Cosmological simulations suggest that mergers were more frequent at higher redshifts compared to fly-by interactions. As the Universe evolved and expanded, the frequency of fly-by interactions increased and become comparable to mergers \citep{Yee-Ellingson1995, Sinha2012A, an.etal.2019}. Both types of interactions are important in the formation and evolution of the galaxies \citep{dimatteo.etal.2007, Cheng-Li2008, Kim2014, Sinha2015B}. Both interactions can induce enhanced star formation, extended tidal features, spiral arms, bars, etc. The results of these interactions depend on various factors such as their initial mass ratio, gas contents, orbital configurations and distance of closest approach (pericenter distance).

\section{Major components of galaxies}
\label{components_of_galaxies}
Elliptical galaxies are featureless and fade smoothly as we move from center to the outer edge. Their one dimensional radial surface brightness along the isophotal semi-major axis R is well represented with S\'ersic profile \citep{Sersic.1968book} defined as
\begin{equation}
    \Sigma(R) = \Sigma_{\rm e}\exp\left[-b_{\rm n}\left\{\left(\frac{R}{R_{\rm e}}\right)^{\frac{1}{n}}-1\right\}\right],
    \label{eqn:elliptical_sersic}
\end{equation}
where R$_{\rm e}$ is the effective radius or half-light radius, $\Sigma_{\rm e}$ is the surface brightness at R$_{\rm e}$, and b$_{\rm n}$ is the function of S\'ersic index `n' and is related with Gamma function ($\Gamma$) and lower incomplete Gamma function ($\gamma$) by the following expression,
\begin{equation}
    \gamma(2n;b_{n}) = \frac{1}{2}\Gamma(2n),
    \label{eqn:gamma_sersic}
\end{equation}
For 1$<$n$<$10, b$_{\rm n}$ cab be approximated as b$_{\rm n}$ $=$ 2n$-$1/3 \citep{Ciotti.Bertin.1999}. S\'ersic index ranges from n $=$ 2 for dim elliptical galaxies to m $=$ 6 for luminous elliptical galaxies. In the middle, n $=$ 4 gives de Vaucouleurs' R$^{1/4}$ profile \citep{de_Vaucouleurs.1948}.

Elliptical galaxies contain very little or no gas and dust. They are generally red in color \citep{Sandage.Visvanathan.1978}, which is an indication of a dominant older stellar population. All or a major fraction of the stars in elliptical galaxies orbit randomly around the center \citep{Bertola.Capaccioli.1975, Illingworth.1977}. The stellar velocity dispersion supports them against gravitational collapse. The fact that elliptical galaxies are dominated by dispersion suggests that they have formed in violent processes. Several theoretical, numerical, and observational works have shown that elliptical galaxies are formed in monolithic collapses or/and in similar mass mergers \citep{Bekki.Shioya.1997, Zepf.1997, De_Lucia.etal.2006, Bournaud.etal.2007, Kormendy.etal.2009, Naab.et.al.2014}.

In contrast, spiral galaxies possess various distinct features such as spiral arms, bars, and bulges. We refer to these structures as the components of (spiral) galaxies. In the following subsections, we give a brief overview of the major galactic components.

\subsection{Disk}
\label{subsec:disk_component}
Most of the baryonic mass of spiral galaxies lies in the form of a flat circular disk. The radial distribution of surface brightness of this disk is well described with an exponential profile \citep{Freeman.1970} defined as
\begin{equation}
    \Sigma(R)=\Sigma_{0}\exp\left(-{\frac{R}{R_{\rm s}}}\right),
    \label{eqn:spiral_disk_face}
\end{equation}
where $\Sigma_{0}$ is the central surface brightness of the disk and $R_{\rm s}$ is the disk scale radius (or scale length). In many cases, due to the presence of other central components, the central surface brightness of the disk is extrapolated to the outer disk during the fitting process. Additionally, the vertical surface brightness of the disk is commonly approximated with a $\sech^{2}$ function. Therefore, the three dimensional (3D) disk surface brightness profile is given by the following expression,
\begin{equation}
    j(R,z)=j_{0}\exp\left(-{\frac{R}{R_{\rm s}}}\right) \sech^{2}\left(\frac{z}{z_{0}}\right),
    \label{eqn:spiral_disk_3D}
\end{equation}
where j$_{0}$ is the central surface brightness at the mid-plane of the disk, R$_{\rm s}$ is disk scale radius, and z$_{0}$ is disk scale height. The disk scale height is nearly constant at each radial position for late type spirals. However, early type galaxies show increasing disk scale height with radius \citep{de_Grijs.Peletier.1997}. 

The majority of stars in spiral galaxies rotate around the galactic center on near circular orbits. The rotation of stars in spiral galaxies provides support against gravitational collapse of the disk \citep{Mo.Bosch.White.2010book}. It is still not fully understood how spiral galaxies develop disks and whether they retain them throughout the cosmic evolution. In $\Lambda$ Cold Dark Matter ($\Lambda$CDM) models of galaxy formation\footnote{also known as hierarchical structure formation model.}, disks form in dark matter halos that have high angular momentum \citep{Dalcanton.etal.1997}.

\subsection{Spiral arms}
\label{subsec:spiral_component}
As the name suggests, spiral galaxies host beautiful spiral arms in their disks. The extent of the spiral arms ranges from the very central region to the edge of the galaxy. The strength of spiral arms is often quantified using the Fourier decomposition of disk surface brightness. For example, in a polar co-ordinate system centered at the nucleus of the galaxy, $\Sigma(\rm R,\phi)$ is the surface brightness of a spiral galaxy at any co-ordinate $(\rm R,\phi)$ and m is the number of spiral arms. Then its surface brightness can be written as follows,
\begin{equation}
    \Sigma(\rm R,\phi) = \Sigma_{0}(\rm R) + \Sigma_{\rm m}(\rm R)\cos\left[m\phi + \phi_{\rm m}(\rm R)\right],
    \label{eqn:spiral_arms_fourier_decomp}
\end{equation}
where $\Sigma_{0}$(R) and $\Sigma_{\rm m}$(R) are respectively amplitudes of the 0$th$ and m$th$ Fourier modes at any radius R. $\phi_{\rm m}$(R) is the phase angle of the m$th$ Fourier mode. Now, the ratio $\Sigma_{\rm m}$(R)/$\Sigma_{0}$(R) provides a quantitative strength of an m-armed spiral pattern present in the disk at any radial position R.

The shape of spiral arms is reasonably well represented by a logarithmic profile. The mathematical form of logarithmic spiral arms is given by the following expression, 
\begin{equation}
    \rm R(\phi) = \rm R_{0} e^{\phi \tan \alpha},
    \label{eqn:spiral_arms_shape}
\end{equation}
where $\rm R_{0} = \rm R(0)$, $\phi$ is the winding angle which constraints the azimuthal extent of the spiral arm, and $\alpha$ is the pitch angle which measures the tightness of the spiral arm. For grand design spirals\footnote{Grand design spirals is a class of spiral arms that shows two-armed spiral pattern clearly traceable to large radial extent.}, pitch angle usually remains constant throughout the galaxy within an error of $\pm5$ degree \citep{Kennicutt.1981, Diaz-Garcia.etal.2019}.

The enhanced star formation in disk galaxies is often seen along their spiral arms which indicates that the over-densities associated with spiral arms act as the sites for star formation. They sweep the inter stellar medium and help in the birth of new stars. Spiral arms also help transferring angular momentum from the inner region to the outer region \citep{Binney.Tremaine.2008}. This angular momentum transfer is useful for the growth of central disk. Furthermore, Spiral arms are thought to be source of wave-like breathing motion seen in the Milky Way \citep{Kumar.etal.2022Oct}.

There are various hypothesis for the formation of spiral arm. However, the fundamental reason for their existence is that the spiral galaxies are dynamically cold stellar systems\footnote{In astronomy, by dynamically cold stellar systems, we mean that these systems are rotation dominating. Conversely, dispersion supported stellar systems are called dynamically hot.} and rotate differentially.

\subsection{Bar}
\label{subsec:bar_component}
Bars are narrow, linear structures observed across the centers of spiral galaxies. As discussed in Section~\ref{sec:classification_of_galaxis}, spiral galaxies with bars are known as barred spirals. Nearly two-third of spirals in the local volume are barred galaxies. Similar to the spiral arms, bars can also be quantified using Fourier decomposition of the disk surface brightness. Since bars have 2-fold symmetry i.e. in one full rotation disk surface brightness looks unchanged twice, we can replace m by 2 in equation~(\ref{eqn:spiral_arms_fourier_decomp}). Now, the surface brightness of barred spirals can be expressed as follows,
\begin{equation}
    \Sigma(\rm R,\phi) = \Sigma_{0}(\rm R) + \Sigma_{2}(\rm R)\cos\left[2\phi + \phi_{2}(\rm R)\right],
    \label{eqn:bars_fourier_decomp}
\end{equation}
where $\phi_{2}$(R) is the phase angle of 2$nd$ Fourier mode. It remains constant in bar region because bars are linear structures.
Now, the ratio $\Sigma_{2}$(R)/$\Sigma_{0}$(R) provides quantitative strength of bar at any radial position R.

Bars play an important role in the secular evolution of spiral galaxies. Bars can shock the gas and ignite star formation in the inner disk region. Funneling of gas along the bar and into the galaxy nucleus helps the growth of central super massive black holes (SMBHs).

\subsection{Bulge}
\label{subsec:bulge_component}
\begin{figure*}[ht!]
    \centering
	\includegraphics[width=0.7\textwidth]{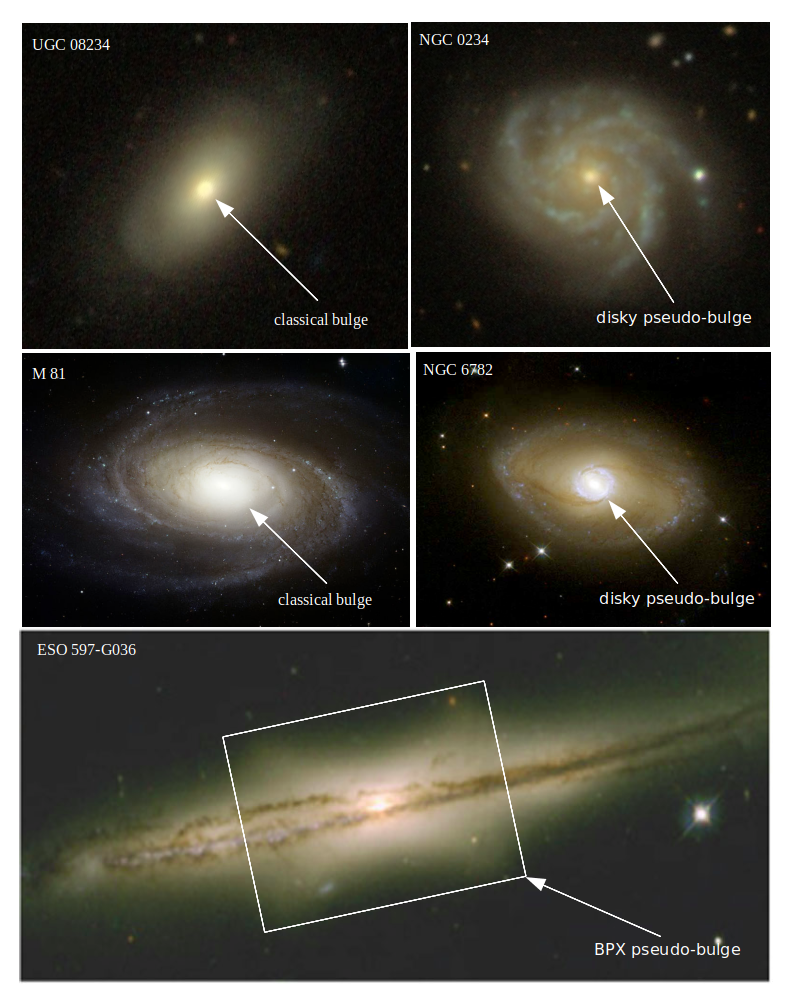}
    \caption{Few examples of different types of bulges. Name of the galaxy and the class of its bulge are mentioned in corresponding panels. Picture courtesy: this collage is made using images from NASA/ESA/Hubble and from Aladin Lite's SDSS9.}
    \label{fig:classification_of_bulges}
\end{figure*}

The dense collection of stars in the center of disk galaxies is knows as the bulge. Moreover, galaxies which do not have any bulge are called bulgeless galaxies. On the basis of morphology and the dynamics of its constituent stars, bulges can be categorized into two types: classical bulges and pseudo-bulges. Classical bulges are nearly spherical in shape and are supported by the rotational motion of the stars, whereas pseudo-bulges are flat structures and show rotational support \citep{Kormendy2006, Drory2007, Athanassoula2008, Fisher2008A, Fisher2008B}. In Fig.~\ref{fig:classification_of_bulges}, we have shown different types of bulges as seen in observed galaxies.

There are many observational similarities between classical bulges and elliptical galaxies which suggests a similar formation mechanism for both systems. For example, their shapes or S\'ersic indices, the orbit families of their stars, their size-luminosity relation, etc \citep{Faber1976, Kormendy1977, Gultekin2009, Savorgnan2013}. On the basis of numerical simulations and theoretical studies, it is well accepted that elliptical galaxies are the remnants of galaxy mergers \citep{Balland1998, Dubinski1998, Burkert2003, Bournaud2007}. Classical bulges are also thought to form in monolithic collapse of gas clouds, and later grow from the accretion of smaller galaxies  \citep{Aguerri2001, Bournaud2005, Brooks2016}. 

The term pseudo-bulge is commonly used to describe two types of bulges: disky pseudo-bulges and boxy/peanut/x-shape pseudo-bulges. Disky pseudo-bulges are circular in shape but in the vertical direction they are as flat as the disks of their host galaxy. Hence, it is nearly impossible to detect them in edge-on galaxies. They are thought to form within the inner disks via star formation \citep{laurikainen.etal.2009}. Boxy/peanut/x-shape bulges (BPX bulges in short) are more extended in the vertical direction and can hence be detected in edge-on galaxies. They are usually associated with bars in disk galaxies \citep{Friedli1990, Debattista2006, Gadotti2011}. Pseudo-bulges, unlike classical bulges, are formed due to the secular evolution of galactic disks, the buckling instability of bars \citep{Combes1981}, or via mergers with gas rich galaxies \citep{Keselman2012}.

\subsection{Dark matter halo}
\label{subsec:dark_matter_component}
Dark matter is an unknown form of the matter and accounts for more than 80\% of the matter in galaxies. It is totally invisible as it does not emit electromagnetic radiation. We have only indirect evidences that dark matter exists from its gravitational interaction with visible baryonic matter. There have been several attempts to detect dark matter and discover its nature but to date no one has succeeded.

\begin{figure*}[ht!]
    \centering
	\includegraphics[width=0.7\textwidth]{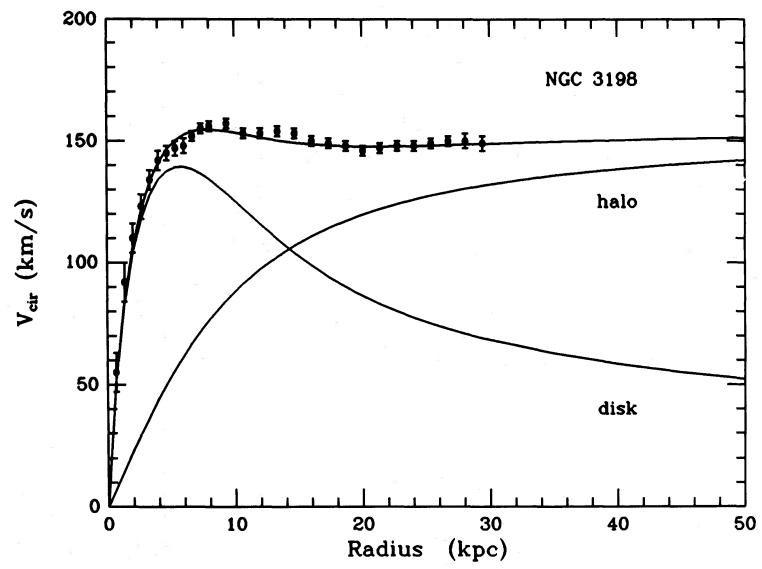}
    \caption{A typical example of rotation curve. Circular velocity increases up to certain radius and then it becomes constant. Here, solid curves show fitted rotation curve and contribution of individual component. Picture courtsey: \citep{van_Albada.etal.1985}.}
    \label{fig:rotation_curve_NGC3198}
\end{figure*}

\cite{Zwicky.1933} was the first who quantitatively showed the presence of dark matter. He calculated the mass of Coma galaxy cluster using kinematics of galaxies within the cluster. Surprisingly, he found that the mass estimated using kinematics is nearly 400 times the mass estimated using the luminosity of all galaxies in the cluster. Later on, \cite{Freeman.1970} and \cite{Rubin.etal.1970} independently showed that the rotation curves\footnote{Rotation curve is the radial variation of orbital velocity as a function of radius from galactic center.} of spiral galaxies become flat in the outer regions (see Fig.~\ref{fig:rotation_curve_NGC3198} for a typical rotation curve). The central rotation curve can be explained with baryonic matter, while the outer rotation curve is hard to explain without assuming extra hidden matter i.e. dark matter. These evidences suggested the need of dark matter in the Universe \citep[e.g., see review on dark matter][]{Swart.etal.2017}.

Dark matter provides gravitational support for the formation of the structure we see today. Early clustering of dark matter results in dark matter halo formation. The huge gravitational potential of these halos helps gas in-flow toward their center, and leads to the formation of proto-galaxies. The merger of small dark matter halos and continuing star formation help the growth of galaxies. Dark matter stabilizes disk galaxies against global instabilities such as bars and spirals.

\section{Our galaxy: The Milky Way}
\label{sec:milkyway}
\begin{figure*}[ht!]
\centering
	\includegraphics[width=0.6\textwidth]{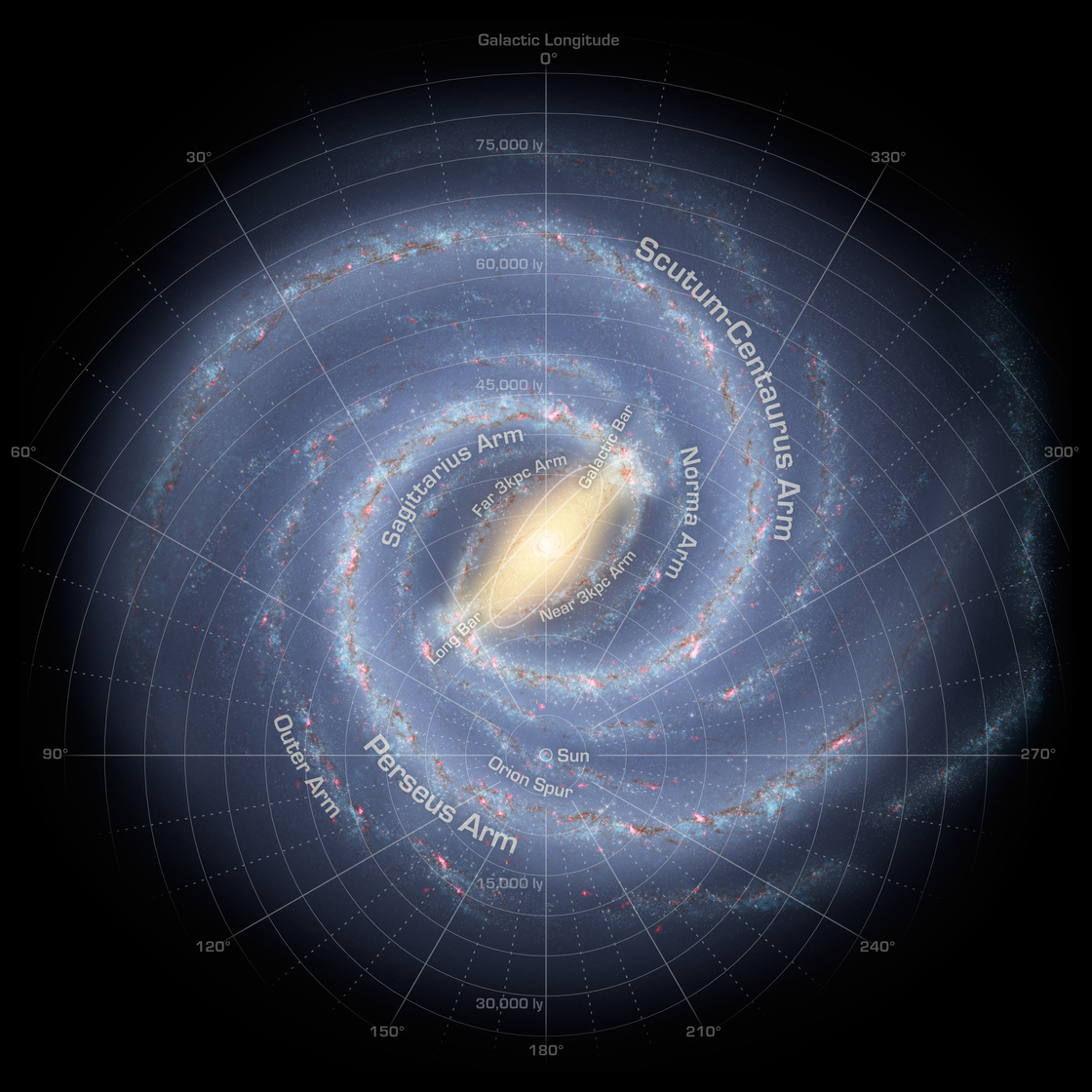}
	\includegraphics[width=0.6\textwidth]{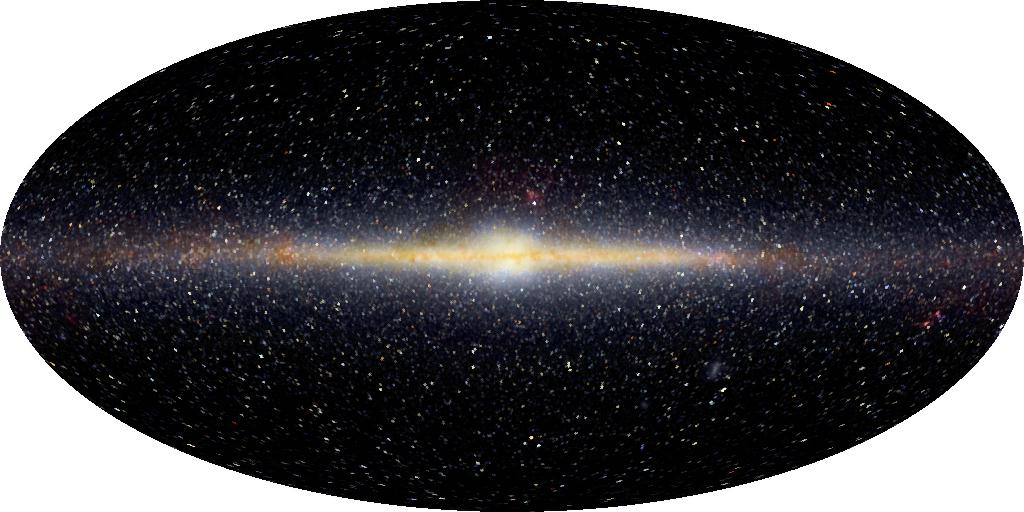}
    \caption{Top: the artistic face-on view of the Milky Way. Two main spiral arms, Perseus and Scutum-Centaurus, are originating from ends of the central bar. Picture courtesy: NASA/JPL-Caltech/R. Hurt (SSC/Caltech). Bottom: the edge-on view of the Milky Way taken from Diffuse Infrared Background Experiment (DIRBE) instrument of Cosmic Background Explorer (COBE) satellite. Picture courtesy: Ned Wright.}
    \label{fig:milkyway}
\end{figure*}
The Milky Way is a barred spiral galaxy comprising of a disk and a BPX bulge \citep{Ness.Lang.2016}. The radius of the Milky Way disk is around 10~kpc which contains most of its stars \citep{Binney.Tremaine.2008}. The top and bottom panels of Fig.~\ref{fig:milkyway} show the face-on and edge-on view of the Milky Way respectively. Face-on view is the detailed artistic impression of the Galaxy. It shows the central bar and the spiral arms in the disk. Edge-on view shows the extent of disk and dense collection of stars in the central region. This image is observed with Diffuse Infrared Background Experiment (DIRBE) instrument onboard Cosmic Background Explorer (COBE) satellite. Here, we cannot clearly distinguish X-shape of the bulge due to our position in the Galaxy. The long axis of the bar makes $\sim$30$^\circ$ angle from our line-of-sight to the Galactic center \citep{Wegg.etal.2015}.

In total, the Milky Way weight is equivalent to 1.2 trillion times the Sun's mass \citep[e.g., see ][]{Wang.Haan2020}. The stellar mass of the Milky way is $\sim$ 6$\times$10$^{10}$M$_{\odot}$ (where M$_{\odot}=$ solar mass) which includes a disk of $\sim$ 5$\times$10$^{10}$M$_{\odot}$ and a bulge of $\sim$ 1$\times$10$^{10}$M$_{\odot}$ \citep{McMillan.2011, Licquia.Newman.2015}. The scale length of the disk is nearly 3~kpc and the half length of bar is 4 to 5~kpc \citep{Cabrera-Lavers.etal.2008, McMillan.2011, Wegg.etal.2015}. The extent of its dark matter halo is not well constrained and expected to lie between 100 and 300~kpc \citep{Dehnen.etal.2006, McMillan.2011, Deason.etal.2020}. The Sun is located at a distance of 8.3 kpc from the center of the Milky Way and completes one orbit around Galactic center every 240 million years at an average speed of 220 km s$^{-1}$ \citep{McMillan.2011, Gillessen.etal.2017}.

The Milky Way is the only galaxy studied in great detail by far. Still, there is a lot to know about it. Recent studies based on enormous 6D data (3 position and 3 velocity) have shown that the Milky Way disk is not perfectly planar. It exhibits a warped disk \citep{Freudenreich.etal.1994, Poggio.etal.2018}, north-south star number asymmetry and wave-like breathing motion \citep{Widrow.etal.2012, Kumar.etal.2022Oct}, corrugation or undulation rings \citep{Xu.etal.2015}, warp in solar neighbourhood \citep{Schonrich.Dehnen.2018}, phase-space spirals \citep{Antoja.etal.2018}.

\section{Thesis layout}
\label{sec:thesi_layout}
Chapter~\ref{chapter1} reviews history of galaxies, their classification, evolution, galactic components, and the Milky Way. Two-body relaxation, theoretical modelling of stellar orbits, collisionless Botlzmann equation, Jeans equations, numerical techniques, simulations, and codes used for this thesis are discussed in chapter~\ref{chapter2}. We have presented our work on the evolution of bulges, disks, and spiral arms in fly-by interactions of galaxies in chapter~\ref{chapter3}. In the chapter~\ref{chapter4}, we have discussed the excitation of wave-like breathing motion by tidally induced spiral arms in fly-by interactions. Chapter~\ref{chapter5} of this thesis talks about the formation of bar and BPX bulges in non-spherical dark matter halos. Chapter~\ref{chapter6} presents our work on the growth of disk-like pseudo-bulges in observed galaxies since $z=0.1$ redshift. Next, chapter~\ref{chapter7} shows our work on bulgeless galaxies in Illustris TNG50 simulations and comparison with local volume observations. Finally, chapter~\ref{chapter8} concludes our work in this thesis along with future directions.

	\begin{savequote}[100mm]
``I had always seen myself as a star; I wanted to be a galaxy.''
\qauthor{\textbf{$-$ Twyla Tharp}}
\end{savequote}

\chapter[Techniques for Numerical Simulations]{Techniques for Numerical Simulations}
\label{chapter2}

\section{Introduction}
\label{sec:intro} 
A significant fraction of the baryonic mass of galaxies lies in the form of stars. The orbits of these stars characterize the morphology, shape and internal kinematics of galaxies. For example, the random stellar orbits in elliptical galaxies make them ellipsoidal, whereas the ordered rotational stellar orbits in spiral galaxies make them resemble a disk. Therefore, to understand the formation and evolution of galaxies, one needs to study how the stars move in galaxies. However, it is challenging to examine stellar orbits analytically, given the large number of stars in galaxies or in other stellar systems. Thus stellar dynamics in galaxies is complex and is mainly investigated using a variety of approximations and numerical techniques. In this chapter we review some analytical and numerical approaches to study the stellar kinematics in galaxies.

\section{Two body relaxation}
\label{sec:relaxation}
\begin{figure*}[ht!]
    \centering
	\includegraphics[width=0.8\textwidth]{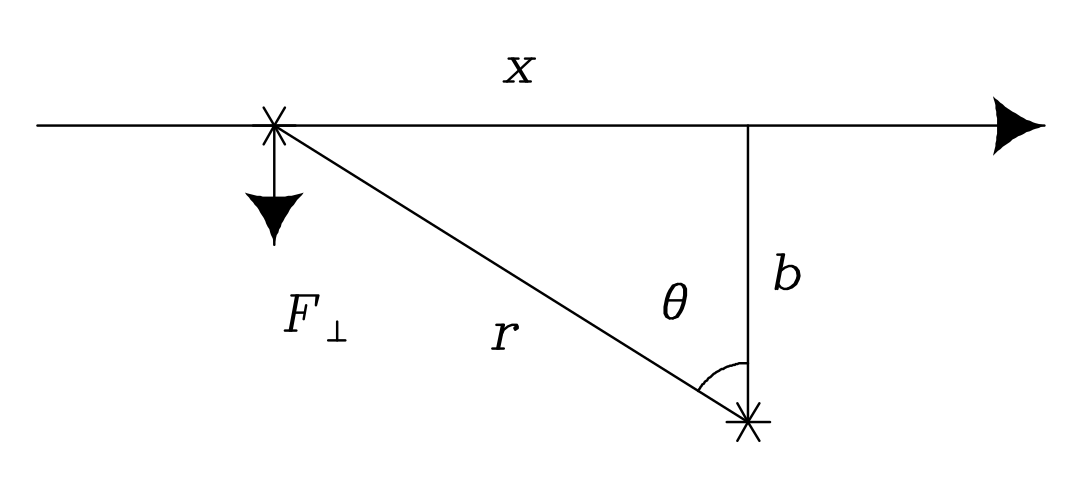}
    \caption{A toy model to measure the effect of an encounter on a subject star assuming a straight line trajectory and impact parameter b. Picture courtesy: \citep{Binney.Tremaine.2008}.}
    \label{fig:two_body_relaxation}
\end{figure*}

Several theoretical studies of galaxies are based on the smooth potential of galaxies. This is a reasonable assumption and is better understood via two-body relaxation processes. Consider the motion of a subject star in a galaxy having $N$ number of identical stars each with mass $m$. Say, the subject star is moving with velocity $\vec{v}$ and passes-by a field star at a distance $b$ (known as the impact parameter). For simplicity, we assume that the field star is stationary during this interaction and the trajectory of a subject star is a straight line (see Fig.~\ref{fig:two_body_relaxation}). In this case, the net force on the subject star along the trajectory will sum up to zero. However, the force perpendicular to the trajectory at any instance will be given by
\begin{align}
    \centering
    F_{\perp} &= \frac{Gm^{2}}{r^{2}}\cos\theta
    = \frac{Gm^{2}}{r^{2}}\frac{b}{r}
    = \frac{Gm^{2}b}{r^{3}}, \notag \\
    F_{\perp} &= \frac{Gm^{2}b}{(b^{2}+x^{2})^{3/2}}.
    \label{eqn:two_boy_relax_1}
\end{align}
Let us assume, time is zero at the closest passage. Now using Newton's law in perpendicular direction,
\begin{align}
    \centering
    m \frac{dv_{\perp}}{dt} &= F_{\perp} \notag \\
    \int m dv_{\perp} &= \int_{-\infty}^{+\infty} F_{\perp} dt \notag \\
    m \delta v_{\perp} &= \int_{-\infty}^{+\infty} F_{\perp} dt = \int_{-\infty}^{+\infty} \frac{Gm^{2}b}{(b^{2}+x^{2})^{3/2}}dt \notag \\
    m \delta v_{\perp} &= \int_{-\infty}^{+\infty} \frac{Gm^{2}b}{(b^{2}+x^{2})^{3/2}} \frac{dx}{v} \notag \\
    m \delta v_{\perp} &= \frac{Gm^{2}b}{v} \int_{-\infty}^{+\infty} \frac{dx}{(b^{2}+x^{2})^{3/2}} \notag \\
    m \delta v_{\perp} &= \frac{Gm^{2}b}{v} \frac{2}{b^2} \notag \\
    \delta v_{\perp} &= \frac{2Gm}{bv}.
    \label{eqn:two_boy_relax_2}
\end{align}
This is the magnitude of deflection in a subject star's velocity due to a single encounter with a field star i.e. $|\delta \vec{v}| = \delta v = \delta v_{\perp}$. For 90$^{\circ}$ deflection ($\delta v = v$), impact parameter will be $b_{90} = 2Gm/v^{2}$.

Suppose $R$ is the size of the galaxy, its number surface density will be of the order of $N/\pi R^{2}$. Therefore, the number of encounters per crossing in the impact parameter range from $b$ to $b+db$ can be written as
\begin{align}
    \centering
    \delta n &\approx \frac{N}{\pi R^{2}} 2\pi b db = \frac{2N}{R^{2}}b db.
    \label{eqn:two_boy_relax_3}
\end{align}
All these encounters will change a subject star's velocity randomly and result in zero net deflection. However, the sum of the squared deflections will be non-zero and can be estimated by integrating over whole range of impact parameter,
\begin{align}
    \centering
    (\Delta v)^{2} &\approx \int_{b_{min}}^{b_{max}} (\delta v)^{2} \delta n \notag \\
    (\Delta v)^{2} &\approx \int_{b_{min}}^{b_{max}} \left(\frac{2Gm}{bv}\right)^{2} \frac{2N}{R^{2}}b db \notag \\
    (\Delta v)^{2} &\approx 8N\left(\frac{Gm}{Rv}\right)^{2} \ln \left(\frac{b_{max}}{b_{min}}\right),  \notag \\
    (\Delta v)^{2} &\approx 8N\left(\frac{Gm}{Rv}\right)^{2} \ln \Lambda,
    \label{eqn:two_boy_relax_4}
\end{align}
where
\begin{align}
    \centering
    \ln \Lambda = \ln \left(\frac{b_{max}}{b_{min}}\right),
    \label{eqn:two_boy_relax_5}
\end{align}
is the Coulomb logarithm. It can be approximated assuming $b_{min} = b_{90}$ and $b_{max} = R$ for most of the stellar systems where $R >> b_{90}$. Now,
\begin{align}
    \centering
    \frac{(\Delta v)^{2}}{v^{2}} &\approx 8N\left(\frac{Gm}{Rv^{2}}\right)^{2} \ln \Lambda.
    \label{eqn:two_boy_relax_6}
\end{align}
We can eliminate $R$ from this equation considering the typical speed of a star in a galaxy to be equal to the circular velocity at the edge of the galaxy i.e. $v^{2} = GNm/R$. 
\begin{align}
    \centering
    \frac{(\Delta v)^{2}}{v^{2}} &\approx \frac{8\ln \Lambda}{N}.
    \label{eqn:two_boy_relax_7}
\end{align}
Every time a subject star crosses the galaxy, its velocity changes by $(\Delta v)^{2}$. Using this expression, we can calculate number of crossings  a subject star will make to change its velocity by the same order. In other words, the number of required crossings to make the star dynamically relax is given by,
\begin{align}
    \centering
    n_{relax} &\approx \frac{N}{8 \ln \Lambda},
    \label{eqn:two_boy_relax_8}
\end{align}
and the relaxation time $t_{relax} = n_{relax} t_{cross}$, where $t_{cross} = R/v$ is the galaxy crossing time. Furthermore, $\Lambda \approx R/b_{90} \approx Rv^{2}/Gm = N$. Therefore,
\begin{align}
    \centering
    t_{relax} &\approx \frac{0.1N}{\ln N} t_{cross}.
    \label{eqn:two_boy_relax_9}
\end{align}
Relaxation time quantifies when the cumulative effect of small encounters significantly changes the orbit of a star. After one relaxation time, a star does not remember its initial conditions. Thus, for an evolution time $t<t_{relax}$, we can consider a system to be collisionless. Also, it is reasonable to neglect the effect of the discreteness of the gravitational potential due to individual stars and consider the potential to be smooth.

\section{Stellar Orbits in spherical potential}
\label{sec:orbit_in_spherical_pot}
The simplest approximation to study stellar orbit is the spherically symmetric potential of a stellar system. In a spherical potential, the star's angular momentum is a conserved quantity, meaning the star orbits in a plane. Let us assume a polar co-ordinate system ($r,\psi$) in the orbital plane positioned at the center of potential $\Phi(r)$. Now, we can write its Lagrangian per unit mass as follows,
\begin{align}
    \centering
    \mathscr{L} = \frac{1}{2}\left[\dot r^2 + (r \dot \psi)^{2}\right] - \Phi(r).
    \label{eqn:orbit_spherical_pot_1}
\end{align}
Equations of motion for this Lagrangian are,
\begin{align}
    \centering
    \frac{d}{dt} \frac{\partial \mathscr{L}}{\partial \dot r} - \frac{\partial \mathscr{L}}{\partial r} &= 0 \implies \ddot r - r \dot \psi^{2} + \frac{d \Phi}{dr} = 0,
    \label{eqn:orbit_spherical_pot_2a} \\
    \frac{d}{dt} \frac{\partial \mathscr{L}}{\partial \dot \psi} - \frac{\partial \mathscr{L}}{\partial \psi} &= 0 \implies \frac{d}{dt}(r^{2} \dot \psi) = 0.
    \label{eqn:orbit_spherical_pot_2b}
\end{align}
Equation~(\ref{eqn:orbit_spherical_pot_2b}) shows that $r^{2} \dot \psi$ is a constant of motion which is nothing but the angular momentum (L) of the star i.e.,
\begin{align}
    \centering
    r^{2} \dot \psi = L.
    \label{eqn:orbit_spherical_pot_3}
\end{align}
This equation can be re-arranged in the form of an operator,
\begin{align}
    \centering
    \frac{d}{dt} = \frac{L}{r^{2}} \frac{d}{d \psi}
    \label{eqn:orbit_spherical_pot_4}
\end{align}
Using this operator, we can replace the time derivative by an angle derivative in the equation~(\ref{eqn:orbit_spherical_pot_2a}) as follows,
\begin{align}
    \centering
     \frac{L}{r^{2}} \frac{d}{d \psi} \left( \frac{L}{r^{2}} \frac{dr}{d \psi} \right) - r \left(\frac{L}{r^{2}}\right)^{2} + \frac{d \Phi}{dr} &= 0, \notag \\
     \frac{L^{2}}{r^{2}} \frac{d}{d \psi} \left( \frac{1}{r^{2}} \frac{dr}{d \psi} \right) - \frac{L^{2}}{r^{3}} + \frac{d \Phi}{dr} &= 0.
    \label{eqn:orbit_spherical_pot_5}
\end{align}
We can simplify equation~(\ref{eqn:orbit_spherical_pot_5}) by substituting,
\begin{align}
    r = \frac{1}{u} \implies \frac{dr}{d \psi} = - \frac{1}{u^{2}} \frac{du}{d \psi}.
    \label{eqn:orbit_spherical_pot_6}
\end{align}
On substitution of these values, equation~(\ref{eqn:orbit_spherical_pot_5}) becomes,
\begin{align}
    -L^{2}u^{2} \frac{d^{2}u}{d \psi^{2}} - L^{2}u^{3} + \frac{d \Phi}{dr} &= 0, \notag \\
    \frac{d^{2}u}{d \psi^{2}} + u = \frac{1}{L^{2}u^{2}} \frac{d \Phi}{dr}.
    \label{eqn:orbit_spherical_pot_7}
\end{align}
To integrate equation~(\ref{eqn:orbit_spherical_pot_7}), multiply both sides by $du/d \psi$, and simplify the right-hand side using equation~(\ref{eqn:orbit_spherical_pot_6}),
\begin{align}
    \frac{du}{d \psi} \frac{d^{2}u}{d \psi^{2}} + u \frac{du}{d \psi} &= - \frac{1}{L^{2}} \frac{d \Phi}{d \psi}, \notag \\
    \frac{1}{2} \left( \frac{du}{d \psi} \right)^{2} + \frac{1}{2} u^{2} &= - \frac{\Phi}{L^{2}} + \frac{E}{L^{2}}, \notag \\
    \left( \frac{du}{d \psi} \right)^{2} + u^{2} + \frac{2\Phi}{L^{2}} &= \frac{2E}{L^{2}},
    \label{eqn:orbit_spherical_pot_8}
\end{align}
where $E/L^{2}$ is an integration constant and E represents total energy of the orbit. Now, we can write equation~(\ref{eqn:orbit_spherical_pot_8}) in polar co-ordinates using equations~(\ref{eqn:orbit_spherical_pot_4}) and (\ref{eqn:orbit_spherical_pot_6}) as follows,
\begin{align}
    \frac{1}{L^{2}} \left( \frac{dr}{dt} \right)^{2} + \frac{1}{r^{2}} + \frac{2\Phi}{L^{2}} &= \frac{2E}{L^{2}}, \notag \\
    \left( \frac{dr}{dt} \right)^{2} &= 2(E - \Phi) - \frac{L^{2}}{r^{2}},
    \label{eqn:orbit_spherical_pot_9}
\end{align}
Using equation~(\ref{eqn:orbit_spherical_pot_9}), we can estimate orbital parameters of the stellar orbit. For example,

\begin{figure*}[ht!]
    \centering
	\includegraphics[width=0.5\textwidth]{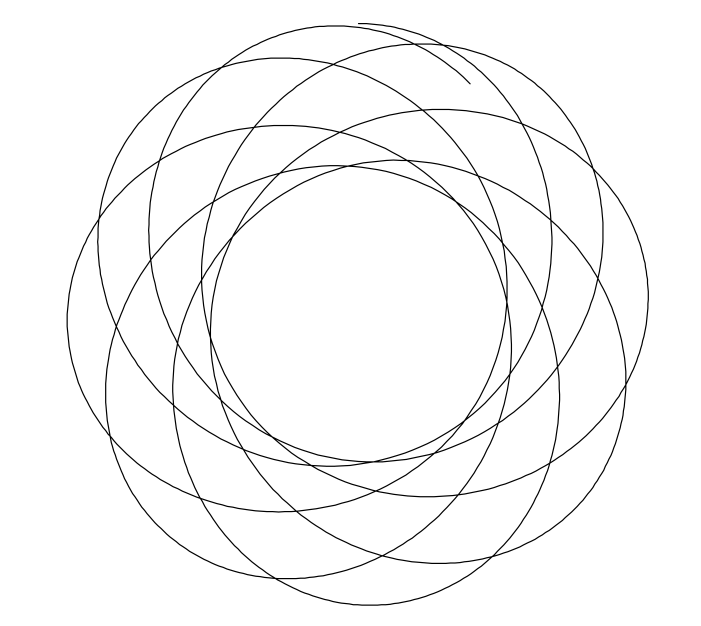}
    \caption{An example orbit in a spherical potential. It looks like a rosette which is typical in spherical potential. Picture courtesy: \citep{Binney.Tremaine.2008}.}
    \label{fig:orbit_in_spherical_pot}
\end{figure*}

\begin{itemize}
    \item {\bf Pericenter and apocenter distances:} From the center of potential, the closest point of orbit is known as perigee and the farthest point is called apogee. Radial distances of perigee and apogee are respectively labeled as pericenter ($r_{p}$) and apocenter ($r_{a}$) distances. At perigee and apogee of the orbit, $dr/dt$ will be zero i.e.,
    \begin{align}
    0 &= 2(E - \Phi) - \frac{L^{2}}{r^{2}}, \notag \\
    r^2 - \frac{L^{2}}{2 \left(E - \Phi \right)} &= 0.
    \label{eqn:orbit_spherical_pot_10}
    \end{align}
    This equation generally has two roots $r_{p}$ and $r_{a}$. A star oscillates between these two radii as it evolves.

    \item {\bf Radial period ($T_{r}$):} It is the time required to orbit from perigee to apogee and back to perigee. Thus,
    \begin{align}
        T_{r} &= 2\int_{r_{p}}^{r_{a}} \frac{dt}{dr} dr, \notag \\
        T_{r} &= 2\int_{r_{p}}^{r_{a}} \frac{dr}{\sqrt{2 \left( E - \Phi \right) - \frac{L^{2}}{r^{2}}}}
        \label{eqn:orbit_spherical_pot_11}
    \end{align}

    \item {\bf Azimuthal period ($T_{\psi}$):} It is the time required to travel $2\pi$ radian in azimuthal direction. Let us first calculate the change in azimuthal angle in one radial period.
    \begin{align}
        \Delta \psi &= 2\int_{r_{p}}^{r_{a}} \frac{d \psi}{dr} dr = 2\int_{r_{p}}^{r_{a}} \frac{L}{r^{2}} \frac{dt}{dr} dr, \notag \\
        \Delta \psi &= 2L\int_{r_{p}}^{r_{a}} \frac{dr}{r^{2} \sqrt{2 \left( E - \Phi \right) - \frac{L^{2}}{r^{2}}}}
        \label{eqn:orbit_spherical_pot_12}
    \end{align}
    Now, azimuthal period can be calculated as,
    \begin{align}
        T_{\psi} = \left| \frac{2\pi}{\Delta \psi} \right| T_{r}
        \label{eqn:orbit_spherical_pot_13}
    \end{align}
    For a close orbit, $\Delta \psi / 2\pi$ should be a rational number which it is not generally. Therefore, the typical orbits in a spherical potential describe a rosette like structure, as can be seen in Fig.~\ref{fig:orbit_in_spherical_pot}. 
\end{itemize}

\section{Stellar Orbits in an axisymmetric potential}
\label{sec:orbit_in_axisymmetric_pot}
The assumption of a spherically symmetric potential completely breaks down for disk galaxies. Thus, disk galaxies are studied considering an axisymmetric potential. Stars moving in the equatorial plane of such a potential feel zero net force in the vertical direction. As a consequence of this the star cannot perceive that it is not moving in a spherical potential. Therefore, stellar orbits in the equatorial plane are described in a similar manner as we discussed in Section~\ref{sec:orbit_in_spherical_pot}.

\begin{figure*}[ht!]
    \centering
	\includegraphics[width=\textwidth]{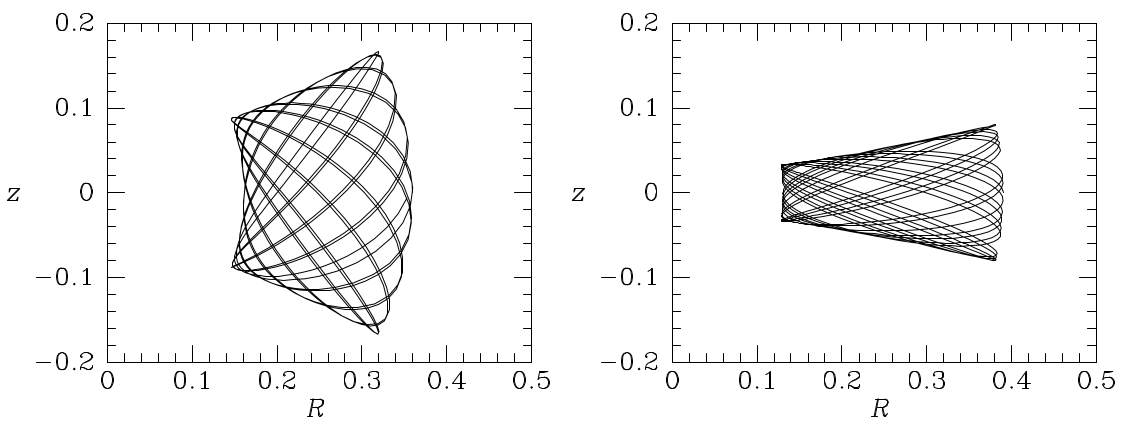}
    \caption{Example orbits in an axisymmetric potential. Both the orbits are shown for the same total energy and the same angular momentum along z-axis in the potential given in equation~(\ref{eqn:orbit_axisymmetric_pot_7}) for $q=0.9$ and $v_{0}=1$. Picture courtesy: \citep{Binney.Tremaine.2008}.}
    \label{fig:orbit_in_axisymmetric_pot}
\end{figure*}

To study typical stellar orbits in axisymmetric potential $\Phi(R,z)$ which is symmetric about $z=0$ plane, let us consider a cylindrical co-ordinate system ($R,\phi,z$) positioned at the center of the potential. Now, we can write its Lagrangian per unit mass as follows,
\begin{align}
    \centering
    \mathscr{L} = \frac{1}{2}\left[\dot R^2 + (R \dot \phi)^{2} + \dot z^{2} \right] - \Phi.
    \label{eqn:orbit_axisymmetric_pot_1}
\end{align}
The momenta for this Lagrangian are,
%
\begin{subequations}
\begin{align}
    \centering
    p_{R} &= \frac{\partial \mathscr{L}}{\partial \dot R} = \dot R
    \label{eqn:orbit_axisymmetric_pot_2a} \\
    p_{\phi} &= \frac{\partial \mathscr{L}}{\partial \dot \phi} = R^{2} \dot \phi 
    \label{eqn:orbit_axisymmetric_pot_2b} \\
    p_{z} &= \frac{\partial \mathscr{L}}{\partial \dot z} = \dot z.
    \label{eqn:orbit_axisymmetric_pot_2c}
\end{align}
\end{subequations}
The corresponding Hamiltonian is,
\begin{align}
    \centering
    \mathscr{H} = \frac{1}{2}\left[p_{R}^2 + \frac{p_{\phi}^{2}}{R^{2}} + p_{z}^{2} \right] + \Phi.
    \label{eqn:orbit_axisymmetric_pot_3}
\end{align}
The equation of motion for this Hamiltonian are,
\begin{subequations}
\begin{align}
    \centering
    \dot p_{R} &= \ddot R = -\frac{\partial \mathscr{H}}{\partial \dot R} = \frac{p_{\phi}^{2}}{R^{3}} - \frac{\partial \Phi}{\partial R} 
    \label{eqn:orbit_axisymmetric_pot_4a} \\
    \dot p_{\phi} &= \frac{d}{dt}(R^{2} \dot \phi) = -\frac{\partial \mathscr{H}}{\partial \dot \phi} = 0 
    \label{eqn:orbit_axisymmetric_pot_4b} \\
    \dot p_{z} &= \ddot z = -\frac{\partial \mathscr{H}}{\partial \dot z} = - \frac{\partial \Phi}{\partial z}.
    \label{eqn:orbit_axisymmetric_pot_4c}
\end{align}
\end{subequations}
Equation~(\ref{eqn:orbit_axisymmetric_pot_4b}) shows that $p_{\phi}$ is a conserved quantity and is, in fact, the component of angular momentum about the z-axis ($L_{z}$), i.e.
\begin{align}
    p_{\phi} = R^{2} \dot \phi = L_{z}.
    \label{eqn:orbit_axisymmetric_pot_5}
\end{align}
Re-writing equations~(\ref{eqn:orbit_axisymmetric_pot_4a}) and (\ref{eqn:orbit_axisymmetric_pot_4c}) using value of $p_{\phi} = L_{z}$,
\begin{subequations}
\begin{align}
    \centering
    \ddot R &= - \frac{\partial \Phi_{eff}}{\partial R} 
    \label{eqn:orbit_axisymmetric_pot_6a} \\
    \ddot z &= - \frac{\partial \Phi_{eff}}{\partial z},
    \label{eqn:orbit_axisymmetric_pot_6b} \\
    \text{where} \notag \\
    \Phi_{eff} &= \frac{L_{z}^{2}}{2R^{2}} + \Phi (R,z).
    \label{eqn:orbit_axisymmetric_pot_6c}
\end{align}
\end{subequations}
is effective potential. Coupled equations~(\ref{eqn:orbit_axisymmetric_pot_6a}) and (\ref{eqn:orbit_axisymmetric_pot_6b}) describe motion of the star in the $R-z$ plane (also known as the meridional plane). These equations can be solved numerically (or analytically for some special potentials) for different initial conditions. In Fig.~\ref{fig:orbit_in_axisymmetric_pot}, we have shown the orbits of a star in the meridional plane of an axisymmetric potential given by the following expression,
\begin{align}
    \Phi(R,z) = \frac{1}{2} v_{0}^{2} \ln \left(R^{2} +\frac{z^{2}}{q^{2}} \right).
    \label{eqn:orbit_axisymmetric_pot_7}
\end{align}
Both the orbits are calculated for the same total energy ($E=-0.8$) and the same angular momentum along the z-axis ($L_{z}=0.2$) in this potential for $q=0.9$ and $v_{0}=1$

\section{Collisionless Boltzmann Equation}
\label{sec:CBE}
In previous sections, we saw that galaxies are collisionless stellar systems and their potentials can be smoothed to study stellar orbits. However, it is not practical and worthwhile to follow the orbits of each star in a galaxy. For practical purpose, we use a distribution function (DF) $f(\vec{x},\vec{v},t)$ to describe galaxies at any given time $t$. Here, $f(\vec{x},\vec{v},t)$ is a phase-space probability density function, where the probability of finding a random star at $\vec{w} = (\vec{x}, \vec{v})$ in volume $dV$ at any time $t$ is given by,
\begin{align}
    p = \int_{dV} f(\vec{w},t) d^{6} \vec{w} = \int_{dV} f(\vec{x},\vec{v},t) d^{3} \vec{x} d^{3} \vec{v}.
    \label{eqn:CBE_1}
\end{align}
By the definition of probability density function,
\begin{align}
    \int_{V} f(\vec{x},\vec{v},t) d^{3} \vec{x} d^{3} \vec{v} = 1.
    \label{eqn:CBE_2}
\end{align}
the DF of the system changes over time as the stars travel through phase-space. Analogous to the continuity equation for mass conservation in fluid dynamics, we can write continuity equation for probability conservation in phase-space as follows,
\begin{align}
    \frac{\partial f}{\partial t} + \frac{\partial }{\partial \vec{w}} \cdot (f \dot{\vec{w}}) = 0
    \label{eqn:CBE_3}
\end{align}
Let us say $\mathscr{H}$ is the Hamiltonian in phase-space $\vec{w} = (\vec{x}, \vec{v})$,
\begin{align}
    \frac{\partial }{\partial \vec{w}} \cdot (f \dot{\vec{w}}) &= \frac{\partial }{\partial \vec{x}} \cdot (f \dot{\vec{x}}) + \frac{\partial }{\partial \vec{v}} \cdot (f \dot{\vec{v}}) \notag \\
    &= \frac{\partial }{\partial \vec{x}} \cdot (f \frac{\partial \mathscr{H}}{\partial \vec{v}}) - \frac{\partial }{\partial \vec{v}} \cdot (f \frac{\partial \mathscr{H}}{\partial \vec{x}}) \notag \\
    &= \frac{\partial f}{\partial \vec{x}} \cdot \frac{\partial \mathscr{H}}{\partial \vec{v}} - \frac{\partial f}{\partial \vec{v}} \cdot \frac{\partial \mathscr{H}}{\partial \vec{x}} \notag \\
    &= \frac{\partial f}{\partial \vec{x}} \cdot \dot{\vec{x}} + \frac{\partial f}{\partial \vec{v}} \cdot \dot{\vec{v}}
    \label{eqn:CBE_4}
\end{align}
Substituting the value of equations~(\ref{eqn:CBE_4}) in equation~(\ref{eqn:CBE_3}) gives,
\begin{align}
    \frac{\partial f}{\partial t} + \dot{\vec{x}} \cdot \frac{\partial f}{\partial \vec{x}} + \dot{\vec{v}} \cdot \frac{\partial f}{\partial \vec{v}} = 0.
    \label{eqn:CBE_5}
\end{align}
This partial differential equation is knows as collisionless Boltzmann equation (CBE) or Vlasov equation. The left-hand side of equation~(\ref{eqn:CBE_5}) is actually a perfect time derivative of $f$. Thus, simply $df/dt = 0$ is the collisionless Boltzmann equation.

The collisionless Boltzmann equation is based on the conservation of probability density function. We can modify it for the creation and destruction of stars in a galaxy as follows,
\begin{align}
    \frac{\partial f}{\partial t} + \dot{\vec{x}} \cdot \frac{\partial f}{\partial \vec{x}} + \dot{\vec{v}} \cdot \frac{\partial f}{\partial \vec{v}} = C - D,
    \label{eqn:CBE_6}
\end{align}
where $C(\vec{x},\vec{v},t)$ and $D(\vec{x},\vec{v},t)$ are respectively creation and destruction rates of stars per unit phase-space volume.

\section{Jeans equations}
\label{sec:jeans_equations}
One can calculate various observational quantities (velocity moments) using the DF. Nonetheless, finding the DF is not that simple. Here, we review how to estimate various velocity moments using the gravitational potential of the system. Let us say $\Phi (\vec{x})$ is the potential of galaxy, we can the write collisionless Boltzmann equation~(\ref{eqn:CBE_5}) as follows,
\begin{align}
    \frac{\partial f}{\partial t} + \vec{v} \cdot \frac{\partial f}{\partial \vec{x}} - \frac{\partial \Phi}{\partial \vec{x}} \cdot \frac{\partial f}{\partial \vec{v}} = 0.
    \label{eqn:Jeans_equations_1}
\end{align}
Integrating this equation over the whole velocity space and writing it in a summation convention \footnote{Suppose a scalar product $\vec{A} \cdot \vec{B} = \Sigma_{i} A_{i} B_{i}$, where i = 1, 2, 3 represents x-, y-, z-components, respectively. In summation convention, we can simply write it as $\vec{A} \cdot \vec{B} = A_{i} B_{i}$. It automatically assumes summation over dummy index i.},
\begin{align}
    \int \frac{\partial f}{\partial t} d^{3}\vec{v} + \int v_{i} \frac{\partial f}{\partial x_{i}} d^{3}\vec{v} - \int \frac{\partial \Phi}{\partial x_{i}} \frac{\partial f}{\partial v_{i}} d^{3}\vec{v} = 0, \notag \\
    \int \frac{\partial f}{\partial t} d^{3}\vec{v} + \int v_{i} \frac{\partial f}{\partial x_{i}} d^{3}\vec{v} - \frac{\partial \Phi}{\partial x_{i}} \int \frac{\partial f}{\partial v_{i}} d^{3}\vec{v} = 0.
    \label{eqn:Jeans_equations_2}
\end{align}
Since $f$ is the function of $\vec{x}$, $\vec{v}$, and $t$, the partial derivatives in the first and second terms can be taken outside. The volume integration in the third term can be converted into surface integration using divergence theorem. At large velocity, $f=0$ as there are no stars with infinitely high velocities. Therefore, surface integration will become zero. Now we get,
\begin{align}
    \frac{\partial \mu}{\partial t} + \frac{\partial (\mu \overline{v}_{i})}{\partial x_{i}} = 0,
    \label{eqn:Jeans_equations_3}
\end{align}
where
\begin{subequations}
\begin{align}
    \mu (\vec{x}, t) &= \int f(\vec{x}, \vec{v}, t) d^{3} \vec{v}, 
    \label{eqn:Jeans_equations_4a}\\
    \overline{v}_{i} (\vec{x}, t) &= \frac{1}{\mu (\vec{x}, t)} \int f(\vec{x}, \vec{v}, t) v_{i} d^{3} \vec{v}.
    \label{eqn:Jeans_equations_4b}
\end{align}
\end{subequations}
Now, multiply equation~(\ref{eqn:Jeans_equations_1}) by $v_{j}$ and integrate again over the whole velocity space,
\begin{align}
    \int v_{j} \frac{\partial f}{\partial t} d^{3}\vec{v} + \int v_{i} v_{j} \frac{\partial f}{\partial x_{i}} d^{3}\vec{v} - \frac{\partial \Phi}{\partial x_{i}} \int v_{j} \frac{\partial f}{\partial v_{i}} d^{3}\vec{v} = 0, \notag \\
    \frac{\partial (\mu \overline{v}_{j})}{\partial t} + \frac{\partial (\mu \overline{{v}_{i}{v}_{j}})}{\partial x_{i}} - \frac{\partial \Phi}{\partial x_{i}} \int v_{j} \frac{\partial f}{\partial v_{i}} d^{3}\vec{v} = 0
    \label{eqn:Jeans_equations_5}
\end{align}
The last term of this equation can be simplified using integration by parts and the divergence theorem,
\begin{align}
    \frac{\partial \Phi}{\partial x_{i}} \int v_{j} \frac{\partial f}{\partial v_{i}} d^{3}\vec{v} &= - \frac{\partial \Phi}{\partial x_{i}} \int \frac{\partial v_{j}}{\partial v_{i}} f d^{3}\vec{v} \notag \\
    &= - \frac{\partial \Phi}{\partial x_{i}} \int \delta_{ij} f d^{3}\vec{v} \notag \\
    &= - \delta_{ij} \frac{\partial \Phi}{\partial x_{i}} \mu \notag \\
    &= - \mu \frac{\partial \Phi}{\partial x_{j}}.
    \label{eqn:Jeans_equations_6}
\end{align}
Substituting it in equation~(\ref{eqn:Jeans_equations_5}) gives,
\begin{align}
    \frac{\partial (\mu \overline{v}_{j})}{\partial t} + \frac{\partial (\mu \overline{{v}_{i}{v}_{j}})}{\partial x_{i}} + \mu \frac{\partial \Phi}{\partial x_{j}} &= 0,
    \label{eqn:Jeans_equations_7a} \\
    \mu \frac{\partial \overline{v}_{j}}{\partial t} + \overline{v}_{j} \frac{\partial \mu}{\partial t} + \frac{\partial (\mu \overline{{v}_{i}{v}_{j}})}{\partial x_{i}} + \mu \frac{\partial \Phi}{\partial x_{j}} &= 0.
    \label{eqn:Jeans_equations_7b}
\end{align}
Making use of equation~(\ref{eqn:Jeans_equations_3}) in equation~(\ref{eqn:Jeans_equations_7b}) returns,
\begin{align}
     \mu \frac{\partial \overline{v}_{j}}{\partial t} - \overline{v}_{j} \frac{\partial (\mu \overline{v}_{i})}{\partial x_{i}} + \frac{\partial (\mu \overline{{v}_{i}{v}_{j}})}{\partial x_{i}} + \mu \frac{\partial \Phi}{\partial x_{j}} &= 0.
    \label{eqn:Jeans_equations_8}
\end{align}
Now, let us define the velocity dispersion tensor as,
\begin{align}
     \sigma_{ij}^{2}(\vec{x}, t) &= \frac{1}{\mu (\vec{x},t)} \int (v_{i} - \overline{v_{i}}) (v_{j} - \overline{v_{j}}) f(\vec{x}, \vec{v}, t) d^{3} \vec{v} \notag \\
     &= \overline{v_{i} v_{j}} - \overline{v}_{i} \overline{v}_{j}.
    \label{eqn:Jeans_equations_9}
\end{align}
Substituting $\overline{v_{i} v_{j}}$ from equation~(\ref{eqn:Jeans_equations_9}) in equation~(\ref{eqn:Jeans_equations_8}),
\begin{align}
     \mu \frac{\partial \overline{v}_{j}}{\partial t} + \mu \overline{v}_{i} \frac{\partial \overline{v}_{j}}{\partial x_{i}} + \frac{\partial (\mu \sigma_{ij}^{2})}{\partial x_{i}} + \mu \frac{\partial \Phi}{\partial x_{j}} &= 0, \notag \\
     \mu \frac{\partial \overline{v}_{j}}{\partial t} + \mu \overline{v}_{i} \frac{\partial \overline{v}_{j}}{\partial x_{i}} &= - \mu \frac{\partial \Phi}{\partial x_{j}} - \frac{\partial (\mu \sigma_{ij}^{2})}{\partial x_{i}}.
    \label{eqn:Jeans_equations_10}
\end{align}
This equation is similar to Euler’s equation in fluid dynamics. Equations~(\ref{eqn:Jeans_equations_3}) and (\ref{eqn:Jeans_equations_10}) is known as Jeans equations for stellar dynamics.

\section{Numerical simulations}
\label{sec:numerical_sims}
Although the analytical description of galactic dynamics is very useful and valuable, it cannot provide us complete information about galaxy formation and evolution. Therefore, numerical simulations come into picture as indispensable tools to study galaxies, thanks to the revolutionary development of computers. In the following subsections, we discuss various methods used in numerical simulations.

\subsection{Force calculation}
\label{subsec:force_calculation}
\begin{itemize}
    \item {\bf Particle-Particle (PP) algorithm:} Consider a stellar system having $N$ number of identical particles\footnote{A system with $N$ number of particles is commonly knows as $N-$body system.} each with mass $m$. Gravitational force per unit mass on $i$th particle in the system is simply given by,
    \begin{align}
        \vec{F}_{i} = \sum_{j\neq i}^{N} Gm \frac{\vec{x}_{j} - \vec{x}_{i}}{|\vec{x}_{j} - \vec{x}_{i}|^{3}}.
        \label{eqn:PP_algorithm}
    \end{align}
    At any time step of the simulation, a computer will perform an order of $N^2$ calculations to estimate forces on all the particles. This method is fine as long as the number of particles is less than $10^{4}$. For galactic scale simulations, it is computationally a very expensive method.

    \item {\bf Tree algorithm:} Here, we review Tree algorithm discussed in \cite{Barnes.Hut.1986}. It reduces force calculations to order of $N \ln N$. In Tree algorithm, the whole simulation is placed inside a cubical box. Next, we divide this box into eight equal cubical sub-boxes. If any sub-box has more than one particle, we divide it again into eight equal cubical sub-sub-boxes until each box has at most one particle. This hierarchical structure is call oct-tree (or quad-tree in 2D). The original cube is the root, the cubes having more than one particle are the nodes, and cubes with at most one particle are leaves.
    
    For force calculation on any subject particle, we walk the tree starting from root nodes until angles subtended by the nodes are less than the tolerance angle ($\theta_{tol}$)\footnote{Tolerance angle also known as opening angle because it decides whether node is open to walk further or not.}. Suppose a node of size $l$ is located at a distance $r$ from the subject particle, the condition to stop walking the tree further is given by,
    \begin{align}
        \frac{l}{r} < \theta_{tol}.
        \label{eqn:Tree_algorithm}
    \end{align}
    At this point, we can treat particles in the node as a single particle of node mass and located at node's center of mass. In Fig.~\ref{fig:Tree_algorithm}, we have illustrated the Tree algorithm in two dimensions. The left-hand panel of the figure shows the quad-tree for a given distribution of particles. The particle on which we want to compute force is colored green. The right-hand panel of the figure shows the corresponding closed nodes and their equivalent particles in red color for an arbitrary $\theta_{tol}$. The connecting lines show the forces acting after the implementation of the Tree algorithm.
    \begin{figure*}[ht!]
    \centering
	\includegraphics[width=0.4\textwidth]{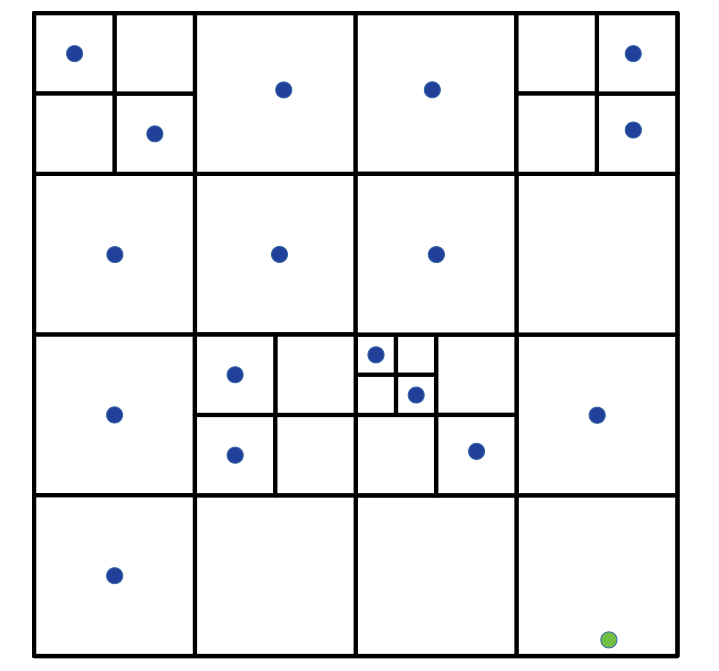}
	\includegraphics[width=0.4\textwidth]{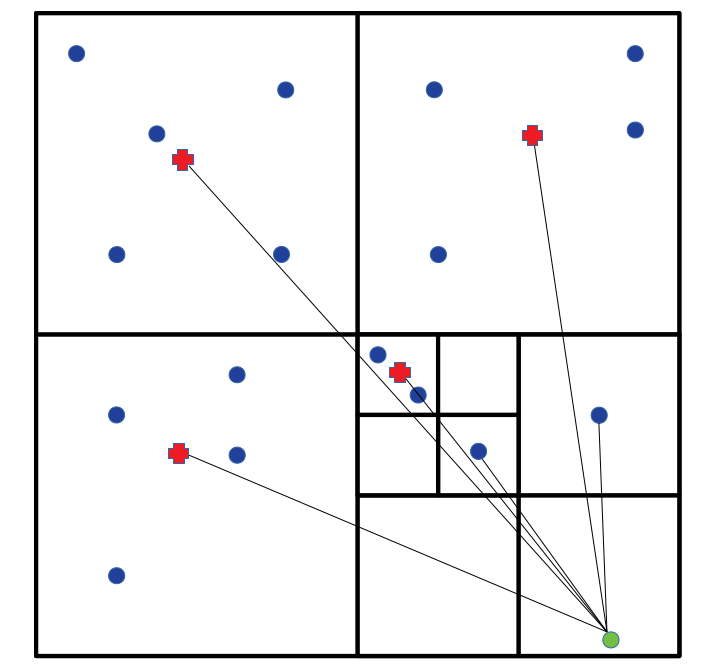}
    \caption{An illustration of the Tree algorithm for force computation in 2D. The left-hand panel shows the quad-tree for a given distribution of particles. The particle on which we want to compute force is colored green. The right-hand panel shows the corresponding closed nodes and their equivalent particles in red color for an arbitrary $\theta_{tol}$. The connecting lines show forces acting after Tree algorithm is applied.}
    \label{fig:Tree_algorithm}
    \end{figure*}

    \item {\bf Particle-Mesh (PM) algorithm:} In the Particle-Mesh algorithm, we first divide the whole simulation box of size $L$ into a grid of $M^{3}$ meshes with fixed size $l=L/M$. Second, we calculate the mass density at mesh-points. Third, we determine the potential on the mesh-points by solving Poisson's equation. Finally, the gradient of the potential at the mesh-points provides the force on the mesh-points \citep{Hockney.Eastwood.1988}.
    
    Consider an $N-$body system with identical particles each with mass $m$. The mass density on a mesh-point $\vec{x}_{p}$ due to particles distribution ($\vec{x}_{i}$, $i=1,2,3,...,N$) can be expressed as follows,
    \begin{align}
        \rho(\vec{x}_{p}) = \frac{m}{L^{3}} \sum_{i=1}^{N} W(\vec{x} - \vec{x}_{p}).
        \label{eqn:Particle-Mesh_algorithm_1}
    \end{align}
    Where W is a normalized smoothing kernel. It is chosen in such a way that the total angular momentum of the system is conserved. For example,
    \begin{align}
        W(\vec{x}) = \frac{1}{\pi l^{3}}
        \begin{cases}
        \frac{3}{4} - R^{2} & \text{($0 \leq R \leq \frac{1}{2}$)} \\
        \frac{1}{2}(\frac{3}{2} - R)^{2} & \text{($\frac{1}{2} < R \leq \frac{3}{2}$)} \\
        0 & \text{otherwise},
        \end{cases}
        \label{eqn:Particle-Mesh_algorithm_2}
    \end{align}
    where $R=|\vec{x}|/l$. Potential $\Phi (\vec{x})$ on a mesh-point can be obtained using Fourier transformation of Poisson's equation as follows,
    \begin{align}
        -k^{2} \Phi(\vec{k}) = 4 \pi G \rho (\vec{k}).
        \label{eqn:Particle-Mesh_algorithm_3}
    \end{align}
    Now, gravitational force per unit mass at the mesh-point can be calculated using the inverse Fourier transformation of
    \begin{align}
        \vec{F}(\vec{k}) = -i \Phi (\vec{k}) \vec{k}.
        \label{eqn:Particle-Mesh_algorithm_4}
    \end{align}
    Finally, the forces on the locations of particles can be obtained using interpolation of forces on mesh-points as follows,
    \begin{align}
        \vec{F}(\vec{x}) = \sum_{i=1}^{N} W(\vec{x} - \vec{x}_{p}) \vec{F}(\vec{x_{p}}).
        \label{eqn:Particle-Mesh_algorithm_5}
    \end{align}

    \item {\bf P$^{3}$M algorithm:} P$^{3}$M is a hybrid algorithm that uses Particle-Particle (PP) algorithm and Particle-Mesh (PM) algorithm simultaneously. In P$^{3}$M algorithm, Long-range forces are computed using the PM algorithm, whereas the short-range forces are calculated using the PP algorithm. The PP algorithm is usually implemented in the regions where particles are highly clustered. If there are many particles within $2l$ distance, the PP algorithm will be initiated otherwise the PM algorithm is used.

    \item {\bf TreePM algorithm:} TreePM is also a hybrid algorithm that uses Tree algorithm and Particle-Mesh (PM) algorithm simultaneously. In TreePM algorithm, long-range forces are computed using PM algorithm, whereas short-range forces are calculated using the Tree algorithm.
    
\end{itemize}

\subsection{Gravitational softening}
\label{subsec:grav_soft}
Sometimes, during the evolution of simulation, particles come so close that the force calculated using equation~(\ref{eqn:PP_algorithm}) diverges. In a collisionless system, such encounters are totally unphysical. To avoid such collisions in the simulation, we soften the gravitational force as follows,
\begin{align}
    \vec{F}_{i} = \sum_{j\neq i}^{N} Gm \frac{\vec{x}_{j} - \vec{x}_{i}}{(|\vec{x}_{j} - \vec{x}_{i}|^{2} + \epsilon^{2})^{3/2}},
    \label{eqn:gravitational_softening}
\end{align}
where $\epsilon$ is a constant known as gravitational softening length or Plummer softening length. Introducing constant $\epsilon$ in the denominator of force equation does not affect force calculation when $|\vec{x}_{j} - \vec{x}_{i}| > \epsilon$. However, it avoids numerical artifacts when two particles are very close. In such cases, the denominator becomes $\epsilon$.

\subsection{Smoothed-particle hydrodynamics (SPH)}
\label{subsec:SPH}
Smoothed-particle hydrodynamics (SPH) is one of the well-known techniques for studying gas dynamics in numerical simulations. It is based on the Lagrangian formulation of fluid dynamics where a fluid is treated as a set of N particles. In SPH, we follow the motion of individual particle whose mass $m$ is constant during the evolution \citep{Monaghan.1992}.

To obtain the smoothed form of any fluid field $A(\vec{x})$, such as density ($\rho$) or temperature ($T$), in SPH simulations, we convolve the field with some smoothing kernel $W$ as follows,
\begin{align}
    A_{s}(\vec{x}) = \sum_{i=1}^{N} \frac{m_{i}}{\rho_{i}} A(\vec{x}_{i}) W(\vec{x}-\vec{x}_{i}, h),
    \label{eqn:SPH_1}
\end{align}
where $h$ is the characteristic radius of the smoothing kernel. It defines the extent of the kernel, and is typically chosen in such a way that $W$ becomes zero at a distance $\eta h$, where $\eta$ is generally an order of unity. A commonly used smoothing kernel in SPH is given by,
\begin{align}
    W(\vec{x},h) = \frac{1}{\pi h^{3}}
    \begin{cases}
    1 - \frac{3R}{2} + \frac{3 R^{3}}{4} & \text{($0 \leq R \leq 1$)} \\
    \frac{1}{4}(2 - R)^{3} & \text{($1 < R \leq 2$)} \\
    0 & \text{otherwise},
    \end{cases}
    \label{eqn:SPH_2}
\end{align}
where $R=|\vec{x}|/h$ and $\eta$ is 2.

\subsection{Time integration}
\label{subsec:time_integration}
Time integration is a crucial part of any numerical simulation. Various simulations adopt the leapfrog integration scheme to predict the next co-ordinates of a particle based on its initial co-ordinates and for a given time-step. Let us say $\vec{x}(t_{n})$, and $\vec{v}(t_{n})$ are respectively the position and velocity co-ordinates of a particle at any time $t_{n}$. If $\vec{F}(t_{n})$ is the force per unit mass acting on the particle at time $t_{n}$, the next position and velocity of this particle after $\Delta t_{n}$ time-step are calculated using following steps,
\begin{subequations}
\begin{align}
    \vec{x}(t_{n+1/2}) = \vec{x}(t_{n}) + \frac{\Delta t_{n}}{2} \vec{v}(t_{n}),
    \label{eqn:time_integration_1a} \\
    \vec{v}(t_{n+1/2}) = \vec{v}(t_{n}) + \frac{\Delta t_{n}}{2} \vec{F}(t_{n}).
    \label{eqn:time_integration_1b}
\end{align}
\end{subequations}
These are the position and velocity co-ordinates at half time-step. We can calculate force at the intermediate position as follows,
\begin{align}
    \vec{F}(t_{n+1/2}) = -\nabla \Phi|_{\vec{x}(t_{n+1/2})}.
    \label{eqn:time_integration_2}
\end{align}
Now the position and velocity co-ordinates at next half time-step are
\begin{subequations}
\begin{align}
    \vec{x}(t_{n+1}) = \vec{x}(t_{n+1/2}) + \frac{\Delta t_{n}}{2} \vec{v}(t_{n+1/2}),
    \label{eqn:time_integration_3a} \\
    \vec{v}(t_{n+1}) = \vec{v}(t_{n+1/2}) + \frac{\Delta t_{n}}{2} \vec{F}(t_{n+1/2}).
    \label{eqn:time_integration_3b}
\end{align}
\end{subequations}
These are the co-ordinate of the particle after $\Delta t_{n}$ time-step and new simulation time becomes
\begin{align}
    t_{n+1} = t_{n} + \Delta t_{n}.
    \label{eqn:time_integration_4}
\end{align}

\section{Existing simulations}
\label{sec:existing_simulations}
Several groups of researcher across the globe have performed cosmological simulations that lead to structure formation and galaxy evolution. These simulations have specific science goals. Some simulations adopt large box sizes to understand large scale structure formation and clustering. On the other hand, some simulations focus on small scale clustering and galaxy formation using small box sizes. These simulation can be simply dark matter only or can include sophisticated baryonic physics. On the basis of these factors, simulations can be categorized into four classes:

\begin{itemize}
    \item {\bf Dark matter only large volume simulations:} These simulations consider all the matter as dark matter and evolve it in a large cosmological box. A typical purpose of these simulations is to investigate the structure formation and gravitational clustering of matter at large scales. They generally have course mass resolution. Bolshoi simulations, Dark Sky, Millennium, Millennium-II, Millennium-XXL are examples of dark matter only large volume simulations.
    
    \item {\bf Dark matter only zoom-in simulations:} These simulations are refined versions of dark matter only large volume simulations in the sense that they evolve a small region of interest in large volume simulations at much higher mass resolution. They are useful to study small scale clustering and dark matter halo formation in cosmological environment. Via Lactea, ELVIS, Phoenix, GHALO, Aquarius are the examples of dark matter only zoom-in simulations.
    
    \item {\bf Dark matter + baryon large volume simulations:} These simulations treat dark matter and baryonic matter as separate entities. They also include baryonic physics to probe its impact on large scale structure and clustering of baryons along dark matter.  EAGLE, Romulus25, Horizon-AGN, Massiveblack-II, Simba, Magneticum, Illustris, IllustrisTNG are examples of (dark matter + baryon) large volume simulations.
    
    \item {\bf Dark matter + baryon zoom-in simulations:} Similar to the dark matter only zoom-in simulations, these simulations also evolve a small region of interest from the large volume simulation at much better resolution and include sophisticated baryonic physics. These simulations help us understand galaxy formation, the effect of baryonic physics, dark matter halo abundance, and several other small scale properties in the cosmological scenario. Eris, Latte/FIRE, APOSTLE, NIHAO, Auriga are the examples of dark matter + baryon zoom-in simulations.
    
\end{itemize}

\section{Codes, tools, and data used}
\label{sec:codes_tools_data}
Every simulation requires initial realization of the system we are interested in. Next, this realization is evolved in time to understand the dynamics of the system over time. Here, we list the simulation codes, analysis tools, and data used in this thesis work.

\begin{itemize}
    \item {\bf GALIC:} The initial realizations of the model galaxies are generated using a publicly distributed code GALIC \citep{Yurin2014}. GALIC uses predefined analytic density distribution functions to populate the particles and searches for the stable solution of the collisionless Boltzmann Equation (CBE). For a stable solution, it randomly changes velocities of the particles keeping their positions fixed in space. Some components of the Schwarzschild's method and made-to-measure technique are employed in GALIC code.

    \item {\bf GADGET-2:} All the initial realizations were evolved in time using a publicly distributed code GADGET-2 \citep{Springel2001, Springal2005man, Springel2005}. GADGET-2 is a massively parallelized and adaptive TreeSPH code. It uses $N-$body approach to treat collisionless dynamics and SPH technique to describe hydrodynamics. It makes use of hybrid TreePM algorithm to compute gravitational forces during evolution. The time integration in GADGET-2 is performed using leapfrog method with adaptive time-steps.

    \item {\bf GALFIT:} For the photometric analysis of simulated galaxies, we used the latest version of the GALFIT code which is a two-dimensional multi-component decomposition tool \citep{Peng2003man, Peng2011}. It uses pre-defined surface brightness profiles and fits them using the Levenberg–Marquardt algorithm. The  best fit solution is determined using chi-square minimization. It is an efficient and powerful tool to perform multi-component fitting simultaneously.

    \item {\bf IllustrisTNG50:} We have also used the simulated data product from the state-of-art gravo-magnetohydrodynamical simulation, IllustrisTNG50. It is one of the highest resolution runs of IllustrisTNG simulations, a successor of the Illustris project. IllustrisTNG comprises three cosmological volumes, namely TNG50, TNG100, and TNG300 having box sizes approximately 50, 100, and 300 Mpc respectively. All the TNG simulations are performed with the moving-mesh and massively parallel code AREPO \citep{Springel.2010} starting from the redshit z=127 to z=0. Now, all three runs are publicly available in their entirety \citep{Nelson.etal.May2019}.
    
\end{itemize}

	\begin{savequote}[100mm]
``Your hand touching mine. This is how galaxies collide.''
\qauthor{\textbf{$-$ Sanober Khan}}
\end{savequote}

\chapter[Galaxy Flybys: Evolution of the Bulge, Disk, and Spiral Arms]{Galaxy Flybys: Evolution of the Bulge, Disk, and Spiral Arms}
\label{chapter3}

\section{Introduction}
\label{sec:intro} 
There are basically two types of galaxy interactions; mergers and flybys. It is well known that both phenomena play a major role in the hierarchical growth of the large scale structure \citep{Aarseth1979, Frenk1985}, as well as produce morphological changes in galaxies \citep{martin.etal.2018}. In the case of flybys, although the galaxies finally separate without merging, the interaction can have a strong effect on the galaxy disks \citep{arp.1966}. Flybys can be further categorized into major and minor interactions, depending upon the primary (host) to secondary (satellite) galaxy mass ratio. In general, if the ratio is greater than 3, it is classified as a minor interaction otherwise it is called a major interaction \citep{Stewart2009}. At early epochs, the universe was smaller in size and mergers were more frequent, hence they played an important role in the growth of galaxies \citep{Kauffmann1993, VanDenBergh1996, Murali2002, Stewart2008}. However, recent studies have shown that at later times, as the universe expanded, flybys became more common than mergers \citep{Sinha2012A}. Thus, although mergers may appear more important for changing galaxy morphology, the cumulative effect of flyby interactions may also be important since they are more frequent than mergers in the cosmic history of galaxies \citep{an.etal.2019}.

Early simulations clearly show the importance of flybys on the evolution of galaxies \citep{ToomreToomre1972, Barnes1992}. Later studies have found that flybys play a crucial role in producing tidally induced bars \citep{berentzen.etal.2004, lang.etal.2014, lokas.etal.2014}. The associated mass inflow and bar evolution can result in the formation of pseudobulges \citep{weinzirl.etal.2009} as well as kinematically decoupled cores \citep{derijcke.etal.2004}. The tidal interaction of flybys have an even stronger effect on the disks of the major galaxy, inducing the formation of spiral arms \citep{dubinski.etal.1999, renaud.etal.2015, Moreno.etal.2019} that result in strong star formation \citep{duc.etal.2018}. If the gas infall towards the center is large enough, then these effects can ultimately lead to starbursts \citep{mihos.hernquist.1994} and the triggering of active galactic nuclear (AGN) activity in galaxies \citep{combes.etal.2001}. Studies have shown that the tidal interaction can also result in the formation of warps in galaxy disks \citep{Kim2014}, as well as increase the stellar velocity dispersion, resulting in disk thickening \citep{Reshetnikov1997}. Other effects of flybys on stellar disks are the formation of stellar streams and tidal bridges \citep{duc.renaud.2013}. Studies suggest that flybys will also affect galaxy spin, although the effect is dependent on several factors such as orbit direction and the nature of the galaxies \citep{choi.yi.2017}.

In this study we investigate the effect of flybys on the two important stellar components of galaxies, their bulges and their disks. Bulges can be broadly classified into two types based on their morphology and the dynamics of their constituent stars; classical bulges that are approximately spherical in shape, and pseudobulges that appear disky and flatter in shape \citep{Kormendy2006, Drory2007, Athanassoula2008, Fisher2008A, Fisher2008B}. Classical bulges are dynamically hotter systems compared to pseudobulges that have more disky orbits \citep{Kormendy1993, Andredakis1994}. Classical bulges are thought to form from the monolithic collapse of gas clouds or clumps at early epochs, and then grow from the accretion of smaller galaxies \citep{Aguerri2001, Bournaud2005, Brooks2016}.

Pseudobulges can be further divided into two categories: disky pseudobulges and boxy/peanut pseudobulges. Disky pseudobulges are circular in shape but in the vertical direction they are as flat as the disks of their host galaxy. Hence, it is nearly impossible to detect them in edge-on galaxies. They are thought to form within the inner disks via star formation \citep{laurikainen.etal.2009}. Boxy/peanut bulges are more extended in the vertical direction and can hence be detected in edge-on galaxies. They are usually associated with bars in disk galaxies \citep{Friedli1990, Debattista2006, Gadotti2011}. Boxy/peanut pseudobulges, unlike classical bulges, are formed due to the secular evolution of galaxy disks, the buckling instability of bars \citep{Combes1981}, or via mergers with gas rich galaxies \citep{Keselman2012}. Using observations of galaxies in our local universe at distances less than 11 Mpc, \citet{Fisher2011} found that 80$\%$ of the galaxies with a stellar mass more than $10^{9} M_{\odot}$ contain either a pseudobulge or no bulge. This domination of pseudobulges in the local universe challenges the hierarchical models of galaxy formation, according to which the majority of galaxies should have classical bulges \citep{Frenk1985}.

In the literature, there have been several observational and numerical studies on the formation and evolution of bulges \citep{Kormendy2004, Athanassoula2005, Gadotti2009, Laurikainen2016}. It is clear that both secular processes and violent ones such as galaxy mergers are important for galaxy evolution \citep{tonini.etal.2016}. But most of the literature is biased towards only one kind of galaxy interaction -- galaxy mergers. The other kind of galaxy interaction, flybys, are largely ignored. This could be due to two reasons. The first is that the effect of flybys is lower than that of mergers. However, recent cosmological simulations have shown that galaxy flybys are as frequent as mergers in the low redshift universe \citep{Yee-Ellingson1995, Sinha2012A} and they may play an equally prominent role as mergers in the evolution of galaxies \citep{dimatteo.etal.2007, Cheng-Li2008, Kim2014, Sinha2015B}. The second reason is that bulges lie deep within the potential of a galaxy and the more massive ones make disks extremely stable against bar instability \citep{kataria.das.2018,K&D2020}. The spherical non-rotating classical bulge cannot be torqued effectively in flyby interactions unless there is some instability e.g. bar \citep{Kanak.etal.2012, Kanak.etal.2016}. Bars can be torqued in flyby interactions \citep{lokas.etal.2014} and so  boxy/peanut pseudobulges can also be torqued because they are essentially a part of the bar. The tidal torque of the satellite galaxies may thus change the kinematics of bars and their associated pseudobulges.

The tidal forces on galaxy disks due to flybys, are known to result in extended spiral arms. They can be similar to grand design spirals such as M51 \citep{dobbs.etal.2010}, or result in extremely large spiral arms such as in NGC 6872 \citep{horellou.koribalski.2007}. The interaction can lead to the formation of tidal tails, bridges and warped disks, but the overall morphology depends on many factors such as the pericenter distances, the galaxy orbits, and the relative masses of the galaxies \citep{oh.etal.2008}. Flybys may also affect the spin of galaxy disks, especially when the integrated effect of several interactions is included \citep{lee.etal.2018}. However, although the effect of tidal forces on galaxy disks has been explored in several earlier numerical studies \citep{walker.etal.1996}, it has not been quantified in terms of disk perturbation parameters, such as the Fourier components A$_2$/A$_0$, which is useful for measuring the effect of the tidal interaction, and is widely adopted in studies of bars \citep{kataria.das.2019}.

Thus, although there are several numerical studies of the tidal effect of flybys on galaxy disks, not much attention has been paid to its effect on the bulges or on quantifying the effects. For example, examining whether there are any changes in bulge S\'ersic index or the strength of the spiral arms. In this chapter we try to measure the changes in bulge and disk properties for different pericenter distances. We also include both classical and boxy pseudobulges in our models, disky pseudobulges will be studied later as their evolution possibly involves star formation. Although galaxy interactions do have a large parameter space, for simplicity in this study we focus on one of the parameters that causes the largest change in galaxy morphology -- the pericenter distance. 

In this study we have used a simple N-body approach; the effect of gas will be included in a future study. In the following sections we first discuss our simulation method which includes creating galaxy disks with classical and pseudobulges, describe how we have analysed the bulges and spiral arms, present our results and then discuss the implications of our work for observational studies.

\section{Simulations And Analysis}
In our simulations, we have fixed the mass ratio of the interacting galaxies at  5:1 and 10:1, as these ratio are similar to a wide range of minor flyby interactions in realistic scenarios. We have also restricted the disk angular momentum vectors to be perpendicular to the orbital plane of the interacting galaxies. The choice of parallel directions of the disk angular momenta of the major and minor galaxies was done to ensure maximum resonance (quasi-resonance) between the angular velocities of stars in the major galaxy and the peak angular velocity of the minor galaxy \citep{D'Onghia.etal.2010}.

\subsection{Galaxy Model: Halo and Disk}
\label{sec:model_holo_disk} 
We generated our model disk galaxies using the publicly available code GalIC \citep{Yurin2014}. Each galaxy has a live dark matter halo, and a stellar disk. In our models, the dark matter halo has a spherically symmetric Hernquist density distribution \citep{Hernquist1990},
\begin{equation}
    \rho_{dm}(r)=\frac{M_{dm}}{2\pi}\frac{a}{r(r+a)^3}
    \label{eqn:halo}
\end{equation}
where `$a$' is the scale radius of the dark matter halo and is related to the concentration parameter `$c$' of a corresponding NFW halo \citep{NFW1996} of mass $M_{dm}=M_{200}$ in the following manner,
\begin{equation}
    a=\frac{r_{200}}{c}\sqrt{2\left[\ln{(1+c)-\frac{c}{(1+c)}}\right]}
    \label{eqn:scale_radius}
\end{equation}
where $r_{200}$ is the virial radius of the galaxy (which is defined as the radius within which the mean density is 200 times the critical density of the universe) and $M_{200}$ is mass within the virial radius.

The stellar disk density has an exponential form in the radial direction and a $\sech^{2}$ form in the vertical direction
\begin{equation}
    \rho_{d}(R,z)=\frac{M_{d}}{4\pi z_{0} R_{s}^{2}}\exp\left(-{\frac{R}{R_{s}}}\right) \sech^{2}\left(\frac{z}{z_{0}}\right)
    \label{eqn:disk}
\end{equation}
where $M_{d}$ is the total mass of the disk, $z_{0}$ is the scale height of the disk and $R_{s}$ is the scale radius of the disk.

\subsection{Galaxy Model: Bulge}
\label{sec:model_bulge}
To understand the effect of flyby events on the bulges and disks of galaxies, we have generated two types of bulges in the disk of the larger galaxy, which we call the major galaxy. The first is the classical bulge. Its density is derived from a spherically symmetric Hernquist potential \citep{Hernquist1990},
\begin{equation}
    \rho_{b}(r)=\frac{M_{b}}{2\pi}\frac{b}{r(r+b)^{3}}
    \label{eqn:bulge}
\end{equation}
where $M_{b}$ is the total mass of the bulge and `$b$' is the scale radius of the bulge. 

To obtain a pseudobulge in the disk of the major galaxy, we generated it naturally through the bar formation in a bulgeless galaxy. At a fixed disk scale height to scale radius ratio, we varied the angular momentum fraction of the disk to halo so that it can form a bar which can spontaneously buckle to form a pseudobulge (boxy/peanut bulge) after evolution. We evolved this model upto 10~Gyrs and calculated the bar strength using the ratio of m=2 to m=0 Fourier modes. The amplitudes of the $m$th Fourier mode at a radius $R$ is given by,
\begin{align}
    a_{m}(R) &= \sum_{i=1}^{N} m_{i}(R) \cos(m\phi_{i}), m=0,1,2,3, ... \\
    b_{m}(R) &= \sum_{i=1}^{N} m_{i}(R) \sin(m\phi_{i}), m=0,1,2,3, ... \\
    \intertext{where $m_{i}(R)$ and $\phi_{i}$ are the mass and the azimuthal angle of $i$th particle at radius $R$ respectively, and $N$ is the total number of particles at the same radial position. The strength of a bar is defined as}
    \frac{A_{2}}{A_{0}} &= max \Big[\frac{\sqrt{[a_{2}(R)]^2+[b_{2}(R)]^2}}{\sum_{i=1}^{N} m_{i}(R)} \Big]
    \label{eqn:bar_strenght}
\end{align}
\begin{figure*}
	\centering
	\includegraphics[width=0.7\textwidth]{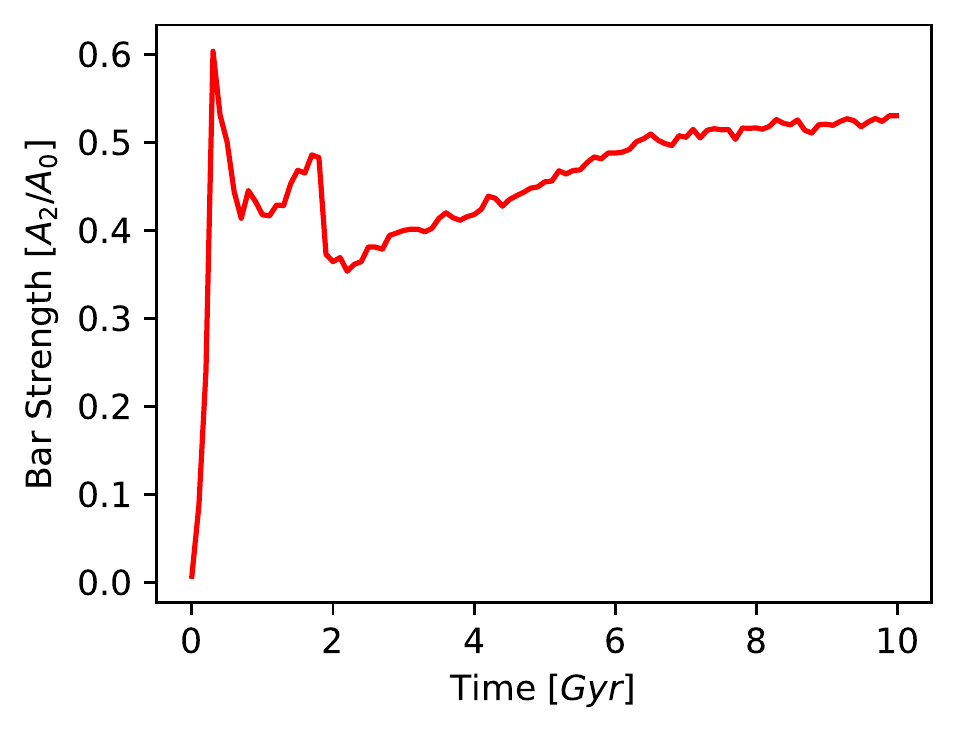}
    \caption{The evolution of bar strength in the pseudobulge model until 10~Gyrs. Due to the buckling of the bar (the 2 peaks in the  $\frac{A_{2}}{A_{0}}$ value), the bar strength changes in the initial stage of evolution. After 6~Gyrs, it shows a nearly constant strength. This is where we take our initial model for a galaxy with a pseudobulge.}
    \label{fig:bar_strength}
\end{figure*}
Fig.~\ref{fig:bar_strength} shows the evolution of bar strength for the major galaxy with a pseudobulge ($PG_{PB}$ model) upto 10~Gyrs. The fluctuations in the bar strength at early epochs of the evolution is the indicator of bar buckling. The bar strength remains approximately constant after 6~Gyrs of evolution. This is where we take the initial conditions for our pseudobulge model i.e. we took the galaxy at 6~Gyr as the initial model for simulations of galaxy flybys with pseudobulges. The formation of the pseudobulge is graphically represented in fig.~\ref{fig:pb_formation} as a  time sequence of the galaxy evolution, using the edge-on view of the disk and bar.
\begin{figure*}
	\centering
	\includegraphics[width=0.7\textwidth]{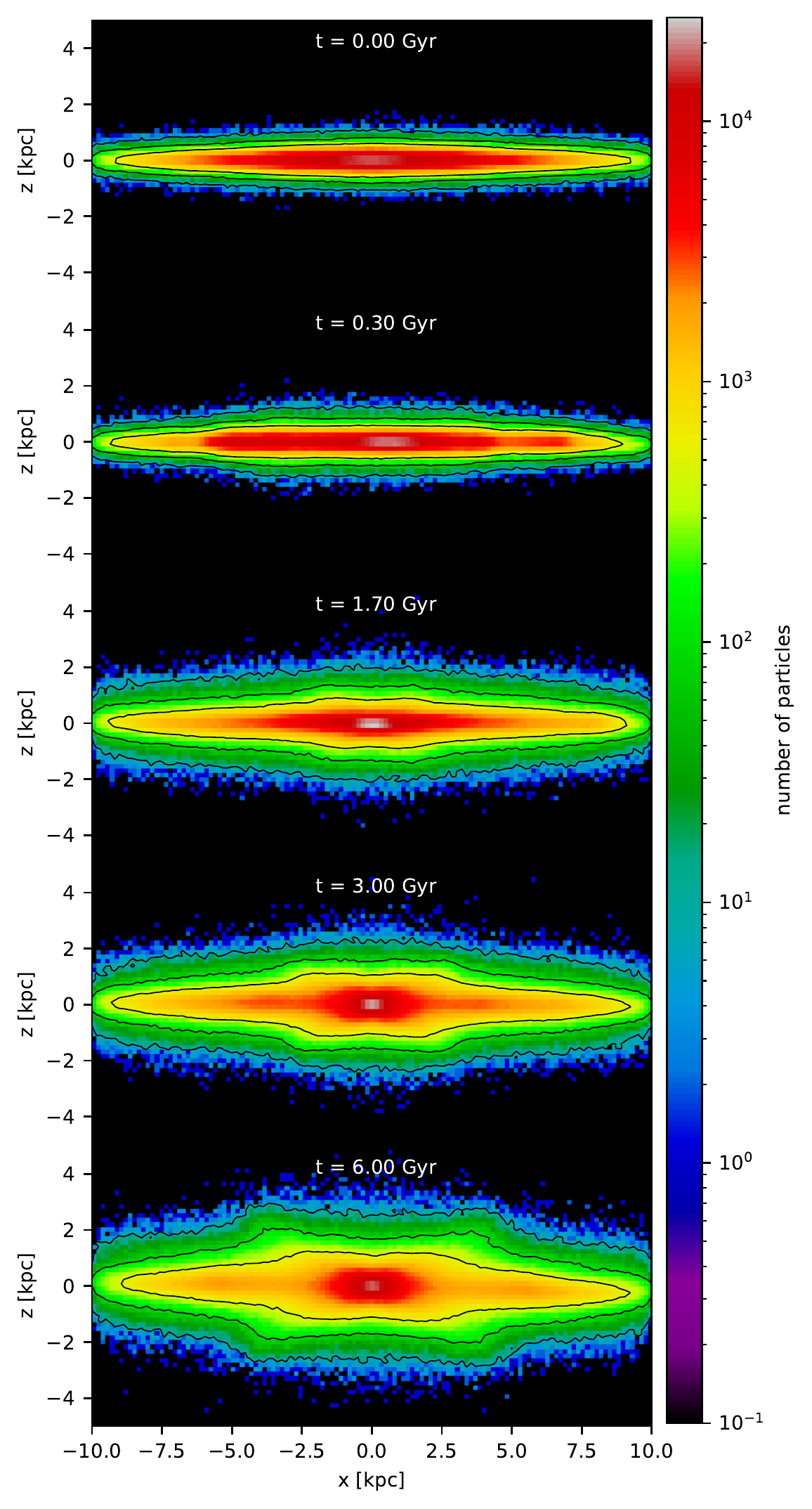}
    \caption{ The visual illustration of the pseudobulge formation. The top to bottom panels shows the time sequence of the edge-on view of the major galaxy, perpendicular to the bar, together with similar isodensity contours. From top to bottom, the five panels show the snapshots at t=0.0, 0.3 (1st peak in the bar strength), 1.7 (2nd peak in the bar strength), 3.0 (after the buckling), and 6.0 Gyr (beginning of the constant bar strength) respectively. The bottom panel displays the initial conditions for the flyby models of galaxies with a pseudobulge .}
    \label{fig:pb_formation}
\end{figure*}

\subsection{Choice of Model Parameters for Simulations}
\label{sec:sim_param}

\begin{table*}
\centering
\caption{Initial parameters of the primary (major) and secondary (minor) galaxies. In this table, $PG_{CB}$ = Primary galaxy with classical bulge, $PG_{PB}$ = Primary galaxy with pseudobulge, $SG_{5}$ = Secondary galaxy with 1/5 mass of primary galaxy, and $SG_{10}$ = Secondary galaxy with 1/10 mass of primary galaxy. (units $M_{\odot}$ = solar mass, $kpc$ = kiloparsec).}
\label{tab:initial_parameters}
\begin{threeparttable}
\begin{tabular}{lcccr}
\hline
Parameters & $PG_{CB}$ & $PG_{PB}$ & $SG_{5}$ & $SG_{10}$ \\
\hline
Total mass ($M$) & 1.2 $\times$ $10^{12} M_{\odot}$ & 1.2 $\times$ $10^{12} M_{\odot}$ & 2.4 $\times$ $10^{11} M_{\odot}$ & 1.2 $\times$ $10^{11} M_{\odot}$ \\
\hline
Halo spin parameter($\lambda$) & 0.035 & 0.035 & 0.035 & 0.035 \\
\hline
Halo concentration ($c$) & 10 & 10 & 11 & 11 \\
\hline
Disk mass fraction & 0.025 & 0.03 & 0.01 & 0.01 \\
\hline
Bulge mass fraction & 0.005 & 0.0 \tnote{a} & 0.002 & 0.002 \\
\hline
Disk spin fraction & 0.03 & 0.022 & 0.01 & 0.01 \\
\hline
Disk scale radius ($R_{s}$) & 3.80~kpc & 2.30~kpc & 1.95~kpc & 1.55~kpc \\
\hline
Disk scale height ($z_{0}$) & 0.38~kpc & 0.23~kpc & 0.195~kpc & 0.155~kpc \\
\hline
Halo particles($N_{Halo}$) & 2.5$\times 10^{6}$ & 2.5$\times 10^{6}$ & 5.0$\times 10^{5}$ & 2.5$\times 10^{5}$ \\
\hline
Disk particles($N_{Disk}$) & 1.5$\times 10^{6}$ & 2.5$\times 10^{6}$ & 3.0$\times 10^{5}$ & 1.5$\times 10^{5}$ \\
\hline
Bulge particles($N_{Bulge}$) & 1.0$\times 10^{6}$ & $0$ \tnote{a} & 2.0$\times 10^{5}$ & 1.0$\times 10^{5}$ \\
\hline
Total particles($N_{Total}$) & 5.0$\times 10^{6}$ & 5.0$\times 10^{6}$ & 1.0$\times 10^{6}$ & 5.0$\times 10^{5}$ \\
\hline
\end{tabular}
\begin{tablenotes}\footnotesize
\item [a] Initially this model does not have any bulge particles. A pseudobulge grows from the disk itself after a bar forms and buckles.
\end{tablenotes}
\end{threeparttable}
\end{table*}

Table~\ref{tab:initial_parameters} summarises the initial parameters of the primary (major) and secondary (minor) galaxies. The mass of the major galaxy is similar to that of the Milky Way galaxy in all the models, as suggested by many studies (see \cite{Wang.Haan2020} and reference therein). The halo spin parameter ($\lambda$) of the galaxies is chosen to be 0.035 which corresponds to the peak of the halo spin distribution given in \cite{Bullock2001}. The values of concentration parameter ($c$) lie well within the mass-concentration relation given by \cite{Wang.Bose2020}. The stellar mass fraction of the galaxies are chosen from the stellar to halo mass relation (SHMR) \citep{Behroozi2013, Moster2013}. The choice of disk scale length for the major galaxy is made such that the model $PG_{CB}$ does not show any bar feature after evolution but the model $PG_{PB}$ develops a bar which then buckles to form a pseudobulge.

\begin{figure*}
	\centering
	\includegraphics[width=0.9\textwidth]{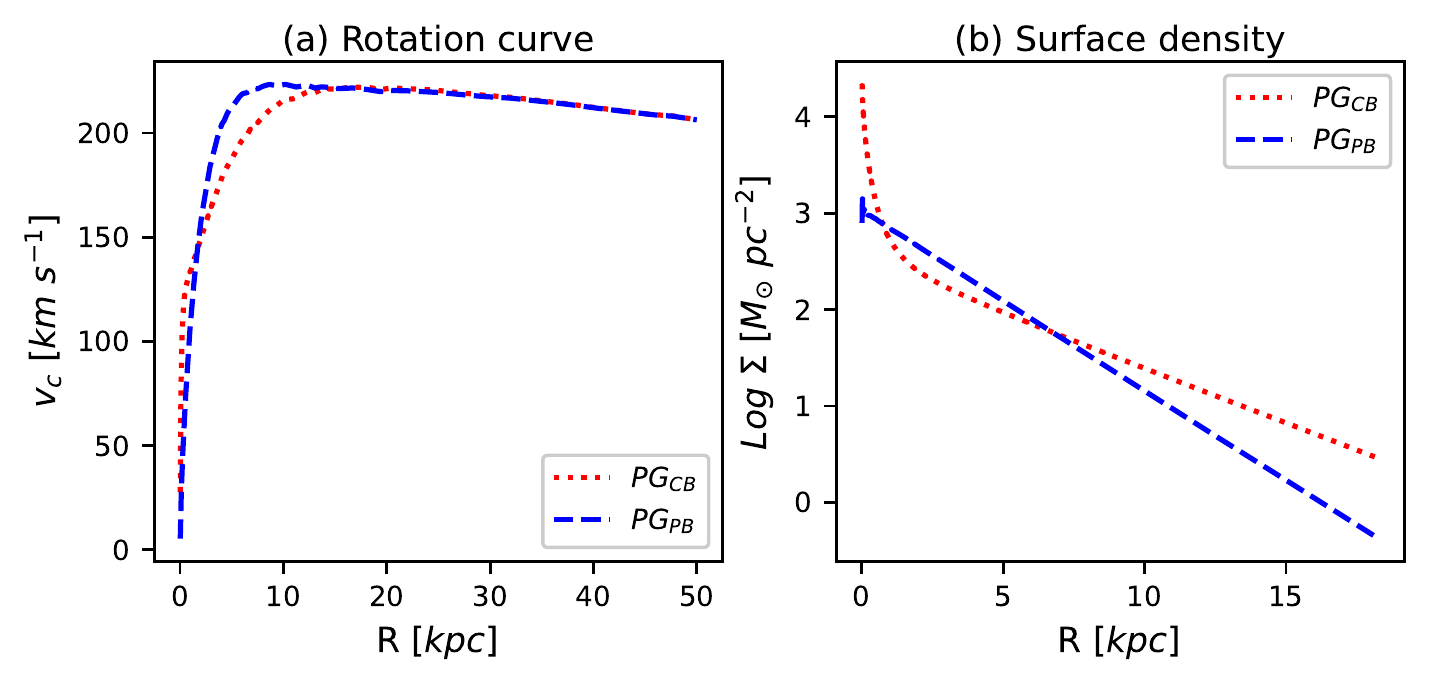}
    \caption{The left panel shows the total rotation curves of initial models of the major galaxy. The panel on the right shows the corresponding stellar surface densities of the galaxy. The red dotted curve and the blue dashed curve represent the $PG_{CB}$ and $PG_{PB}$ models respectively.}
    \label{fig:rot_curve_surf_den}
\end{figure*}

To generate a bar stable major galaxy (hereafter referred to as an unbarred galaxy) and bar unstable major galaxy (hereafter referred to as a pseudobulge galaxy), we tuned the angular momentum fraction of the disk to halo, at a fixed disk scale height to scale radius ratio. This tuning of angular momentum decides the size of the disk (see \cite{Mo.Mao.White.1998} for the galaxy models). The unbarred galaxy models did not show any signatures of bar formation throughout the simulation time of 5~Gyrs but the barred galaxy forms the bar and buckles to form a pseudobulge. In fig.~\ref{fig:rot_curve_surf_den}, we have shown the rotation curves (left panel) and surface densities (right panel) of the major galaxy models.

The choice of the number of particles in the major galaxy is based on the region of interest within the galaxies. Since we are interested in the bulge and the disk of the major galaxy, we selected the number of particles in the major galaxy in such a way that the two body relaxation time is greater than the time of interest (i.e. the simulation time) in the central region of the galaxy \citep{Power2003}. The relaxation time is 5~Gyr for the models $PG_{CB}$ and $PG_{PB}$ at the radii 0.14~kpc and 0.27~kpc respectively. To confirm that our results do not depend on mass resolution of the simulations, we have also performed four high mass resolution simulation runs (one isolated and one $r_{p}=40~kpc$ for both $PG_{CB}$ and $PG_{CB}$ models) with twice as many particles as listed in the Table~1.

\begin{figure*}
	\centering
	\includegraphics[width=0.9\textwidth]{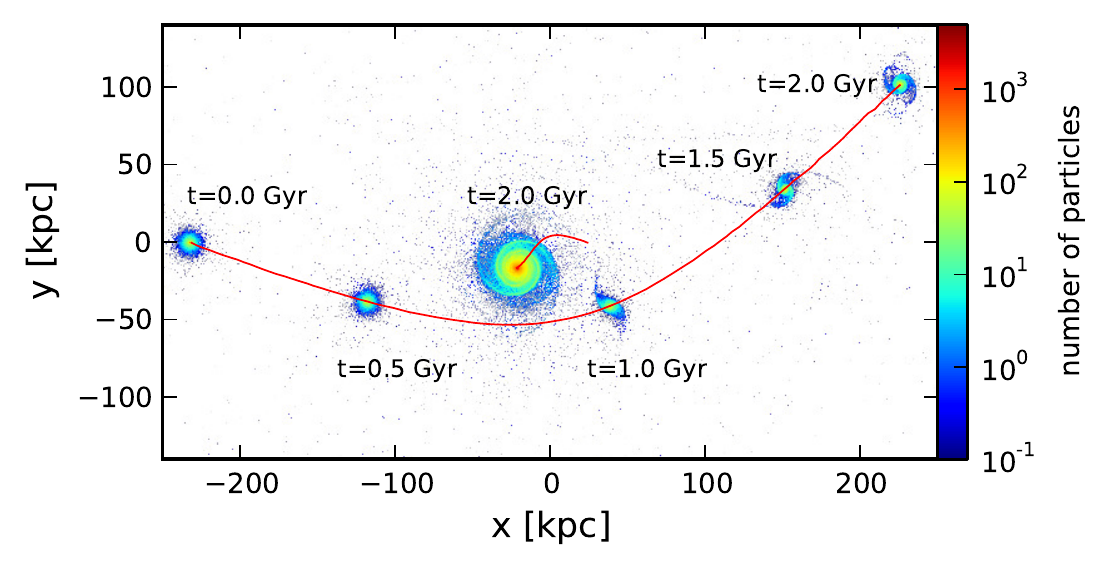}
    \caption{The orbits of two disk galaxies of mass ratio 10:1 undergoing a flyby interaction with pericenter distance of 40~kpc. The red solid curves trace the orbits of the galaxies during flyby. The minor galaxy is shown in time steps of 0.5~Gyr and the major galaxy is shown at the time 2~Gyr. The mass weighted particle number density is represented by the color on the log scale. In this illustration, only the stellar particles are shown.}
    \label{fig:orbits_of_galaxies}
\end{figure*}

After generating the major and minor galaxies, we put them on hyperbolic orbits of eccentricity $e=1.1$ as illustrated in fig.~\ref{fig:orbits_of_galaxies} so that the minor galaxy does not decay by dynamical friction of the major galaxy. In this figure, the minor galaxy is shown at time intervals of 0.5~Gyr and the major galaxy is shown at time t=2.0~Gyr. The solid red curves traces the actual paths of the center of mass of the two galaxies in one of our simulations. The initial separation of the galaxies was chosen to be 255~kpc which is the sum of the virial radii ($r_{200}$) of the major and minor galaxies. We evolved the galaxies for 4~Gyr using the publicly available massively parallel code Gadget-2 \citep{Springel2001, Springal2005man, Springel2005}. The evolution time is set to 4.0~Gyr because galaxies are well separated after evolving for this much amount of time. In all of our simulations, the gravitational softening is 0.02~kpc for the stellar particles and 0.03~kpc for the halo particles. The change in total angular momentum is within 0.15$\%$ for all the simulations in 4~Gyrs of evolution, which ensures an output with minimum numerical errors. 

To study the effect of flyby interactions on the bulge and the disk of the major galaxy, we varied the distance of closest approach i.e. the pericenter distance of the galaxies ($r_{p}$) from 40~kpc to 80~kpc, keeping all the other parameters fixed. Hence, we have two sets of simulation models for each pericenter distance ($r_{p}$), which are : (1) a major galaxy with a classical bulge and (2) a major galaxy with a pseudobulge bulge. Each model is further simulated for 10:1 and 5:1 galaxy mass ratios. As a control model, we have also evolved the major galaxy in isolation for the same time period of 4~Gyr. The isolated model provides the benefit of removing the effect of secular evolution and the effect of discreteness, if any.


For naming the models, we have used pericenter distances determined assuming 2-body systems. For e.g. $r_{p}=40$ denotes the model with pericenter distance = 40~kpc as determined from a two body system. However, the real N-body systems always deviates from these pre-determined pericenter distances. Note that all the quantities are in units of the Hubble parameter '$h$' (where the Hubble constant is $H_{0}= 100$ $h$ $km$ $s^{-1} Mpc^{-1}$ and $h=0.67$ from \cite{Planck.Collaboration2020}) and all the plots are for the major galaxy unless otherwise specified.

\subsection{Analysis}
\label{sec:analysis}

We used the latest released version of GALFIT, which is a widely used two dimensional galaxy image decomposition tool, for the bulge-disk decomposition of the simulated galaxies \citep{Peng2003man, Peng2011}. Since we are interested in the effect of the flyby on the major galaxy or host galaxy only, we did the two dimensional image decomposition of the major galaxy and not the minor galaxy. Also, since GALFIT requires the galaxy images in `fits' format as an input, we first made `fits' format images of the galaxies using simulation snapshots at time steps of 0.2~Gyr. For the galaxy decomposition, we did not consider any background sky and hence the point spread function (psf) is a delta function. We used a simple exponential profile for the two dimensional fitting of a face-on disk,
\begin{equation}
    \Sigma_{d}(R)=\Sigma_{d0}\exp\left(-{\frac{R}{R_{s}}}\right)
    \label{eqn:galfit_disk_face}
\end{equation}
where $\Sigma_{d0}$ is the central surface density of the disk and $R_{s}$ is the disk scale radius. For the two dimensional fitting of an edge-on disk, we used following profile
\begin{equation}
    \Sigma_{d}(R,z)=\Sigma_{d0}\left(\frac{R}{R_{s}}\right)K_{1}\left(\frac{R}{R_{s}}\right)\sech^{2}\left(\frac{z}{z_{0}}\right)
    \label{eqn:galfit_disk_edge}
\end{equation}
where $K_{1}$ is a Bessel function and $z_{0}$ is the scale height of the disk. We used a S\'ersic profile for the two dimensional fitting of the bulge
\begin{equation}
    \Sigma_{b}(R)=\Sigma_{b0}\exp\left[-b_{n}\left\{\left(\frac{R}{R_{e}}\right)^{\frac{1}{n}}-1\right\}\right]
    \label{eqn:galfit_sersic}
\end{equation}
where $R_{e}$ is effective radius of the bulge, $\Sigma_{b0}$ is the surface density of the bulge at $R_{e}$, and $b_{n}$ is a function of the S\'ersic index $n$. Previous studies of galaxy bulges suggest that the S\'ersic index $n=2$ is a good proxy for distinguishing between classical bulges and pseudobulges \citep{Fisher2006, Fisher2008B}. Generally, classical bulges have a S\'ersic index $n>2$ while pseudobulges have a S\'ersic index $n<2$. We have applied this classification scheme for bulges in our study.

\begin{figure*}
	\centering
	\includegraphics[width=0.85\textwidth]{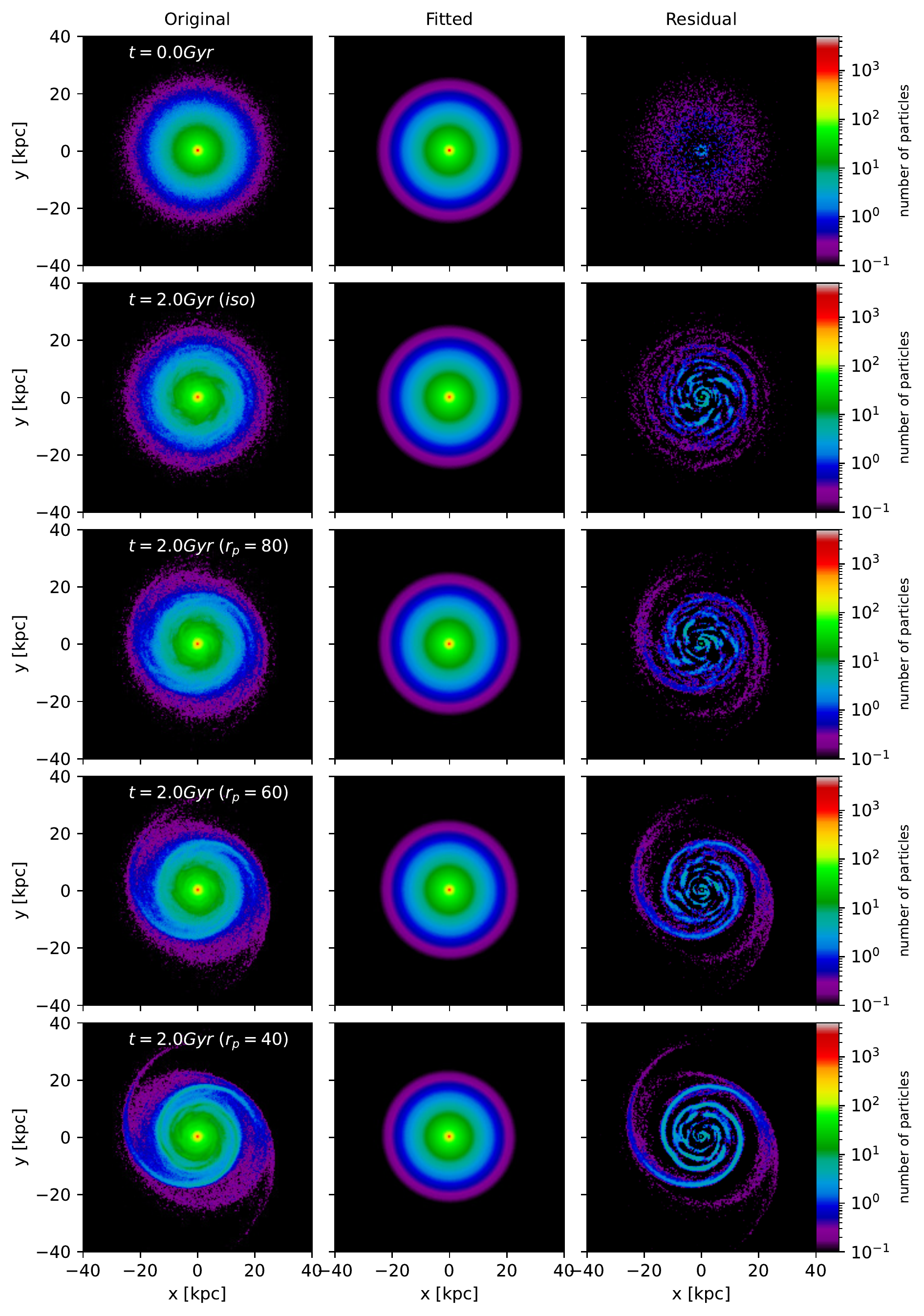}
    \caption{The two dimensional decomposition of the major galaxy with a classical bulge. The left, middle and, right columns show the original images, fitted images, and residual images respectively. The mass weighted particle number density is represented by the color on the log scale. For the sake of visualization, only positive pixels are shown in the residual images. From top to bottom, the five rows show the model galaxy at t=0.0~Gyr, $isolated$ model at t=2.0~Gyr, $r_{p}=80$ model at t=2.0~Gyr, $r_{p}=60$ model at t=2.0~Gyr, and $r_{p}=40$ model at t=2.0~Gyr respectively}
    \label{fig:2d_decomposition}
\end{figure*}

For improving the goodness of fit, we did several trials by setting different box sizes, bin sizes (pixel size), and convolution box sizes. We chose all these parameters in such a way that there is minimum effect of over-fitting and under-fitting. Therefore, we centered the major galaxy in a box of 80~kpc $\times$ 80~kpc size with a pixel size of 0.08~kpc $\times$ 0.08~kpc. For each fitting, the initial guess parameters of the galaxy components were chosen visually. See fig.~\ref{fig:2d_decomposition} for a visual illustration of the fitting procedure in face-on galaxies. The original, fitted and residual distributions of the simulated galaxies with classical bulges are shown in the first, second and third column respectively. Each row of the figure represents the two dimensional bulge-disk decomposition of a model galaxy at a given time as written in the legend of the first column. The figure suggests that GALFIT fits our simulated galaxies quite well with minimum residuals.

\begin{figure*}
	\centering
	\includegraphics[width=0.6\textwidth]{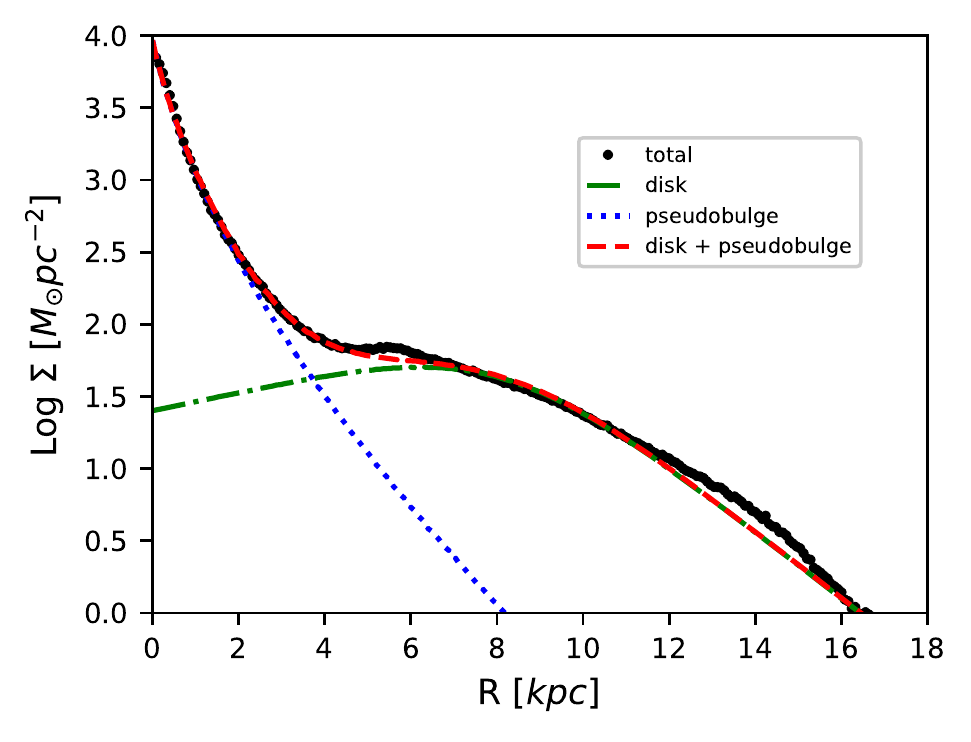}
    \caption{One dimensional density profiles of pseudobulge galaxy model and fit obtained from its two dimensional decomposition. The black circle-solid curve, green dash-dotted curve, blue dotted curve, and red dashed curve demonstrate the total, disk, pseudobulge, and disk + pseudobulge profiles respectively. The disk profile is smoothly truncated in the inner region because most of the matter is swept by the bar. Our model pseudobulge fits well with $n=1.4$ S\'ersic index.}
    \label{fig:decomp_profile}
\end{figure*}


In our simulations, a pseudobulge is formed from the buckling of the bar. So the bar and pseudobulge are essentially the same component of the galaxy. Fig.~\ref{fig:decomp_profile} shows the 1-d density profiles of the pseudobulge galaxy model and the fits obtained from its 2d decomposition using GALFIT. It also includes the individual profiles of the disk and pseudobulge components. The disk profile is smoothly truncated in the inner region of the galaxy because most of the central mass is swept up by the bar. Our model pseudobulge fits well with $n=1.4$ S\'ersic index which verifies its flat nature. However, the morphological decomposition of a galaxy reveals only the projected distribution of mass or light in the different components. This is fine for a spherical bulge but a flat bulge requires more detailed decomposition. Therefore, we explored kinematic methods to decompose the pseudobulges \citep{abadi.etal.2003}.

\begin{figure*}
	\centering
	\includegraphics[width=0.7\textwidth]{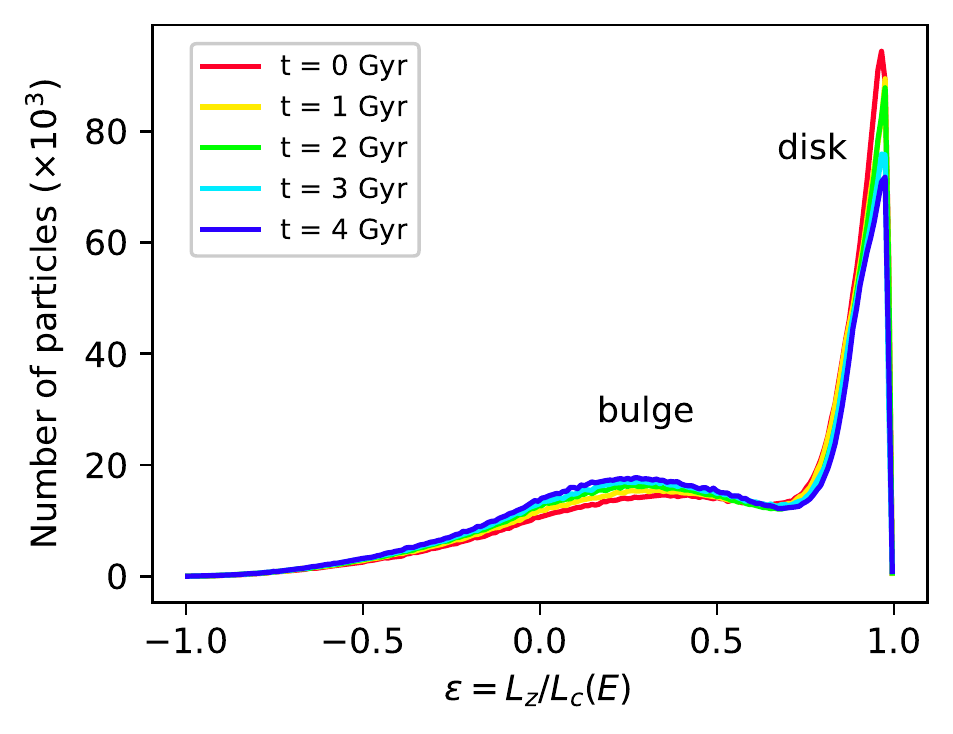}
    \caption{ An example of the kinematic decomposition of the bulge (hot) and disk (cold) components of the galaxy with a pseudobulge (model $PG_{PB}$). The X-axis denotes the ratio of stars' angular momentum perpendicular to the disk plane ($L_{z}$) and the maximum circular angular momentum ($L_{c}(E)$) corresponding to the stars' energy E. The peak around $\epsilon=1.0$ shows the rotation dominated component and the peak away from $\epsilon=1.0$ shows the dispersion dominated component. The different colors show the bulge-disk decomposition of the $isolated$ model with a pseudobulge ($PG_{PB}$) at time intervals of 1.0~Gyr.}
    \label{fig:kinematic_decomp}
\end{figure*}

In the kinematic method, we used the angular momentum of the particles perpendicular to the galaxy plane to separate the `cold' disk particles and `hot' bulge particles that had larger velocity dispersion. Let $L_{z}$ be the angular momentum of a particle perpendicular to the galaxy plane, let $E$ be it's total energy, and $L_{c}(E)$ be the maximum angular momentum for the particle with energy $E$ in the galaxy potential. Now the ratio $|L_{z}/L_{c}(E)|$ will be close to one for cold/disk particles and close to zero for hot/bulge/halo particles (see the fig.~\ref{fig:kinematic_decomp}). For the calculation of $L_{c}(E)$, we divided all the particles in $\sim$200 equal size energy bins and found the particle with maximum $L_{z}$ in each bin. These $L_{z}$ are the $L_{c}(E)$ in the corresponding energy bins. After that, the $L_{c}(E)$ of all the particles are calculated using $3rd$ order spline interpolation of these $\sim$200 values.

All of our models show a two armed spiral structure during the flyby interactions, which are radially symmetric to each other (see the fig.~\ref{fig:2d_decomposition}). The spiral arms produced in flybys are very close to the logarithmic shape. So we used the Fourier analysis method as discussed in \cite{Sellwood1984} and \cite{Sellwood1986} to calculate the strength and pitch angle of the spiral arms. The Fourier coefficients are given by the following expression,
\begin{equation}
    A(m,p)=\frac{1}{N}\sum_{j=1}^{N} \exp\left[i\left(m\phi_{j}+p\ln R_{j}\right)\right]
    \label{eqn:spiral_strenght}
\end{equation}
where $m$ is the number of spiral arms ($m=2$ for our models), $p$ is a real number related to the pitch angle ($\alpha$) of the spiral arms, $N$ is the number of stars in a given annular region from radii $R_{min}$ to $R_{max}$, and $(R_{j},\phi_{j})$ are the polar coordinates of the $j^{th}$ star. The range $R_{min}$ to $R_{max}$ is chosen in such a way that the maximum spiral arm strength lies in this range and no extra feature (e.g. bar and tidal arms) falls within this range. Then we calculated $A(m,p)$ for $p\in[-50,+50]$ at a step $dp=0.25$ and determine the parameter $p_{max}$ such that the value of $A(m,p)$ maximizes at $p=p_{max}$ \citep{Puerari.etal.2000}. So then the pitch angle is given by $\alpha=\arctan(m/p_{max})$ \citep{Sang-Hoon.etal.2015, Semczuk.etal.2017} and the spiral arm strength is given by $|A(m,p_{max})|$. For comparison, we also calculated the spiral arm strengths using the method as described for bars in equation~\ref{eqn:bar_strenght}. Both the methods give similar spiral arm strengths.

\section{Results}
\label{sec:results}
We have simulated flyby interactions for galaxies with mass ratio 10:1 and 5:1, and for two types of bulges in the major galaxy: the classical bulge and the pseudobulge. The pericenter distances are set at 40~kpc, 60~kpc, and 80~kpc. As a control case, we have evolved isolated galaxy models for the same length of time. In the following subsections we discuss our results of the effect of galaxy flybys on the disks and bulges. As mentioned earlier, we focus on the properties of the major galaxy only, and not the minor one.

\subsection{Disk Scale Radius}
\label{sec:disk_radius}

\begin{figure*}
	\centering
	\includegraphics[width=\textwidth]{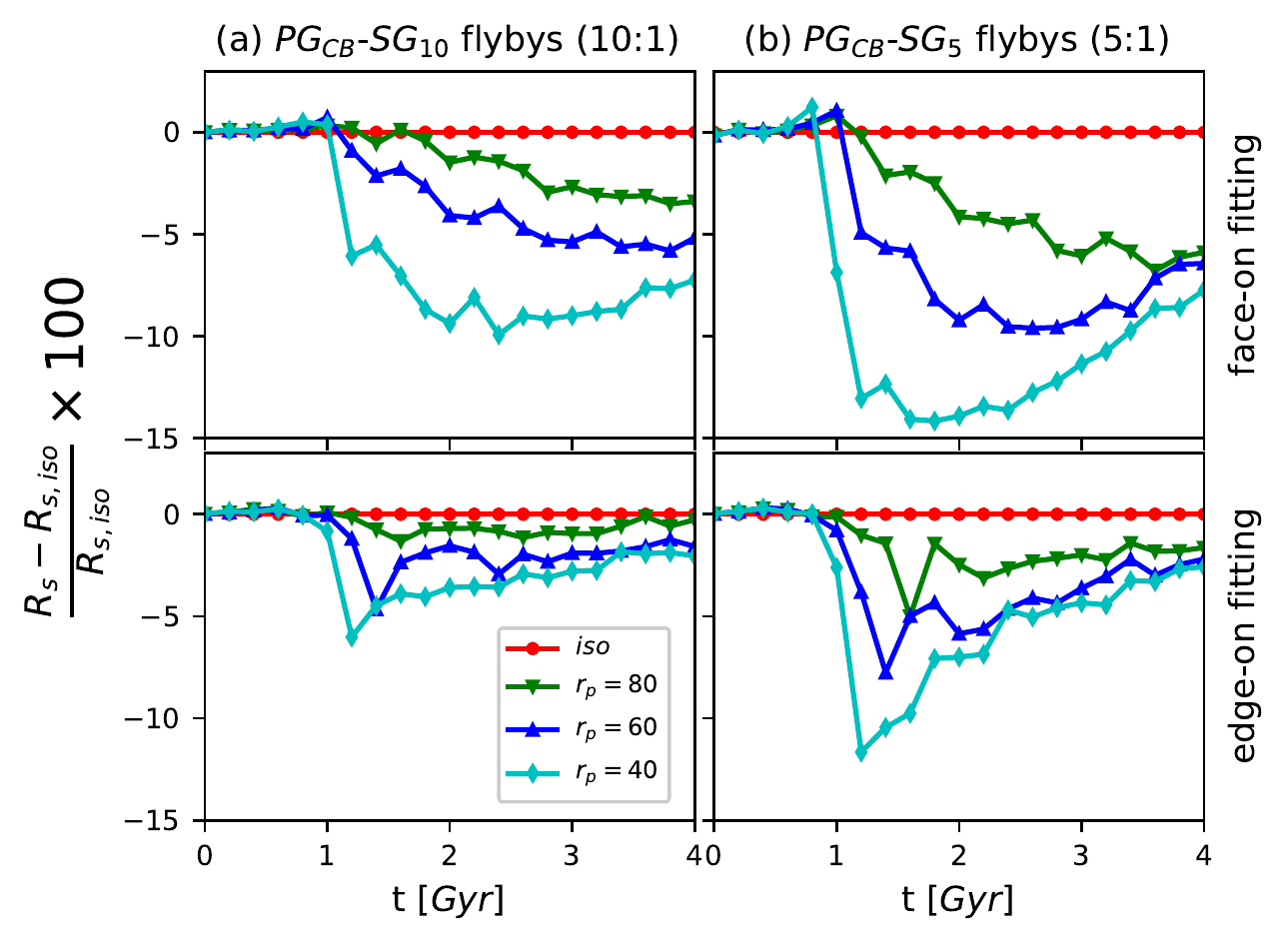}
    \caption{The evolution of the percentage change in the disk scale radius of the unbarred, classical bulge galaxy due to flybys. The left and right columns show the flyby simulations of $1/10$ and $1/5$ galaxy mass ratios respectively. The top and bottom rows represent the outputs from face-on and edge-on galaxy fitting respectively. The red circle-solid curves, the green down triangle-solid curves, the blue upper triangle-solid curves, and the cyan diamond-solid curves represent the models $isolated$, $r_{p}=80~kpc$, $r_{p}=60~kpc$, and $r_{p}=40~kpc$ respectively.}
    \label{fig:disk_radius}
\end{figure*}

During close flybys, galaxies experience strong tidal forces due to each others gravity. This tidal force pulls out the stellar mass from the galaxies and results in the formation of strong spiral arms and tidal streams as shown in the third column of fig~\ref{fig:2d_decomposition}. Since the spiral arms form from the stellar particles of the pristine disk, there will be some resultant changes in the disk surface densities. The question is how does the formation of spiral arms affect the pristine disk? To answer this question, we examined the time evolution of the percentage change in the disk scale radius ($R_{s}$) of the unbarred, classical bulge galaxies in flybys with 10:1 and 5:1 mass ratios, relative to a control isolated model, as shown in column (a) and column (b) of fig~\ref{fig:disk_radius}. The top and bottom rows are the corresponding face-on and edge-on fits of the galaxies. We have performed the edge-on fitting of the galaxies for two perpendicular viewing angles to minimize the bias of the viewing angle on the morphology of the galaxy. The bottom row shows the percentage change in the average disk scale radius for the edge-on fitting. At the beginning of the simulation, the face-on and edge-on fittings give the disk scale radius to be 3.65~kpc and 3.78~kpc respectively. The scale radius from edge-on fitting is nearly equal to the theoretical scale radius (3.8~kpc) but the face-on fitting gives slightly smaller disk scale radius. The cause of this mismatch is the discreteness of the matter distribution in the simulations, which lacks a smooth distribution of particles in the outermost region. Therefore the outermost particles are excluded in the face-on fitting but edge-on fitting does not have this drawback. However, this small mismatch is not a problem for studying the disk evolution in our simulations.

From this figure (fig~\ref{fig:disk_radius}) it is very clear that the scale radius of the unbarred, classical bulge galaxy decreases with time. The magnitude of the decrease in disk scale radius depends on the pericenter distances of the galaxies. The smaller the pericenter distance, the larger is the change in disk scale radius. Note that the isolated models do not show any change in the scale radius because they are control models and all the change are measured with respect to them. The closest flyby models, which experience the largest tidal forces, show the greatest change in the scale radius because they also have the strongest spiral features. Hence the changes in the disk scale radii are correlated with the formation of the spiral arms (which are discussed later in this section). All the models show a significant change in the disk scale radius just after the pericenter passage, at approximately t=1.0~Gyr. Before pericenter passage, all the models show no significant change in the disk scale radius. In both 10:1 and 5:1 flybys, the face-on fittings show a larger change compared to the edge-on fitting. This is due to the presence of spiral arms which remain in the residual in the face-on fits. However, in the edge-on fits a significant fraction remains in the disk. For a given pericenter distance ($r_{p}$), the 5:1 simulations show more changes than the 10:1 simulations. Similarly, for a given major to minor galaxy mass ratio, the closest pericenter passage shows more change than the relatively farther pericenter passage. The effects of close flybys are short lived. As the minor galaxy goes away, the matter pulled out due to the flyby falls back into the major galaxy and starts building the disk again.

\subsection{Disk Scale Height or Thickness}
\label{sec:disk_height}

\begin{figure*}
	\centering
	\includegraphics[width=\textwidth]{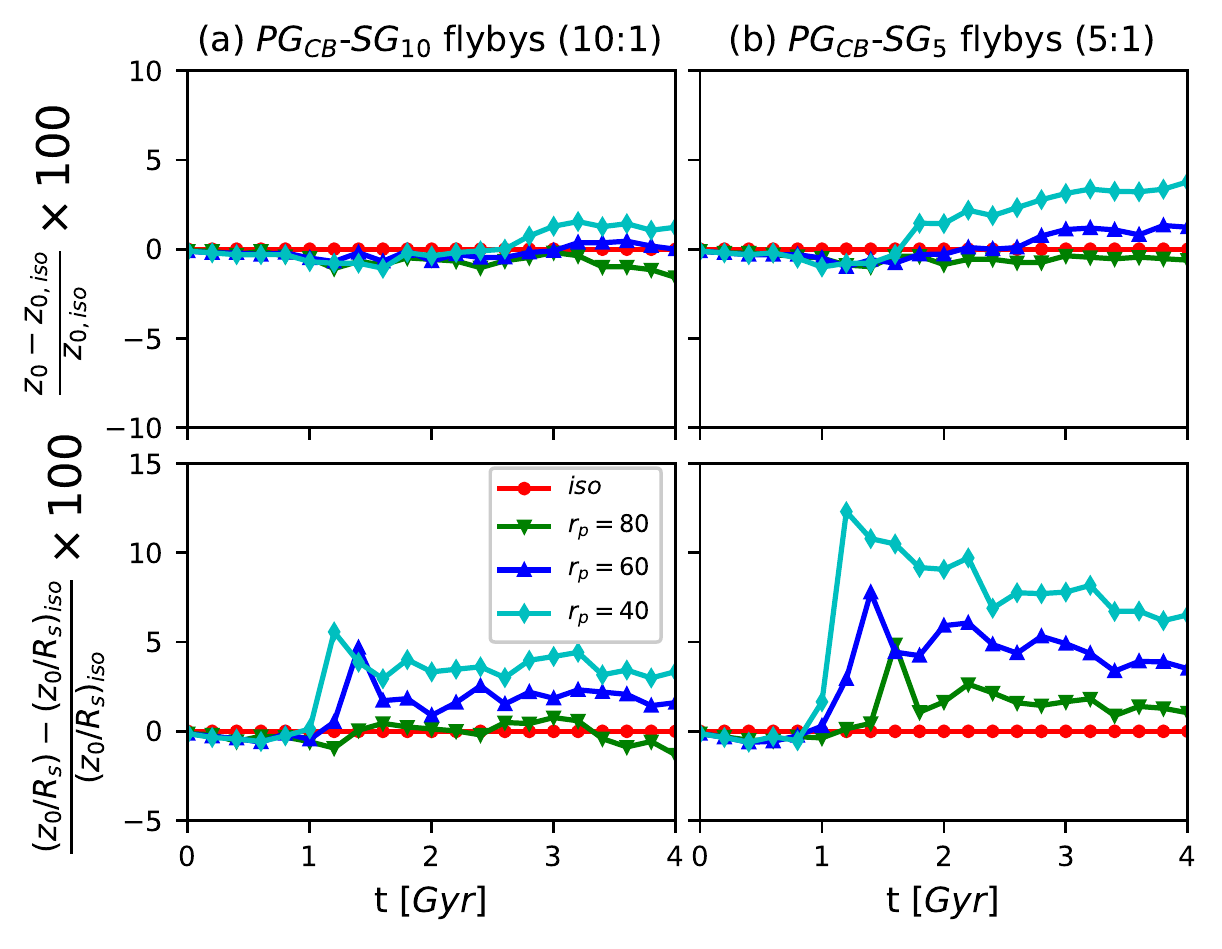}
    \caption{Evolution of the percentage change in the disk height or thickening in unbarred galaxies with bulges due to flybys. Left and right columns show the flyby simulations of galaxies with $1:10$ and $1:5$ mass ratio respectively around the major galaxy with classical bulge. Top and bottom rows represent the change in the disk scale height and the ratio of the disk scale height to the disk scale radius respectively. The red circle solid curves, green down triangle solid curves, blue upper triangle solid curves, and cyan diamond solid curves represent the models $isolated$, $r_{p}=80~kpc$, $r_{p}=60~kpc$, and $r_{p}=40~kpc$ respectively.}
    \label{fig:disk_height}
\end{figure*}

The minor flyby interactions of galaxies heats up the minor galaxy dynamically and can also induce disk instabilities (e.g. bar formation and disk warping) in the minor galaxy which results in the thickening of the disk \citep{lokas.etal.2014, Gajda.etal.2018, Lokas.etal.2018}. Although the minor galaxy exerts a much weaker tidal force on the major galaxy, the perturbation may dynamically heat up the major galaxy disk as well. To see if there is any disk thickening for our simulations of flybys with unbarred, classical bulge galaxies, we have plotted the percentage change in the disk scale height ($z_{0}$) and the ratio of disk scale height to disk scale radius ($\frac{z_{0}}{R_{s}}$) relative to the control isolated model, as a function of time in the top and bottom rows of the fig.~\ref{fig:disk_height} respectively. The left (a) and right (b) columns of the figure represent the galaxies in 10:1 and 5:1 mass ratio flybys respectively. To reduce the bias of the viewing angle on the morphology of the galaxies, we have plotted average disk scale heights and the ratio of the average disk scale height to average scale radius of two fitted models at perpendicular viewing angles. From the first row of the figure, we can see that both the models show very small or insignificant disk thickening due to the dynamical heating by the minor flybys. The model $r_{p}=40~kpc$ for a mass ratio of 5:1 shows a maximum of $\sim 4\%$ increase in the disk scale height due to the flyby. These results indicate that minor flybys cannot heat or thicken the disks of the major galaxy significantly. Only the very close and nearly equal mass flybys (major flyby) can heat the disk of the major galaxy significantly.

However, the disk thickness can also be measured relative to the disk scale radius at the time of measurement as shown in the bottom row of the figure (fig.~\ref{fig:disk_height}). To bring out the effect of flybys on the disk thickening, we have shown the percentage change in the ratio of the disk scale height to the disk scale radius relative to that of control isolated model. At the beginning of the simulation, the disk scale height is $0.1R_{s}$ which is equal to the theoretical value we started with. As we go forward in time, all the models irrespective of the flyby mass ratio follow the same track until nearly t=1.0~Gyr i.e. before close approach. After that, all the models move on different tracks. In the 10:1 models, the disk thickening remains nearly the same after the passage of the minor galaxy, but the 5:1 models show sudden disk thickening just after the flyby, and then thinning due to increasing disk scale length as discussed in subsection~\ref{sec:disk_radius}. From both panels of the bottom row, it is clear that the closest approach produces the largest effect on disk thickening at a given major to minor galaxy mass ratio. Similarly, the major flyby produces the greatest effect on disk thickening at a given pericenter passage.

\subsection{Spiral Arms Strength}
\label{sec:spiral_arms}

\begin{figure*}
	\centering
	\includegraphics[width=\textwidth]{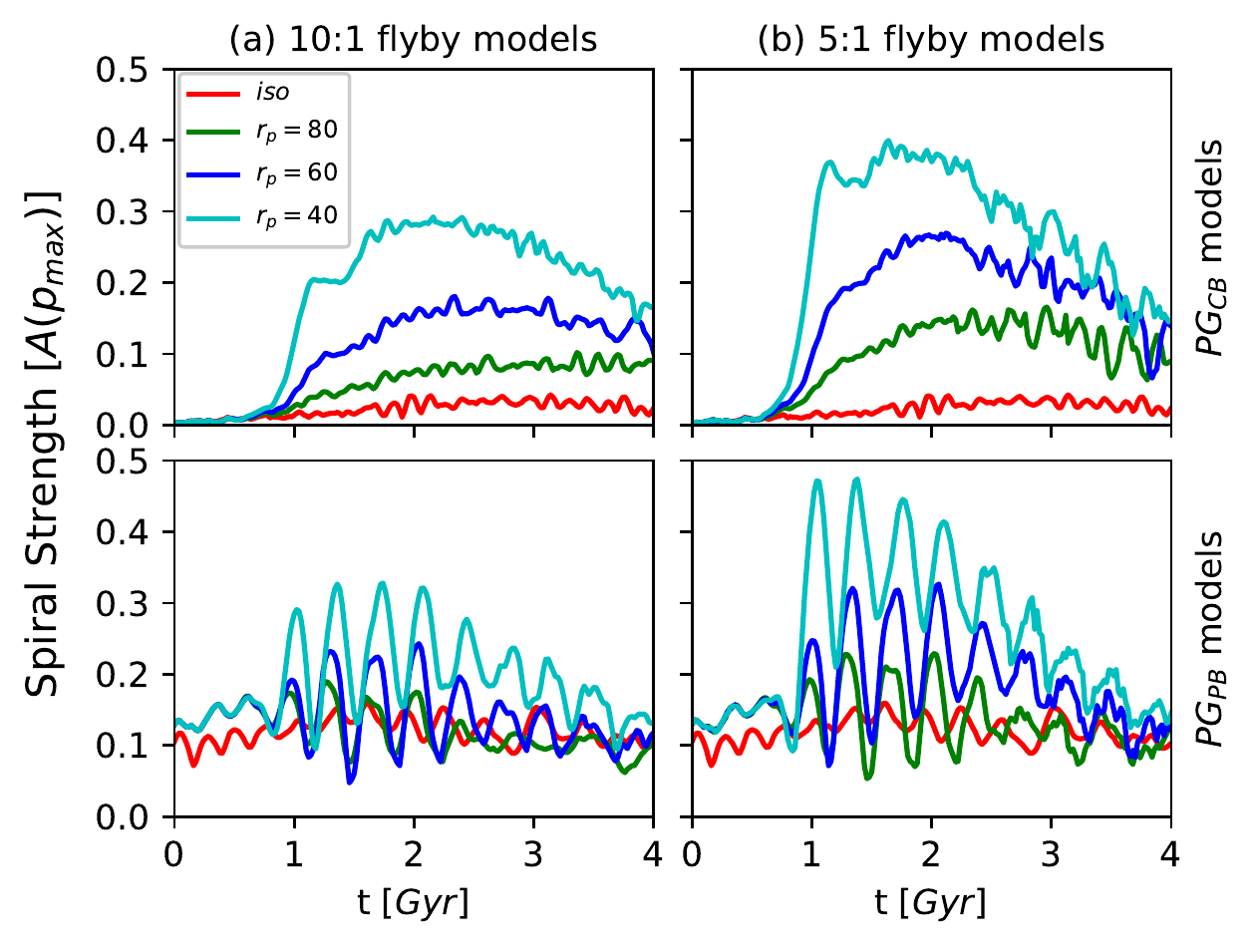}
    \caption{Time evolution of the spiral arms strength. The left and right columns show the flyby simulations of mass ratios 10:1 and 5:1 respectively. The top and bottom rows represent the major galaxy model of classical bulge and pseudobulge respectively. The red, green, blue, and cyan curves represent the models $isolated$, $r_{p}=40~kpc$, $r_{p}=60~kpc$, and $r_{p}=80~kpc$ respectively.}
    \label{fig:spiral_arms}
\end{figure*}

The tidal force of the minor galaxy pulls out stellar particles from the major galaxy. At the same time, the resonance between the stellar orbits and the orbit of the minor galaxy helps the major galaxy to develop spiral arms. In our {simulations, the} galaxies move in a prograde-prograde configuration which is the most favourable orientation for the formation of spiral arms. All of our models develop two spiral arms during the interactions which are radially symmetric to each other (see the fig.~\ref{fig:2d_decomposition}). The strength of spiral arms in a galaxy represents how well the spiral features are distinguishable from the host disk. Hence the amplitude of the m=2 Fourier mode will be a good estimator of the spiral arm strength at any given radius (see equation~\ref{eqn:spiral_strenght}).

In fig~\ref{fig:spiral_arms}, we have plotted the strength of the spiral arms for our models. The left and right columns of the figure show the strength of spiral arms in the flyby models with mass ratios 10:1 and 5:1 respectively. The top and bottom rows represent the spiral arms strength in galaxies with a classical bulge and a pseudobulge respectively. From all the panels of the figure, one can see that the major galaxy in all the flybys shows a sudden increase in the spiral strength at approximately t=1.0~Gyr which is the time of pericenter passage of the galaxies. There is no change or negligible change in the strength of the spiral arms for the control or isolated models throughout the simulation. This shows the importance of flyby interactions in the formation of strong spiral arms.

The time evolution of the spiral arms strength in the classical bulge models shows that the strength increases as the satellite crosses the pericenter, which is at approximately 1.0~Gyr. After some time, the magnitude of the arm strength reaches its maximum and starts decreasing slowly. As expected the value of the peak strength increases with decreasing pericenter distance of the galaxies. Also, at a given pericenter distance, the 5:1 flyby models show higher peak strength than the 10:1 flyby models. The closer flybys attain peak strength earlier than the farther flybys. However, although the closest flyby interactions show the most rapid growth in spiral arms strength, they also show rapid decay in the spiral arms strength. As a result all the flyby models reach the same strength level after some time. This result shows the importance of satellite galaxies or massive clusters as spiral strength boosters. They help the major galaxies to maintain well defined and long lasting spiral arms.

Also, for the close flybys of the unbarred, classical bulge models, the spiral arms can be traced all the way to the center of the galaxy. For example, in the fig.~\ref{fig:2d_decomposition}, the $r_{p}=40~kpc$ model shows a grand design spiral structure after the flyby. The two spiral arms are distinct all the way to the center. It is well known that the strength of global disk instabilities in disks (e.g. bars, spiral arms) depend on the mass and the concentration of the bulge \citep{shen.etal.2004, Athanassoula.Lambert2005, kataria.das.2018}. We see this effect in our simulations as well, since both the strength and the extent of the spiral arms formed in the flybys depends on the distance of closest approach and the presence of the bulge mass.

The evolution of spiral arms strength in the pseudobulge models is even more interesting as can be seen from the bottom row of the fig~\ref{fig:spiral_arms}. Both the 10:1 and 5:1 flyby models exhibit an oscillatory nature on top of the time-varying spiral arms strength, whose nature is similar to those for classical the bulge models. After the pericenter passage, the crest values of these oscillations vary in an approximately similar manner as the spiral arm strength of the classical bulge models. They first increase, reach a peak value and then start decreasing slowly. At the end of the interaction, at 4~Gyr, all the models show similar spiral arm strengths which are equal to the control or isolated model. The isolated models have some initial arms strength $A(p_{max})=0.1$ which is because we have taken the initial pseudobulge models after the evolution of bar strength becomes nearly time independent (see the fig.~\ref{fig:bar_strength}), and by then these models have grown weak spiral arms. All the flyby models show the same oscillation frequency, irrespective of the flyby mass ratio and pericenter distances. This indicates that these oscillations are intrinsic to the host galaxy and the flyby interactions amplify the amplitudes of these oscillations. Barred galaxies have intrinsic resonances \citep{Binney.Tremaine.2008}. The oscillations in the effective potential and the positions of equilibrium points near the co-rotation resonances in barred galaxies has been studied by \cite{Wu.etal.2016}.

\subsection{S\'ersic Index of the Classical Bulge}
\label{sec:sersic_index}

\begin{figure*}
	\centering
	\includegraphics[width=\textwidth]{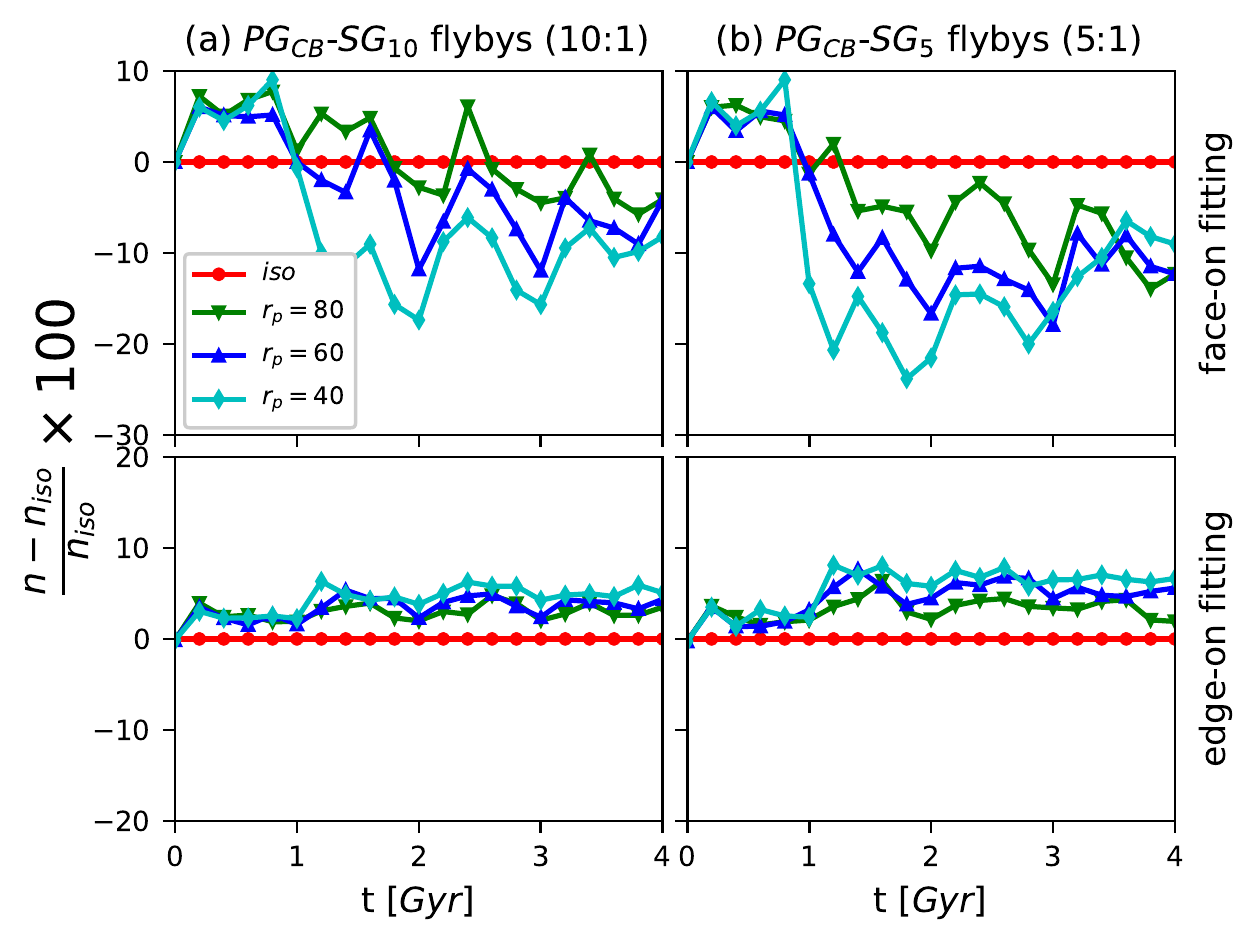}
    \caption{ Evolution of the percentage change in the S\'ersic index of the bulge in unbarred galaxies due to flybys. Left and right columns show the flyby simulations of $1/10$ and $1/5$ mass minor galaxy respectively around the major galaxy with classical bulge. Top and bottom rows represent the outputs from face-on and edge-on galaxy fitting respectively. The red circle solid curves, green down triangle solid curves, blue upper triangle solid curves, and cyan diamond solid curves represent the models $isolated$, $r_{p}=80~kpc$, $r_{p}=60~kpc$, and $r_{p}=40~kpc$ respectively.}
    \label{fig:sersic_index}
\end{figure*}

The evolution of the S\'ersic indices of the classical bulges relative to the control isolated model, derived from the two dimensional decomposition of the major galaxy, is shown in fig.~\ref{fig:sersic_index}. As in the previous figures, the left and right columns represent flybys of 1:10 and 1:5 mass ratios. The output from the face-on and edge-on bulge-disk fittings are shown in the top and bottom rows respectively. Our classical bulge models have initial S\'ersic index values ($n$) greater than 2. It should be noted that the value of $n$ strongly depends on the surface density of the host galaxy disk i.e. the background of the bulge. For example, the classical bulge without a disk has a S\'ersic index of $n=3.6$ but when the disk is included the face-on fit gives the value $n=2.1$. However, with the edge-on disk, the S\'ersic index remains approximately equal to that without a disk. In our study, we fitted both components of the galaxy simultaneously because that is how it is done in observational studies.

\begin{figure*}
	\centering
	\includegraphics[width=0.8\textwidth]{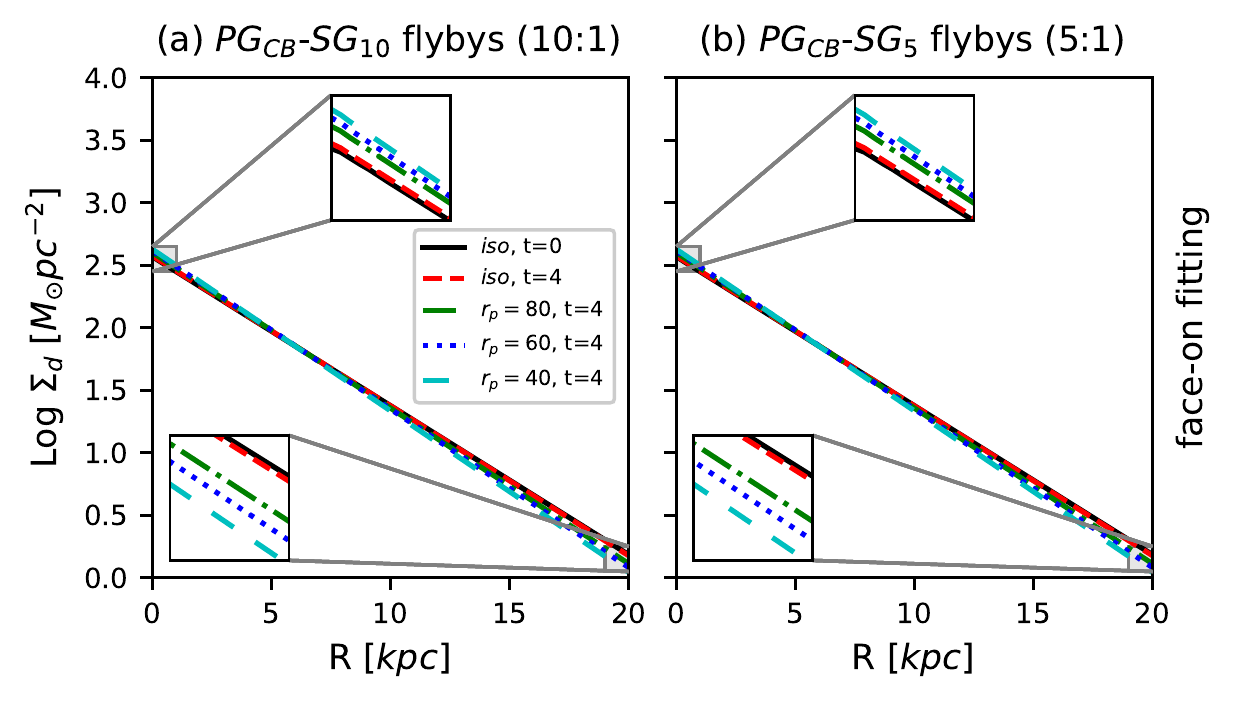}
    \caption{The left and right panels show the initial and final disk surface densities from face-on decomposition of the unbarred galaxies of 10:1 and 5:1 flyby models respectively. The black solid curves show the initial profiles and red dashed curves, green dash-dotted curves, blue dotted curves, and cyan dash-spaced curves represent the final profiles of the $isolated$, $r_{p}=80~kpc$, $r_{p}=60~kpc$, $r_{p}=40~kpc$ models respectively. The rectangular insets show the 5 times zoom in regions.}
    \label{fig:disks_density}
\end{figure*}

In the face-on fittings, the value of the S\'ersic index increases until $t\sim$1.0~Gyr for all the models (top row of the fig.~\ref{fig:sersic_index}). This small increment in the S\'ersic index is due to the steepening of the surface density of the inner disk region \citep{Hohl1971, Laurikainen2001, Debattista2006} which contributes to the mass of the spherical/classical bulge. After t=1.0~Gyr, the value of the S\'ersic index decreases for all the models. The decrement of the S\'ersic index after 1.0~Gyr, which is the time at when the minor galaxy passes the pericenter, is due to the formation of strong spiral arms. The spiral arms reduce the surface density of the outer disk region. For the sake of visualisation, we have plotted the final and initial disk surface densities from the face-on decomposition of unbarred, classical bulge galaxies of 10:1 and 5:1 flybys models in the left and right panels of fig.~\ref{fig:disks_density} respectively. The density change in the inner and outer disk regions causes the bulge-disk fitting to take some bulge particles into the disk component, thus reducing the value of the S\'ersic index. In both flyby models, the change in the S\'ersic index just after pericenter passage is the largest for the closest flyby. But it become nearly equal in the end for all the models of the given galaxy mass ratio. In the end, the 5:1 simulations show more change than the 10:1 simulations. Hence it appears that closer and major flybys are more effective in transforming classical bulges into pseudobulges compared to more distant and minor flybys.

\begin{figure*}
	\centering
	\includegraphics[width=\textwidth]{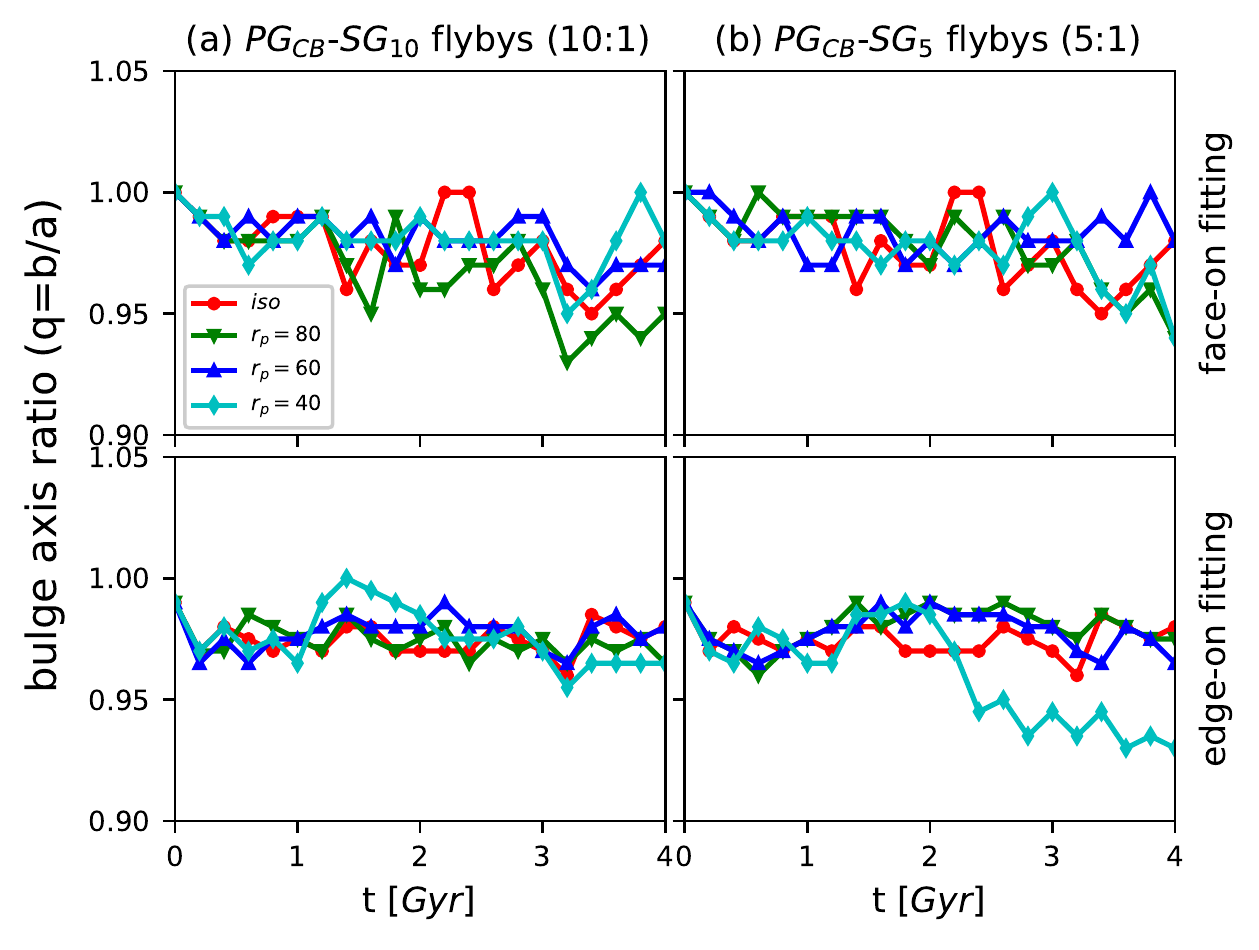}
    \caption{Evolution of the minor to major axis ratio ($q=b/a$) of the classical bulges in unbarred galaxies. Left and right columns show the 10:1 and 5:1 flybys simulations respectively. Top and bottom rows represent the outputs from face-on and edge-on galaxy fitting respectively. The red circle solid curves, green down triangle solid curves, blue upper triangle solid curves, and cyan diamond solid curves represent the models $isolated$, $r_{p}=80~kpc$, $r_{p}=60~kpc$, and $r_{p}=40~kpc$ respectively.}
    \label{fig:bulge_axis_ratio}
\end{figure*}

However, this apparent decrement in the S\'ersic index due to flyby interactions does not really represent the flattening of the classical bulge. This is because if the bulge became more disky or flattened in the z direction, it would be effectively turning into a pseudobulge or it would have a more oval shape. However, we could not detect any such change in the bulge shape in any of the classical bulge models. Fig.~\ref{fig:bulge_axis_ratio} illustrates this result. It shows the time evolution of the minor to major axis ratio ($q=b/a$) of the classical bulges. The left and right columns represent the 10:1 and 5:1 flyby models respectively. The top and bottom rows show the outputs of face-on and edge-on fitting respectively. All the panels show nearly constant axis ratios for the bulges. But the two dimensional decomposition of the face-on galaxy gives a lower S\'ersic index which corresponds to the flattening of the bulge. This lowering of the S\'ersic index, however, is due to the changing mass distribution in the disk caused by the flyby interaction as discussed earlier in this subsection. Hence, we conclude that classical bulge shapes are not changing due to flyby interactions.

In the edge-on fittings (see the bottom row of the fig.~\ref{fig:sersic_index}), in contrast to the face-on fittings, the value of the S\'ersic index for all the flyby models is always greater than or equal to the corresponding isolated model. Hence, the flybys are not decreasing the S\'ersic index of the classical bulge (or flattening the spherical bulge). Therefore, the real indicator of bulge type/shape is the edge-on decomposition of the galaxy.

\subsection{Mass of the Classical Bulge}
\label{sec:bulge_mass}

\begin{figure*}
	\centering
	\includegraphics[width=\textwidth]{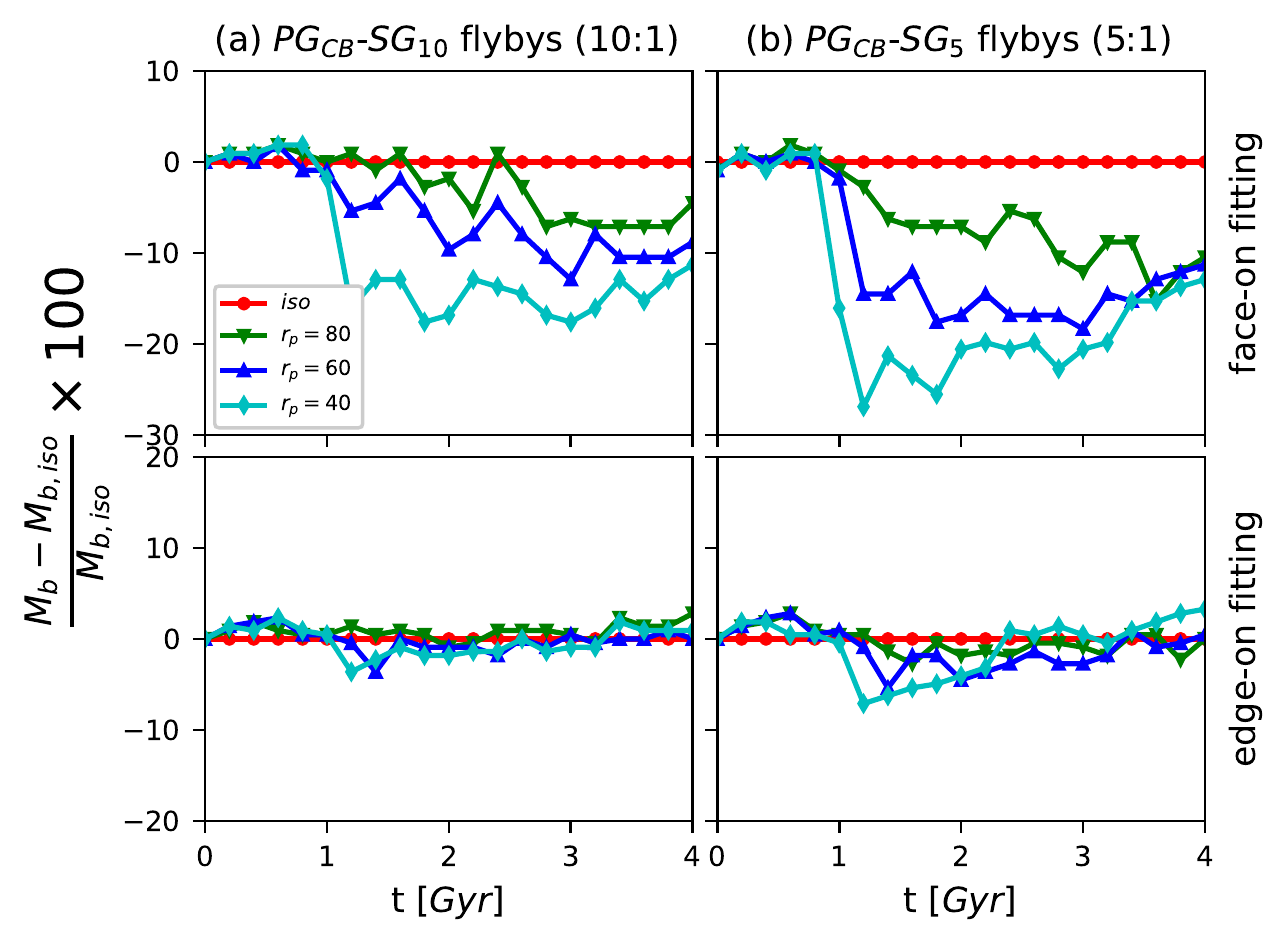}
    \caption{Evolution of the percentage change in the mass of the bulge in unbarred galaxies due to flybys. Left and right columns show the flyby simulations of $1/10$ and $1/5$ mass minor galaxy respectively around the major galaxy with classical bulge. Top and bottom rows represent the outputs from face-on and edge-on galaxy fitting respectively. The red circle solid curves, green down triangle solid curves, blue upper triangle solid curves, and cyan diamond solid curves represent the models $isolated$, $r_{p}=80~kpc$, $r_{p}=60~kpc$, and $r_{p}=40~kpc$ respectively.}
    \label{fig:bulge_mass}
\end{figure*}

To see whether flybys affect the mass of classical bulges or not, we calculated the bulge mass at time steps of 0.2~Gyr using the integrated magnitudes of the bulges obtained from the output file of GALFIT. The left and right columns of fig.~\ref{fig:bulge_mass}, show the time evolution of the percentage change in the bulge mass of unbarred, classical bulge galaxies for the 10:1 and 5:1 flyby models respectively. The output from face-on and edge-on fittings are shown in the top and bottom rows of the figure respectively. The initial mass of the bulge in the face-on fitting (0.42 $\times 10^{10}M_{\odot}$) is smaller than the edge-on fitting (0.55 $\times 10^{10}M_{\odot}$). The bulge mass from edge-on fitting is much closer to the initial value of 0.6 $\times 10^{10}M_{\odot}$. In the face-on fittings, there are clear evolutionary trends which are similar to the evolutionary trends of the S\'ersic indices. There is a marginal increase in the bulge masses until the time t=1.0~Gyr of the evolution. This marginal increment is due to the steepening of the surface density of the inner disk region \citep{Hohl1971, Laurikainen2001, Debattista2006} which contributes to the mass of the bulge. Then there is some decrease in the bulge masses for both types of flyby models after 1.0~Gyr of the evolution. This decrease in the bulge mass is due to the formation of strong spiral arms after the flyby interaction. The formation of spiral arms reduces the surface density in the outer region of the disk (see fig.~\ref{fig:disks_density}). Hence, the redistribution of disk particles causes some of the bulge particles to fit into the disk which changes the mass of the bulge in the 2d bulge-disk decomposition of the face-on galaxies. For a given pericenter distance, the 5:1 flybys show more change than the 10:1 flybys. Similarly, the closest pericenter passage shows the more change at a given major to minor galaxy mass ratio. But at the end of the simulations of a given galaxy mass ratio, the change in all the models appear to converge to the same value. On the other hand, edge-on fittings do not show any change in the bulge mass for both the flyby models as seen in the bottom row of the figure. Thus we conclude that flybys do not effect the mass of the classical bulges in the major galaxy.

\subsection{Angular Momentum of the Classical Bulge}
\label{sec:classical_bulge_Lz}

\begin{figure*}
	\centering
	\includegraphics[width=0.8\textwidth]{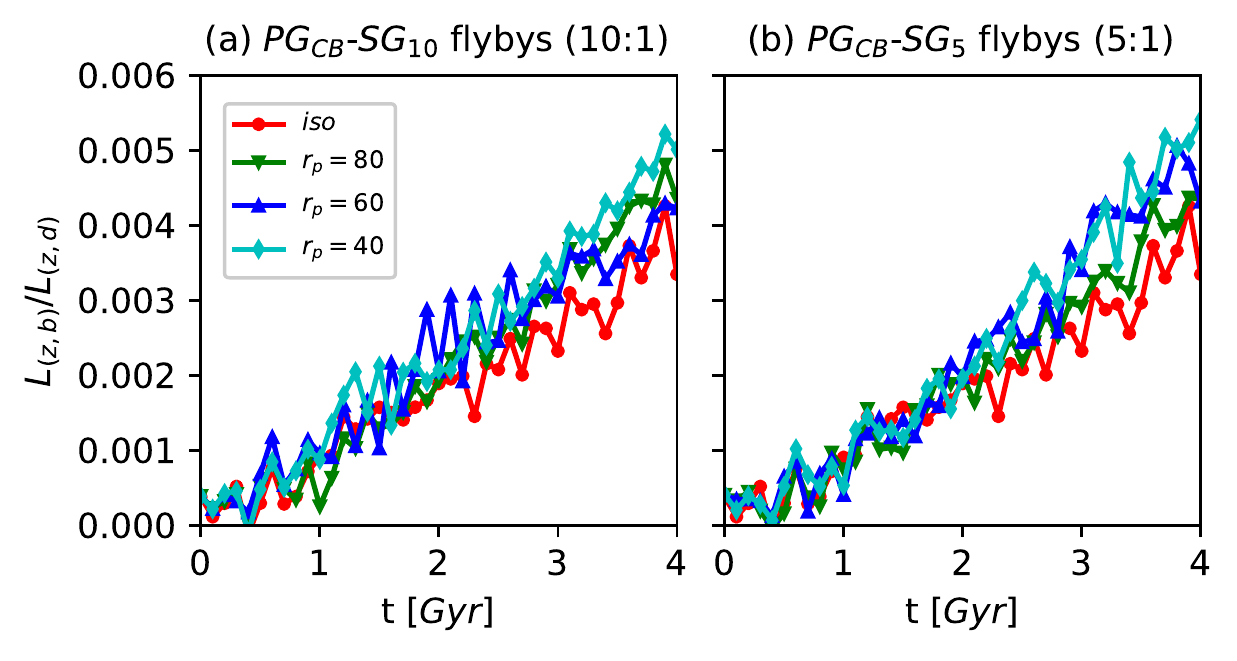}
    \caption{Evolution of the bulge to disk angular momentum ratio within a sphere of 5 bulge effective radius ($5R_{e}$) in unbarred galaxies. Left and right panels show the 10:1 and 5:1 flyby models respectively. The red circle solid curves, green down triangle solid curves, blue upper triangle solid curves, and cyan diamond solid curves represent the models $isolated$, $r_{p}=80~kpc$, $r_{p}=60~kpc$, and $r_{p}=40~kpc$ respectively.}
    \label{fig:classical_bulge_Lz}
\end{figure*}

Flybys play an important role in the angular momentum transfer between galaxies. They can spin-up or spin-down the galaxies depending on the configuration of the flyby, such as prograde-prograde, prograde-retrograde, retrograde-prograde, or retrograde-retrograde directions of the galaxies orbits \citep{Bett.Frenk.2012}. But how do the bulges of the galaxies respond to the angular momentum transfer in flybys? To answer this question, we have calculated the z-component of the angular momentum of the bulge and disk within a sphere of radius 5 times the bulge effective radius ($5R_{e}$). In fig.~\ref{fig:classical_bulge_Lz}, we have shown the time evolution of the bulge to disk angular momentum ratio in unbarred, classical bulge galaxies. The left and right panels of this figure show the 10:1 and 5:1 flyby models respectively. We have plotted only the ratio of the z-component of the angular momenta because all of our model galaxies lie in the x-y plane.

From the two panels of the fig.~\ref{fig:classical_bulge_Lz}, we can clearly see that the ratio of bulge to disk angular momentum is increasing with time in all the models. This is the indication of the angular momentum transfer from the disk to bulge. However, until 2~Gyr, all the flyby models show a gain similar to that of the isolated model. There is a small angular momentum gain in the flyby models after 2~Gyr but it is not significant enough to clearly distinguish an angular momentum gain in the flyby models compared to the corresponding isolated model. Although it is possible that multiple prograde flybys may add up to a significant transfer of angular momentum to the bulge, it must however be noted that in nature, flybys are not always on prograde orbits. Hence, classical bulges do not gain significant angular momentum in flybys, but instead they could gain angular momentum due to the rotating disk \citep{Pedrosa.Tissera.2015}. The classical bulges in galaxies reside in the deepest part of the potential and there in these models there is no bar type instability in their disks. Hence they do not gain much angular momentum. However, the case is very different when the galaxies are barred as there can be significant angular momentum gain by the bulge \citep{kataria.das.2018}.

\subsection{Evolution of the pseudobulge}
\label{sec:pb_evolution}

\begin{figure*}
	\centering
	\includegraphics[width=0.5\textwidth]{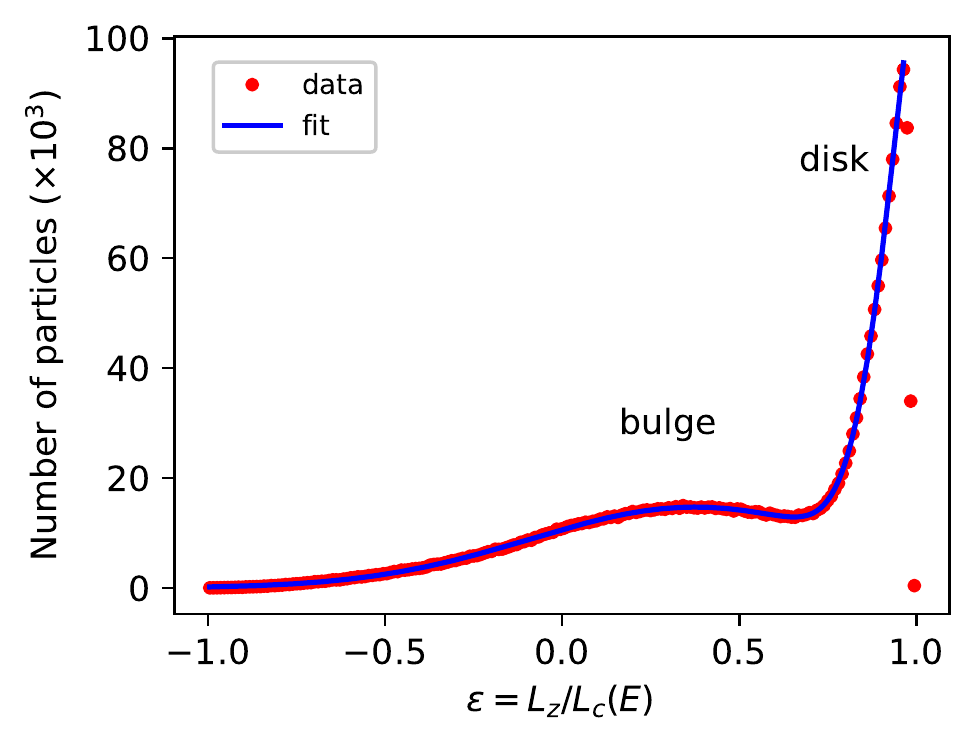}
    \caption{ An example of two generalized Gaussian fitting (up to the disk peak) to the kinematic decomposition of bulge and disk components of the galaxy with pseudobulge ($PG_{PB}$). The X-axis shows the ratio of stellar particle angular momentum perpendicular to the disk plane ($L_{z}$) and the maximum circular angular momentum ($L_{c}(E)$) corresponding to stellar energy E. The peak around $\epsilon=1.0$ shows the rotation dominated component and the peak away from $\epsilon=1.0$ shows the dispersion dominated component. The red filled circles and blue continuous curve show the data points and best fit curve respectively.}
    \label{fig:kinematic_decomp_fit}
\end{figure*}

As discussed before in the subsection~\ref{sec:analysis}, the existence of the bar and the boxy/peanut pseudobulge makes the 2d decomposition of the galaxy very difficult. Therefore, we have used the kinematic decomposition method for the models with pseudobulges ($PG_{PB}$) \citep{abadi.etal.2003}. For the kinematic decomposition of the bulge, stellar particles within 15~kpc from the center of the galaxy are used because the bulge resides in the center of the galaxy. We simultaneously fitted the peaks corresponding to the bulge and the disk in the $\epsilon=L_{z}/L_{c}(E)$ distribution plot using two generalized Gaussians (for example see the fig.~\ref{fig:kinematic_decomp_fit}). We used the position ($\epsilon_{mean}$) and full width at half maximum ($\epsilon_{fwhm}$) of the generalized Gaussian corresponding to the bulge component as the indicator of bulge evolution. The idea is that the $\epsilon_{mean}$ of the bulge will move toward $\epsilon=0$ on increasing the random motion of the stellar particles, which represents the dynamical heating of the bulge, and vice-versa for dynamical cooling which is due to increasing ordered motion of the stellar particles. Similarly, the $\epsilon_{fwhm}$ will be larger for bulges that are less distinguishable from the disk compared to small $\epsilon_{fwhm}$ values which represent bulges that are dominant in disks.

\begin{figure*}
	\centering
	\includegraphics[width=\textwidth]{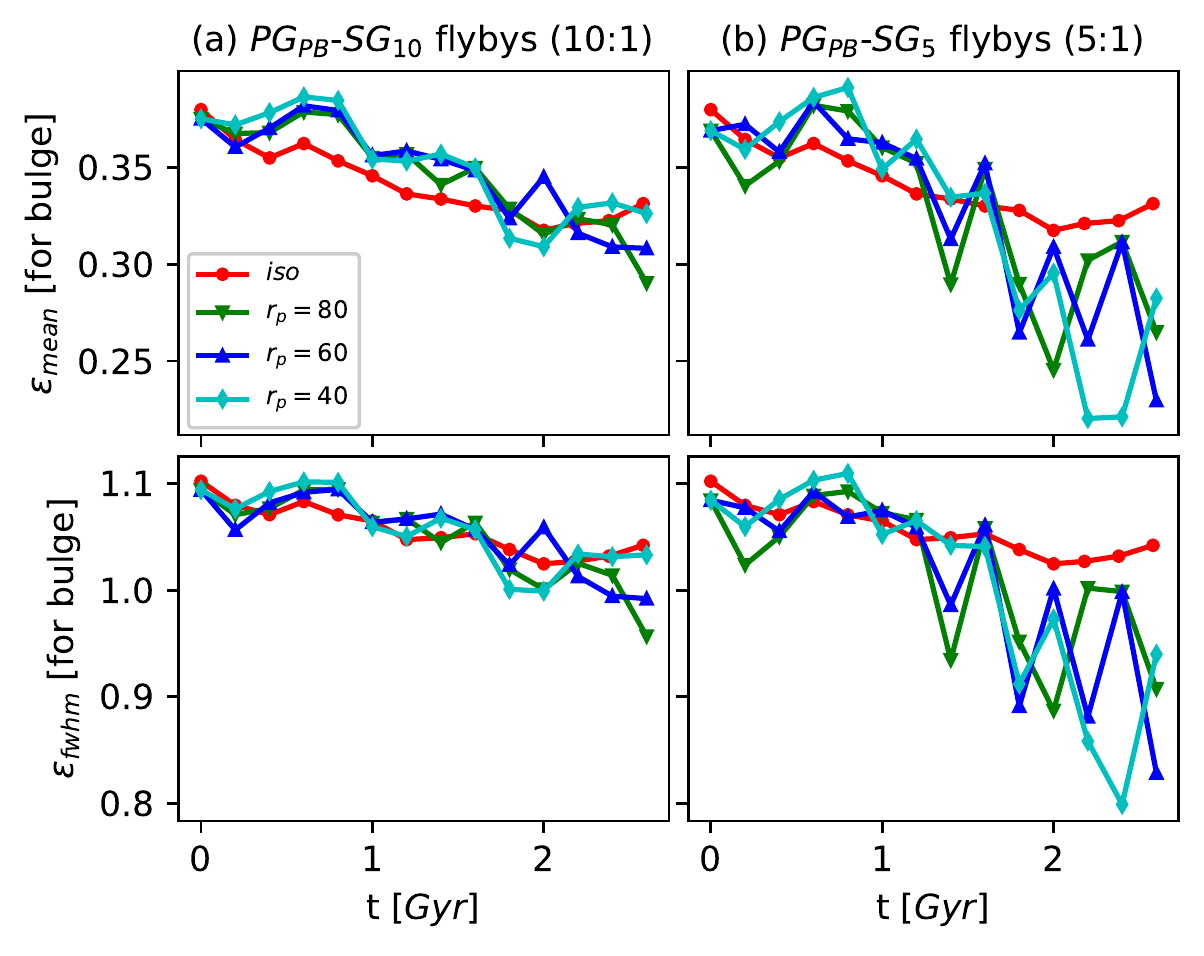}
    \caption{The top and bottom rows show the time evolution of the position ($\epsilon_{mean}$) and full width at half maximum ($\epsilon_{fwhm}$) of the generalized Gaussian fitted to the pseudobulge component. Left and right columns represent the flyby simulations of $1/10$ and $1/5$ mass minor galaxy respectively around the major galaxy with pseudobulge. The red circle solid curves, green down triangle solid curves, blue upper triangle solid curves, and cyan diamond solid curves represent the models $isolated$, $r_{p}=80~kpc$, $r_{p}=60~kpc$, and $r_{p}=40~kpc$ respectively.}
    \label{fig:pb_mean_fwhm}
\end{figure*}

The top and bottom rows of fig.~\ref{fig:pb_mean_fwhm} demonstrate the time evolution of the mean position ($\epsilon_{mean}$) and full width at half maximum ($\epsilon_{fwhm}$) of the bulge component in galaxies with pseudobulges ($PG_{PB}$) respectively. The left and right columns show the outputs from 10:1 and 5:1 flyby models respectively. From the left column of this figure, one can easily interpret that there is no effect of the 10:1 flyby interactions on the pseudobulge of the major galaxy. The time evolution of $\epsilon_{mean}$ and $\epsilon_{fwhm}$ in the 10:1 flyby models are approximately similar to that of the corresponding isolated models. Although these flyby models do show some increase in the $\epsilon_{mean}$ during the closest pericenter passage, it does not last long. Hence, we can conclude that flybys with relatively low mass satellites cannot affect the evolution of the pseudobulge of the host galaxy.

In contrast to the 10:1 flyby simulations, the 5:1 flyby simulations show significant deviation from the corresponding control isolated models, as can be seen from the right column of the fig.~\ref{fig:pb_mean_fwhm}. The values of $\epsilon_{mean}$ and $\epsilon_{fwhm}$ decrease relative to the corresponding isolated models after the passage of the minor galaxy. The decrease in  $\epsilon_{mean}$ indicates increasing random motion in the bulge i.e the bulge becomes kinematically hotter. On the other hand, decrease in the $\epsilon_{fwhm}$ value indicates that the bulge is becoming more distinct from the disk. These results show that flybys with relatively large mass ratios (such as 1:5) can increase the random motion of stars in the pseudobulge of the host galaxy.

\begin{figure*}
	\centering
	\includegraphics[width=\textwidth]{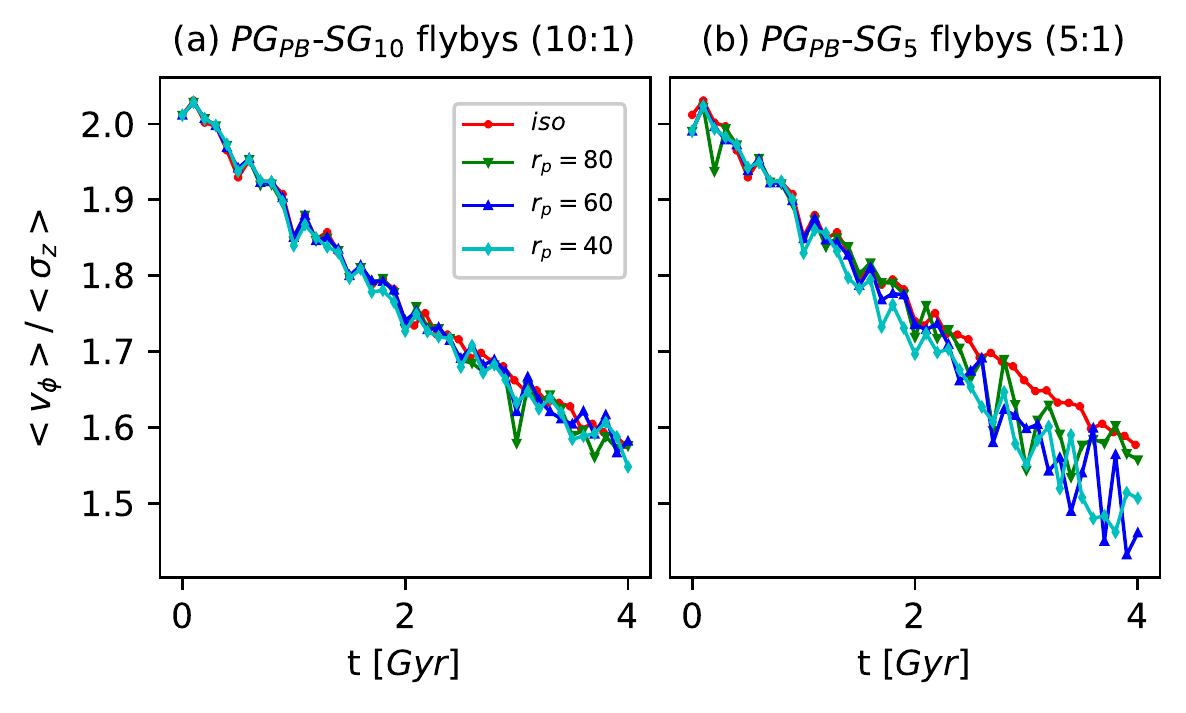}
    \caption{ The time evolution of the ratio of the mean azimuth velocity to mean vertical dispersion ($<v_{\phi}>/<\sigma_{z}>$) within 15~kpc. Left and right columns represent the flyby simulations of $1/10$ and $1/5$ mass minor galaxy respectively around the major galaxy with pseudobulge. The red circle solid curves, green down triangle solid curves, blue upper triangle solid curves, and cyan diamond solid curves represent the models $isolated$, $r_{p}=80~kpc$, $r_{p}=60~kpc$, and $r_{p}=40~kpc$ respectively.}
    \label{fig:pb_v_by_sigma}
\end{figure*}

The above mentioned kinematic decomposition of bulge and disk component requires the prior knowledge of the lesser know dark matter halo distribution. Therefore, we have also shown the evolution of the cold and hot component fractions of the galaxy in terms of observable quantities; mean azimuth velocity ($<v_{\phi}>$) and mean vertical dispersion ($<\sigma_{z}>$). Fig.~\ref{fig:pb_v_by_sigma} shows the time evolution of the ratio of the mean azimuth velocity to mean vertical dispersion ($<v_{\phi}>/<\sigma_{z}>$) within 15~kpc. The left and right panels of the figure represent the 10:1 and 5:1 mass ratio flyby simulations respectively. The left panel clearly shows that the 10:1 flyby models follow the same trend as that of the control isolated model. There is no effect of 10:1 flyby interactions on the evolution of the pseudobulge. But, the 5:1 simulations show a different trends from the control isolated model as can be seen from right panel of the figure. There is more vertical dispersion in the flyby model than the isolated model. This supports the previous results that the flyby interactions with larger mass ratios can heat the pseudobulge more significantly than smaller mass ratios in minor flyby interactions.

\subsection{Mass and Angular Momentum of the Pseudoulge}
\label{sec:pb_mass_angmom}

\begin{figure*}
	\centering
	\includegraphics[width=\textwidth]{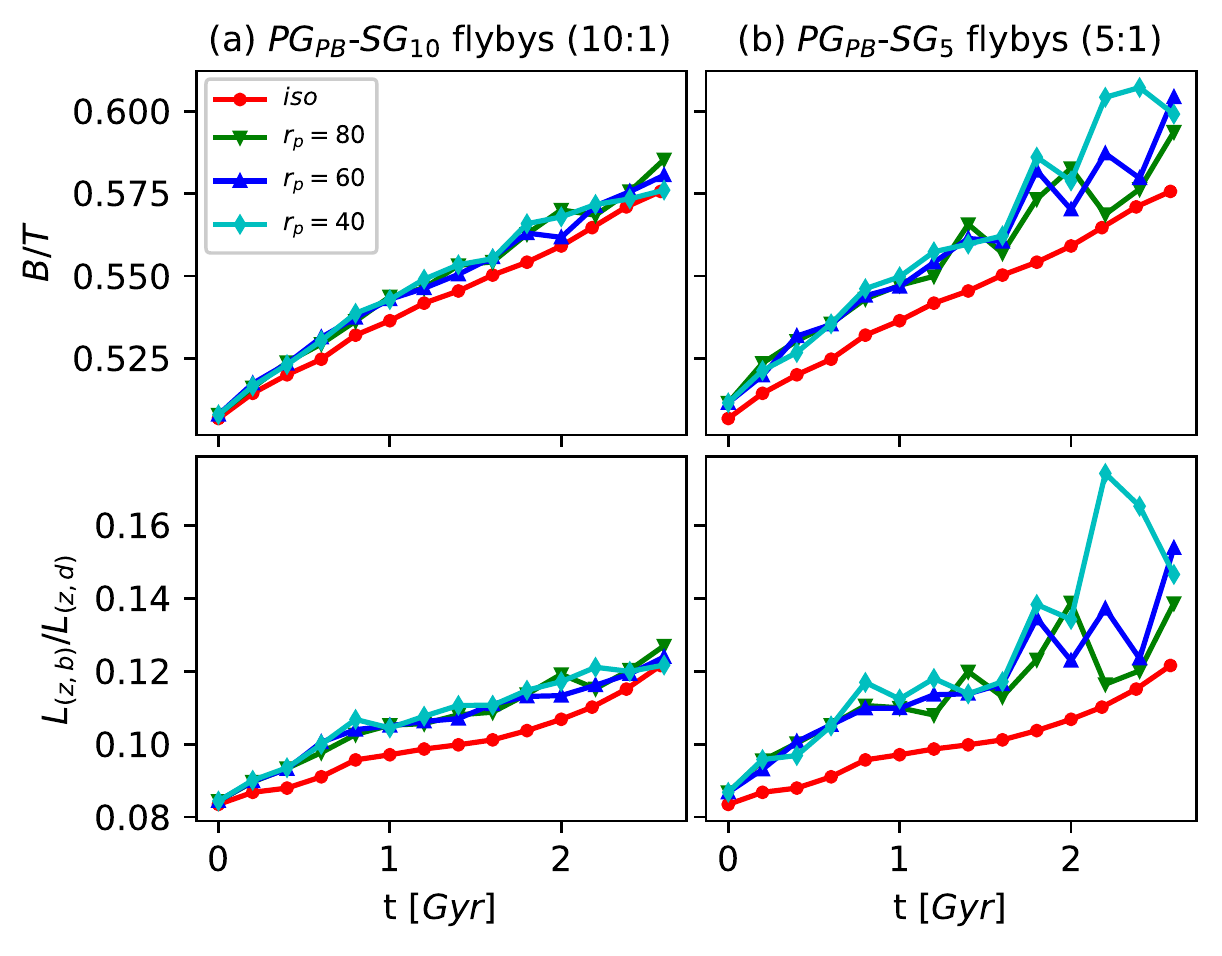}
    \caption{The top and bottom rows show the time evolution of the bulge to total mass ratio ($B/T$) and bulge to disk angular momentum ratio ($L_{(z,b)}$/$L_{(z,d)}$) within 15~kpc respectively. The left and right columns represent the flyby simulations of $1/10$ and $1/5$ mass minor galaxy respectively around the major galaxy with pseudobulge. The red circle solid curves, green down triangle solid curves, blue upper triangle solid curves, and cyan diamond solid curves represent the models $isolated$, $r_{p}=80~kpc$, $r_{p}=60~kpc$, and $r_{p}=40~kpc$ respectively.}
    \label{fig:pb_mass_angmom}
\end{figure*}

In the previous subsection, we have discussed the effects of flyby interactions on the kinematic heating and separability of a pseudobulge from the disk. In this section we investigate its effect on the mass and angular momentum of the pseudobulge. To obtain the mass and angular momentum of the bulge using kinematic decomposition, we have used a widely accepted fixed value of $\epsilon=0.7$ as a separator of the bulge and disk particles \citep{Peebles2020}. The particles with $\epsilon<0.7$ are considered as bulge particles and particles with $\epsilon \geq 0.7$ are considered as disk particles.

The time evolution of the bulge to total stellar mass ratio ($B/T$) and the disk to bulge angular momentum ratio ($L_{(z,b)}$/$L_{(z,d)}$) within 15~kpc (the maximum radius for kinematic decomposition) are shown in the top and bottom rows of fig.~\ref{fig:pb_mass_angmom} respectively. The left and right columns represent the flybys of 1/10 and 1/5 mass ratios for a major galaxy with a pseudobulge. Both panels of the left column show the slow increase in the mass and angular momentum of the bulge in flybys relative to the control isolated model. But after the flyby ends, all the models converge to values similar to that of the isolated model. Thus, the 10:1 flyby models do not show any net effect of the interaction on the mass and angular momentum of the pseudobulge. Hence, flyby interactions with relatively low mass satellites show only small changes in mass and angular momentum, and the effect disappears after the satellite has moved away.

On the other hand, the effect of 5:1 flyby interactions are very significant, as can be seen from comparing the adjacent columns of fig.~\ref{fig:pb_mass_angmom}. The mass and angular momentum gained by the pseudobulge remains always larger than the control isolated model. At time t$\sim$2~Gyr, there is approximately 20$\%$ angular momentum gain in flyby interactions relative to the isolated evolution. The trends of mass and angular momentum gain are approximately similar. Therefore, there is a continuous transfer of mass from the disk to the pseudobulge. This is a clear indication of the kinematic heating of the pseudobulge and bulge growth. Hence, relatively massive satellites can help the pseudobulges in flybys to gain mass and angular momentum from the disk. It is possible that the effect of multiple flyby interactions can strongly affect the evolution of psudobulges.

\subsection{Effect of Flyby on the Buckling of the Bar}
\label{sec:bar_buckling}

\begin{figure*}
	\centering
	\includegraphics[width=0.7\textwidth]{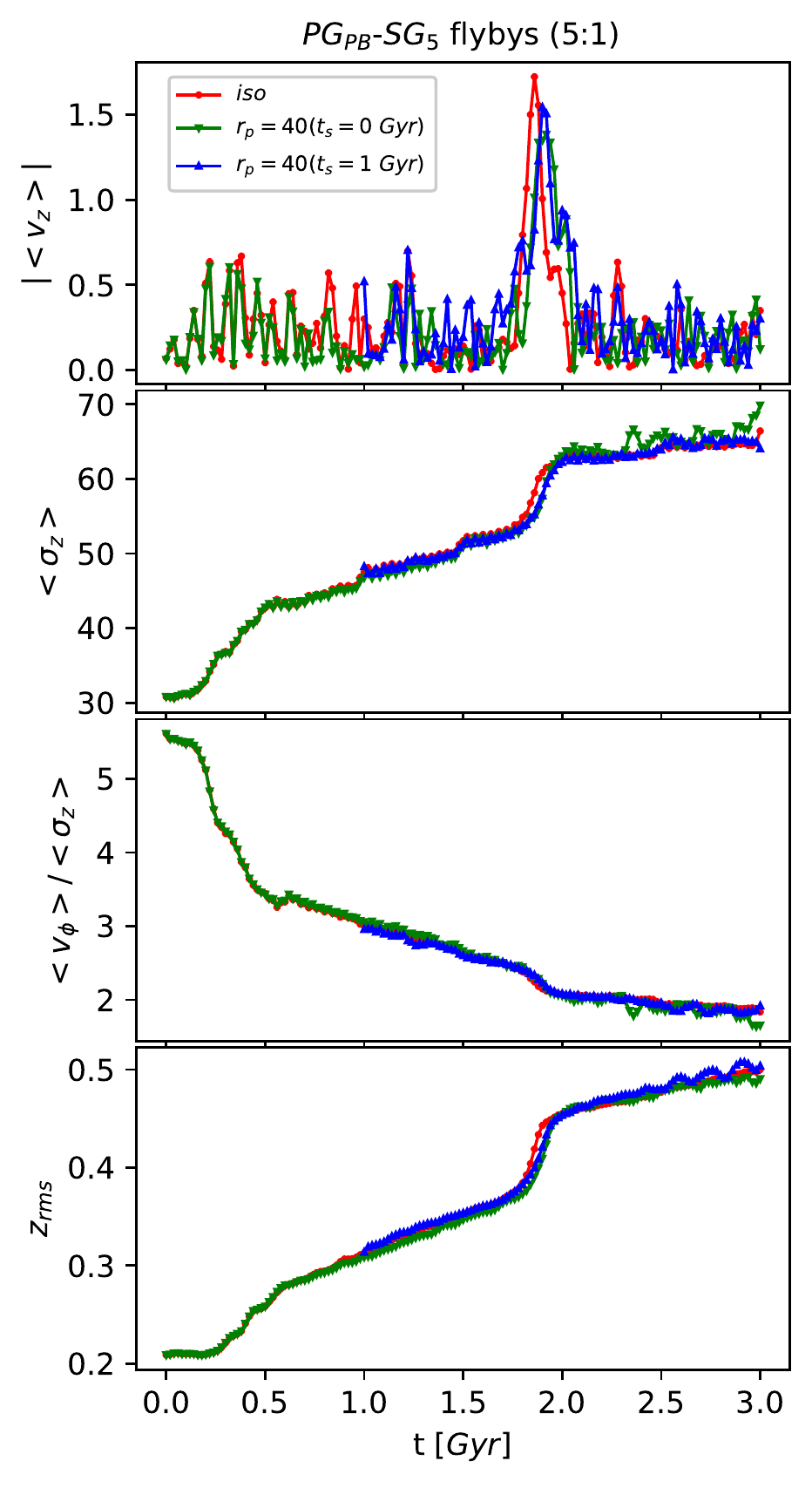}
    \caption{ The time evolution of absolute value of mean vertical velocity($|<v_{z}>|$), mean vertical dispersion ($<\sigma_{z}>$), ratio of the mean azimuth velocity to mean vertical dispersion ($<v_{\phi}>/<\sigma_{z}>$), and root mean square value of vertical coordinates ($z_{rms}$) in the top to bottom panels respectively. The red circle solid curve, green down triangle solid curve, and blue upper triangle solid curve represent the $isolated$, $r_{p}=40~kpc$ starts at $t_{s}=0$ Gyr, and $r_{p}=40~kpc$ starts at $t_{s}=1$ Gyr models respectively.}
    \label{fig:bar_buckling}
\end{figure*}

We have previously discussed the response of pseudobulges to flyby interactions with 10:1 and 5:1 mass ratios. The origin of the pseudobulge discussed in this chapter is the buckling of the bar. An important question is then how does bar buckling respond to flyby interactions? This is a relevant question as bar buckling modifies the kinematics, rendering the system kinematically hotter. To understand the effect of the flyby interactions on bar buckling we have simulated two 5:1 mass ratio flyby interactions, one starting before the first peak in the bar strength and other starting before the second peak in the bar strength (see fig.~\ref{fig:bar_strength}). The choice of 5:1 mass ratio is made simply because the 10:1 mass ratio does not seem to effect pseudobulges as much as the larger mass ratio.

To quantify the strength of the buckling of the bar, we have adopted some commonly used physical quantities in the literature \citep{Ciambur.etal.2017, Lokas2019}. In  fig.~\ref{fig:bar_buckling}, we have shown the time evolution of absolute value of mean vertical velocity($|<v_{z}>|$), mean vertical dispersion ($<\sigma_{z}>$), ratio of the mean azimuth velocity to mean vertical dispersion ($<v_{\phi}>/<\sigma_{z}>$), and the root mean square value of the vertical positions ($z_{rms}$) in the top to bottom panels respectively. All these quantities are calculated within the 7.5~kpc radius of the galaxies (half of that used in the kinematic decomposition) which is large enough to include the bar region throughout the evolution. The red circle solid curve, green down triangle solid curve, and blue upper triangle solid curve represent the control isolated model, flyby stating at $t_{s}=0$ Gyr from control model, and flyby stating at $t_{s}=1$ Gyr from control model respectively.

From the top panel of the fig.~\ref{fig:bar_buckling}, one can see that all the models show a sharp peak at the same time in the evolution of $|<v_{z}>|$. These peaks occur at the time of the second peak in the bar strength. The peak in the evolution of mean $v_{z}$ signifies the fraction of stars going out of the disk plane. The height of the peak quantifies the magnitude of bar bending out of the disk plane. Though the bending is not strong but it seems that the control isolated model bends the most and flyby interactions reduce the bar bending. The second panel from the top shows the evolution of mean $\sigma_{z}$ which quantifies the the vertical height. All the models show similar evolution of $<\sigma_{z}>$. This result can be further confirmed from the evolution of the other two quantities. The evolution of $<v_{\phi}>/<\sigma_{z}>$, and $z_{rms}$ in the flyby models follows a similar trend as that of the control isolated model. Hence, flyby interactions do not seem to affect the time and strength of the bar buckling in our galaxy model.

\subsection{Effect of Numerical Resolution on Our Results}
\label{sec:numerical_resolution}

\begin{figure*}
	\centering
	\includegraphics[width=0.7\textwidth]{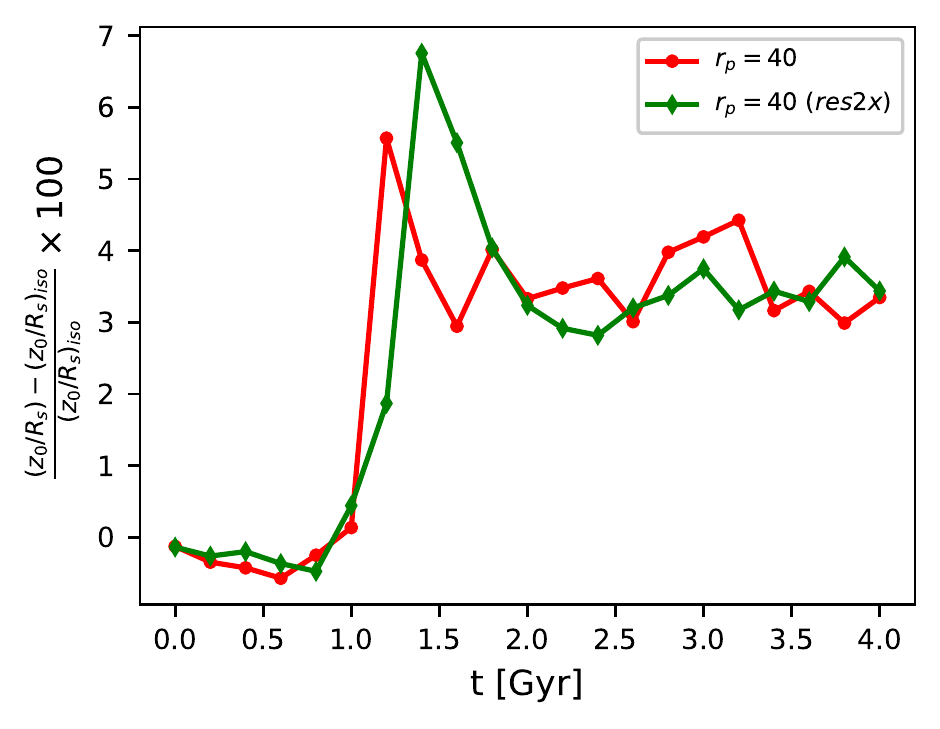}
    \caption{Effect of numerical resolution on the evolution of the percentage change in the disk scale height to disk scale radius ratio in 10:1 flyby model of classical bulge galaxy. The green diamond solid curve represents the simulation of $r_{p}=40~kpc$ model with twice as many particles as red circle solid curve. Both the curves show nearly similar trends.}
    \label{fig:scale_height_res2x}
\end{figure*}

All the results of the flyby interactions, discussed in this chapter, are relative to the control isolated model. The flyby interactions include secular evolution of the galaxies which can only be removed by comparing with the evolution of an isolated galaxy. On the other hand, numerical simulations always suffer from the discreteness of particle distribution. Some effects of this discreteness on the results of flyby simulations can be removed by subtracting the secular evolution from the evolution in interactions. This is the reason why we compared our results with the corresponding isolated models. We chose the number of particles for our simulation using an analytical study of two body relaxation \citep{Power2003}. We also performed a few high resolution simulations with twice as many particles as given in the Table~\ref{tab:initial_parameters}. We find that the trends as discussed in the results section (section~\ref{sec:results}) are consistent with these high resolution simulation. For comparison, in  fig.~\ref{fig:scale_height_res2x}, we have shown the time evolution of the percentage change in the disk scale height to disk scale radius ratio in the 10:1 flyby model of a classical bulge galaxy with respect to the corresponding isolated model. The green diamond solid curve represents the simulation of $r_{p}=40~kpc$ model with twice as many particles as red circle solid curve. Both curve show more or less same evolutionary trends. This indicates that our results are consistent with high resolution.

\section{Implications}
\label{sec:implication}
In the previous sections, we have presented a qualitative and quantitative study of the effect of flyby interactions on the disks, spiral arms, classical bulges (in unbarred galaxies) and pseudobulges in galaxies. The results of our simulations are very significant for observations of interacting galaxies because in our local universe the rate of flybys is larger at lower redshifts (z$<$2) \citep{Sinha2012A}. In the following paragraphs we discuss the implications of our study for observations of flyby interactions in the local universe.

The gravitational force of a flyby produces strong perturbations in the outer disk regions of the major galaxy. As a result, significant amounts of stellar mass is pulled out from the disk in the form of spiral arms/ tidal tails. This results in the decrease of the disk scale radius ($R_{s}$) and the increase of the disk scale height ($z_{0}$). Although the increase in $z_{0}$ is small, the ratio of disk scale height to disk scale radius ($\frac{z_{0}}{R_{s}}$) shows a significant increase \citep{Reshetnikov1996, Reshetnikov1997}. Since a galaxy experiences multiple flybys during its evolution, the resultant vertical thickening or heating of the host galaxies will be significant and perhaps comparable to that produced by minor mergers.

Flyby interactions stimulate the formation of strong spiral arms in the host galaxies just after the pericenter passage. However, the maximum strength of the spiral arms attained in the flyby does not last. It decays slowly with time and all the flyby models show nearly the same strength in their spiral arms after 4~Gyrs. Our N-body simulations thus show the slowly decaying nature of the spiral arms. This supports the idea that spiral arms are not static density waves as discussed in observations \citep{Masters.2019}. These results will be very helpful in understanding the nature of spiral arms in interacting and isolated galaxies. The spiral arms induced by a closely bound satellite galaxy may survive for a long period of time because a bound satellite will always perturb its host galaxy. The effect of multiple satellite galaxies orbiting around their host on different orbital configurations can give a detailed insight into the origin of the spiral structure.

Strong spiral arms are formed in both the unbarred, classical bulge galaxies and the pseudobulge galaxies. But in contrast to the unbarred, classical bulge galaxies, the flyby induced spiral arms in the pseudobulge host galaxies show oscillations on the top of the time-varying strength of the spiral arms. These oscillations in the spiral arms strength $A_{2}/A_{0}$ could be due to existing small amplitude transients near the resonances associated with the bar \citep{Binney.Tremaine.2008, Wu.etal.2016}, which have been detected in earlier simulations of barred galaxies. The oscillations may be amplified by the energy input due to the flyby interaction. These transient spiral waves may have important implications for star formation and are triggered by the flyby. A good way to test this is by comparing the star formation rates in flybys with host galaxies that have bars and those that do not have bars.

Our overall results of flybys with classical bulges suggests that the inner regions of such galaxies are very stable. Neither the bulge morphology nor the bulge angular momentum changes significantly during flybys. The photometric bulge-disk decomposition of the face-on unbarred, classical bulge galaxies shows a decrease in the mass and S\'ersic index of the classical bulge. This apparent decrease in the S\'ersic index of the classical bulge can be mistakenly considered as evidence for the transformation of the classical bulge into the pseudobulge. So if we take the S\'ersic index as a parameter for distinguishing between classical and pseudobulges, a significant fraction of classical bulges will be mistakenly classified as pseudobulges in face-on galaxies. As a consequence, this classification will provide more weight to the rotation dominated universe and challenge the hierarchical nature of the structure formation \citep{Fisher2011}. So it is clearly important to examine the edge-on decompositions of galaxies whenever possible for S\'ersic index studies of bulges.

Our simulations also show that although the pseudobulges are supported by the ordered motion of the stars, they are very stable against the small perturbations in the outer disk regions of the galaxies. In the 10:1 mass ratio flybys, the pseudobulges do not show any noticeable change except during pericenter passage. There are signatures of dynamical heating, mass growth and angular momentum gain only in the 5:1 mass ratio flybys. Thus only similar mass flybys or large mass ratio minor flybys ($>$1:5) are efficient enough to perturb the evolution of pseudobulges.

Our simulations do not account for the presence of gas in the galaxies. Therefore our findings are valid only for dry or gas poor galaxies. To make a rough estimate of the difference that gas inflow during a flyby would make to the bulge mass, consider the following. Let us assume that the central gas mass fraction of the major galaxy is 5-10$\%$ of the total stellar mass and the star formation efficiency (SFE) is $\sim$10$\%$. During the evolution if the gas is converted to stars, it will contribute 3-6$\%$ of the bulge mass in the center of the galaxy during the flyby. Hence, a gas rich galaxy with a high star formation rate (SFR) may show an increase in bulge mass during such flyby interactions but it will not significantly affect our results.

\section{Conclusions}
\label{sec:conclusions}
We have simulated 10:1 and 5:1 mass ratio flyby interactions of an unbarred, classical bulge galaxy and a barred galaxy with a pseudobulge. We study the effect of the flyby on the major galaxy only. The galaxies are on prograde-prograde orbits, with pericenter distances (distance of closest approach) varying from 40 to 80~kpc. We evolved the flyby for 4~Gyrs and then quantified the evolution of the bulge, disk and spiral arms of the major galaxy using bulge-disk decomposition and Fourier analysis techniques. The main findings of our study are as follows.

1. The disk scale radius ($R_{s}$) of the galaxies decreases in the flyby interactions. The change in the disk scale radius is strongly correlated with the pericenter distance and mass ratio of the galaxies. For a given mass ratio, the close flyby shows the most decrease but for a given pericenter distance, the major flyby (5:1) shows the most decrease. Due to presence of the spiral arms, this change is more pronounced in face-on decomposition than egde-on decomposition. The final disk scale radius seems to converge to the same value for all pericenter distances for a given flyby mass ratio.

2. The absolute disk scale height ($z_{0}$) of the galaxies is marginally affected by the flyby interactions. There is a maximum increase of $\sim 4\%$ in the disk scale height for our set of simulations. However, the ratio of disk scale height to disk scale radius ($z_{0}/R_{s}$) depends significantly on the pericenter distance. The galaxies with smaller pericenter distance show larger changes in this ratio, which is similar to observations of interacting galaxies \citep{Reshetnikov1996,Reshetnikov1997}. But our simulations also show that this is the effect of reducing disk scale length.

3. The minor flyby interactions reduce the disk size but at the same time result in the formation of the grand design spiral arms. The simulations of unbarred, classical bulge galaxies suggests that grand design spiral arms can form in minor flyby interactions when the major galaxy has a concentrated bulge and/or low surface density disk. Both features prevent bar formation in the disk. Hence, grand design spiral galaxies form only in flyby interactions of bar-stable, unbarred galaxies with satellite galaxies.

4. All the models show that the spiral arm strength increases with decreasing pericenter distance passage. However, at the end of the flyby simulation (at 4~Gyrs), all the flyby models show approximately similar spiral arms strength values for a given galaxy mass ratio model, indicating that the final effect does not depend on pericenter distance. The barred, pseudobulge galaxies show small oscillations in the strength of the spiral arms on top of the overall spiral arm strength. The amplitude of these oscillations increases with decreasing pericenter distance. The frequency of these oscillations is approximately independent of the pericenter distances and galaxy mass ratios.

5. The classical bulge of the major galaxy remains generally unaffected by the flyby interactions. In the face-on bulge-disk decomposition, the apparent change in the bulge S\'ersic index is instead due to the steepening of the inner disk surface density and the formation of spiral arms during the flyby. But the edge-on fittings and visualizations of the bulge indicates that the bulge morphology remains unchanged. The kinematic decomposition of the pseudobulges shows some heating and mass growth in the 5:1 simulations. This growth comes at the cost of the disk heating/thickening.

6. There is only a very minute increment in the angular momentum of the classical bulges. This small change is not due to the flyby interactions but due to the rotational motion of the disk particles. The change in the angular momentum of the classical bulges is nearly similar for all the flyby models and isolated models. The evolution of the classical bulges in dry (of gas poor) galaxies remain unaffected of flyby interactions. Pseudobulges show around 20$\%$ gain in the bulge angular momentum after 2~Gyr relative to the secular evolution. But most of it comes from the conversion of disk particles to bulge particles by vertical heating of the disk.

7. The tidal force of the orbiting satellite can induce bar formation in the host galaxy \citep{lang.etal.2014, Lokas.etal.2018}. Tidally induced bars can also buckle with different different strength \citep{Lokas2019}. In our study we find that flyby interactions affect neither the time nor the strength of bar buckling in our galaxy models.

	\begin{savequote}[100mm]
``The Milky Way is nothing else but a mass of innumerable stars planted together in clusters.''
\qauthor{\textbf{$-$ Galileo Galilei}}
\end{savequote}

\chapter[Excitation of Vertical Breathing Motion in Disc Galaxies by Tidally-induced Spirals in Fly-by Interactions]{Excitation of Vertical Breathing Motion in Disc Galaxies by Tidally-induced Spirals in Fly-by Interactions}
\label{chapter4}

\section{Introduction}
\label{sec:intro}
The second data release from the \gaia\ mission (hereafter \gaia\ DR2) has revealed the presence of large-scale bulk vertical motions ($\sim 10 \kms$ in magnitude) and the associated bending and breathing motions for stars in the Solar vicinity and beyond \citep{Gaia.Collaboration.2018}. The presence of such breathing motions, i.e., stars on both sides of the Galactic mid-plane moving coherently towards or away from it, has also been reported in various past Galactic surveys, for example, the SEGUE (Sloan Extension for Galactic Understanding and Exploration) survey \citep{Widrow.etal.2012}, the LAMOST (Large Sky Area Multi-Object Fibre Spectroscopic Telescope) survey  \citep{Carlin.etal.2013}, and the RAVE (Radial Velocity Experiment) data \citep{Williams.etal.2013}. The existence of such non-zero bulk vertical motions in the Milky Way raises questions about the plausible driving mechanism(s), since, in an axisymmetric potential, the bulk radial and vertical motions should be zero \citep[e.g.][]{BT08}.
\par
Much of the understanding of the excitation of breathing motions in Milky Way-like galaxies have been gleaned from numerical simulations. Using semi-analytic models and test-particle simulations, \citet{Faureetal2014} was the first to show that a strong spiral can drive large-scale vertical motions ($|\avg{v_z}| \sim 5-20 \kms$). The amplitude of such breathing motion increases at first with height from the mid-plane, and then starts to decrease after reaching its maximum value at a certain height. Also, using self-consistent $N$-body simulation, \citet{Debattista.2014} showed that a \textit{vertically-extended} spiral feature can drive strong large-scale breathing motions, with amplitude increasing monotonically from the mid-plane. The relative sense of these bulk motions, whether compressing or expanding, changes across the corotation resonance (hereafter CR) of the spiral \citep{Faureetal2014,Debattista.2014}. Furthermore, using a {\it self-consistent}, high-resolution simulation with star formation,  \citet{Ghosh.etal.2020} studied the age-dependence of such vertical breathing motions excited by spiral density waves. They showed that, at fixed height, the amplitude of such vertical breathing motion decreases with stellar age. They showed a similar age-variation in the breathing amplitude in the \gaia\ DR2, thereby supporting the scenario that the breathing motion of the Milky Way might well be driven by spiral density waves \citep{Ghosh.etal.2020}. Instead, \citet{Monari.etal.2015} showed that a stellar bar can also drive such breathing motion in disc galaxies. However, the resulting amplitudes of the breathing motions are small ($|\avg{v_z}| \sim 1 \kms$) when compared to the spiral-driven breathing amplitudes. 
\par
In the Lambda cold dark matter ($\Lambda$CDM) paradigm of hierarchical structure formation, galaxies grow in mass and size via major mergers and/or multiple minor mergers, and cold gas accretion \citep{WhiteandRees1978,Fall1980}. During the evolutionary phase, a galaxy also experiences multiple tidal interactions with satellites and/or passing-by companion galaxies. The frequency of such fly-by encounters increases at lower redshifts \citep[e.g., see][]{SinhaandBockelmann2015}, and their cumulative dynamical impact on the morphology as well as on the dynamics of the host galaxies can be non-negligible \citep[e.g., see][]{Anetal2019}.
 Fly-by encounters can excite an $m=2$ bar mode \citep[e.g., see][]{Noguchi1987,Lokasetal2016,Martinez-Valpuestaetal2017,Ghoshetal2021a}, off-set bars with a one-arm spiral \citep[e.g,][]{Pardyetal2016}, and an $m=1$ lopsidedness in the stellar disc \citep{Bournaudetal2005,Mapellietal2008,Ghoshetal2021b}. They can also trigger star formation \citep[e.g.,][]{Ducetal2018}, and if  the gas inflow towards the centre is exceedingly large, this can lead to a starburst \citep{MihosandHernquist1994} as well as triggering of AGN activity \citep{Combes2001}. Furthermore, the role of fly-bys has been investigated in the context of forming warps \citep{Kimetal2014,Semczuketal2020}, disc heating and disc thickening \citep{ReshetnikovandCombes1997, Kumar.etal.2021}, tidal bridges and streams \citep[e.g., see][]{ToomreToomre1972,DucandRenaud2013}, altering the galaxy spin \citep{ChoiandYi2017}, and the evolution of classical and pseudo-bulges \citep{Kumar.etal.2021}. 
 Our Galaxy has also experienced such a tidal interaction with the Sagittarius (Sgr) dwarf galaxy \citep[e.g., see][]{Majewskietal2003}. Recent studies have indicated that such a tidal interaction could excite a `snail-shell' structure (phase-space spiral), bending motions in the Solar neighbourhood \citep[for details see, e.g.,][]{Widrowetal2014,Antojaetal2018,Gaia.Collaboration.2018}. 
\par
A tidal encounter with another galaxy can excite spiral features, as was first proposed in the seminal work of \citet{Holmberg1941}. Later, pioneering  numerical work of \citet{ToomreToomre1972} showed that a tidal interaction can excite tidal tails, bridges, and spiral features in the disc of the host galaxy for a wide variety of orbital configurations. Following that, numerical simulation has become an indispensable tool to study the dynamical effect of galaxy interactions. Several past studies attempted to understand the role of tidal encounters in the context of excitation of spirals as well as to understand the longevity and nature of the resulting spirals \citep[e.g. see][]{Sundeliusetal1987,DonnerandThomasson1994,SaloandLaurikainen2000,dobbs.etal.2010,Pettittetal2018}. 
Furthermore, a recent study by \citet{Pettittetal2017} investigated star formation and the properties of interstellar medium in tidally-induced spirals. 
\par
Tidal interactions can also induce vertical distortions and oscillations in the disc of the host galaxy \citep[e.g., see][]{HunterandToomre1969,Araki1985,Mathur1990,Weinberg1991,VesperiniandWeinberg2000,Gomezetal2013,Widrowetal2014,Donghiaetal2016}. \citet{Widrowetal2014} proposed a dynamical scenario where a satellite galaxy, while plunging into the disc, can excite both bending and breathing motions. Interestingly, such tidal  interactions also excited a strong spiral response within the disc in their model (see Fig.~10 there). Therefore, it is still unclear whether the tidal interactions are `directly' responsible for driving breathing motion, or the tidally-induced spirals are driving the breathing motions.
\par
 We aim to test this latter hypothesis in detail in this chapter. We study a set of $N$-body models of galaxy fly-by interactions while varying the orbital parameters. We investigate the generation of the spiral features due to such fly-by encounters, and quantify the nature and longevity of such spirals in different fly-by models. We closely follow the generation of the vertical breathing motions and their subsequent evolution. In particular, we look for evidence that the generation and evolution of the vertical breathing motions is correlated with the temporal evolution of the tidally-induced spirals.
\par
 The rest of the chapter is organized as follows. Section~\ref{sec:sim_setup} provides the details of the simulation set-up and the fly-by models. Section~\ref{sec:spiral_quant} presents the results of the tidally-induced spirals, the density wave nature of spirals, and the temporal evolution of their strength. Section~\ref{sec:breathing} measures the properties of the vertical breathing motions driven by these tidally-induced spirals. Section~\ref{sec:discussion} discusses a few limitations of this work while section \ref{sec:conclusion} summarizes the main findings of this work.  

\section{Simulation set-up of galaxy Fly-by models}
\label{sec:sim_setup}

\begin{table*}
\centering
\caption{Key galaxy parameters for the equilibrium models of the host and the perturber galaxies.}
\resizebox{\columnwidth}{!}{%
\begin{tabular}{ccccccccccccc}
\hline
Galaxy & $M$$^{(1)}$ & $\lambda$$^{(2)}$ & $c$$^{(3)}$ & $f_{\rm disc}$$^{(4)}$ & $f_{\rm bulge}$$^{(5)}$ & $j_{\rm d}$$^{(6)}$ &  $R_{\rm d}$$^{(7)}$ & $z_0$$^{(8)}$ & $N_{\rm halo}$$^{(9)}$ & $N_{\rm disc}$$^{(10)}$ & $N_{\rm bulge}$$^{(11)}$ & $N_{\rm tot}$ $^{(12)}$\\
& ($\times 10^{12} M_{\odot}$) &&&&&& ($\kpc$) & ($\kpc$) & ($\times 10^{6}$) & ($\times 10^{6}$) & ($\times 10^{6}$) & ($\times 10^{6}$)\\
\hline
Host & 1.2 & 0.035 & 10 & 0.025 & 0.005 & 0.03 & 3.8 & 0.38 & 2.5 & 1.5 & 1 & 5 \\
Perturber & 0.24 & 0.035 & 11 & 0.01 & 0.002 & 0.01 & 1.95 & 0.195 & 0.5 & 0.3 & 0.2 & 1\\
\hline
\end{tabular}%
}
\centering
{ \hspace{0.9 cm} (1) total mass (in $M_{\odot}$); (2) halo spin; (3) halo concentration; (4) disc mass fraction; (5)  bulge mass fraction; (6) disc spin fraction; (7)  disc scale length (in $\kpc$); (8) disc scale height (in $\kpc$); (9) total DM halo particles; (10) total disc particles; (11) total bulge particles; (12) total number of particles used. }
\label{tab:initial_parameters}
\end{table*}

To motivate our study, we construct a set of $N$-body models of galaxy fly-bys where the host galaxy experiences an unbound interaction with a perturber galaxy. The mass ratio of the perturber and the host galaxy is set to 5:1, and is kept fixed for all the models considered here. A prototype of such a galaxy fly-by model is already presented in  \citet{Kumar.etal.2021}. Here, we construct a suite of fly-by models varying the orbital configuration (e.g., angle of interaction, orientation of the orbital spin vector). The details of the modelling and the simulation setup is discussed in  \citet{Kumar.etal.2021}. For the sake of completeness, here we briefly mention the equilibrium model of the galaxies as well as the orbital configurations of the galaxy interaction.

\subsection{Equilibrium models}
\label{sec:equilibrium_model}

The initial equilibrium model of each galaxy (host and the perturber) consists of a classical bulge, a stellar disc, and a dark matter (hereafter DM) halo. Each of the galactic components is treated as live, thereby allowing them to interact with each other. The DM halo is assumed to be spherically symmetric, and is modelled with a Hernquist density profile \citep{Hernquist1990} of the form 
\begin{equation}
    \rho_{\rm dm}(r)=\frac{M_{\rm dm}}{2\pi}\frac{a}{r(r+a)^3}\,,
    \label{eqn:halo}
\end{equation}
\noindent where $M_{\rm dm}$ and $a$ are the total mass and the scale radius of the DM halo, respectively. The scale radius of the Hernquist halo is related to the concentration parameter `$c$' of NFW DM halo \citep{NFW1996}. For an NFW DM halo with mass $M_{200}=M_{\rm dm}$, this relation is given by the following equation,
\begin{equation}
    a=\frac{r_{200}}{c}\sqrt{2\left[\ln{(1+c)-\frac{c}{(1+c)}}\right]},
    \label{eqn:scale_radius}
\end{equation}
where $r_{200}$ represents the radius of an NFW halo\footnote{It is defined as the radius from the centre of the halo inside which the mean density is 200 times the critical density of the Universe.}, and the mass within this radius is defined as $M_{200}$. The classical bulge is also assumed to be spherically symmetric and is modelled with another Hernquist density profile \citep{Hernquist1990} of the form
\begin{equation}
    \rho_{\rm b}(r)=\frac{M_{\rm b}}{2\pi}\frac{b}{r(r+b)^{3}}\,,
    \label{eqn:bulge}
\end{equation}
where $M_{\rm b}$ and $b$ represent the total bulge mass and the bulge scale radius, respectively.  The initial radial surface density of the stellar disc follows an exponential fall-off and has a $\mathrm {sech}^2$ profile along the vertical direction, thereby having the form 
\begin{equation}
   \rho_{d}(\rm R,\rm z)=\frac{M_{\rm d}}{4\pi z_{0} R_{\rm d}^{2}}\exp\left(-{\frac{R}{R_{\rm d}}}\right) \sech^{2}\left(\frac{z}{z_{0}}\right), 
    \label{eqn:disc}
\end{equation}
\noindent where $M_{\rm d}$ is the total mass, $R_{\rm d}$ is the exponential disc scale length, and  $z_{0}$ is the scale height. The corresponding values of the structural parameters, used to model the host as well as the perturber galaxy, are listed in Table~\ref{tab:initial_parameters}.
\par
\begin{figure*}
  \begin{multicols}{2}
	\includegraphics[width=0.95\linewidth]{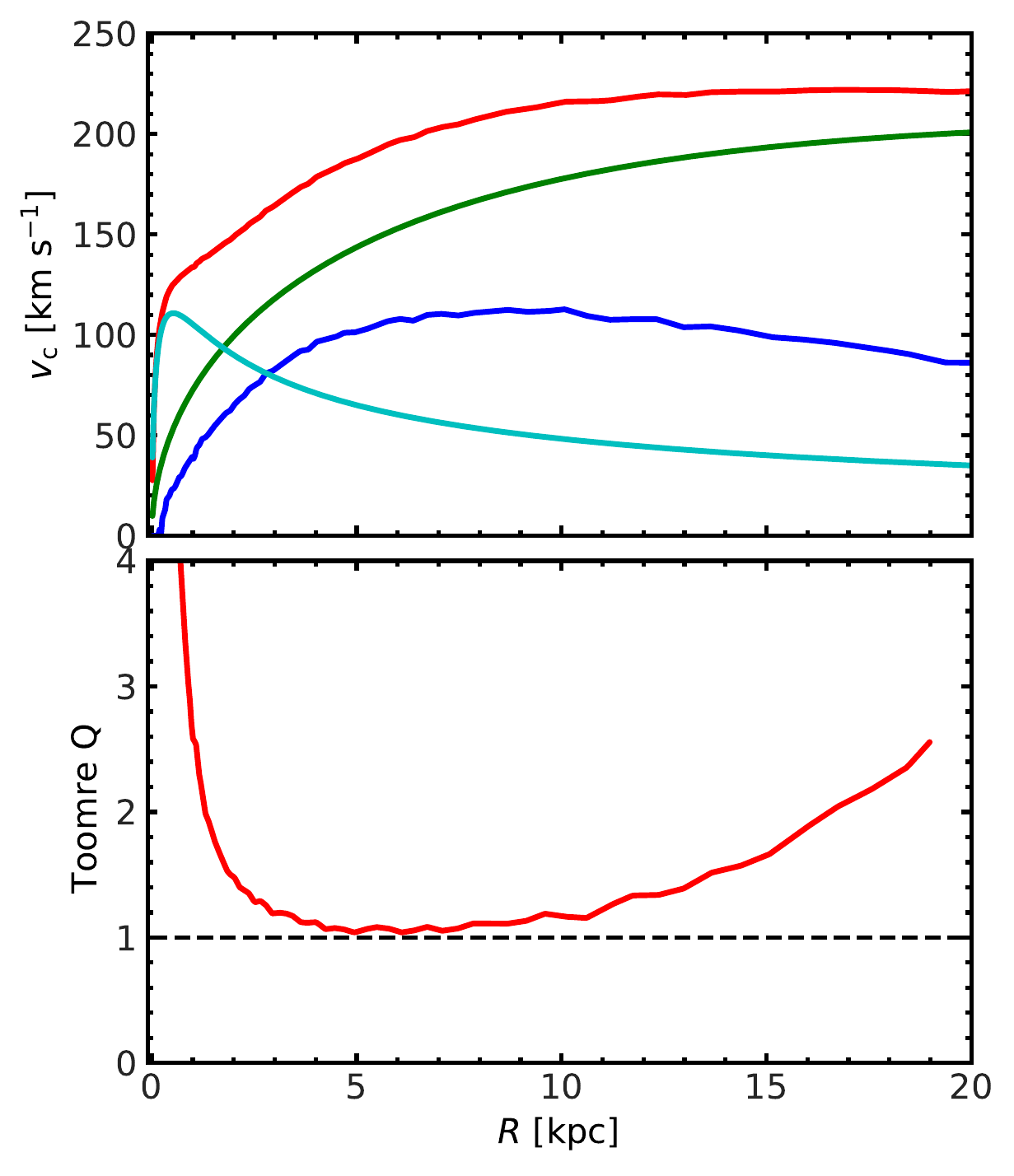}\par
	\includegraphics[width=0.95\linewidth]{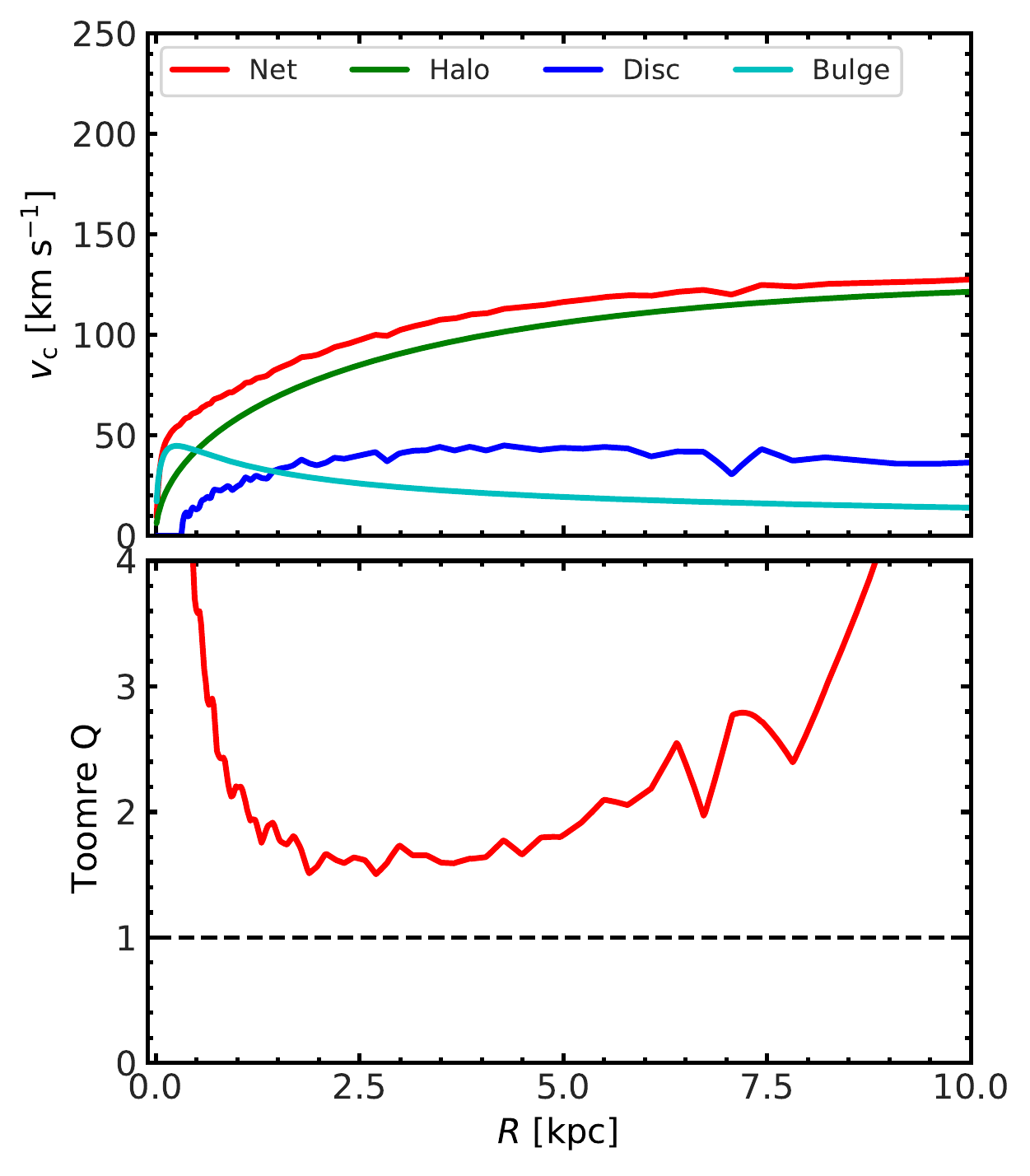}\par
	\end{multicols}
    \caption{Radial profiles of the circular velocity ($v_{\rm c}$) and the Toomre $Q$ parameter are shown for equilibrium models of the host (left-hand panels) and the perturber galaxy (right-hand panels). In the top panels, the blue line denotes the contribution of the stellar disc while the DM halo contribution is shown by the green line. The bulge contribution is shown by the cyan solid line whereas the red line denotes the total/net circular velocity.}
    \label{fig:equilibrium_dynamics}
\end{figure*}

The equilibrium models for the host as well as the perturber are generated using the publicly distributed code {\sc galic} \citep{Yurin2014}. This code uses elements of the Schwarzschild's method and the made-to-measure method to search for a stable solution of the collisionless Boltzmann Equation (CBE) for a set of collisionless stellar particles, initialized by some predefined analytic density distribution functions \citep[for details, see][]{Yurin2014}. A total of $5 \times 10^6$ particles are used to model the host galaxy whereas a total of $1 \times 10^6$ particles are used to model the perturber galaxy. The number of particles used to model each of the galaxy components of the host and the perturber galaxy are also listed in Table~\ref{tab:initial_parameters}. The stellar particles have gravitational softening $\epsilon = 20 \pc$ while the DM halo particles have $\epsilon = 30 \pc$. Fig.~\ref{fig:equilibrium_dynamics} shows the corresponding radial profiles of the circular velocity ($v_{\rm c}$) and the Toomre $Q$ parameter for the host and the perturber galaxy at $t =0$.

\subsection{Set-up of galaxy fly-by scenario}
\label{sec:flyby_setup}

\begin{table}
\centering
\caption{Key orbital parameters for the galaxy fly-by models.}
\label{tab:sims_name}
\begin{threeparttable}
\begin{tabular}{lccccr}
\hline
Model\tnote{(a)} & $r_{\rm p}\tnote{(b)}$ & $i^{\degrees}\tnote{(c)}$ & $t_{\rm p}\tnote{(\rm d)}$ & $t_{\rm enc}\tnote{(e)}$ & $T_{\rm p}\tnote{(f)}$ \\
& (kpc) & & (Gyr) & (Gyr)\\
\hline
$RP40i00pro$ & 53.09 & 0 & 0.88 & 0.1298 & -4.1348\\
$RP40i30pro$ & 52.34 & 30 & 0.85 & 0.1272 & -4.1162\\
$RP40i60pro$ & 52.14 & 60 & 0.85 & 0.1265 & -4.1113\\
$RP40i90pro$ & 52.23 & 90 & 0.85 & 0.1267 & -4.1133\\
$RP40i00ret$ & 53.07 & 0 & 0.85 & 0.1298 & -4.1342\\
\hline
\end{tabular}
\centering
{(a) Galaxy fly-by model; (b) pericentre distance (in kpc); (c) orbital angle of interaction (in degree); (d) time of pericentre passage (in Gyr); (e) encounter time (in Gyr); (f) tidal parameter.}

\end{threeparttable}
\end{table}

To simulate the unbound galaxy fly-by scenario, we place our galaxy models on a hyperbolic orbit with eccentricity, $e=1.1$ so that the orbit of the perturber galaxy remains unbound throughout the interaction. We avoid choosing a parabolic orbit as the dynamical friction of the host galaxy decays the orbit of the perturber galaxy and puts the perturber galaxy on a bound elliptical orbit. Our choice of hyperbolic orbit avoids a bound fly-by interaction. We place the galaxies at an initial separation of $255 \kpc$ before the start of the simulation. For different models with different orbital configurations, we vary the distance of their closest approach (the pericentre distance) assuming the two-body Keplerian orbit.  
For further details of the orbital configuration and the geometry of the unbound fly-by scenario, the reader is referred to \citet{Kumar.etal.2021}. A total of five such galaxy fly-by models are used for this study. The angle of interaction, and the pericentre distance for these models are listed in Table~\ref{tab:sims_name}.
\par
All the simulations are run using the publicly available code {\sc{Gadget-2}} \citep{Springel2001, Springal2005man, Springel2005} for a total time of $6 \Gyr$, with a tolerance parameter $\theta_{\rm tol} = 0.5$ and an integration time-step $0.4 \Myr$. The maximum error in the angular momentum of the system is well within $0.15$ percent throughout the evolution for all models considered here.
\par
Following \citet{dimatteo.etal.2007}, we also calculate the encounter time ($t_{\rm enc}$), and tidal parameter ($T_{\rm p}$) for the host galaxy by using 
 \begin{equation}
    t_{\rm enc} = \frac{r_{\rm p}}{v_{\rm p}},\\
    T_{\rm p} = \log_{10} \left[\frac{m_{\rm per}}{M_{\rm host}} \left(\frac{R_{\rm d}}{r_{\rm p}} \right)^{3}\right],
\end{equation}
\noindent where $r_{\rm p}$ is pericentre distance and $v_{\rm p}$ is the corresponding relative velocity. $M_{\rm host}$ and $m_{\rm per}$ are the masses of the host and the perturber galaxy, respectively. The corresponding values of $t_{\rm enc}$, and $T_{\rm p}$ for all the models consider here, are listed in Table~\ref{tab:sims_name}. We note that the quantities mentioned here will be in units of the dimensionless parameter `$h$', defined via the Hubble constant ($H_{0}=100 h$ km s$^{-1}$ Mpc$^{-1}$) and can be scaled to observed values.
 \par
 Each model is referred as a unique string given by `{\sc [pericentre distance][angle of interaction][orbital spin]}' where {\sc [pericentre distance]} denotes the pericenter distance, obtained using the standard two-body formalism. {\sc [angle of interaction]} denotes the angle at which the perturber encounters with the host galaxy while {\sc [orbital spin]} denotes the orbital spin vector (`pro' for prograde and `ret' for retrograde orbits). We follow this scheme of nomenclature throughout the chapter.  As an example, RP40i30pro denotes a fly-by model where  the perturber galaxy interacts with the host galaxy at an angle of $30 \degrees$ in a prograde orbit, and the calculated pericentre distance between these two galaxies is $40 \kpc$, obtained by using the standard two-body  Keplerian orbit.

\section{Quantification of tidally-induced spirals in galaxy fly-bys}
\label{sec:spiral_quant}
%
\begin{figure*}
    \centering
	\includegraphics[width=\textwidth]{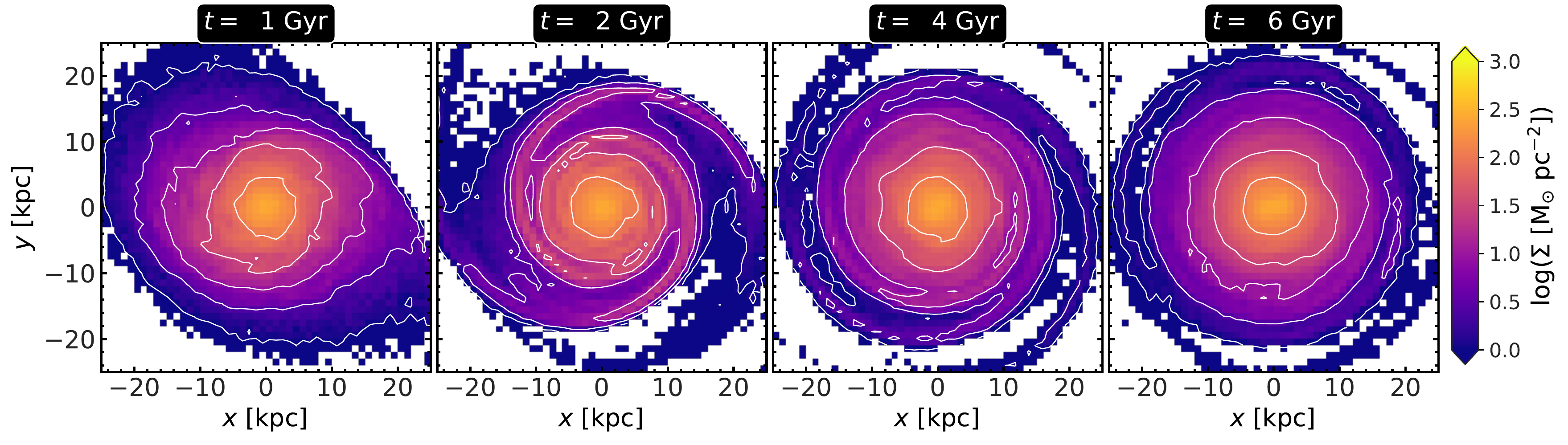}
	\includegraphics[width=\textwidth]{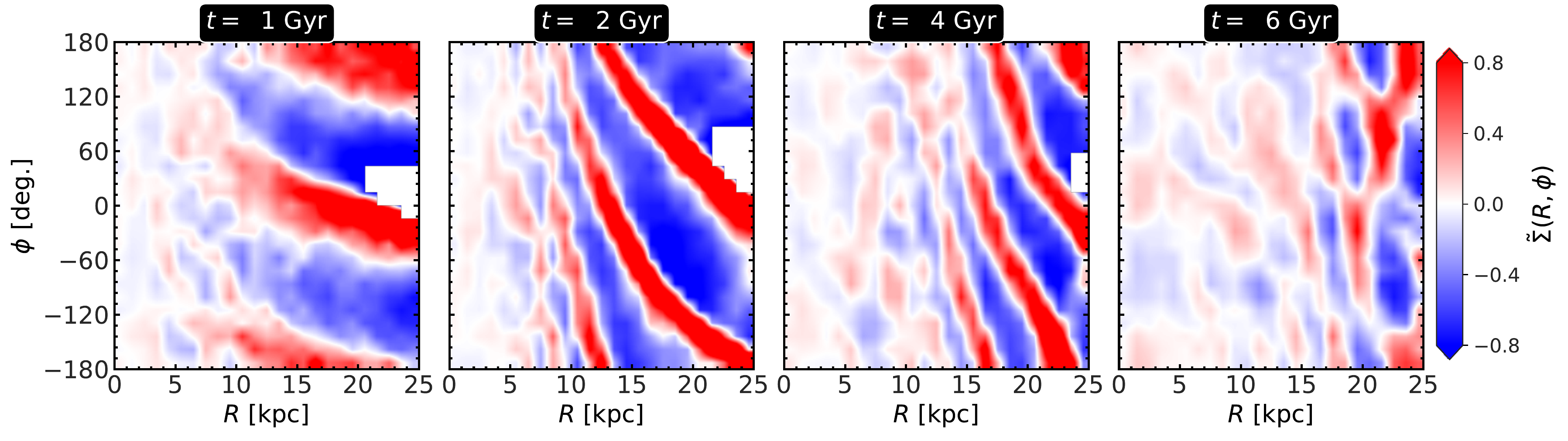}
    \caption{ Top: the density distribution of the stars of the host galaxy in the face-on projection, at four different times after the pericentre passage, for the model  $RP40i00pro$. White solid lines show the contours of constant surface density. A strong spiral feature is excited in the disc of the host galaxy after it experiences a tidal encounter with the perturber galaxy. Bottom: the corresponding distribution of the residual surface density ($\tilde \Sigma (R, \phi)$), calculated using Eq.~\ref{eqn:res_dens}, at the same four different times. The sense of rotation is towards increasing $\phi$.}
    \label{fig:dens_map}
\end{figure*}
Here, we investigate the excitation of spiral structure in the host galaxy as a consequence of the tidal interaction with the perturber galaxy. We first study the model $RP40i00pro$, where the host galaxy experiences an in-plane, unbound fly-by interaction with the perturber galaxy (with mass $1/5$th that of the host galaxy). During the pericentre passage, the perturber exerts a strong tidal pull on the host galaxy. Fig.~\ref{fig:dens_map} (top panels) shows the face-on density distribution of the stars of the host galaxy at four different times, after the tidal interaction occurs. A visual inspection reveals that after the interaction happens, a strong spiral feature is excited in the outer parts of the host galaxy (e.g., at $t = 2 \Gyr$). The spiral features are also seen at later times. However, by the end of the simulation ($t = 6 \Gyr$), there is no discernible, strong spiral features left in the host galaxy.
To study this trend further, we calculate the residual surface density ($\tilde \Sigma(R, \phi)$) in the $(R, \phi)$-plane using 
\begin{equation}
    \tilde \Sigma (R, \phi) = \frac{\Sigma (R, \phi) - \Sigma_{\rm avg} (R)}{\Sigma_{\rm avg} (R)},
    \label{eqn:res_dens}
\end{equation}
\noindent where $\Sigma_{\rm avg} (R)$ is the azimuthally-averaged surface density of the disc at radius $R$. This is shown in  Fig.~\ref{fig:dens_map} (bottom panels). As seen clearly, after the interaction happens, a strong, coherent spiral feature, denoted by the presence of a periodic over- and under-density, is excited in the outer region ($R \geq 10 \kpc$) of the host galaxy. At the end of the simulation ($t = 6 \Gyr$), the corresponding residual density distribution does not exhibit any coherent spiral structure in the disc of the host galaxy.

\subsection{Strength and temporal evolution of spirals}
\label{sec:strength_spiral}

\begin{figure}
    \centering
	\includegraphics[width=0.7\textwidth]{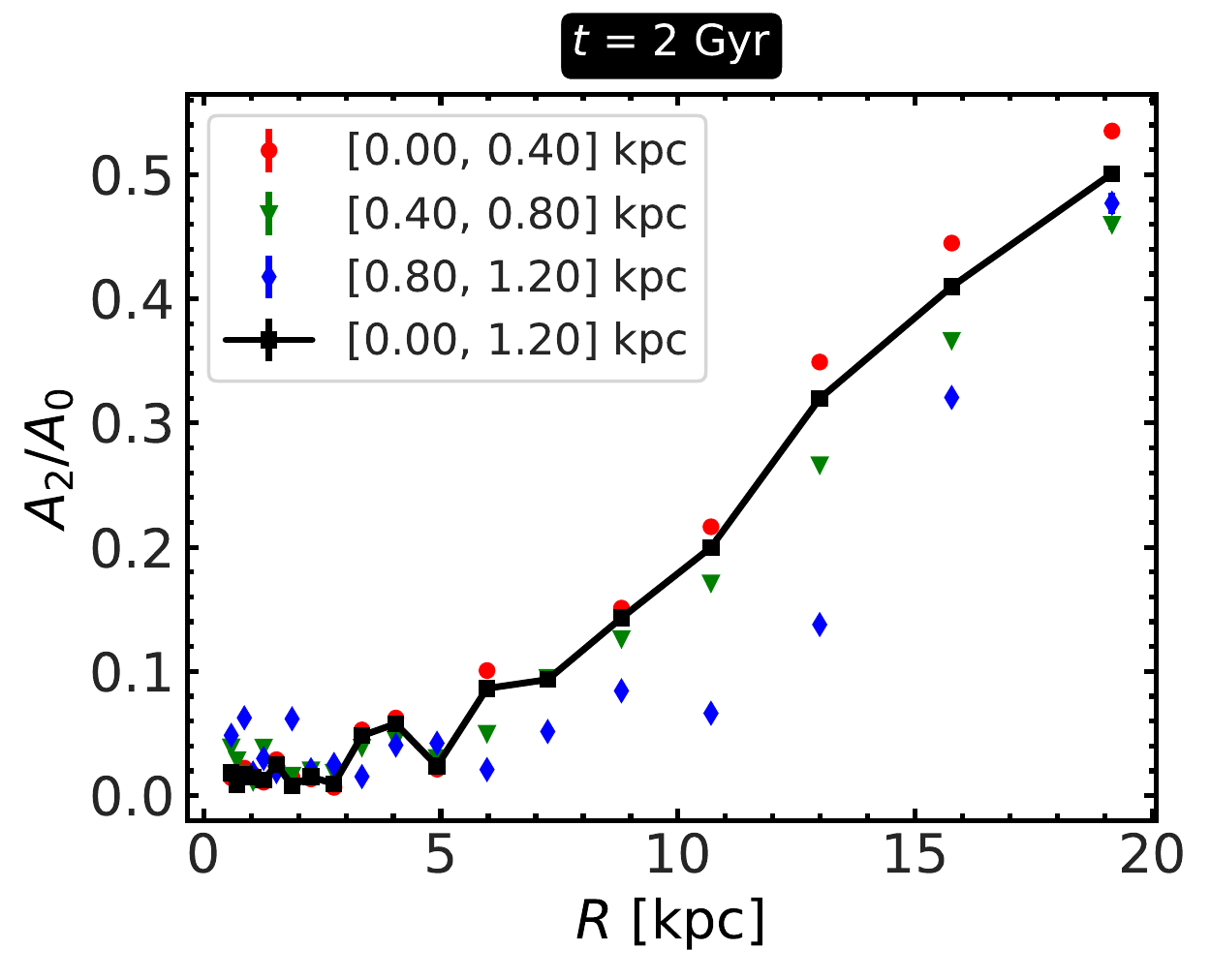}
    \caption{Radial variation of the Fourier coefficient of the $m=2$ Fourier component (normalised by the $m=0$ component) is shown at $t = 2 \Gyr$ for the model $RP40i00pro$ (see the black solid line). The same quantity is then measured for stars in different vertical layers (as indicated in the legend).}
    \label{fig:m2_strength_dz_0.4}
\end{figure}

In the previous section, we have shown that a tidal interaction with a perturber galaxy excites a prominent spiral feature in the outer disc region of the host galaxy for the model $RP40i00pro$. Next, we quantify the strength of the spiral and follow its temporal evolution. For this, we first calculate the radial variation of the Fourier moment of the surface density of the stellar particles of the host galaxy using 
\begin{align}
A_m/A_0 (R)= \left|\frac{\sum_j m_j e^{im\phi_j}}{\sum_j m_j}\right|\,,
    \label{eqn:fourier_mode}
\end{align}
\noindent where $A_m$ is the coefficient of the $m$th Fourier moment of the density distribution, $m_j$ is the mass of the $j$th particle \citep[e.g., see][]{Kumar.etal.2021, Kumar.etal.2022}. Fig.~\ref{fig:m2_strength_dz_0.4} shows the corresponding radial variation of the $m=2$ Fourier moment at $t = 2 \Gyr$ for the model $RP40i00pro$. In the outer disc regions, there are less particles when compared to the inner disc regions. Therefore, using a \textit{linearly-spaced} radial binning when calculating the Fourier coefficient (using Eq.~\ref{eqn:fourier_mode}) introduces noise in the calculation for the outer regions. To avoid that, we employ a logarithmic binning in the radial direction. As seen clearly, in the outer parts ($R \geq 10 \kpc$), the values of the coefficient $A_2/A_0$ are non-zero, indicating the presence of a strong spiral structure. Are these tidally-induced spirals mostly confined to the disc mid-plane or are they vertically-extended? To investigate this further, we calculate the radial variation of the same Fourier coefficient $A_2/A_0$, but for stars in different vertical layers of thickness $400 \pc$. The resulting radial variations are also shown in Fig.~\ref{fig:m2_strength_dz_0.4}. As seen clearly from Fig.~\ref{fig:m2_strength_dz_0.4}, the Fourier coefficient $A_2/A_0$ shows non-zero values even for stars at the largest heights from the mid-plane ($|z| = [0.8, 1.2] \kpc$). Also, at a certain radius $R$ within the extent of the spirals, the values of the Fourier coefficient $A_2/A_0$ decreases monotonically as one moves farther away from the mid-plane. We checked this variation of the Fourier coefficient $A_2/A_0$ with height at other time-steps as well, and found that this trend remains generic whenever the tidally-induced spirals are strong in the disc of the host galaxy. This demonstrates that the tidally-induced spirals in our model $RP40i00pro$ is \textit{vertically-extended}, similar to what was reported in \citet{Debattista.2014} and \citet{Ghosh.etal.2020}. 
\par
\begin{figure}
    \centering
	\includegraphics[width=0.9\textwidth]{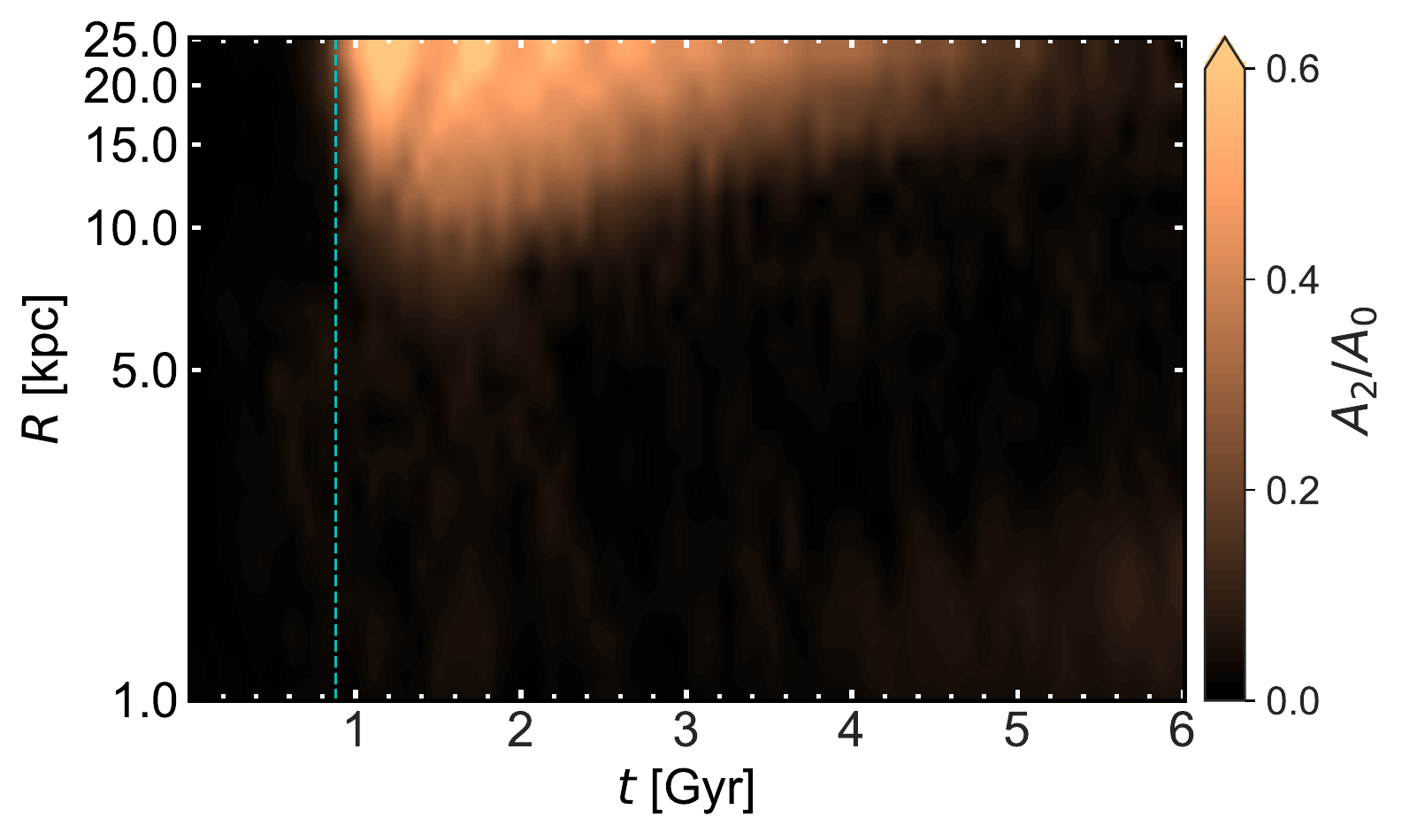}
    \caption{The evolution of the coefficient of the $m=2$ Fourier moment ($A_2/A_0$) in the $R-t$ space for the model $RP40i00pro$. A logarithmic binning is employed along the radial direction, for details see text. The colour bar shows the values of $A_2/A_0$. The vertical dashed line (in cyan) denotes the time of the pericentre passage of the perturber.}
    \label{fig:spiral_strength_t-R}
\end{figure}

To further investigate the spatio-temporal evolution of the tidally-induced spirals in model $RP40i00pro$, we calculate the Fourier coefficient $A_2/A_0 (R)$ at different radial locations for the whole simulation run-time (a total of $6 \Gyr$). This is shown in Fig.~\ref{fig:spiral_strength_t-R}. The tidal interaction with the perturber excites a strong spiral feature in the disc of the host galaxy after $t \sim 1.1 \Gyr$ or so. These spirals remain mostly in the outer regions of the host's disc ($R \geq 10 \kpc$), as the values of $A_2/A_0$ in the inner part ($R \leq 10 \kpc$) are almost zero (see Fig.~\ref{fig:spiral_strength_t-R}). After $t \sim 3 \Gyr$ or so, the spiral starts to decay, as shown by the decreasing values of the $A_2/A_0$. By the end of the simulation run, the values of the $A_2/A_0$ become almost zero, implying that the tidally-induced spirals have wound up almost completely. 
\par
\begin{figure}
    \centering
	\includegraphics[width=0.7\textwidth]{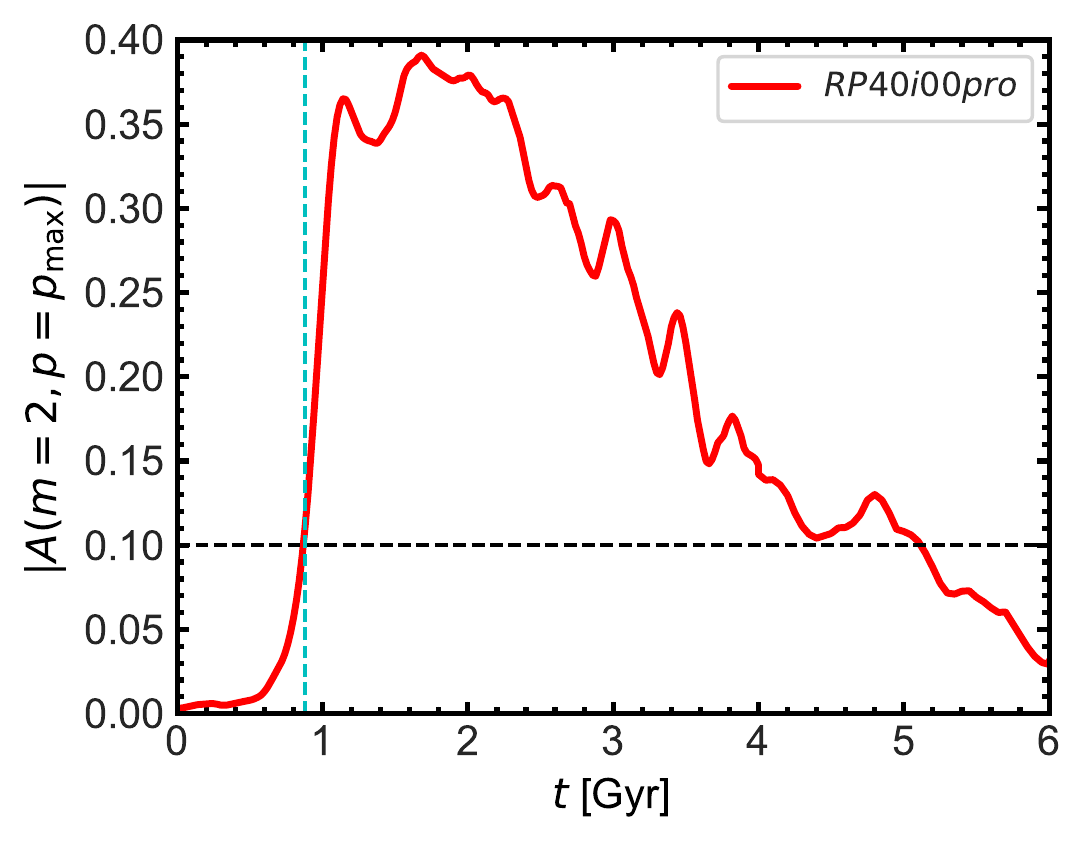}
    \caption{Temporal evolution of the spiral strength ($|A(m=2, p =p_{\rm max})|$), calculated using Eq.~\ref{eqn:spiral_strenght} for the model $RP40i00pro$. The tidally-induced spirals decay by the end of the simulation run. The horizontal black dotted line denotes  $|A(m=2, p =p_{\rm max})| = 0.1$, used as an operational definition for the onset of the spirals, see text for details. The vertical dashed line (in cyan) denotes the time of the pericentre passage of the perturber.}
    \label{fig:spiral_strength_temporal}
\end{figure}

Next, we quantify the temporal evolution of the strength of the tidally-induced spirals in our fly-by model $RP40i00pro$. Following \citet{Sellwood1984}, \citet{Sellwood1986}, \citet{Puerari.etal.2000}, we define
\begin{equation}
    A(m,p)=\frac{\sum_{j}^{N} m_{j} \exp[i(m\phi_{j}+p\ln R_{j})]}{\sum_{j}^{N} m_{j}},
    \label{eqn:spiral_strenght}
\end{equation}
\noindent where $|A(m, p)|$ is the amplitude, $m_{j}$ is the mass of $j^{th}$ star, $m$ is the spiral arm multiplicity, $(R_{j},\phi_{j})$ are the polar coordinates of the $j^{th}$ star in the plane of the disc, and $N$ is the total number of stellar particles in the annulus $R_{\rm min} \leq R \leq R_{\rm max}$ within which the spiral feature exists and/or is most prominent. Here, we take $R_{\rm min} = 3R_{\rm d}$, and $R_{\rm max} = 6 R_{\rm d}$, where $R_{\rm d} = 3.8 \kpc$. For this annular region, we estimate $A(m,p)$ as a function of $p$ for $p\in[-50, 50]$, with a fixed step of $dp=0.25$  \citep[as suggested by][]{Puerari.etal.2000} for different values of $m$. Then, we evaluate the parameter $p_{\rm max}$ which corresponds to the maximum value of $|A(m,p)|$.  We find that the amplitude $|A(m,p = p_{\rm max})|$ shows a maximum value for $m=2$,   indicating that the $m=2$ spiral is the strongest. Therefore, at a certain time $t$,  we define the amplitude $|A(m=2, p =p_{\rm max})|$ as the strength of spirals \citep{Sang-Hoon.etal.2015, Semczuk.etal.2017, Kumar.etal.2021}. The resulting temporal evolution of the strength of spirals for the model $RP40i00pro$ is shown in Fig.~\ref{fig:spiral_strength_temporal}. As seen from Fig.~\ref{fig:spiral_strength_temporal}, the tidally-induced spirals grow for some time after the interaction happens, then remain stable for about $1 \Gyr$ before weakening from around $t = 3 \Gyr$. By the end of the simulation, the spirals' strength becomes almost zero. For quantifying the longevity of the spirals, we define $|A(m=2, p =p_{\rm max})| = 0.1$ as the onset of the strong spiral perturbation\footnote{The threshold value of 0.1 is used purely as an operational definition for the onset of the spirals.}. Therefore, the spirals persist for a time of $\sim 4.2 \Gyr$ (after their formation) for the model $RP40i00pro$ (also see Fig.~\ref{fig:spiral_strength_temporal}).
\begin{figure}
    \centering
	\includegraphics[width=0.5\textwidth]{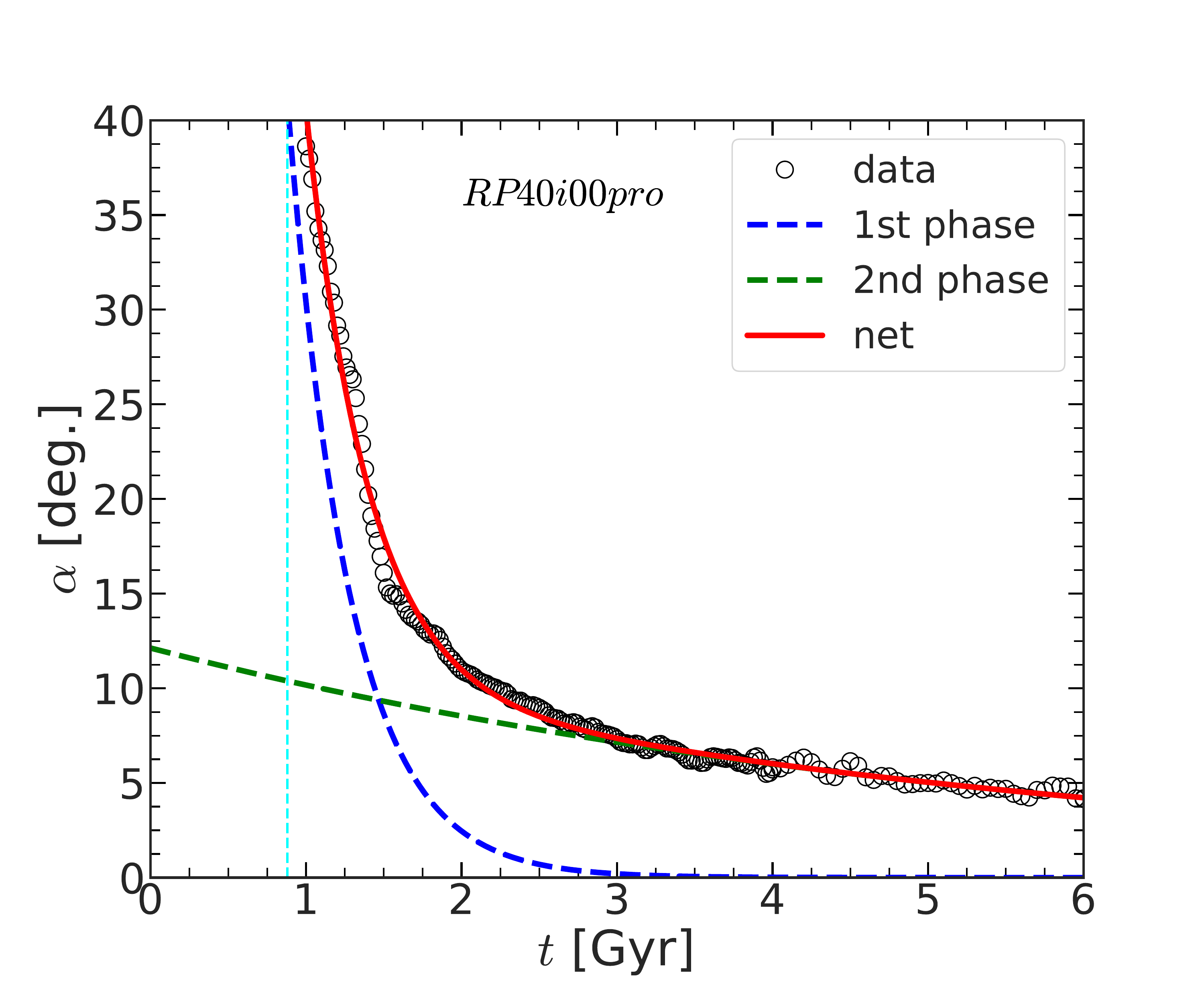}
    \caption{Temporal evolution of the pitch angle ($\alpha$) of the tidally-induced spirals is shown for the model $RP40i00pro$. A double-exponential profile (see Eq.~\ref{eq:winding_modelling}) is fitted to model the temporal evolution. The blue dashed line denotes the initial rapid winding phase whereas the green dashed line denotes the subsequent slow winding phase, for details see text. The red solid line denotes the best-fit double-exponential profile. The vertical dashed line (in cyan) denotes the time of the pericentre passage of the perturber.}
    \label{fig:winding_spiral}
\end{figure}

\subsection{Winding of the tidally-induced spirals}
\label{sec:winding_spiral}

Lastly, we investigate the winding of the tidally-induced spirals in our fly-by model $RP40i00pro$. As shown in past $N$-body simulations of galactic discs \citep[e.g. see][]{oh.etal.2008, Struck.etal.2011, Kumar.etal.2021}, a spiral arm can wind up with time. Following \citet{Sang-Hoon.etal.2015}, and \citet{Semczuk.etal.2017}, at a certain time $t$, we define the pitch angle, $\alpha$, as $\alpha=\tan^{-1}(m/p_{\rm max})$ using Eq.~\ref{eqn:spiral_strenght}. The resulting temporal evolution of the pitch angle for the model $RP40i00pro$ is shown in Fig.~\ref{fig:winding_spiral}. As revealed in Fig.~\ref{fig:winding_spiral}, the temporal evolution of the pitch angle displays two distinct phases, namely, the initial rapid winding phase where the pitch angle decreases sharply, and the subsequent slow winding phase where the pitch angle decreases less drastically. To model the temporal evolution of the pitch angle, we fit a double-exponential profile having the form
\begin{equation}
\alpha = \alpha_{1} \exp \left[-\lambda_1 (t - t_0) \right] + \alpha_{2} \exp \left[-\lambda_2 (t - t_0) \right]\,,
\label{eq:winding_modelling}
\end{equation}
\noindent where $\alpha_1$, $\alpha_2$, $\lambda_1$, and $\lambda_2$ are free parameters. Here, $t_0 = 1 \Gyr$, and denotes the time of the spirals' formation. The fitting is performed via the {\sc scipy} package {\sc curvefit} which uses the  Levenberg-Marquardt algorithm. The resulting best-fit double exponential profile is shown in Fig.~\ref{fig:winding_spiral}. We define the winding time-scale, $\tau_{\rm wind}$, as $\tau_{\rm wind} = \alpha/|\dot \alpha|$ where $|\dot \alpha|$ denotes the (absolute) rate of change of the pitch angle with time. For the initial rapid winding phase, $\tau_{\rm wind} \simeq 1/\lambda_1$ whereas for the subsequent slow winding phase $\tau_{\rm wind} \simeq 1/\lambda_2$ (see Eq.~\ref{eq:winding_modelling}). We find the best-fit values as $\lambda_1 =  2.51 \pm 0.05$ Gyr$^{-1}$ , and $\lambda_2 = 0.18 \pm 0.01$ Gyr$^{-1}$ which translate to a winding time-scale $\tau_{\rm wind} = 0.4 \Gyr$ for the initial rapid winding phase, and a winding time-scale $\tau_{\rm wind} = 5.7 \Gyr$ for the subsequent slow winding phase for the model $RP40i00pro$.

\subsection{Nature of tidally-induced spirals}
\label{sec:nature_spirals}

%
\begin{figure}
    \centering
	\includegraphics[width=0.6\textwidth]{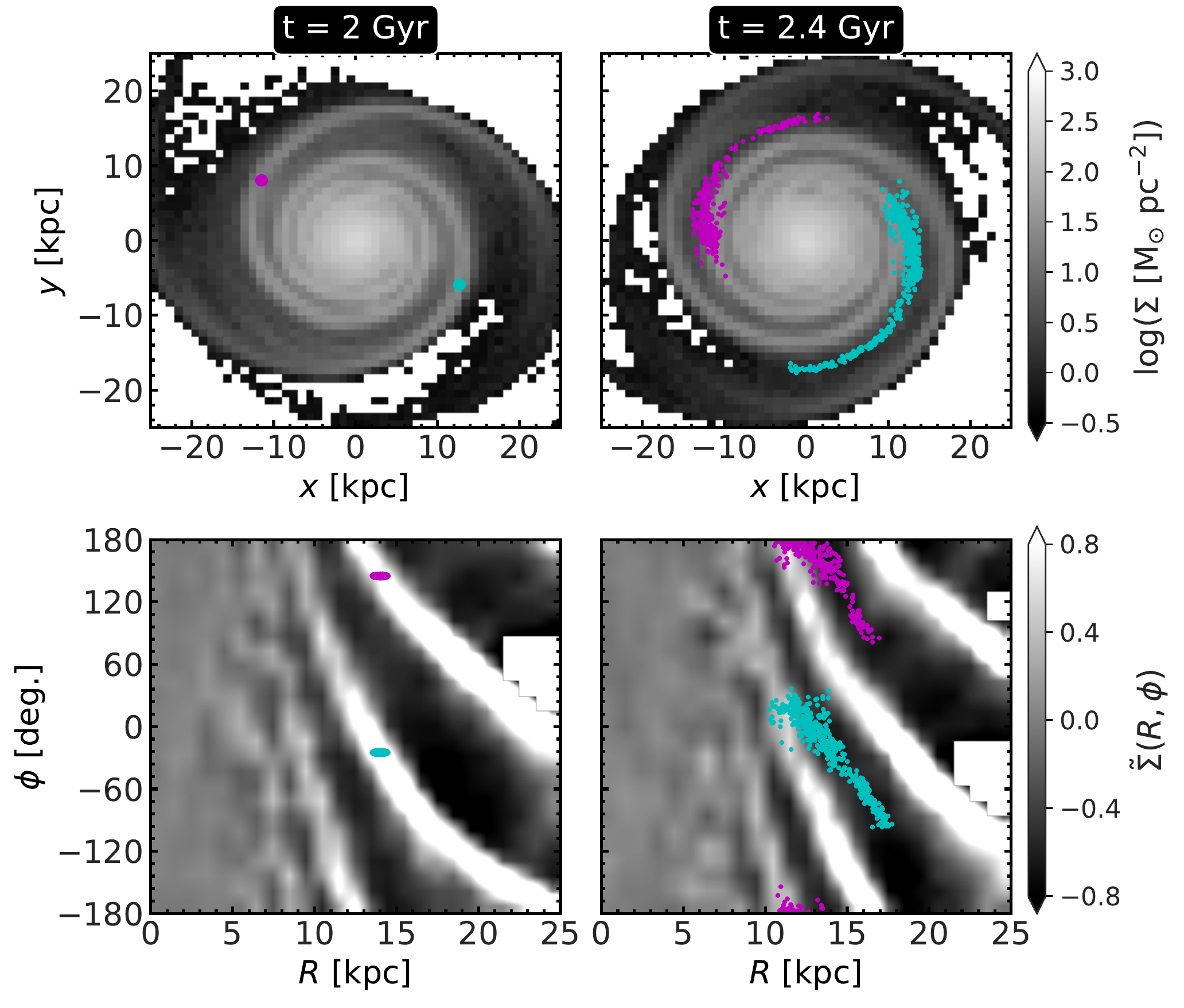}
    \caption{\textit{Nature of spirals:} top and bottom panels of the first column show the surface density and residual density maps of $RP40i00pro$ model at $t=2 \Gyr$ over-plotted with two patches (in magenta and cyan dots) of stellar particles in fully developed spiral arms. Right column is the analogous to the left column, but for $t=2.4 \Gyr$. Both patches of stars, initially associated along the spirals, have subsequently sheared out by the differential rotation, and left the spirals.}
    \label{fig:spiral_nature}
\end{figure}

So far, we have demonstrated that a fly-by interaction with a perturber induces a strong spiral feature in the outer disc of the host galaxy. However, the question remains whether these spirals are density waves or material arm in nature. For a comprehensive review on the nature of the spirals in disc galaxies, the reader is referred to \citet{BT08}, and \citet{DobbsandBaba2014}.
\par
For the set of the $N$-body models we are using here, we  do not have the age information of the stellar particles. This, in turn, restricts us from dividing stellar particles into different age-bins to trace the existence of spirals in different stellar population with different ages, as previously done in \citet{Ghosh.etal.2020}. Therefore, following \citet{Grand.etal.2012, D'Onghia.etal.2013}, we test the nature of the tidally-induced spirals by following the stars which are located on spirals arms at a certain time. We chose a time, say $t = 2 \Gyr$ when the spirals are fully-developed for the model $RP40i00pro$, and select two small patches (shown in magenta and cyan) of stars along the arms. This is shown in Fig.~\ref{fig:spiral_nature} (top left-hand panel). Now, if the spirals are of material arm in nature, then the stars would not leave the spiral arm at a subsequent time. To check that, we follow the selected stars at a later time, $t = 2.4 \Gyr$ (see top right-hand panel of Fig.~\ref{fig:spiral_nature}). As seen clearly from Fig.~\ref{fig:spiral_nature}, the stars, initially concentrated in small patches along the spiral arm, have sheared out due to the underlying differential rotation. This shearing out of the stars is more prominent in the distribution shown in the $(R, \phi)$-plane (see bottom panels of Fig.~\ref{fig:spiral_nature}). Interestingly, the stars, initially contained in the patches have left the spirals at a subsequent time, and the pitch angle of the selected stars is different from that of the spirals at $t = 2.4 \Gyr$. Furthermore, we calculate the pattern speed of the spirals at different radial locations in the disc. We find that the pattern speeds of the spirals are very different from the values of the circular frequency ($\Omega$), calculated at different radial locations. For brevity, these are not shown here. This shows that the spirals present in our fly-by model $RP40i00pro$ are density waves in nature.

\subsection{Dependence on orbital parameters}
\label{sec:orbital_variation}
%
\begin{figure}
    \centering
	\includegraphics[width=0.7\textwidth]{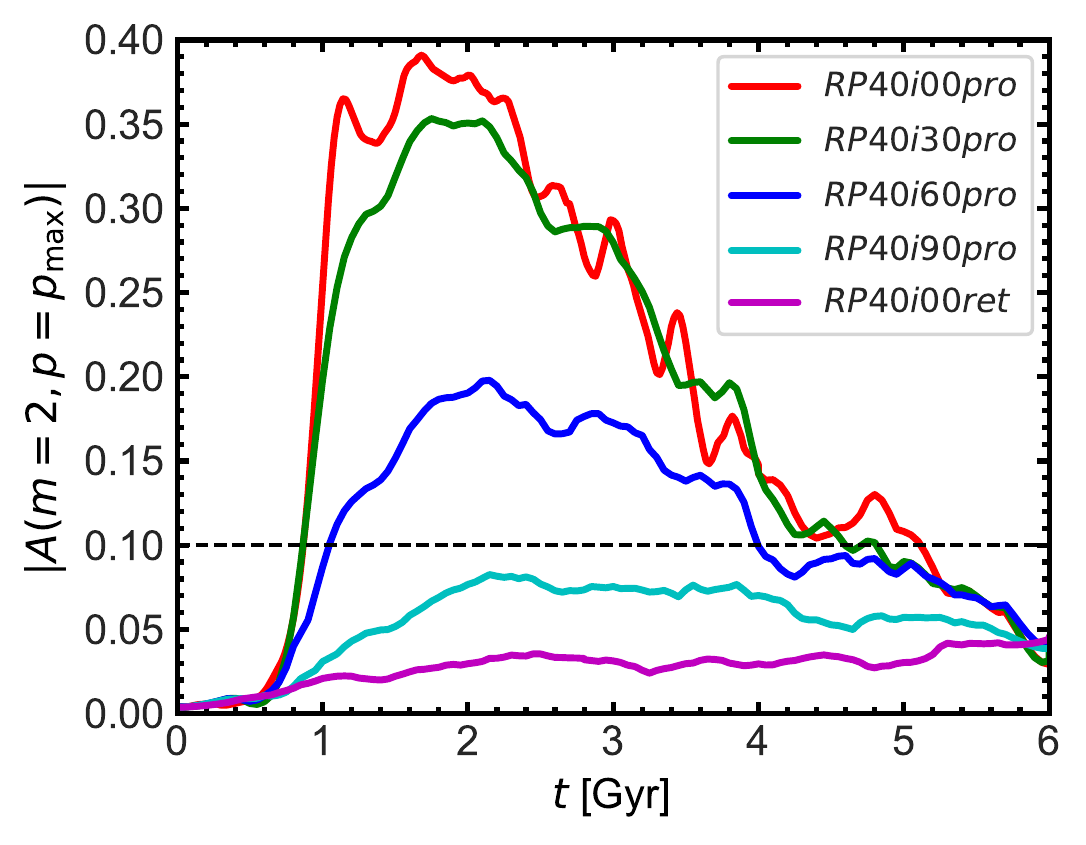}
    \caption{Temporal evolution of the spiral strength ($|A(m=2, p =p_{\rm max})|$), calculated using Eq.~\ref{eqn:spiral_strenght} is shown for models with different angle of interaction, and orbital spin vector. For comparison, we kept the model $RP40i00pro$ here (see red solid line). For models with prograde configuration, the strength of spirals decreases monotonically from the co-planar ($i =0 \degrees$) to polar ($i =90 \degrees$) orbital configuration. The horizontal black dotted line denotes  $|A(m=2, p =p_{\rm max})| = 0.1$, used as an operational definition for the onset of the spirals.}
    \label{fig:spiral_strength}
\end{figure}
Here, we explore the generation of the spirals in the host galaxy due to a tidal interaction with the perturber galaxy for different angles of interaction of the tidal encounter as well as for different orbital spin vectors (prograde or retrograde). Intuitively, the response of the host galaxy will be different for polar ($i =90 \degrees$) and co-planar ($i =0 \degrees$) orbital configurations. First we consider the other prograde models, with angles of interaction $30 \degrees$ and $60 \degrees$ (for details see section~\ref{sec:flyby_setup}). Both the models exhibit a similar trend of excitation of tidally-induced spirals in the disc of the host galaxy, as in model $RP40i00pro$. Shortly after the tidal encounter happens, the disc of the host develops a spiral feature which grows for a certain time, and then starts decaying. We calculate the strength of the spirals using Eq.~\ref{eqn:spiral_strenght} at different times for these models. The resulting temporal variations of the strength of the spirals for these three models are shown in Fig.~\ref{fig:spiral_strength}. The maximum strength of the tidally-induced spirals decreases monotonically with larger angle of interaction. Next, we use $|A(m=2, p =p_{\rm max})| = 0.1$ for defining the onset of the strong spiral perturbation. We find that the spirals persist for a time-scale of $\sim 2.9 - 4.2 \Gyr$, depending on the angle of interaction (also see Fig.~\ref{fig:spiral_strength}). For the polar ($i =90 \degrees$) configuration, the spirals are very weak as the values of $|A(m=2, p =p_{\rm max})|$ remain below $0.1$ (see Fig.~\ref{fig:spiral_strength}).

Next, we study the strength of the spirals in a galaxy fly-by model where the perturber interacts with the host galaxy in a retrograde orientation (with $i = 0 \degrees$) (for details see section~\ref{sec:flyby_setup}). A visual inspection of the face-on distribution of the stars (in the host galaxy) does not reveal any prominent spirals. We again calculate  the strength of the spirals using Eq.~\ref{eqn:spiral_strenght} for the model $RP40i00ret$, and show this also in Fig.~\ref{fig:spiral_strength}. The value of the amplitude $|A(m=2, p =p_{\rm max})|$ remains close to zero throughout the evolution of the model $RP40i00ret$, indicating that no prominent spirals are triggered/excited by the tidal encounter with the perturber galaxy for this model.
\par
Furthermore, we have checked that the tidally-induced spirals display a similar winding, namely, an initial rapid winding phase, followed by a slow winding phase, as seen for the model $RP40i00pro$ (see Fig.~\ref{fig:winding_spiral}). Also, we have checked the nature of the resulting spirals for these other models, using the same technique employed in section~\ref{sec:nature_spirals}. We find that for the models showing prominent spirals, the resulting spirals show a density wave nature, similar to the model $RP40i00pro$. For the sake of brevity, we do not show these results here.

\section{Breathing motions excited by tidally-induced spirals}
\label{sec:breathing}
%
\begin{figure}
    \centering
	\includegraphics[width=0.75\textwidth]{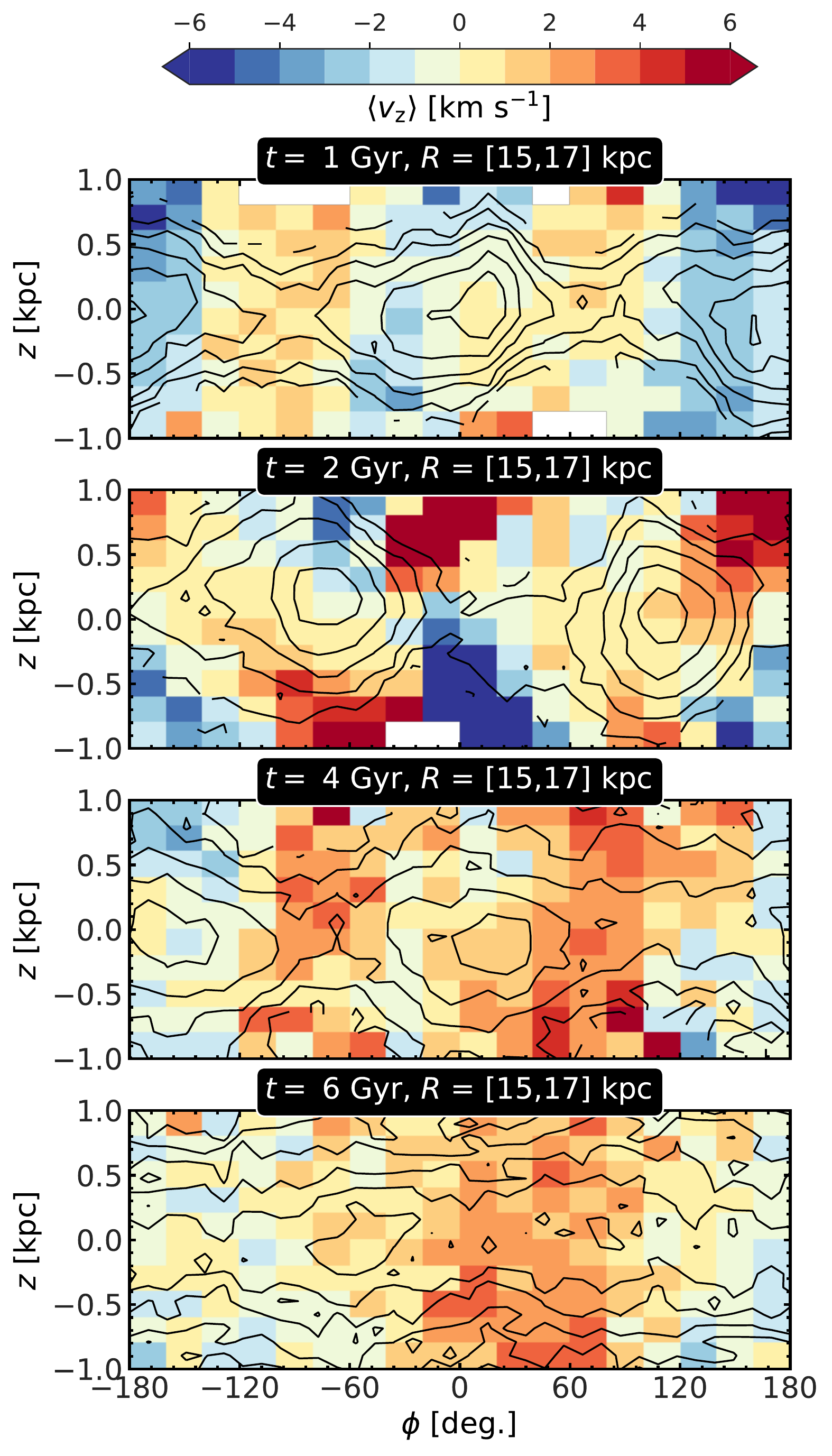}
    \caption{Distribution of the mean vertical velocity, $\avg{v_z}$, in the $(\phi, z)$-plane, for stars in the radial annulus $15 \leq R/\kpc \leq 17$ are shown at four different times for the model  $RP40i00pro$. The solid black lines denote the contours of constant density. The presence of large-scale, non-zero vertical velocities for stellar particles are seen at all four time-steps, for details see text.}
    \label{fig:vertical_velocity}
\end{figure}
\begin{figure*}
    \centering
	\includegraphics[width=0.9\textwidth]{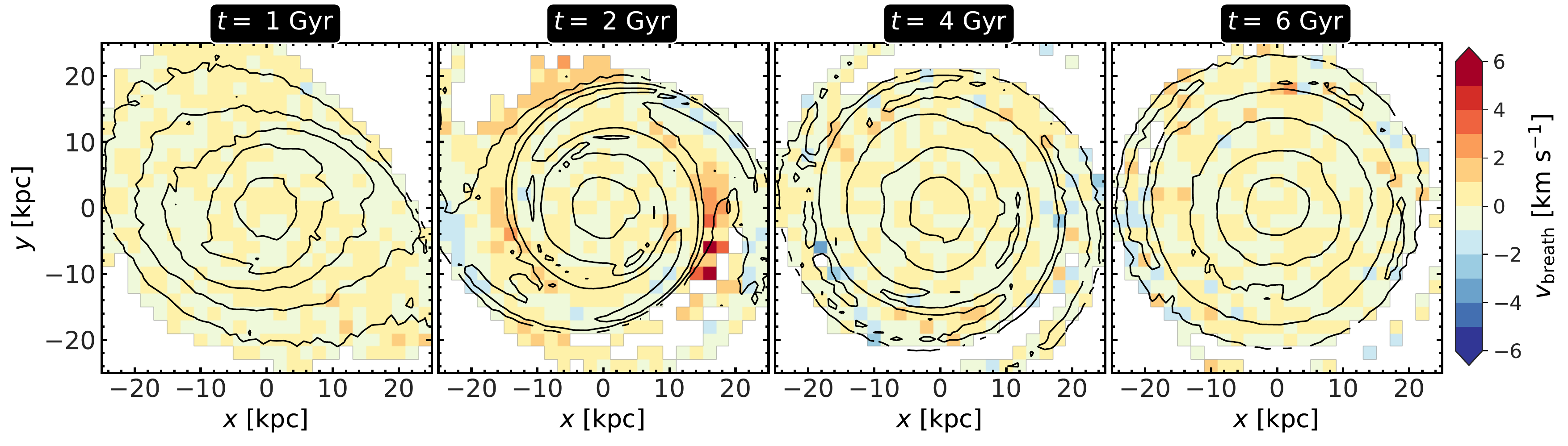}
	\medskip
    \includegraphics[width=0.9\textwidth]{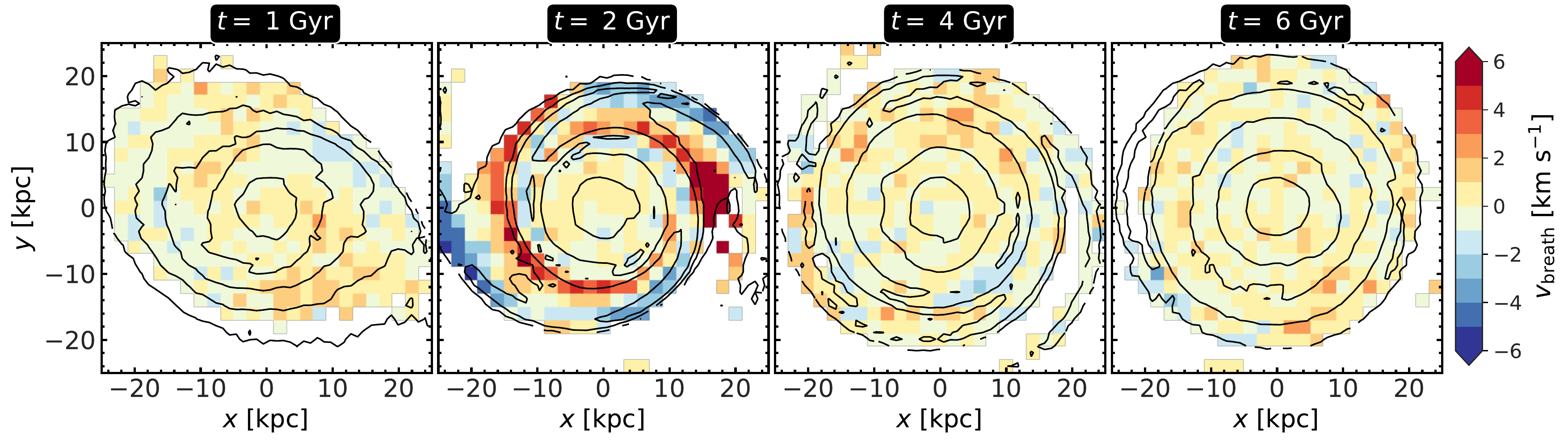}
    \caption{Distribution of the breathing velocity, \vb \ (Eq.~\ref{eq:breathing_velocity}) at different vertical distances from the mid-plane, at four different times for the model  $RP40i00pro$. The solid black lines denote the contours of constant density. The {\it top panels} show stars at $|z| = [0, 400]~\pc$ whereas the \textit{bottom panels} show stars at $|z| = [400, 1200]~\pc$, respectively.}
    \label{fig:breathing_map}
\end{figure*}
In the previous section, we have demonstrated that a tidal interaction with a perturber galaxy can excite prominent spirals in the disc of the host galaxy. Here, we investigate the dynamical impact of these tidally-induced spirals on the bulk vertical motions of the host galaxy. We first choose the fly-by model $RP40i00pro$ which harbours a strong spiral after the interaction. We choose the radial extent $15 \leq R/\kpc \leq 17$ where the spirals are prominent at later times in this radial annulus, and we calculate the mean vertical velocity ($\avg{v_z}$) in the $(\phi, z)$-plane at four different times, namely, at $t = 1, 2, 4,$ and $6 \Gyr$ (same as in Fig.~\ref{fig:dens_map}). This is shown is Fig.~\ref{fig:vertical_velocity}. During our chosen time interval, spirals' strength varies from strong to weak  (for details see section~\ref{sec:strength_spiral}).  At $t = 1 \Gyr$, just after the fly-by encounter, the distribution of the bulk vertical velocity ($\avg{v_z}$) in the $(\phi, z)$-plane predominantly shows bending motions, i.e., stellar particles on both sides of the mid-plane are moving coherently in the same direction. At this time, a prominent spiral is yet to form in the host galaxy (see Fig.~\ref{fig:spiral_strength_t-R}). However, by $t = 2 \Gyr$, the host galaxy shows a prominent spiral (see Fig.~\ref{fig:spiral_strength_t-R}), and the distribution of the bulk vertical velocity ($\avg{v_z}$) in the $(\phi, z)$-plane changes drastically. Now, the stellar particles on both sides of the mid-plane are moving coherently towards or away from it, indicating vertical breathing motion dominates. The relative sense of the $\avg{v_z}$ varies as a function of the azimuthal angle. However, at $t = 4 \Gyr$ when the tidally-induced spirals have weakened substantially, the distribution of the $\avg{v_z}$ is again seen to be dominated by the bending motions. By the end of the simulation  ($t = 6 \Gyr$), the spiral has wound up, and the distribution of $\avg{v_z}$ remains dominated by the bending motions of the stars.
\par
To quantify the breathing motions, we define the breathing velocity, $\vb $, as \citep{Debattista.2014, Gaia.Collaboration.2018, Ghosh.etal.2020}
\begin{equation}
\vb(x, y) = \frac{1}{2}\left[\avg{v_z(x,y, \Delta z)} - \avg{v_z(x,y, -\Delta z)}\right]\,,
    \label{eq:breathing_velocity}
\end{equation}
\noindent where $\avg{v_z(x,y, \Delta z)}$ is the mean vertical velocity at position $(x,y)$ in the galactocentric cartesian coordinate system, averaged over a vertical layer of thickness $\Delta z$ \citep[for details see][]{Gaia.Collaboration.2018, Ghosh.etal.2020}. A positive breathing velocity ($\vb > 0$) implies that the stars are coherently moving away from the mid-plane (expanding breathing motion), while $\vb < 0$ implies that the stars are moving coherently towards the mid-plane (compressing breathing motion). We find that at larger heights, the particle resolution of our selected fly-by model is not well-suited to compute meaningful values of $\vb$. Therefore, we calculate the distribution of $\vb$ in the $(x,y)$-plane for two vertical layers, namely, $|z| = [0, 400] \pc$ and $|z| = [400, 1200] \pc$.
The resulting distributions of $\vb$ at the same four times (as in Fig.~\ref{fig:vertical_velocity}) are shown in Fig.~\ref{fig:breathing_map}. The breathing velocity is close to zero near the mid-plane, for all four times considered. However, at $t = 2 \Gyr$ when the strong spirals are present in the host galaxy, the upper vertical layer ($|z| = [400, 1200] \pc$) shows significant, coherent breathing velocity ($\sim 6-8 \kms$, in magnitude). This trend is similar to what is shown for spiral-driven breathing motions \citep[e.g., see][]{Debattista.2014,Ghosh.etal.2020}, and is also similar to the breathing motions seen from the \gaia\ DR2 \citep{ Gaia.Collaboration.2018}. A visual inspection also reveals that at $t = 2 \Gyr$, the compressing breathing motions are closely associated with the spiral arms whereas the expanding breathing motions arise in the inter-arm regions (see Fig.~\ref{fig:breathing_map}), similar to the findings of \citet{Faureetal2014}, \citet{Debattista.2014} and \citet{Ghosh.etal.2020}. However, after $t = 4 \Gyr$ when the spirals are either significantly weaker or completely absent, the host galaxy does not show any prominent, coherent breathing motion in the upper layer ($|z| = [400, 1200] \pc$). To probe this further, we calculate the distribution of $\vb$ at two vertical layers in every $200 \Myr$ from $t = 2 \Gyr$ to $t = 6 \Gyr$ for the model  $RP40i00pro$. For the sake of brevity, this is not shown here. We find that the incidence of a prominent, coherent breathing motion is strongly related with the presence of a strong spiral feature in the host galaxy. This, together with the amplitude of the breathing velocity increasing with height \citep[a signature of spiral-driven breathing motion, see][]{Ghosh.etal.2020} indicate that the breathing motions present in the host galaxy are driven by the tidally-induced spirals.
\begin{figure}
    \centering
	\includegraphics[width=0.75\textwidth]{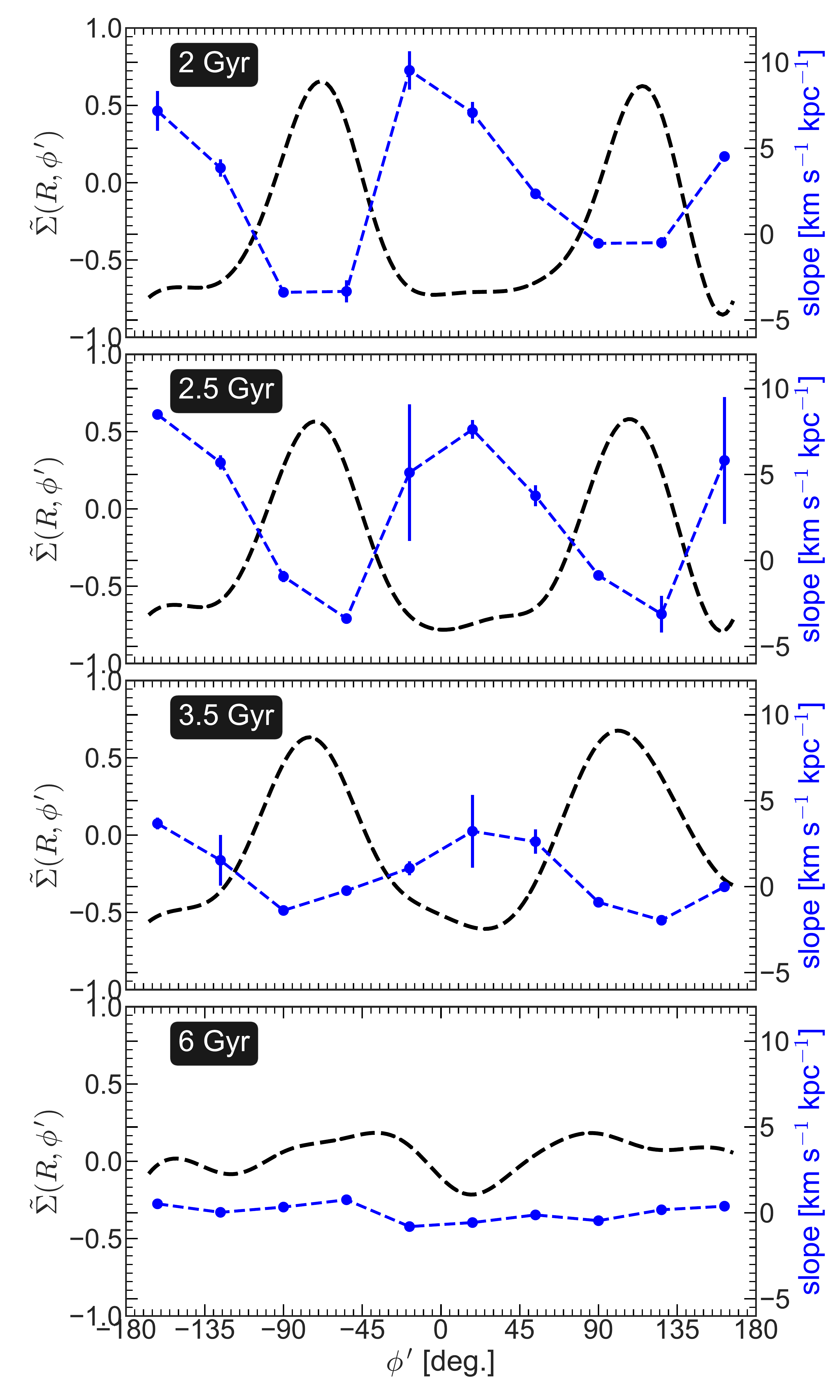}
    \caption{Variation of the slopes as a function of the rotated azimuthal angle ($\phi'$) at four different times (having different spiral strength) for the model $RP40i00pro$ (see blue dashed lines). Only the stellar particles in the radial extent $15 \leq R/\kpc \leq17 $ are chosen here. The black dashed line denotes the residual surface density ($\tilde \Sigma (R, \phi')$) as a function of the azimuthal angle, calculated in this chosen radial extent. The particles have first been binned in $1~\kpc$ annuli and then azimuthally rotated relative to each other so that the minimum in $\tilde \Sigma (R, \phi)$) in each annulus is coincident, and then the mean vertical velocity ($\avg{v_z}$) is calculated, amounting to stacking different radial ranges while unwinding the spiral. For details, see text. The sense of rotation is towards increasing $\phi'$.}
\label{fig:meanvz_resden_azimuthal_variation}
\end{figure}
\par
Finally, we study the azimuthal variation of the breathing motions and their connection with the peak(s) of the spirals. Figs.~\ref{fig:vertical_velocity} and \ref{fig:breathing_map} already demonstrated that the breathing motions in model  $RP40i00pro$, is associated with the spiral arm and the inter-arm regions. Here, we explore this further. Following \citet{Ghosh.etal.2020}, we quantify these breathing motions whose amplitude increases with height from the mid-plane, by fitting a straight line. In this formalism, the presence of a prominent breathing motion would result in a significantly \textit{non-zero} slope. Furthermore, an expanding (positive) breathing motion will yield a positive slope whereas a compressing (negative) breathing motion will yield a negative slope. On the other hand, the non-zero value of the intercept of the best-fit straight line indicates the presence of a bending motion. For details, the reader is referred to \citet{Ghosh.etal.2020}. A similar approach was also used in \citet{Widrowetal2014}. We consider the same radial extent $15 \kpc \leq R \leq 17 \kpc$ where a prominent spiral is present at $t = 2 \Gyr$. Since the phase-angle of the $m=2$ Fourier mode ($\phi_2$) varies as a function of radius, indicating that the azimuthal locations of the density peaks vary as a function of radius, to obtain a stronger signal of the slope, we first rotate the stellar particles in two different $1\kpc$-wide radial bins, in such a way that the density peaks in our chosen radial extent coincide. Then, we recalculate the slope of the breathing velocity as a function of the rotated azimuthal angle ($\phi'$). The resulting variation of the slope as well as the residual surface density ($\tilde \Sigma(R, \phi')$) with rotated azimuthal angle ($\phi'$) are shown in Fig.~\ref{fig:meanvz_resden_azimuthal_variation} for four different times. As seen clearly from Fig.~\ref{fig:meanvz_resden_azimuthal_variation}, at $t = 2 \Gyr$ when the spirals are quite strong ($|A(m=2, p = p_{\rm max})| = 0.38$), the compressive breathing motions are associated with the peak(s) of the spiral whereas the expanding  breathing motions are associated with the density minima. This trend is consistent with the signature of a spiral-driven breathing motions as shown in \citet{Debattista.2014} and \citet{Ghosh.etal.2020}. We show that in our model, the amplitudes of the expanding breathing motions are higher than that of the compressing breathing motions -- a trend attributed to the fact of having a more abrupt density variation as the stellar particles leave the spirals compared with when they enter the spirals \citep[see][]{Debattista.2014}. At $t = 3.5 \Gyr$, when the spirals are decaying ($|A(m=2, p = p_{\rm max})| = 0.21$), the corresponding (absolute) values of the slope also decrease, thereby indicating that the breathing motions also weaken. By $t = 6 \Gyr$, the spirals get wound up (almost) completely ($|A(m=2, p = p_{\rm max})| = 0.04$), and the (absolute) values of the slope are close to zero, indicating the absence of breathing motion. This further strengthens the case for the breathing motions being driven by the spirals, and not the \textit{direct} dynamical consequence of a fly-by interaction. The breathing motions are seen to last for $\sim 1.5 -2 \Gyr$ after their generation for the model $RP40i00pro$.
\par
We repeated this whole set of analyses for all the other models, to investigate the breathing motions and their connection with the incidence and strength of spirals present in the model. We find that a strong spiral always drives a prominent breathing motion with expanding breathing motions associated with the inter-arm region whereas the compressing breathing motions are associated with the peak(s) of the spirals. As the spirals get wound up, the breathing motions also cease to exist in the fly-by models. These trends are similar to what we have found for the model $RP40i00pro$. Therefore, for brevity, these are not shown here. In the fly-by model with retrograde orbital configuration ($RP40i00ret$), the spiral structure itself is weaker compared to the other models in prograde orbital configuration (see Fig.~\ref{fig:spiral_strength}). We checked that no prominent breathing motion is excited by this feeble spiral structure.

\section{Discussion}
\label{sec:discussion}
Our fly-by interactions excite strong spirals in the outer regions of the host galaxy's disc. The spirals show a variation in their maximum strength depending on the angle of interaction, and the orbital spin vector. Here, we compare the  strength, location, and nature of the tidally-induced spirals in our models with past studies. The numerical simulations of \citet{Oh.etal.2015} showed that in their models, a stronger tidal encounter induces prominent spirals in the inner regions ($ 5 \leq R/\kpc \leq 10$) of the host galaxy's disc (in addition to the tidal tails in the outer parts). The arm strength of the spirals vary in the range $\sim 0.1 - 0.18$, depending on the values of the relative tidal force, and the relative imparted momentum. Moreover, the spirals are of kinematic density wave in nature. Also, the simulations of \citet{Semczuk.etal.2017} showed the generation of transient spirals due to the tidal force exerted by the potential of a cluster; the spirals appear with each pericentre passage, followed by a fast decay. These spirals appear in the outer disc region ($ 12 \leq R/\kpc \leq 17$), with the maximum arm strength varying in the range $\sim 0.4 - 0.75$. In comparison, the spirals in our fly-by models are most prominent in the outer regions ($R \geq 10 \kpc$). The maximum arm strength varies in the range $\sim 0.15-0.38$, depending on the angle of interaction, and the orbital spin vector (see Fig.~\ref{fig:spiral_strength}). The spirals appear shortly after the pericentre passage of the perturber, grow rapidly, followed by winding up of those spirals. This winding of spirals show two distinct phases, namely, the initial rapid winding phase, followed by a slow winding phase (see Fig.~\ref{fig:winding_spiral}).
\par
The amplitude of the spiral-driven breathing motions also merits some discussion. In our fly-by models, when the spirals are most prominent, the values of the best-fit slope (used as a proxy for the breathing amplitude) vary from $\sim -5 \kms$ kpc$^{-1}$ to $\sim 10 \kms$ kpc$^{-1}$ (see Fig.~\ref{fig:meanvz_resden_azimuthal_variation}).  In comparison, the fly-by model (with spirals present) of \citet{Widrowetal2014} showed the (absolute) values of the best-fit slope $\sim 10 \kms$  kpc$^{-1}$. As for the spiral-driven breathing motions where the spirals arise due to internal instability, \citet{Ghosh.etal.2020} reported the values of the slope varying from $\sim - 2.5 \kms$ kpc$^{-1}$ to $\sim 3 \kms$ kpc$^{-1}$. Furthermore, the strong spirals present in the models used by \citet{Faureetal2014,Debattista.2014} can drive large breathing motions ($|\avg{v_z}|~ \sim 5-20 \kms$).
\par
We have considered only $N$-body models of an unbound, fly-by interaction, excluding the interstellar gas. It is well known that a disc galaxy contains a finite amount of gas \citep[e.g., see][]{ScovilleandSanders1987}. Additionally, in the $\Lambda$CDM galaxy formation scenario, a galaxy can accrete cold gas \citep[e.g.,][]{BirnboimDekel2003,Keresetal2005,DekelBirnboim2006,Ocvriketal2008} either during the merger-phase or at a later stage. Past studies have shown the dynamical importance of the interstellar gas in the context of cooling the stellar disc and facilitating the generation of fresh spiral waves \citep{SellwoodCarlberg1984}, and in maintenance of spiral density waves in infinitesimally-thin discs \citep{GhoshJog2015,GhoshJog2016} as well as in a galactic disc with finite thickness \citep{GhoshJog2021}. For gas rich galaxies undergoing such fly-by interactions, the vertical breathing motion may be important for increasing the turbulence in the gas where star formation is insignificant \citep{Stilp.etal.2013}. This is because as the stars are in the breathing motion, the vertical potential will change with time and the gas distribution will be affected. However, since the vertical stellar velocity induced by the breathing motion is fairly small (of the order of a few km s$^{-1}$), this effect may not be very significant, especially if compared to the much larger kinematic effect of supernova explosions and stellar winds \citep{Krumholz.etal.2018, Yu.Brian.etal.2021}. In addition, galaxy bulges play an important role in maintaining spiral density waves in the disc for a longer time \citep{SahaandElmegreen2016}. Although, our galaxy models have a classical bulge, we have not varied the contribution of the bulges in our models.
\par
Lastly, our models are specifically designed in a way that the unperturbed disc galaxy is not bar unstable and only forms weak spirals; the strong spirals that form therefore are a dynamical result of the fly-by encounter.
However, in reality, the Milky Way also harbours a stellar bar \citep[e.g., see][]{LisztandBurton1980,Binneyetal1991,Weinberg1992,Binneyetal1997,BlitzandSpergel1991,Weiland.etal.1994,Dwek.etal.1995,Freudenreich.1998,Hammersleyetal2000,WegandGerhard2013}. Furthermore, the bulk vertical motions in the Solar neighbourhood and beyond, display both bending and breathing motions \citep[e.g.,][]{Gaia.Collaboration.2018,Carrilloetal2018}. This simultaneous presence of bending and breathing motions could well be collectively manifested due to a combination of internal (spiral and/or bar-driven) and external driving mechanisms (tidal encounters), as previously investigated by \citet{Carrilloetal2019}. We stress that the aim of this work is to clarify whether the excitation of breathing motions are `directly' related to tidal interactions or whether they are driven by the tidally-induced spirals (as also mentioned in section~\ref{sec:intro}), and not to replicate the observed dynamical state of the Milky Way.

\section{Summary }
\label{sec:conclusion}
In summary, we investigated the dynamical impact of an unbound, single fly-by interaction with a perturber galaxy on the generation of the tidally-induced spiral features and the associated excitation of vertical breathing motions. We constructed a set of $N$-body models of fly-by encounter, with mass ratio kept fixed to 5:1 while varying different orbital parameters.
Our main findings are :\\

\begin{itemize}

\item{Fly-by interactions trigger a strong spiral structure in the disc of the host galaxy. The spirals grow rapidly in the initial times, followed by a slow decay. The generation and the strength of these tidally-induced spirals depend strongly on the angle of interaction as well as on the orbital spin vector. For the same orbital energy and the angle of interaction, the models in prograde configuration are  more efficient at driving strong spirals when compared to models in retrograde configuration.}

\item{The tidally-induced spirals in the host galaxy can survive for $\sim 2.9 - 4.2 \Gyr$ after their formation. The pitch angle of the resulting spirals display two distinct phases of winding, namely, a fast winding phase ($\tau_{\rm wind} \sim 0.4 \Gyr$) and a subsequent slow winding phase ($\tau_{\rm wind} \sim 5.7 \Gyr$). }

\item{When the tidally-induced spirals are strong, they drive coherent, large-scale vertical breathing motions whose amplitude increases with height from the mid-plane. Furthermore, the azimuthal locations of the compressing breathing motions ($\vb <0$) are associated with the peaks of the spirals whereas the azimuthal locations of the expanding breathing motions ($\vb > 0$) coincide with the density minima of the spirals. These trends are in agreement with the signatures of spiral-driven breathing motions.}

\item{The temporal evolution of these breathing motions follow closely the temporal evolution of the strengths of the spirals. A stronger spiral drives breathing motions with larger amplitudes. These breathing motions, excited by tidally-induced spirals, can persist for $\sim 1.5-2 \Gyr$ in the disc of the host galaxy.}

\end{itemize}

Thus, the results presented in this chapter demonstrate that a strong spiral structure can drive large, coherent vertical breathing motions irrespective of their formation scenario, i.e., whether induced by tidal interactions (as shown here) or generated via internal disc gravitational instability \citep[e.g.,][]{Faureetal2014,Debattista.2014,Ghosh.etal.2020}. Furthermore, our results highlight the cautionary fact that although in past studies, the tidal interactions are considered as the `usual suspect' for driving the vertical breathing motions, it is indeed the tidally-induced spirals which drive the breathing motions, and the dynamical role of such tidal encounters remains only ancillary.

\section{Appendix: Evolution in isolation}
\begin{figure*}
    \centering
	\includegraphics[width=0.6\textwidth]{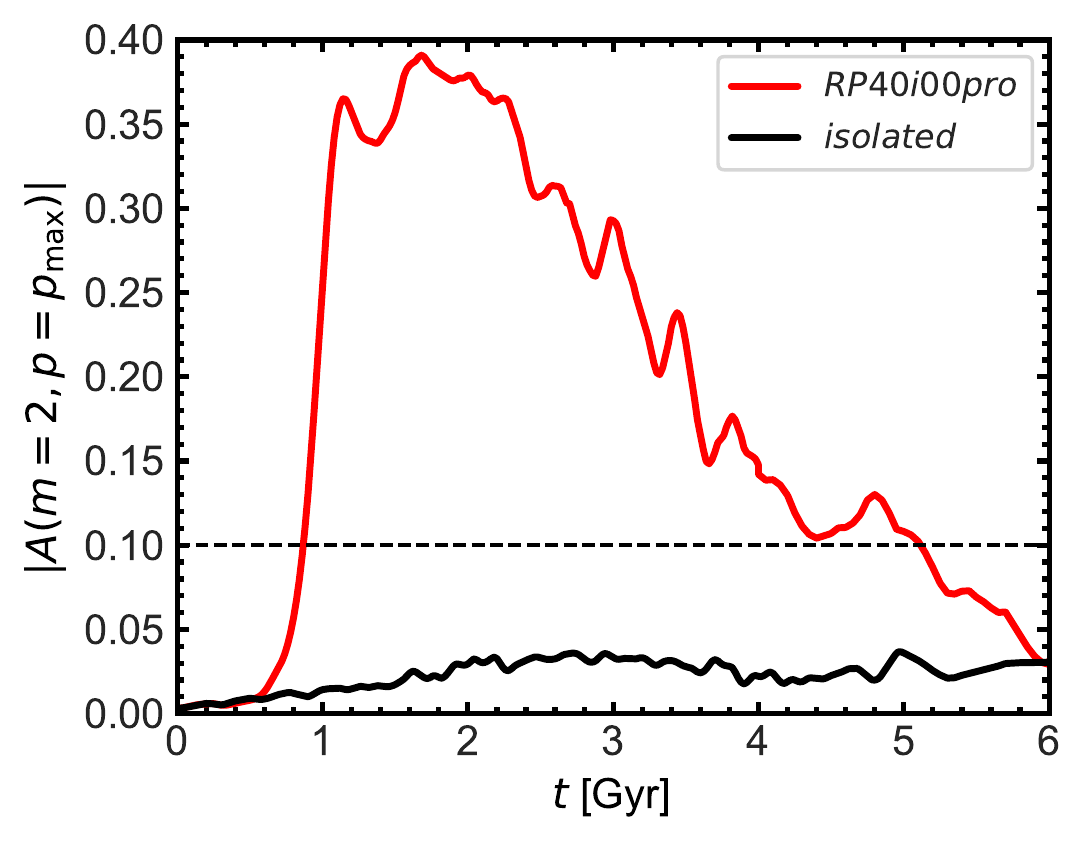}
    \caption{Temporal evolution of the spiral strength ($|A(m=2, p =p_{\rm max})|$), calculated using Eq.~\ref{eqn:spiral_strenght} is shown for the isolated host galaxy model. For comparison, we kept the model $RP40i00pro$ here (see red solid line). The isolated model does not show any prominent spirals throughout the simulation run. }
    \label{fig:spiral_strength_isolated}
\end{figure*}

\begin{figure*}
    \centering
	\includegraphics[width=\textwidth]{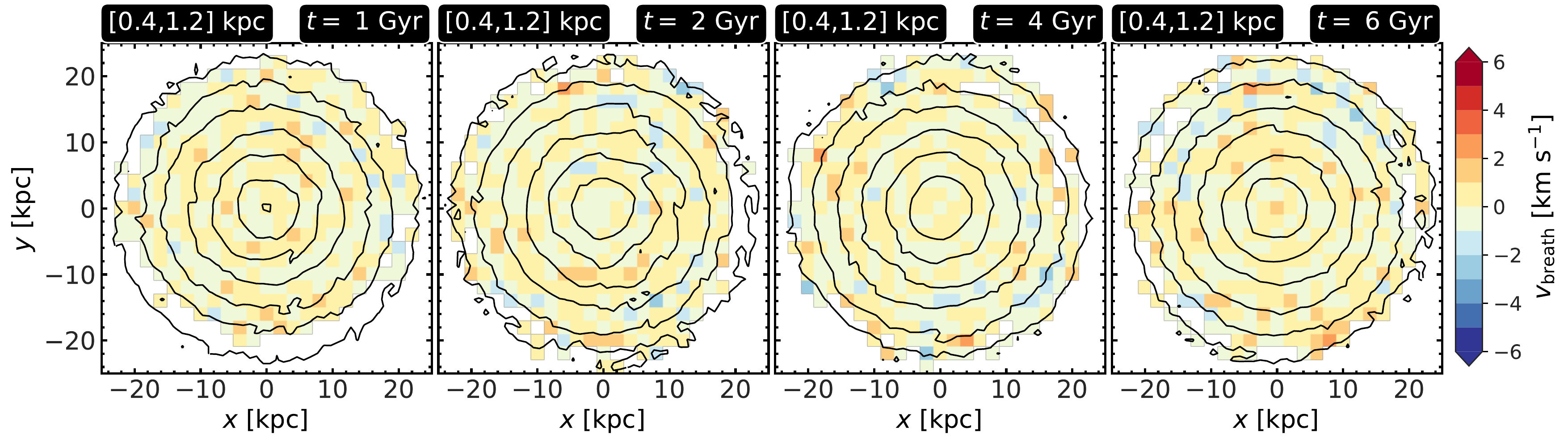}
    \caption{Evolution in isolation: distribution of the breathing velocity, \vb, for stars at $|z| = [400, 1200]~\pc$ at four different times for the isolated host galaxy model. The solid black lines denote the contours of constant density. No prominent vertical breathing motion is seen for the isolated evolution, in sharp contrast with the fly-by models (e.g., compare with Fig.~\ref{fig:breathing_map}).}
    \label{fig:breathing_map_isolated}
\end{figure*}

So far, we have shown that in our fly-models, a prominent spiral appears shortly after the interaction happens, and this spiral drives a coherent vertical breathing motion in the host galaxy. However, it remains to be investigated whether the host galaxy, when evolved in isolation, could still generate spirals and the associated vertical breathing motions.  We evolve the host galaxy model in isolation, for $6 \Gyr$. A visual inspection of the face-on density distribution of the stellar particles reveals no prominent spirals, throughout the simulation run, (as can be seen in the density contours in Fig.~\ref{fig:breathing_map_isolated}). Following the methodology described in Section~\ref{sec:strength_spiral}, we measure the values of $|A(m=2, p =p_{\rm max})|$ (used to quantify the strength of spirals) for the isolated host galaxy model. The resulting temporal evolution is shown in Fig.~\ref{fig:spiral_strength_isolated}. The values of $|A(m=2, p =p_{\rm max})|$ remain close to zero throughout the simulation run, demonstrating that no prominent spiral arm is generated during the entire isolated evolution of the host galaxy model. 

Furthermore, we calculate the breathing motions ($\vb$), using Eq.~\ref{eq:breathing_velocity}, for both the vertical slices, namely, $|z| = [0, 400] \pc$ and $|z| = [400, 1200] \pc$. As before, the vertical slice $|z| = [0, 400] \pc$ does not show any breathing motion. Interestingly, the upper vertical slice ($|z| = [400, 1200] \pc$) does not show any prominent breathing motion either throughout the simulation run (see Fig.~\ref{fig:breathing_map_isolated}), in sharp contrast with the fly-by models (compare with Fig.~\ref{fig:breathing_map}). This clearly demonstrates that the spirals in the fly-by models are indeed tidally-induced. In other words, the generation of spirals and the associated spiral-driven vertical breathing motions can indeed be attributed to the dynamical impact of a fly-by interaction.

	\begin{savequote}[100mm]
``As serious we are with this Universe. The Universe is also that serious with us.''
\qauthor{\textbf{$-$ Marianne Williamson}}
\end{savequote}

\chapter[The Effect of Dark Matter Halo Shape on Bar Buckling and Boxy/Peanut Bulges]{The Effect of Dark Matter Halo Shape on Bar Buckling and Boxy/Peanut Bulges}
\label{chapter5}

\section{Introduction}
\label{sec:intro} 
The fraction of bars in the observable Universe varies from $30\%$ to $70\%$ \citep{Aguerri.etal.2009, Masters.etal.2011, Diaz-Garcia.etal.2016}. Bars are the major driver of secular evolution in disc galaxies \citep{Weinberg.etal.2007, Athanassoula2013book, Long.etal.2014, Gadotti.etal.2020} and play a crucial role in the re-distribution of the mass and angular momentum among different components of a galaxy \citep{Athanassoula.2002, Athanassoula.2003, Saha.etal.2012, saha.etal.2016, kataria.das.2019}. Studies have shown that during disc evolution bars themselves also evolve, both in length and vertical thickening. The latter can result in the formation of boxy/peanut/x-shape (in short BPX) pseudo-bulges \citep{Friedli1990, Debattista2006, Gadotti2011}.


The origin of the vertical thickening of bars has been studied widely using  numerical simulations \citep{Combes1981, Combes.etal.1990, Raha.etal.1991, Athanassoula.etal.2005}. There are mainly three mechanisms which lead to bar thickening:- (i) bar buckling \citep{Combes.etal.1990, Raha.etal.1991}, (ii) the 2:1 vertical resonance \citep{Quillen.etal.2014}, and (iii) the gradual trapping of stellar orbits into the 2:1 resonance \citep{Sellwood.Gerhard.2020}. Bar buckling is the most violent thickening mechanism during which the bar bends out of the disc plane. The buckling reduces the size and strength of the bar and results in the formation of boxy/peanut bulges \citep{Sellwood.Merritt.1994, Debattista.etal.2005, Martinez-Valpuesta.etal.2004}. In some cases where buckling is slow and less energetic, it can even destroy the bar \citep{Collier.2020}. There are evidences of multiple (recurrent) buckling in numerical simulations \citep{Martinez-Valpuesta.etal.2006, Lokas.2019B}. The primary buckling occurs in the inner part of the bar and persists for less than 1~Gyr; however the secondary buckling take place in the outer region of the bar and remains for few $Gyr$ \citep{Martinez-Valpuesta.etal.2006, Athanassoula2016book}. In this chapter we focus only on bar buckling, which is the most rapid process that produces the vertical thickening of bars. 

Some studies suggest that properties of disc, bulge, and gas affect bar buckling. For example in their N-body simulations, \cite{Friedli1990} found that a small asymmetry about the mid plane of the disc accelerates bar buckling. The presence of a classical bulge in the center of galaxies may also prevent the onset of buckling instability in bars \citep{Smirnov.2019}. The warm gas in the galaxies has also been found to decrease the distortion in a bar which is due to buckling \citep{Lokas.2020}. Buckling time and strength remain unaltered in galaxy flybys \citep{Kumar.etal.2021}.

On a large scale, bar formation and evolution are affected by dark matter halo properties. The presence of live dark matter halo supports the bar formation instability, in contrast to rigid dark matter halo that delay onset of bar formation \citep{Athanassoula.2002, Saha.Naab.2013}. There have been several studies of angular momentum re-distribution between bar and live dark matter halo \citep{Sellwood.1980, Athanassoula.2003, Martinez-Valpuesta.etal.2006, Collier.etal.2019}. The detailed study of the effect of halo triaxiality and gas mass fraction on the formation and evolution of the bars is discussed in \cite{Berentzen.etal.2006} and \cite{Athanassoula.etal.2013}. Recently, \cite{Collier.etal.2018} have studied the evolution of bars in rotating and non-spherical live halos. They noticed multiple (=two) buckling events only in prolate halo. The detailed nature of bar buckling in non-spherical halos and the evolution of the buckling induced boxy/peanut bulges are not well explored.

There have been a few attempts to find the signatures of ongoing buckling in the observable Universe. However, the time period of bar buckling is very short and the presence of a central concentration can halt the buckling instability, so it is not easy to detect ongoing bar buckling in galaxies. The first attempt used the bar isophotes and the kinematic signatures of bar buckling derived from simulations to detect buckling events in NGC 3227 and NGC 4569 \citep{Erwin.Debattista.2016}. Recently, \cite{Xiang.etal.2021} have used kinematic signatures to detect ongoing buckling in face-on galaxies.

In this chapter we show that the non-spherical nature of halos affects bar buckling significantly and has very important implications for the observations of bars. We vary the halo shape from oblate to prolate, keeping the ratio of halo axes equal in the disc plane, which is a fairly good assumption as shown in recent numerical studies \citep{Bett.etal.2010, Liao.etal.2017}. We have also characterized the properties of the buckling induced boxy/peanut bulges in oblate, spherical and prolate halos.

\section{Simulations And Analysis}
We have simulated isolated disc galaxies with different dark matter halo shapes ranging from oblate to prolate including spherical. All of our model galaxies form a bar and undergo buckling instability. As a result of buckling, the bar thickens and forms a boxy/peanut bulge. To trace the properties of the buckling induced boxy/peanut pseudo-bulge, we have evolved all the models until 8~Gyr. The model setup has been described in detail in earlier papers \citep{kataria.das.2018, Kumar.etal.2021}, and so we describe it only briefly below.

\subsection{Model Galaxies:}
\label{sec:galaxy_model} 
We used a publicly available open source code GALIC \citep{Yurin2014} to generate the initial conditions of our model galaxies. GALIC populates the particles according to the given density distribution and finds the equilibrium solution of the collisionless Boltzmann Equation (CBE) by iteratively changing the initial velocities of the particles. This code makes use of Schwarzschild's method and made-to-measure technique (see \cite{Yurin2014}).

Each of our model galaxies incorporates a stellar disc and a dark matter halo. Since we are interested in bar buckling and the resulting pseudo-bulges, we did not include the bulge component in our models because the presence of a spherical potential in the center of a galaxy slows down bar formation and hence hinders the buckling of bars \citep{kataria.das.2018}. To populate the particles of a spherical dark matter halo, we used the Hernquist density profile \citep{Hernquist1990} defined as,
\begin{equation}
    \rho_{dm}(r)=\frac{M_{dm}}{2\pi}\frac{a}{r(r+a)^3}
    \label{eqn:halo}
\end{equation}
where $M_{dm}$ is the total mass of the dark matter halo and `$a$' is its scale radius and is related to the concentration parameter `$c$' of the NFW halo \citep{NFW1996} with mass $M_{200}=M_{dm}$ by the relation,
\begin{equation}
    a=\frac{r_{200}}{c}\sqrt{2\left[\ln{(1+c)-\frac{c}{(1+c)}}\right]},
    \label{eqn:scale_radius}
\end{equation}
where $r_{200}$ is the radius of the NFW halo. It is defined as the radius of the sphere within which the average density is 200 times the critical density of the Universe and $M_{200}$ is the mass within this radius.

The non-spherical halos (oblate and prolate) are generated by linearly distorting the spherical halo along the z-axis perpendicular to the disc plane. Assuming an ellipsoid has $a_{x} = b_{y}$, and $c_{z}$ as its three axes, then the ratio $q=c_{z} / a_{x}$ defines the shape of the halo. An oblate halo has $q<1$, a spherical halo has $q=1$, and a prolate halo has $q>1$. The density profile of such non-spherical halos is given by
\begin{equation}
    \bar \rho_{dm}(R,z,q)=\frac{1}{q} \rho_{dm}\left(\sqrt{R^{2}+\frac{z^{2}}{q^{2}}}\right)
    \label{eqn:non_spherical_halo}
\end{equation}
where $\rho_{dm}$ is the Hernquist density profile as shown in equation (~\ref{eqn:halo}). This new profiles keeps the total mass of the halo invariant. 

The distribution of particles in the stellar disc is represented by the exponential profile in the radial direction and the $\sech^{2}$ profile in the direction perpendicular to the disc. So the net density profile is given as,
\begin{equation}
   \rho_{d}(R,z)=\frac{M_{d}}{4\pi z_{0} R_{s}^{2}}\exp\left(-{\frac{R}{R_{s}}}\right) \sech^{2}\left(\frac{z}{z_{0}}\right), 
    \label{eqn:disc}
\end{equation}
where $M_{d}$ is the total disc mass, $z_{0}$ is the disc scale height and $R_{s}$ is the disc scale radius.

\begin{table}
\centering
\caption{Initial parameters of the model galaxies.}
\label{tab:initial_parameters}
\begin{threeparttable}
\begin{tabular}{lr}
\hline
Total mass ($M$) & 6.4 $\times$ $10^{11} M_{\odot}$ \\
\hline
Halo spin parameter ($\lambda$) & 0.035 \\
\hline
Halo concentration parameter ($c$) & 20 \\
\hline
Disc mass fraction & 0.10 \\
\hline
Disc scale radius ($R_{s}$) & 2.90~kpc \\
\hline
Disc scale height ($z_{0}$) & 0.58~kpc \\
\hline
Halo particles ($N_{Halo}$) & 2.0$\times 10^{6}$ \\
\hline
Disc particles ($N_{Disc}$) & 2.0$\times 10^{6}$ \\
\hline
Total particles ($N_{Total}$) & 4.0$\times 10^{6}$ \\
\hline
\end{tabular}
\end{threeparttable}
\end{table}

\begin{figure*}
    \centering
	\includegraphics[width=\textwidth]{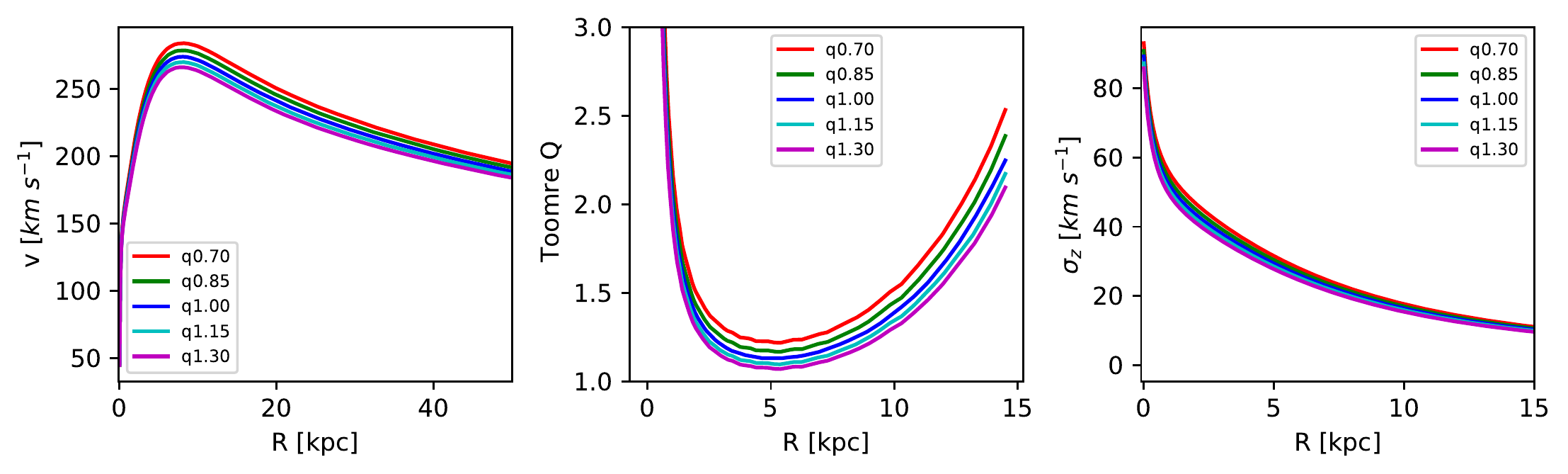}
    \caption{Initial condition of model galaxies. Panels from left to right show the rotation curve, Toomre instability parameter, and vertical velocity dispersion of model disc galaxies respectively. Different colors in the legend represent the models with different the halo shape parameters.}
    \label{fig:init_cond}
\end{figure*}

We have simulated five models of non-spherical dark matter halos in isolated disc galaxies for the halo shape parameter $q \in \{0.70, 0.85, 1.00, 1.15, 1.30\}$. Each of our model galaxies has $2\times 10^{6}$ dark matter particles and $2\times 10^{6}$ stellar particles making a total of $4\times 10^{6}$ particles. For testing purposes, we have also simulated all the models with $2\times 10^{6}$ total particles and found similar results as that for $4\times 10^{6}$ total particles. The total mass of each galaxy is set to $6.4\times 10^{11} M_{\odot}$, where each galaxy contains $90\%$ dark matter mass and $10\%$ stellar mass. Table~\ref{tab:initial_parameters} summarises the initial parameters of our model galaxies which are standard in all the models. The rotation curve $v(R)$, Toomre Q parameter, and the vertical velocity dispersion $\sigma_{z}$ are shown in Fig.~\ref{fig:init_cond}. Since the equilibrium model of the galaxy is generated by iteratively solving CBE for disc and halo, the rotation velocity, Toomre Q, and the velocity dispersion of oblate halos are always a little higher than the respective prolate halos.

After generating initial realizations of model galaxies, we evolved them using the open source code Gadget-2 \citep{Springel2001, Springal2005man, Springel2005} upto 8~Gyr. Gadget-2 is a massively parallelized and adaptive (in both space and in time) code. The gravitational softening for stellar particles and dark matter particles are set to 0.02~kpc and 0.03~kpc respectively. The maximum percentage change in the total angular momentum of the galaxy is well within 0.1$\%$ for all model galaxies throughout the evolution.

We used the shape parameter $'q'$ of the halos for naming the models. For example 'q1.15' represents the initial galaxy model with shape parameter $q=1.15$. However, the shape of the halo may change after evolution of the models. It should be noted that all the quantities discussed here will be in units of the dimensionless Hubble parameter $'h'$ where the Hubble constant is defined as $H_{0}= 100$ $h$ $km$ $s^{-1} Mpc^{-1}$.

\subsection{Analysis:}
\label{sec:analysis}
The bar strength is usually determined using the amplitude of the m=2 Fourier mode relative to m=0 Fourier mode. The amplitude of the $m$th Fourier mode at a cylindrical radius $R$ is given by,
\begin{align}
    A_{m}(R) &= \left| \sum_{j=1}^{N} m_{j} \exp{\left(i m \phi_{j}\right)} \right|,
    \label{eqn:bar_strenght}
\end{align}
where $m_{j}$ is the mass and $\phi_{j}$ is the azimuth angle of the $j$th particle at a radius $R$, and $N$ is the total number of particles at a radius $R$. The strength of the bar is defined as $A_{bar} = max \left[ \frac{A_{2}}{A_{0}}(R) \right]$.

For the quantification of the buckling instability, we have  adopted the commonly used expressions in the literature \citep{Debattista2006, Xiang.etal.2021}, i.e. the $m=2$ Fourier mode weighted by the vertical velocity
\begin{align}
    A_{buck,v_{z}}(R) &= \left| \frac{\sum_{j=1}^{N} v_{z_j} m_{j} \exp{\left(2 i \phi_{j}\right)}}{\sum_{j=1}^{N} m_{j}} \right|,
    \label{eqn:buck_strenght_vz}
\end{align}
and the $m=2$ Fourier mode weighted by the vertical height
\begin{align}
    A_{buck,z}(R) &= \left| \frac{\sum_{j=1}^{N} z_{j} m_{j} \exp{\left(2 i \phi_{j}\right)}}{\sum_{j=1}^{N} m_{j}} \right|.
    \label{eqn:buck_strenght_z}
\end{align}
Equation~\ref{eqn:buck_strenght_vz} is a very useful relation for quantifying buckling in face-on galaxies where stellar kinematic is known while equation~\ref{eqn:buck_strenght_z} quantifies the buckling in edge-on galaxies where vertical stellar morphology is known. We have also used median vertical height ($z_{med}$) for the quantification of buckling. The evolution of the boxy/peanut/x-shape structure is traced using the root mean square vertical height ($A_{BPX} = z_{rms}$). For all these calculations, we have considered only those stellar particles which lie above and below the mid plane within 10 disc scale height ($10 z_{0}$). This choice reduces the discreteness noise from the measurement when some particles reach higher vertical distances.

\section{Results}
\label{sec:results}
We have evolved isolated disc galaxy models with varying dark matter halo shapes ranging from q=0.70 (oblate) to q=1.30 (prolate). All the models are bar unstable and show bar buckling instability after reaching maximum bar strength. In the following subsections, we  discuss how bar buckling is affected by halos shape and how it affects the final product, the boxy/peanut pseudo-bulge.

\subsection{Dependence of bar formation time and bar strength on halo oblateness:}
\label{sec:bar_strength}
In Fig.~\ref{fig:bar_strength}, we have shown the time evolution of the bar strength in our model galaxies. We have smoothed the bar strength with the Savitzky–Golay filter \citep{Savgol.1964} using window size = 5, and polynomial order = 3 for the filter parameters. The main motive for using this filter was to remove the noise and reveal the global evolution of the bar instability. All the models show a quick rise in bar strength. After reaching maximum strength, the bar strength decreases a little and again starts increasing. Later in this section we will show that this drop in bar strength is the result of bar buckling (or bending) event which weakens the bar. Finally, the strength of the bar saturates in all the models.
\begin{figure*}
    \centering
	\includegraphics[width=0.7\textwidth]{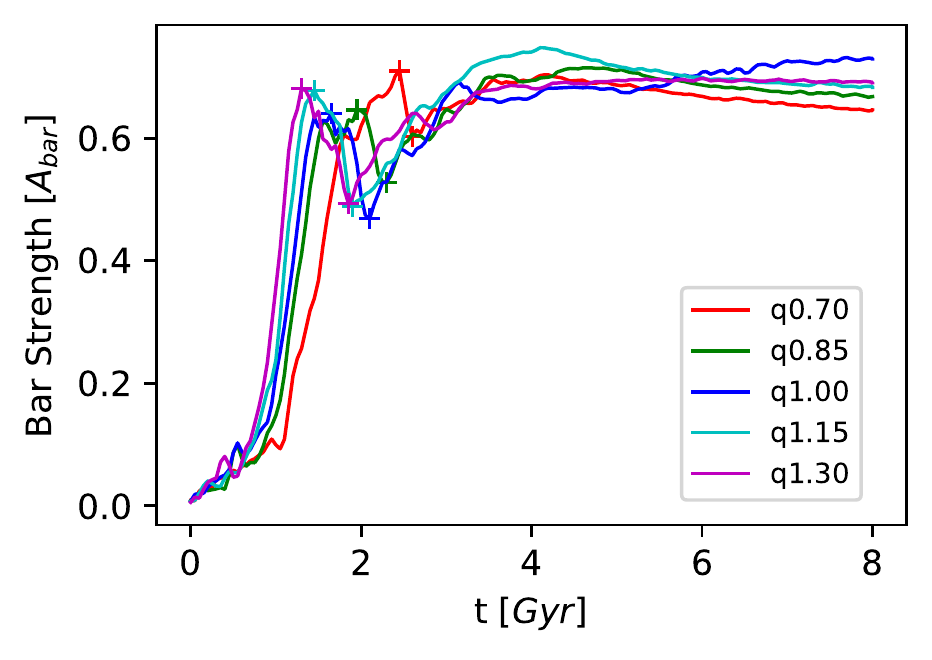}
    \caption{Evolution of bar strength in non-spherical dark matter halos with different shape parameters $'q'$. The positions of bar formation and re-growth after buckling are marked with '+' symbol.}
    \label{fig:bar_strength}
\end{figure*}

The effect of non-spherical dark matter halos on the bar formation is clearly visible. Prolate halos promote early bar formation, whereas oblate halos delay bar formation (where we consider the bar formation time to be the time from the beginning of simulation to the peak bar strength just before the decrease due to bar buckling). If we consider the galaxy with a spherical halo as the control model, oblate halos delay bar formation by more amount of time compared to the corresponding time taken for prolate halos to promote early bar formation. 

At the time of bar formation, the amplitude of bar strength is always higher in non-spherical halos. The decrease in bar amplitude due to buckling is highest for prolate halos and lowest for oblate halos. The change in bar amplitude is 0.19 for the $q=1.30$ model and 0.11 for the $q=0.70$ model. The difference between the bar formation time and the re-growth time (by re-growth time, we mean the time when bar starts recovering from the weakening caused due to buckling) is longer for prolate halos as compared to oblate halos. This difference is 0.55~Gyr for the $q=1.30$ model halo and 0.15~Gyr for the $q=0.70$ model.

\subsection{Bar buckling in oblate and prolate halos:}
\label{sec:bar_buckling}
\begin{figure*}
    \centering
	\includegraphics[width=\textwidth]{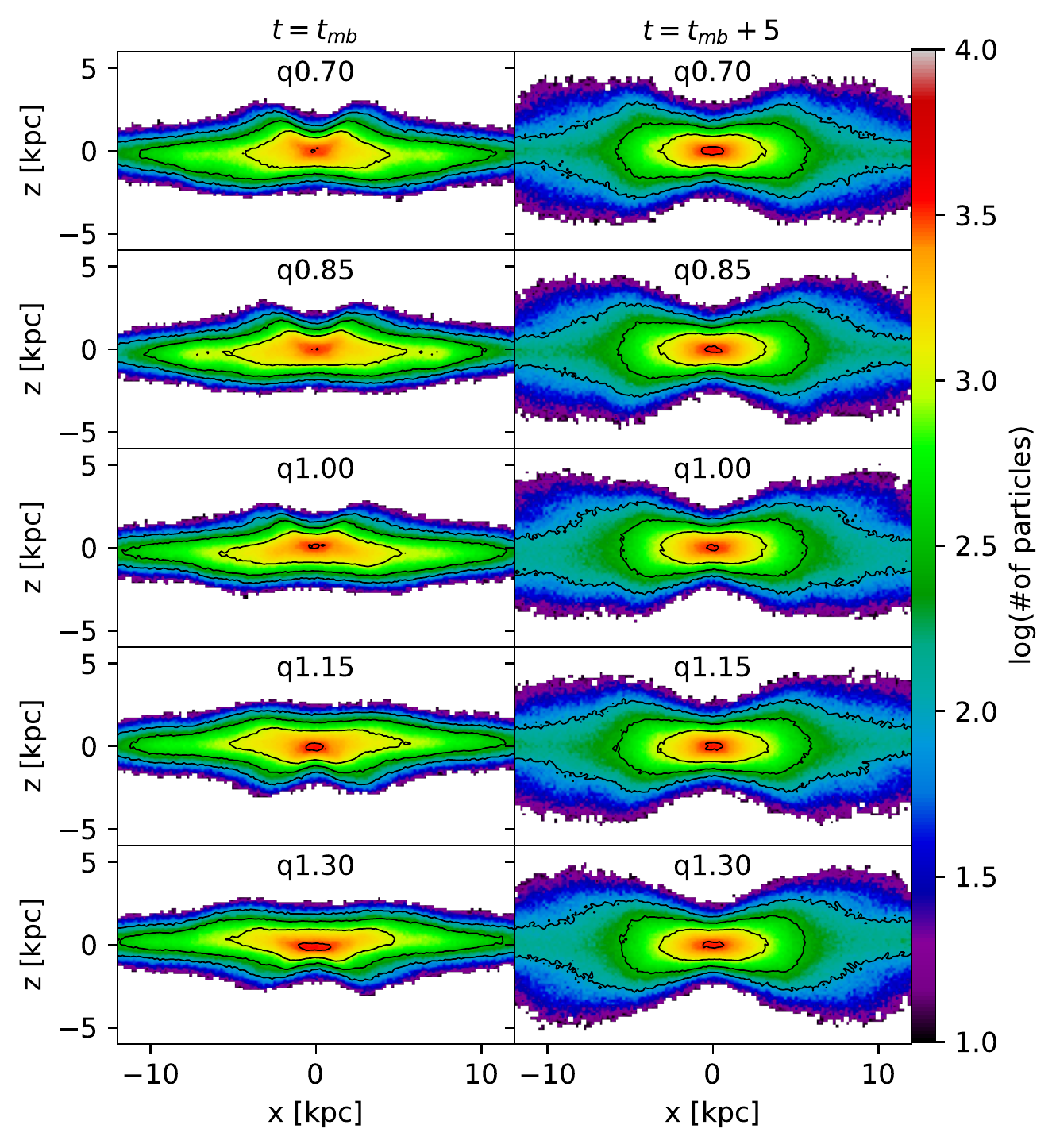}
    \caption{Bar buckling in non-spherical dark matter halos along with iso-density contours. Left column shows the side-on view (perpendicular to bar) of the disc at the time of maximum buckling amplitude ($t_{mb}$) and right column shows the side-on view after 5~Gyr of maximum buckling. }
    \label{fig:bar_buckling_qual}
\end{figure*}
All of our model galaxies go though the buckling event during which the bar bends out of the disc plane and and becomes overall weaker. To verify that  the buckling instability takes place, we have shown the edge-on view of the bars in the left column of Fig.~\ref{fig:bar_buckling_qual} at the time of maximum buckling/bending amplitude (hereafter, we use $t_{mb}$ to represent this time) calculated using equation~\ref{eqn:buck_strenght_vz}. The right column of this figure shows the buckling induced boxy/peanut/x-shape pseudo-bulge after 5~Gyr of the maximum buckling amplitude ($t_{mb}$+5~Gyr). The direction of the bending depends on the mass asymmetry around the mid plane of the disc before buckling, and so the vertical direction can vary from model to model \citep{Friedli1990}.

\begin{figure*}
    \centering
	\includegraphics[width=0.85\textwidth]{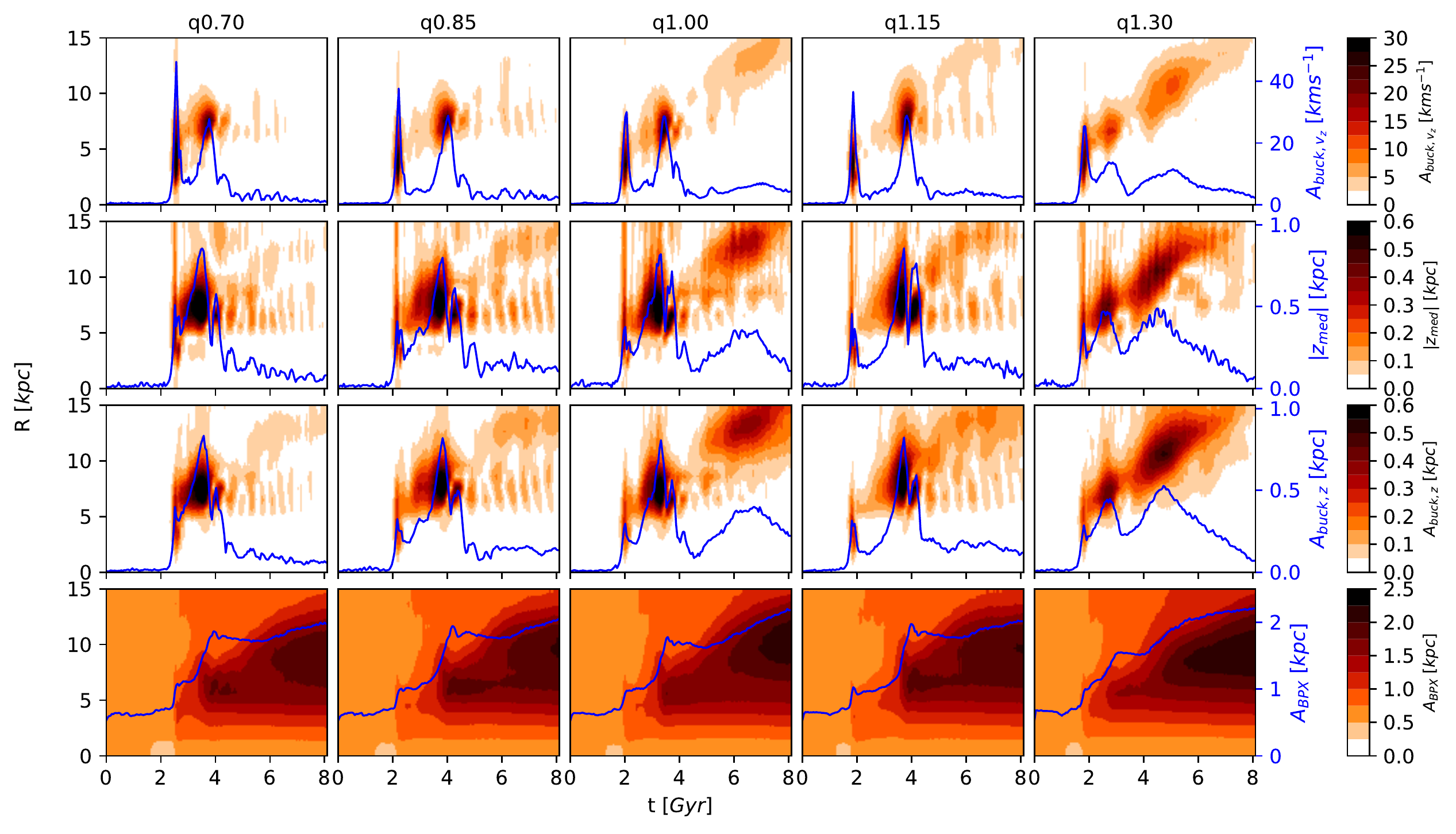}
    \caption{Quantification of bar buckling and resultant BPX structure in non-spherical dark matter halos. First row panels show the buckling amplitude using stellar kinematics while second and third rows panels represent buckling amplitude using stellar morphology. Last row panels show BPX strength. Each column shows a different galaxy model. The temporal variation of maximum strength is shown by over-plotted blue curve.}
    \label{fig:bar_buckling_quant}
\end{figure*}
In Fig.~\ref{fig:bar_buckling_quant}, we have shown the color coded radial and temporal distribution of buckling amplitude (panels in top three rows) and boxy/peanut strength (panels in bottom row) in oblate, spherical and prolate halos. We have also over-plotted each panel with the temporal distribution of the maximum strength. Each column of this figure represents a model galaxy as shown at the top. The panels in the first row show the kinematic amplitude of bar buckling events while the second and third rows quantify morphological asymmetry caused in the bar due to buckling. Here all the high amplitude (i.e. distinct and measurable) peaks correspond to a buckling event during evolution. The fourth row at the bottom demonstrate the strength of boxy/peanut/x-shape structure induced by bar buckling.

As, in Fig.~\ref{fig:bar_strength}, we see a variation in the bar formation time with halo shape. The time of the bar buckling also varies with halo shape. From the first peaks of blue curves in the top row of Fig.~\ref{fig:bar_buckling_quant}, One can notice that the buckling timescale increases with decreasing halo shape parameter $q$. Prolate halos buckle earlier and oblate halos buckle later with respect to spherical halo. The kinematic signature of buckling indicates that oblate halos start buckling just after bar formation, whereas their prolate counterparts take some time to buckle after bar formation. On the other hand, the first buckling event attains its maximum amplitude just before bar re-growth time. These two effects result in closely similar buckling period during first buckling as can be interpreted from the width of the first peak in all the panels of the first row.

\begin{figure*}
    \centering
	\includegraphics[width=0.75\textwidth]{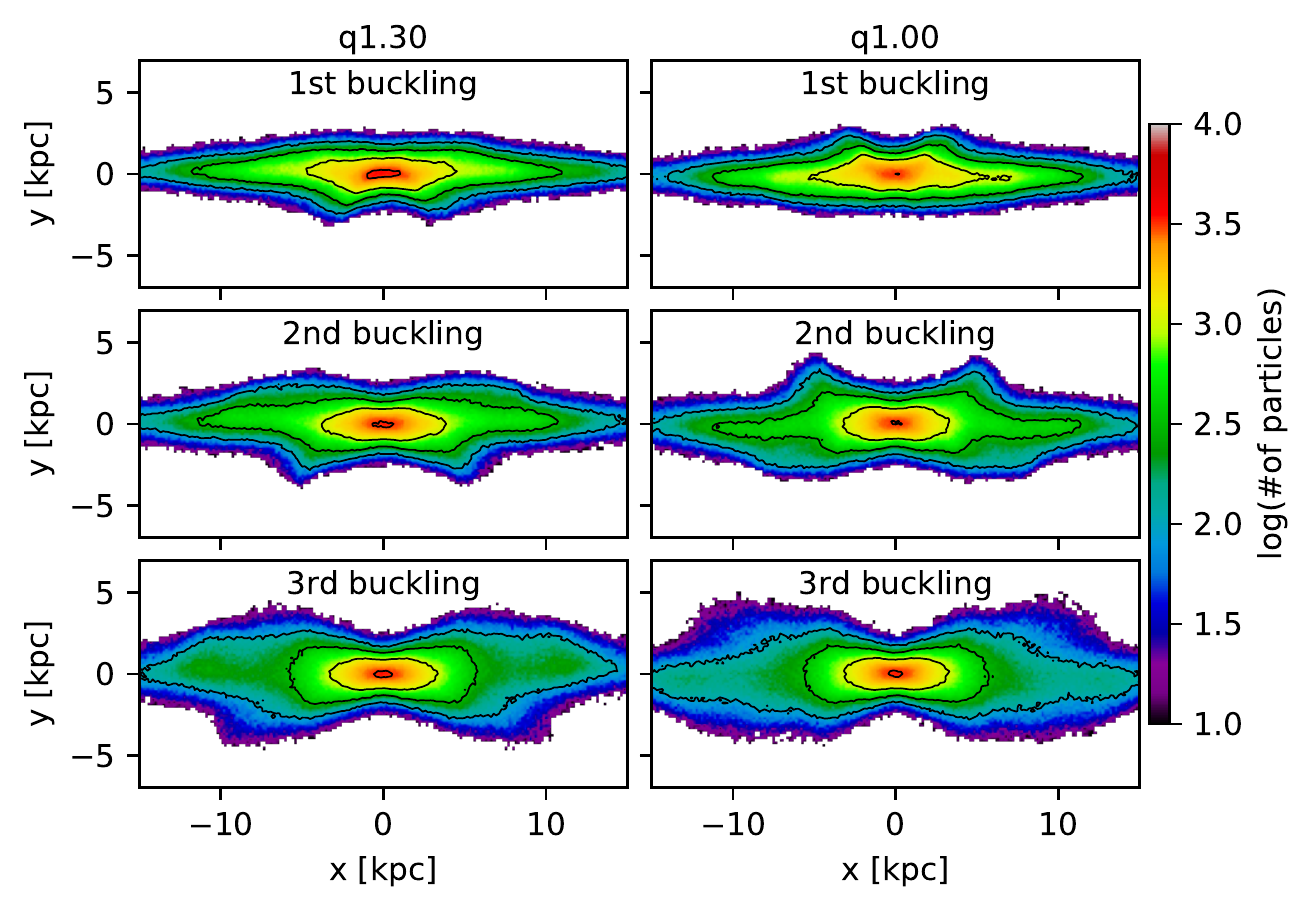}
    \caption{Triple bar buckling events in q=1.30 halo along with iso-density contour compared with the q=1.00 halo. Each panel is shown at the time of distinct peaks in $A_{buck,v_{z}}$ as shown by blue curves in the first row of Fig.~\ref{fig:bar_buckling_quant}. A clear bending of the iso-density contour can be noticed in the outer bar of the prolate halo model during 3rd buckling event.}
    \label{fig:triple_buckling}
\end{figure*}

By comparing the panels of the first row in Fig.~\ref{fig:bar_buckling_quant}, one can clearly notice that the kinematic signature of first buckling is always higher in amplitude than the succeeding buckling events. However, the morphological signatures have higher amplitude during the second buckling event as can be seen in the second and third row panels of the figure. When moving from oblate to prolate halos in any of the first three rows, we can spot a clear difference of increasing buckling signature at outer radial positions. This implies that prolate halos help the bar to buckle in the outer parts of the bar while oblate halos suppress the buckling at outer edges. At the outer edge of the bar, the buckling strength of the q1.15 model is slightly weaker than the q1.00 model. But it is strongest for the q1.30 model which indicates that the small deviation from spherical shape towards prolateness decreases the buckling amplitude but more deviation increases it. Prolate halo for q=1.30 shows an explicitly distinct buckling strength at the outer edges of the bar in both kinematic and morphological quantification which qualifies it to be a third buckling event. The spherical halo with q=1.00 also shows a weak signature of the third buckling event. In Fig.~\ref{fig:triple_buckling}, we have shown the third buckling event in our most prolate halo and compared it with the spherical model. During the third buckling event, the iso-density contours in the prolate halo show clear noticeable distortion in the disc while the corresponding distortion in the spherical halo is very weak. Hence, the most prolate halo in our set of models has a halo shape that promotes the onset of three buckling events in the bar.

Seeing the widths of the peaks in each panel of the top three rows of Fig.~\ref{fig:bar_buckling_quant}, one can easily interpret that the duration of a preceding buckling is always smaller than the succeeding one. Since, the signatures of first buckling are short lived so, it has a lower probability to be detected in observations of galaxies. Instead it is the second buckling event which has a higher probability of being detected as an ongoing buckling event. In our models, the prolate halo of q=1.30 show a remarkable third episode of buckling, as can be seen in the last column of the figure. This suggests that in observations of galaxies, buckling events are more likely to be associated with prolate halos rather than oblate or spherical halos.

\subsection{Effect of halo shape on the boxy/peanut bulge:}
\label{sec:bpx_bulge}
The last row of Fig.~\ref{fig:bar_buckling_quant} shows the evolution of the boxy/peanut pseudo-bulge in distorted halos. There is one to one correlation between the bar buckling event and the steep (or sudden) rise in BPX strength. After each buckling event, the bar quickly gets thicker as can be seen in each panel of the bottom row. There is no significant difference in the inner bar BPX strength for different halos. But, the outer part of the bar in prolate halos gets more thicker than the oblate halos. The same result can be interpreted from the closer look at iso-density contours in the right column of Fig.~\ref{fig:bar_buckling_qual}. It shows that oblate halos try to restrict bar/disc against vertical thickening, whereas prolate halos promote vertical heating by the mean of continuous buckling.

\begin{figure*}
    \centering
	\includegraphics[width=0.65\textwidth]{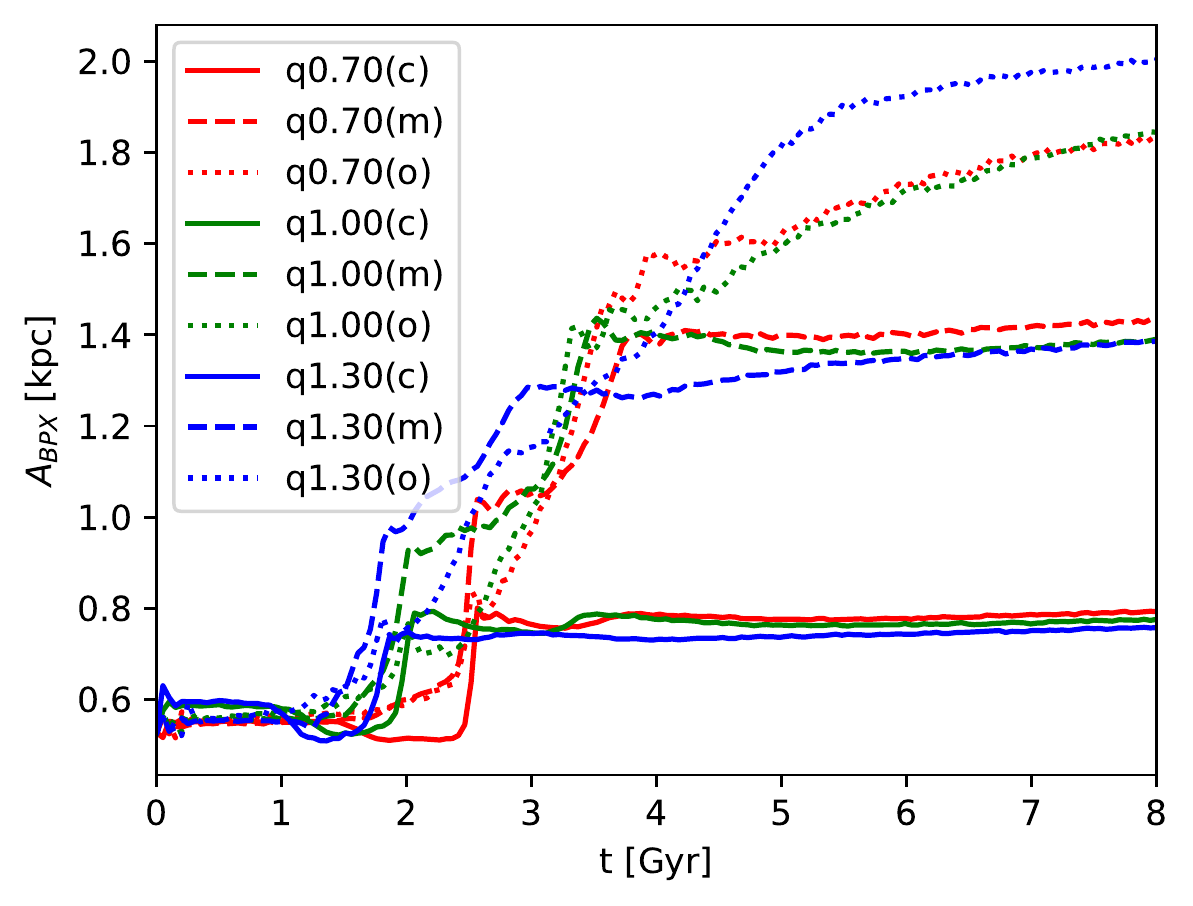}
    \caption{ Evolution of boxy/peanut/x-shape (BPX) structure in three different regions of the disc. For the sake of better visualization, only three extreme models (q0.70, q1.00, q1.30) are shown with different colors. Here, solid curves show central region $R \in [0,R_{s})$, dashed curves represent middle region $R \in [R_{s},2R_{s})$, and dotted curves demonstrate outer region $R \in [2R_{s},3R_{s})$ as marked with symbols 'c', 'm', and 'o' respectively in the legend.}
    \label{fig:bpx_strength}
\end{figure*}

The blue curves in the bottom row panels of Fig.~\ref{fig:bar_buckling_quant} represent the time evolution of the maximum BPX strength in the shown region of the disc. Thus, for the sake of better visualization, we have also shown the strength of BPX pseudobulges in three different parts of the disc for the three extreme models q0.70, q1.00 and q1.30 in Fig.~\ref{fig:bpx_strength}. In the figure the solid curves show the central region $R \in [0,1R_{s})$, dashed curves represent middle region $R \in [1R_{s},2R_{s})$, and dotted curves demonstrate outer region $R \in [2R_{s},3R_{s})$. These regions are marked with symbols 'c', 'm', and 'o' respectively in the legend of the figure. The solid curves of Fig.~\ref{fig:bpx_strength} clearly show that the central part of the bar that lies within disk scale radius thickens least and the amplitude is fixed in all the models. Similarly, the middle part of the bar shows nearly equal BPX amplitude in all the models after evolution but its magnitude is higher than that of the central part of the bar. The outer bar region shows more thickening than the central and middle regions for all the models. But the outer region also shows the maximum thickening for the prolate halo as compared to the spherical and oblate halos.

\section{Conclusions and Discussion}
\label{sec:conclusions}
We have performed N-body simulations of bar buckling in isolated disc galaxies with different dark matter halo shapes ranging from oblate to prolate. The Fourier analysis techniques are used for the quantification of the bar and buckling strength. The main findings of this work are listed below.

(i) Oblate halos delay the bar formation and provide support to the bar against the weakening due to buckling instability, whereas prolate halos helps in early bar formation and this bar weakens more during buckling. But, the re-growth of the bar does not leave any signature of halo shape in bar amplitude. Bar strength become same in both types of halos after recovering from first buckling.

(ii) Though the bar in the prolate halo takes a longer time to re-grow than the bar in an oblate halo, the time period of first buckling event remains closely similar. This is because prolate halos take extra time to buckle after bar formation but oblate halos buckle just after bar formation. On the other hand, introducing small prolateness in the spherical halo suppresses the third buckling event that is seen in the spherical halo but, increasing more prolateness promotes the third buckling event while oblate halos always suppress third bar buckling. In our models, all the oblate halos show only two distinct buckling events. However, the most prolate halo in our set of simulations, q=1.30, displays three noticeable well defined buckling events. 

(iii) The buckling induced boxy/peanut/x-shape structure (or BPX bulge) remains closely similar in the inner part of the bar irrespective of the dark matter halo shape. But, outer edge of the bar thickens more in prolate dark matter halos as compared to oblate and spherical dark matter halos. Prolate halos make bar/disc more thicker than the oblate and spherical dark matter halos.

Oblate halos provide more gravitational potential near the disc plane as compared to prolate halos (see Fig.~\ref{fig:init_cond}). As a result, the rotation velocity, Toomre instability parameter, and vertical dispersion of the disc particles increases and bar formation is delayed in oblate halos. In our set of simulations, the most prolate halo with shape parameter q=1.30 shows the noticeable third buckling event and the spherical model shows only a very weak third buckling event. All the oblate halos and the halo with small prolateness suppress the signature of third buckling event. On the other hand, the timescale of first and second buckling events are very small as compared to the third one. This itself explains why the detection of ongoing buckling events in the observations of nearby galaxies is such a small fraction of all barred galaxies. Till date only 8 buckling events have been observed (see Table 2 in \cite{Xiang.etal.2021}) using various existing methodologies.

Our simulations do not account for the presence of gas, classical bulge and interactions with other galaxies. All these factors also affect the bar formation and buckling of the bar. The presence of gas, and classical bulge prevent bar formation whereas flyby interactions have been seen to produce bars \citep{Lokas.etal.2018, Smirnov.2019, Lokas.2020}. Studying the effect of various components of a galaxy on bar buckling and the galaxy environment, may provide constraints on the shapes of dark matter halos. \cite{O'Brien.etal.2010} and many previous studies suggest that dark matter halos shape parameter lies in the range of q=0.1 to q=1.4. A detailed statistical study of halo shapes in observations of galaxies and state of art cosmological simulations may help improve our understanding of bar buckling and the formation of boxy/peanut/x-shape structures.

	\begin{savequote}[100mm]
``We all travel the Milky Way together, trees and men.''
\qauthor{\textbf{$-$ John Muir}}
\end{savequote}

\chapter[Growth of Disc-like Pseudo-bulges in SDSS DR7 Since $z=0.1$]{Growth of Disc-like Pseudo-bulges in SDSS DR7 Since $z=0.1$}
\label{chapter6}

\section{Introduction}
\label{sec:intro}
The galaxy evolution is broadly governed by two kind of processes namely a) gravitational clustering i.e. collapses, merger events and b) internal secular processes like bar, spiral arms etc \citep{Norman.etal.1996, Conselice.2014}. It is well known that internal secular evolution of disc galaxies leads to significant changes in the properties of the bulges \citep{Kormendy.2004, Combes.2009}. Therefore, in order to understand the galaxy formation and evolution processes, the study of bulges is quite insightful. The nature of different type of bulges has been explored in simulations \citep{Athanassoula.2005} as well as in observations \citep{Fisher.Drory.2008, Fisher.Drory.2011, Erwin.et.al.2015}. Morphological studies of galaxies show that the bulge to disc ratio varies from early to late type spiral galaxies in the Hubble sequence \citep{Laurikainen.et.al.2007, Graham.Worley.2008}.

There are broadly two type of bulges given the recent understanding; classical and disc-like bulges. These are differentiated with the help of photometric, kinematic properties and stellar population of the stars they possess \citep{Athanassoula.2005, Fisher.Drory.2008, Athanassoula.2016, Laurikainen.Salo.2016}. Classical bulges are thought to be formed in major mergers \citep{Kauffmann.et.al.1993, Baugh.et.al.1996, Hopkins.et.al.2009, Naab.et.al.2014}, accretion of smaller satellites \citep{Aguerri.2001}, multiple minor mergers \citep{Bournaud.2007, Hopkins.2010}, and monolithic collapse of a primordial cloud \citep{Eggen.1962}. Classical bulges are rounder objects like elliptical galaxies and contains older population star with higher velocity dispersion compare to disc stars \citep{Kormendy.2004}. On the other hand, disc-like bulges are flattened systems like an exponential disc in the nuclear region \cite{Athanassoula.2005}. They are thought to be formed by the inward pulling of gas along the orbits and the consequent star formation \citep{Kormendy.1993, Heller.1994, Regan.2004}. These disc-like bulges are also known as pseudo-bulges \citep{Kormendy.2004} which were conceptualized by \cite{Kormendy.1982,Kormendy.1983}.

Now, the term pseudo-bulge is commonly used to describe disc-like bulges and boxy/peanut structures seen in edge-on galaxies \citep{Athanassoula.2005}. Boxy/peanut structure are vertically thick systems and are dominated by rotational motion of the stars similar to the disc-like bulges. However, it is well-established that Boxy/Peanut structures are just bars seen under different galaxy inclinations \citep{Bureau.Freeman.1999, Lutticke.etal.2000}. They are thought to be formed by disc instability during secular evolution \citep{Kormendy.2004}, vertical heating of the bar due to buckling \citep{Combes.1990, Raha.1991, Martinez.2006, Shen.et.al.2010, kataria.das.2018, Kumar.etal.2022}, or heating of the bar due to vertical resonances \citep{Pfenniger.1990}. Classical bulges are very stable against galaxy flybys but boxy/peanut structures grow significantly in major flybys \citep{Kumar.etal.2021}. Both, disc-like pseudo bulges and box/peanut structures, are very different in terms of physical properties and formation mechanism \citep{Athanassoula.2005, Laurikainen.Salo.2016}. To be more specific, we focus our study on classical bulges and disc-like pseudo-bulges (hereafter, we refer disc-like pseudo-bulge as pseudo bulge in through out the draft.)

Magneto-hydrodynamics zoom-in cosmological simulations "Auriga simulations" with sub-grid physics \citep{Gargiulo.et.al.2019} show that pseudo-bulges are prominent in Milky Way type halos. In the same study, it has been shown that around 75 $\%$ of the bulges in these simulations have in-situ stars which are formed around $z=0$ rather than in accretion events. \cite{Fisher.2006} used the PAHs (polycyclic aromatic hydrocarbons) surface brightness profiles and found that the star formation mechanisms for classical bulges and pseudo-bulges are very different. Star formation in classical bulges is fast episodic, whereas pseudo-bulges show long lasting star formation. These studies resonate with semi-analytical modeling of L-galaxies in Millenium and Millenium II simulations \citep{Izquierdo.et.al.2019} which has shown the quiet merger history for pseudo-bulged galaxies compare to classical one.

\cite{Laurikainen.et.al.2007} has shown that pseudo-bulges are widespread along all the Hubble sequence galaxies like classical bulges. It has been also pointed out that pseudo-bulges are found mostly in galaxies with lower bulge to total mass ratio ($B/T$), and galaxies with $B/T$ $\geq$ 0.5 claimed to have classical bulges surely \citep{Kormendy.Fisher.2005}. Several observational studies \citep{Weinzirl.et.al.2009,Kormendy.et.al.2010} find that the fraction of pseudo-bulges in their nearby galaxies sample is larger than 0.5. This fact has raised questions regarding hierarchical structure formation scenario under standard $\Lambda CDM$ cosmology \citep{Kormendy.et.al.2010}.

As we have seen the evolution of pseudo-bulges with cosmic time still remains a mystery. We are motivated to ask questions such as "How do the pseudo-bulge and classical bulge fraction varies with redshift?", "How do the pseudo-bulge properties varies with redshift?". In this article, we study the evolution of bulges where we want to look at the frequencies of the two types of bulges with redshifts. This will surely lead us to understand the connection between two types of bulges.

The plan of the chapter is as follows. In Section~\ref{sec:data_analysis}, we have mentioned about the complete sample and our selection criterion along with the $N-$body modeling of galaxies. The evolution of the bulges, their shapes, correlation with various photometric parameters of galaxies, and comparison with local volume survey are shown in Section~\ref{sec:results}. The effect of telescope resolution and disc inclination using simulated galaxies is explored in Section~\ref{sec:effect_of_inc_res}. The effect of data selection criteria on our results is described in Section~\ref{sec:effect_of_data_selection}. In Section~\ref{sec:discussion}, implication of this study are discussed. The brief concluding summary is pointed out in Section~\ref{sec:summary}.

\section{Data and Sampling}
\label{sec:data_analysis}
\subsection{Complete Data}
\label{sec:complete_data}
In this work, we use the archival data from \cite{Simard2011}. The detailed information of data and analysis can be found in \cite{Simard2002, Simard2011} but, for the benefit of the reader, we are describing in brief about the data. \cite{Simard2011} provides the two-dimensional decomposition of 1.12 million objects in g- and r-band from Legacy area of Sloan Digital Sky Survey Data Release Seven (SDSS DR7) \citep{Abazajian2009}. Morphologically, these objects are galaxies and have galactic extinction corrected r-band Petrosian magnitude in the range 14 to 18. The structural parameters of the galaxies were determined using 2D decomposition tool GIM2D \citep{Simard2002}. Only two components, S\'ersic bulge and exponential disc, were used for the decomposition of all the galaxies in the sample. The data is available in three formats of the fittings: (1) fixed S\'ersic index $n=4$ + disc fitting, (2) free S\'ersic index from 0.5 to 8 + disc fitting, and (3) free S\'ersic index from 0.5 to 8 fitting. For the calculation of physical parameters of the galaxies, the cosmological parameters $H_{0}=70$ km s$^{-1}$ Mpc$^{-1}$, $\Omega_{\rm m}=0.3$, $\Omega_{\Lambda}=0.7$ are used.

We also used the archival data from \cite{Salo2015}. The detailed information of data and methodology is described in \cite{Munoz-Mateos.etal.2015, Salo2015}. They provide the two-dimensional decomposition of 2352 galaxies at $3.6~\mu m$ (mid-infrared) form Spitzer Survey of Stellar Structure in Galaxies ($S^4G$) \citep{Sheth2010}. The structural parameters of the galaxies were calculated using 2D decomposition tool GALFIT \citep{Peng2002, Peng2010}. Five components, exponential disc, edge-on disc, S\'ersic bulge, ferrer bar, and unresolved central component psf (point spread function), were used to decompose the galaxies whenever required. The data is available in two formats of the fitting: (1) multi-component fitting, and (2) free S\'ersic index from 0.3 to 19 fitting.

\subsection{Our Criteria}
\label{sec:our_criteria}
\begin{figure*}
    \centering
    \includegraphics[width=0.85\textwidth]{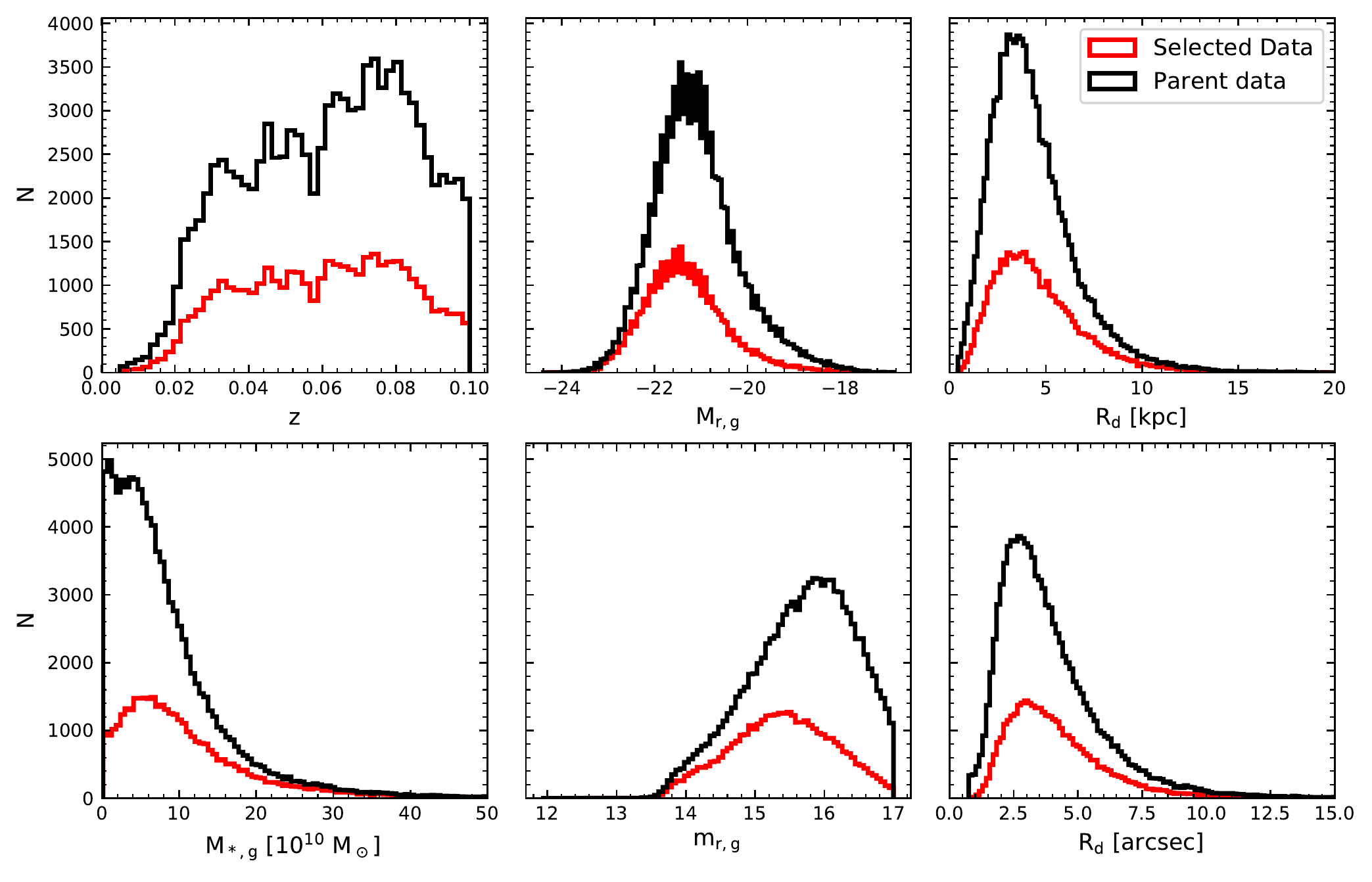}
    \caption{Complete data of galaxies and selected data for this study from \protect\cite{Simard2011}. Data distribution from left to right, top row: spectroscopic redshift ($z$), r-band absolute magnitude of galaxies (M$_{\rm r,g}$), physical disc scale radius (R$_{\rm d}$); bottom row: stellar mass of galaxies (M$_{\rm *,g}$), r-band apparent magnitude of galaxies (m$_{\rm r,g}$), and apparent disc scale radius (R$_{\rm d}$). Black and red histograms show complete data and selected data respectively. Large difference in complete data and selected data is the result of our stringent selection criteria which mostly remove faint and low-mass galaxies.}
    \label{fig:sample_sdss}
\end{figure*}

To understand the evolution of bulges in disc galaxies, we will use the free S\'ersic index from 0.5 to 8 + disc fitting of \cite{Simard2011} (hereafter SDSS data) and multi-component fitting from \cite{Salo2015} (hereafter $S^4G$ data). The redshift or distance of nearby galaxies is severely contaminated by their peculiar velocities. Therefore, we have excluded a small fraction of galaxies with $z<0.005$ as suggested in \cite{Shen2003}. We have limited our study to maximum redshift of $z=0.1$, and spectroscopically confirmed galaxies which are best fitted by two components. We have avoided all the galaxies with apparent bulge effective radius and disc scale radius equal or smaller than the psf radius as these galaxies will have more uncertainties in the photometric parameters. Next we rejected fainter (m$_{\rm r}$ > 17) and smaller galaxies (M$_{*}$ < 10$^{9}$ M$_{\odot}$) from our sample. Stellar mass of the galaxy (M$_{*}$) was estimated using color-stellar mass-to-light ratio relation (CMLR). Stellar mass-to-light ratio ($\gamma_{*}$) of galaxies is related to the color of galaxies by the following expression,
\begin{equation}
    \log \gamma_{*}^{\rm j} = a_{\rm j} + b_{\rm j} \times color,
\end{equation}
where $a_{\rm j}$ and $b_{\rm j}$ are two constants calculated for $\rm j^{th}$ imaging band at given color index (e.g. $g-r$ in our case). To calculate r-band stellar mass-to-light ratio, We have adopted values of $a_{\rm j}$ and $b_{\rm j}$ from \cite{Bell.etal.2003}. Recently, \cite{Du.McGaugh.2020} have re-calibrated these constants for the galaxy mass consistency in various stellar mass-to-light ratio estimators. Both studies give same constants in r-band for $g-r$ color. After all these constraints, we end-up with a total sample of 105,160 galaxies. Hereafter, we refer this sample to parent data.

To minimize the uncertainties in our results, we have imposed the strict constraint of maximum statistical error of $10\%$ on the magnitude of each component, bulge S\'ersic index, disc inclination, bulge to total light ratio, bulge ellipticity, and size of each component. We removed all the galaxies whose S\'ersic index is 0.5 or 8.0 and bulge ellipticity is 0.0 or 0.7 to avoid the fitting bias in the sample due to presence of bar, close merger, point source etc \citep{Simard2011}. Finally, we constrain our sample to the galaxies where half-light radius of the bulge is smaller than the half-light radius of the disc. These constraints reduce our parent data of 105,160 galaxies to 40,504 galaxies.

Since we are interested in the evolution of the bulges, we need to take care of barred galaxies in the sample. To remove the barred galaxies from our sample, we used the catalog of "Galaxy Zoo 2" classification \citep{Willett.etal.2013, Hart.etal.2016}. Galaxy Zoo is a citizen science project, where images of the galaxies are shown to the interested citizens and they are asked to answer some basic questions related to the galaxy morphology. These answers are used for morphological classification of the galaxies depending on the probability of the voting. Though this method does not removes all the barred galaxies, but it improves our sample. Removal of barred galaxies reduces our sample from 40,504 to 38,996 galaxies. Hereafter, we refer this sample to selected data.

For a comparison between our selected data and the parent data of galaxies, we have shown the number distribution of galaxies in Fig.~\ref{fig:sample_sdss}. From left to right, top row panels: spectroscopic redshift ($z$), r-band absolute magnitude of galaxies (M$_{\rm r,\rm g}$), physical disc scale radius (R$_{\rm d}$); bottom row panels: stellar mass of galaxies (M$_{*,\rm g}$), r-band apparent magnitude of galaxies (m$_{\rm r,\rm g}$), and apparent disc scale radius (R$_{\rm d}$). Physical size of galaxy was calculated using its angular distance as mentioned in \cite{Simard2011}. Our selected data represents the fair sample of the parent data except at faint and low-mass end of the distribution. It is because our strict constraint of maximum $10\%$ statistical error in each photometric parameter. Later, in Section~\ref{sec:effect_of_data_selection}, we will discuss the effect of this strict selection criteria on our results.

In our analysis, we have divided data in several different categories. The disc dominating and bulge dominating galaxies are defined on the basis of bulge to total light ratio $B/T$ in r-band. The disc dominating galaxies are those which have $B/T \leq 0.5$ and the bulge dominating galaxies have $B/T > 0.5$. None of the photometric classification of the bulges provides proper separation between classical bulges and pseudo-bulges. Particularly, $n=2$ S\'ersic index does not show bi-modality in the distribution of the bulges \citep{Graham.2013, Graham.2014, Costantin.etal.2018, Mendez-Abreu.etal.2018, Gao.etal.2020, Kumar.etal.2021}. Some elliptical galaxies, and bulges of S0 and merger built galaxies also show lower S\'ersic indices \citep{Davies.etal.1988, Young.Currie.1998, Eliche-Moral.2011, Querejeta.etal.2015, Tabor.etal.2017}. In Section~\ref{sec:effect_of_inc_res}, We have also verified using numerical models of Milky Way mass galaxies that S\'ersic index is not reliable for bulge classification, given the inclination of the disc and resolution of the telescope. Hence, for a more precise classification of the bulges, we have adopted a combination of S\'ersic index based classification and Kormendy relation based classification, which we call S\'ersic-Kormendy classification. S\'ersic index based classification (hereafter S\'ersic classification) makes use of $n=2$ S\'ersic index. All the bulges having $n \leq 2$ are categorized as pseudo-bulges, whereas bulges with $n > 2$ are categorized as classical bulges \citep{Fisher2006_proce, Fisher.Drory.2008}. On the other hand, Kormendy relation based classification (hereafter Kormendy classification) uses well known Kormendy relation for elliptical galaxies to separate two types of bulges. All the bulges which lie above $3\sigma$ limit of the Kormendy relation are grouped as classical bulges, while the bulges which fall below $3\sigma$ limit are grouped as pseudo-bulges \citep{Gadotti.2009}. For this study, we have taken Kormendy relation from \cite{Lackner.Gunn.2012}. In S\'ersic-Kormendy classification, only those bulges are considered which satisfy both S\'ersic and Kormendy classifications simultaneously. For the sake of comparison, we will also be showing our results using S\'ersic classification and Kormendy classification along with S\'ersic-Kormendy classification.

\subsection{Simulated Galaxy Models}
\label{sec:galaxy_models}
To understand the effect of spatial resolution of telescope, projection of galaxy, and surface density of galaxy on fitting parameters, particularly S\'ersic index, we simulate model galaxies with bulge to disc mass ratio ($B/T$) 0.1, 0.3, 0.5, and 0.7 keeping fixed disc mass and fixed bulge scale radius. There are four disc surface densities for each model. The total mass (stellar bulge + stellar disc + dark halo) of each the model galaxy is similar to that of Milky Way type galaxy. We use the publicly available $N-$body code GALIC \citep{Yurin2014} to generate our model galaxies. In our models, dark matter halo density is represented by Hernquist profile,
\begin{equation}
    \rho_{\rm dm}(r)=\frac{M_{\rm dm}}{2\pi}\frac{a}{r(r+a)^3}
    \label{eqn:halo}
\end{equation}
where `$a$' is dark matter halo scale radius. This is related to the concentration parameter `$c$' of a corresponding NFW halo \citep{NFW1996} of mass M$_{\rm dm}$=M$_{\rm 200}$ by the following expression,
\begin{equation}
    a=\frac{r_{200}}{c}\sqrt{2\left[\ln{(1+c)-\frac{c}{(1+c)}}\right]}
    \label{eqn:scale_radius}
\end{equation}
where r$_{200}$ is the virial radius of galaxy. This is the radius within which the average matter density is 200 times the critical density of the Universe. M$_{200}$ is mass within the virial radius.

The disc density decays exponentially in the radial direction and its vertical distribution is described using $\sech^{2}$ profile
\begin{equation}
    \rho_{\rm d}(\rm R,\rm z)=\frac{M_{\rm d}}{4\pi z_{0} R_{\rm s}^{2}}\exp\left(-{\frac{R}{R_{\rm s}}}\right) \sech^{2}\left(\frac{z}{z_{0}}\right)
    \label{eqn:disc}
\end{equation}
where M$_{\rm d}$ is total disc mass, R$_{\rm s}$ is disc scale radius, and $z_{0}$ is disc scale height.

The bulge density is also modelled by the spherically symmetric Hernquist density profile
\begin{equation}
    \rho_{\rm b}(r)=\frac{M_{\rm b}}{2\pi}\frac{b}{r(r+b)^{3}}
    \label{eqn:bulge}
\end{equation}
where M$_{\rm b}$ represents the total bulge mass and `$b$' represents the bulge scale radius.

We set 1 million particles for each component which results in a total of 3 million particles in each model galaxy. The rotation velocity of each model galaxy is set to be 220 km s$^{-1}$ which corresponds to a total mass $2.47\times10^{12}$M$_{\odot}$ (where M$_{\odot}$ = mass of the Sun). The disc mass is fixed at 0.04 of the total galaxy mass. The gravitational softening is set to be 0.01 $kpc$ for each type of particle.

For the purpose of bulge-disc decomposition of the simulated galaxies, we use the latest version of GALFIT \citep{Peng2002, Peng2010} which is a widely used for 2d decomposition of galaxies. To see the effect of the spatial resolution, the pixel size is set to be $0.025~kpc\times0.025~kpc$ (hereafter high-resolution) and $0.05~kpc\times0.05~kpc$ (hereafter low-resolution). For the decomposition of simulated galaxies, we did not consider any background sky and the point spread function (psf) is taken to be delta function.

\section{Results}
\label{sec:results}
\subsection{Distribution of Bulges with Redshift}
\label{sec:bulge_with_z}
\begin{figure*}
    \centering
    \includegraphics[width=\textwidth]{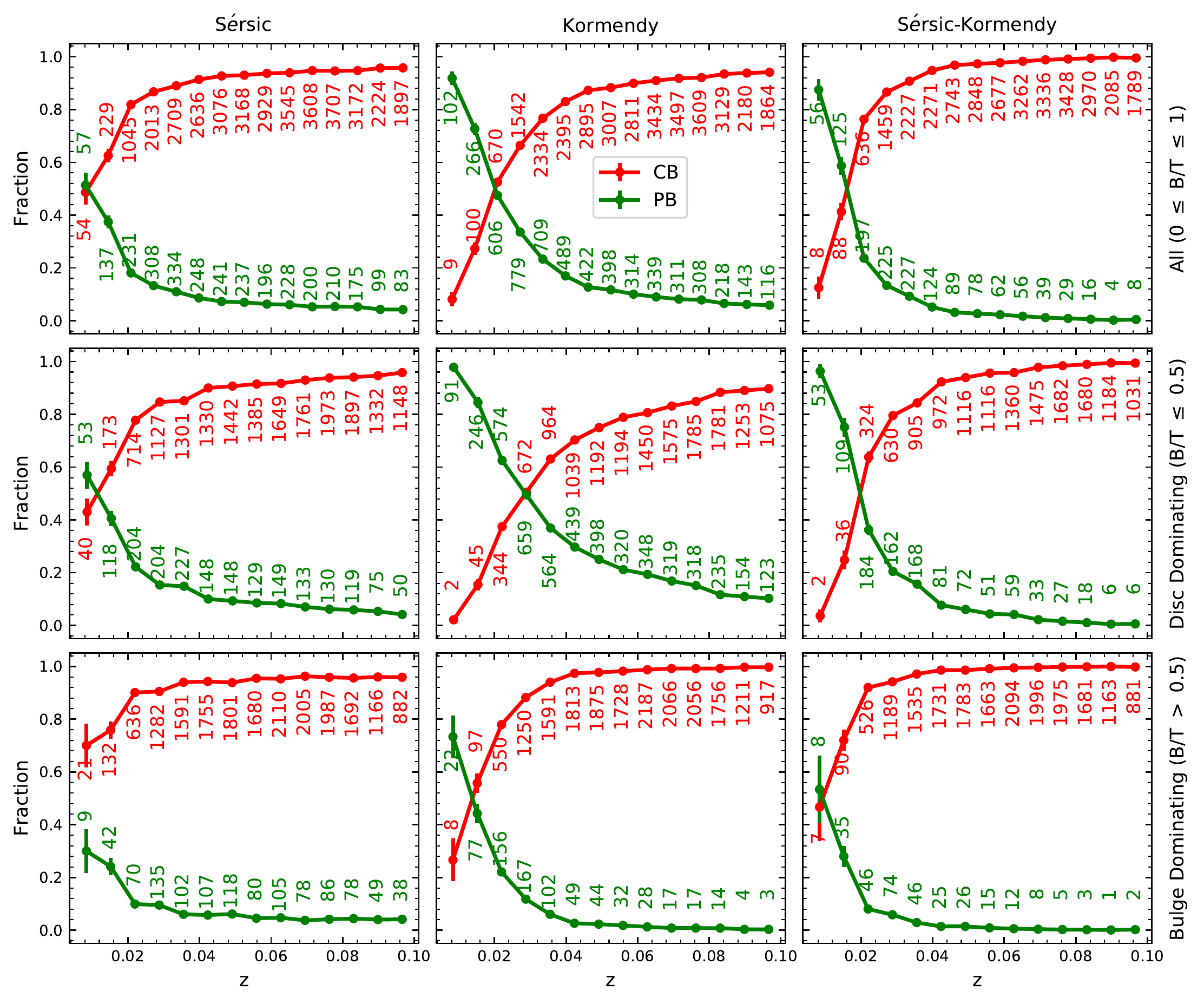}
    \caption{Fractional distribution of the classical and pseudo-bulges with redshift ($z$) in the SDSS survey. Left, middle, and right columns show S\'ersic, Kormendy, S\'ersic-Kormendy classifications respectively. Top, middle, and bottom rows represent all, disc dominating, and bulge dominating galaxies respectively. Red and green curves represent classical bulges (CB) and pseudo-bulges (PB) respectively. The numbers of classical bulge and pseudo-bulge galaxies at any point are shown with their respective colors. All the methods show domination of pseudo-bulges over classical bulges with decreasing redshift.}
    \label{fig:bulge_with_z}
\end{figure*}

The evolution of the classical and pseudo-bulges over the cosmic time from redshift 0.005 to 0.1 is represented in the Fig.~\ref{fig:bulge_with_z} from SDSS survey. The effect of classification method on the distribution of the bulge type is shown in the left, middle, and right columns of the figure which are corresponding to the S\'ersic classification, Kormendy classification, and S\'ersic-Kormendy classification respectively. The contribution of galaxy type on the distribution of the bulge type is shown in the top, middle, and bottom rows of the figure which are the representative of all, disc dominating, and bulge dominating galaxies respectively. Red curves show fraction of classical bulges and green curves show fraction of pseudo-bulges. Statistical uncertainty in the fraction is shown using error bars and is estimated from $\sqrt{[f \times (1-f)]/N}$, where $f$ is the fraction of any point and $N$ is number of objects at that point in the distribution \citep{Sheth.etal.2008}. The numbers of classical bulge and pseudo-bulge galaxies at any point are shown with their respective colors. At many points, error bar is smaller than the size of the symbol. All the panels (except bottom left) of this figure indicate that the local Universe is dominated by pseudo-bulges. The high fraction of the pseudo-bulges is contrary to the prediction of cosmological simulations. According to widely accepted $\Lambda CDM$ cosmological model, we expect more classical bulges instead of pseudo-bulges \citep{White.Rees.1978, Aguerri.2001, Bournaud2005, Baugh.2006, Brooks2016}.

As we go back in the time towards the high-redshift Universe, the fraction of pseudo-bulge decreases and classical bulges start dominating over the pseudo-bulges. In the evolution of the Universe, there comes a time before that Universe was dominated by the classical bulges or spheroids. Since no bulge classification method shows perfect bimodality in two types of bulges, the precise redshift of equality in classical bulges and pseudo-bulges cannot be well constrained. S\'ersic classification shows the point of equality a lower redshift than the Kormendy classification. As a consequence, S\'ersic-Kormendy classification show equality in between the two classification. If we see the distribution of all the sample galaxies in S\'ersic, Kormendy, and S\'ersic-Kormendy classifications, the fraction of the classical and pseudo-bulge becomes equal at $z \approx$ 0.009, 0.020, 0.016 redshifts respectively.

One can see the effect of bulge classification criterion on the distribution of bulge type in all, disc dominating, and bulge dominating galaxies by comparing the panels of top, middle, and bottom row respectively. All three rows show that the fractional distribution of the bulges in S\'ersic classification is very different than the other two classifications. In the low-redshift region, it always shows small fraction of pseudo-bulges as compare to the Kormendy and S\'ersic-Kormendy classifications. But, the trends of distribution are more or less similar in Kormendy and S\'ersic-Kormendy classifications except for bulge dominating galaxies at low-redshift where Kormedy classification show more pseudo-bulges than the S\'ersic-Kormendy classification.

Similarly, the contribution of galaxy type on the distribution of bulge type in S\'ersic, Kormendy, S\'ersic-Kormendy classifications can be seen by comparing the panels of left, middle, and right column respectively. From all three columns, it is clear that the disc dominating galaxies provide more pseudo-bulges than the bulge dominating galaxies in the fractional distribution of the bulges at low-redshift. This implies that most of the pseudo-bulges are low mass relative to their hosting discs. Hence, we conclude that the Kormendy and S\'ersic-Kormendy classifications show quite similar distribution of the bulges but, S\'ersic classification results in a lower fraction of pseudo-bulges at low-redshift. Kormendy classification show slightly higher fraction of pseudo-bulges than S\'ersic-Kormendy classification at low-redshift in bulge domination galaxies. However, local volume remains pseudo-bulge domination irrespective to the classification scheme, and the dominating contribution of pseudo-bulges comes, mostly, from the disc dominating galaxies.

\subsection[]{Distribution of Bulges with R$_{\rm e}$/R$_{\rm hlr}$}
\label{sec:bulge_with_ReRd}
\begin{figure*}
    \centering
    \includegraphics[width=\textwidth]{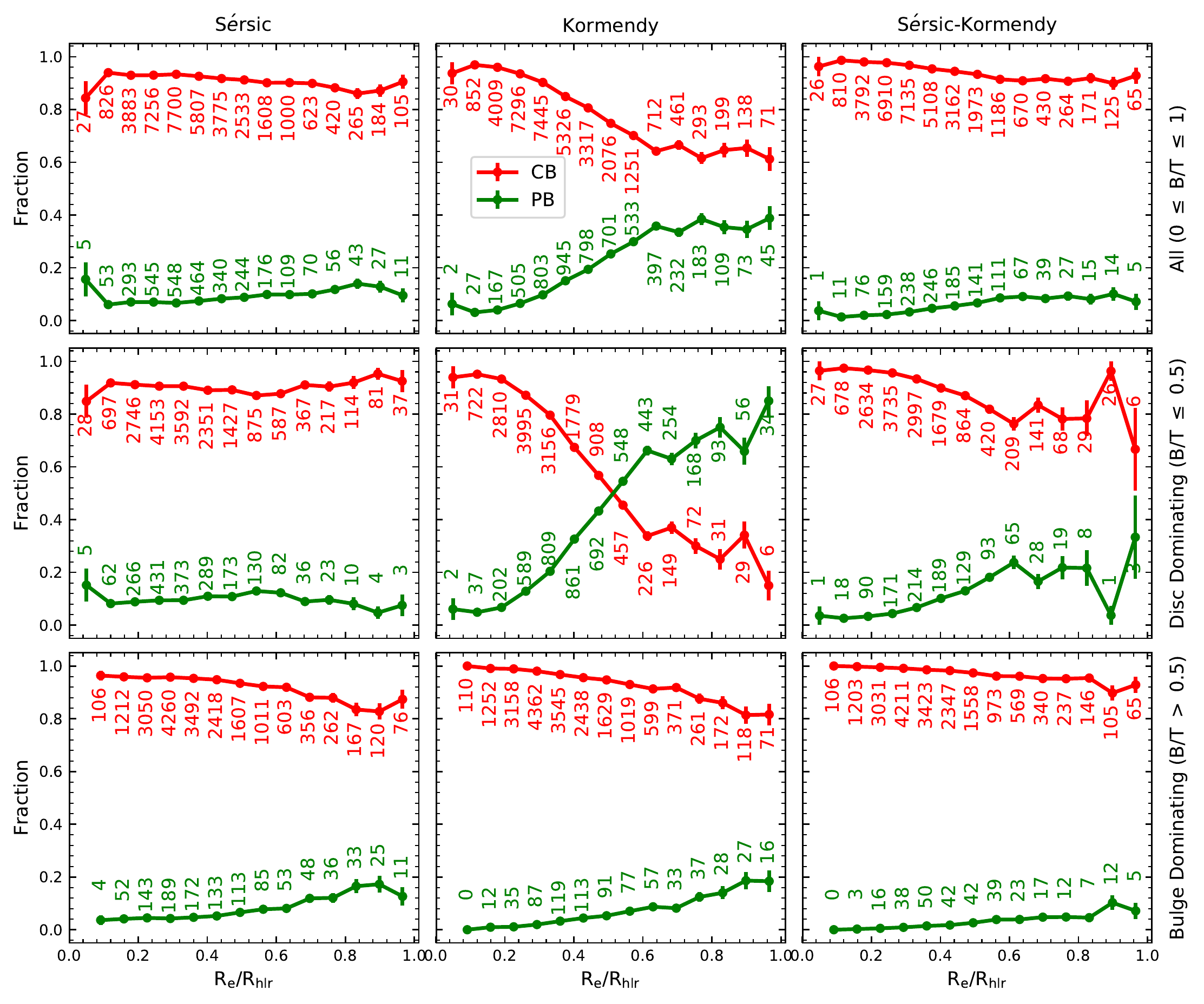}
    \caption{Fractional distribution of classical and pseudo-bulges with the ratio of bulge semi-major half-light radius (R$_{\rm e}$) to disc semi-major half-light radius (R$_{\rm hlr}$) in the SDSS survey. Left, middle, and right columns show S\'ersic, Kormendy and S\'ersic-Kormendy classifications respectively. Top, middle, and bottom rows represent all, disc dominating, and bulge dominating galaxies respectively. Red and green curves represent classical bulges and pseudo-bulges respectively. The numbers of classical bulge and pseudo-bulge galaxies at any point are shown with their respective colors. The fraction of pseudo-bulges increase with increasing bulge to disc size ratio.}
    \label{fig:bulge_with_ReRd}
\end{figure*}

To see the effect of the bulge and disc sizes on the distribution of bulge type, we have calculated the fractional distribution of the classical and pseudo-bulges with the ratio of bulge semi-major half-light radius R$_{\rm e}$ to disc semi-major half-light radius (R$_{\rm hlr}$) from SDSS survey in Fig.~\ref{fig:bulge_with_ReRd}. The ratio of bulge semi-major half-light radius to disc semi-major half-light radius (R$_{\rm e}$/R$_{\rm hlr}$) is better than the absolute bulge or disc size because it removes the error, if any, in the size calculation using distance or redshift of the galaxy. However, in appendix~\ref{app:z_vs_Re_Rd}, we have also shown the absolute sizes of the bulges and discs with redshift and compared them together. In this figure, columns show the effect of the classification method and rows show the effect of galaxy type on the distribution of bulge type. Left, middle, and right columns represent S\'ersic, Kormendy, and S\'ersic-Kormendy classifications respectively. Top, middle, and bottom rows are the representatives of all, disc dominating, and bulge dominating galaxies respectively. From all the panels of the figure, it is clear that a large fraction of the small bulges (small R$_{\rm e}$/R$_{\rm hlr}$) is classical in nature. The fraction of pseudo-bulges increases with increasing R$_{\rm e}$/R$_{\rm hlr}$ or vice-versa the fraction of classical bulges increases with decreasing R$_{\rm e}$/R$_{\rm hlr}$. Unitl R$_{\rm e}$/R$_{\rm hlr} \approx 0.6$, these trends are valid in all the panels.

The effect of bulge classification method on the distribution of bulge type in all, disc dominating, and bulge dominating galaxies can be seen by comparing the panels of the top, middle, and bottom rows respectively. All the three rows of the figure show that the Kormendy classification results in more pseudo-bulges than the other two classification methods for R$_{\rm e}$/R$_{\rm hlr} \geq 0.6$. Particularly, this difference is more distinct in disc dominating galaxies. However, the classification of bulges in bulge dominating galaxies does not shows significant change among three methods. In the same way, one can see the effect of galaxy type on the distribution of bulge type in S\'ersic, Kormendy, and S\'ersic-Kormendy classifications by comparing the panels of left, middle, and right columns respectively. All the columns of the figure clearly indicate that the contribution of disc dominating galaxies in the distribution of pseudo-bulges is always great or equal to that of the bulge dominating galaxies until R$_{\rm e}$/R$_{\rm hlr} \approx 0.6$. Hence, for R$_{\rm e}$/R$_{\rm hlr} \leq 0.6$, a large fraction of pseudo-bulges comes from the disc dominating galaxies. Inversely, a large fraction of classical bulges comes from the bulge dominating galaxies. These conclusions hold for all the classification methods discussed here. When moving towards larger value of R$_{\rm e}$/R$_{\rm hlr}$, disc dominating galaxies start showing increasing classical bulges in S\'ersic and S\'ersic-Kormendy classifications. But, the Kormendy classification still shows increasing pseudo-bulges in disc dominating galaxies. According to Hubble classification, S0 galaxies have massive and large bulges which are comparable to their discs. These bulges are generally classical in nature. Therefore, we expect increase in the fraction of classical bulges close to R$_{\rm e}$/R$_{\rm hlr} = 1$ which is not coming out in Kormendy classification. However, S\'erisc-Kormendy classification captures it well.

For a given bulge to total light ratio ($B/T$), the distribution of the bulges with R$_{\rm e}$/R$_{\rm hlr}$ follows more or less similar trends to that of Fig.~\ref{fig:bulge_with_ReRd}. This implies that the ratio R$_{\rm e}$/R$_{\rm hlr}$ can be treated as an indicator of the ratio of disc to bulge mean surface densities (see the appendix~\ref{app:bulge_disc_bright}). For example, we can say that the bulges with R$_{\rm e}$/R$_{\rm hlr}<0.5$ are more concentrated (or dense) than the bulges with R$_{\rm e}$/R$_{\rm hlr}>0.5$ relative to their hosting discs. From all the panels of the Fig.~\ref{fig:bulge_with_ReRd}, we can see that the preferential condition for the bulges to be pseudo is their low concentration relative to the hosting discs. Concentrated bulges are usually classical in nature. Hence the relative surface densities of the bulge and disc play a crucial role in the formation and evolution of the bulges. A low surface density disc usually lacks from global instabilities in its center \citep{Toomre.1964, Mihos.etal.1997, Mayer.Wadsley.2004, Sodi.etal.2017, Peters.Rachel.2019}. So, most probably, it will also lack from the pseudo-bulge.

\subsection{Distribution of Bulge Ellipticity with Redshift}
\label{sec:ellipticity_with_z}
\begin{figure*}
    \centering
    \includegraphics[width=\textwidth]{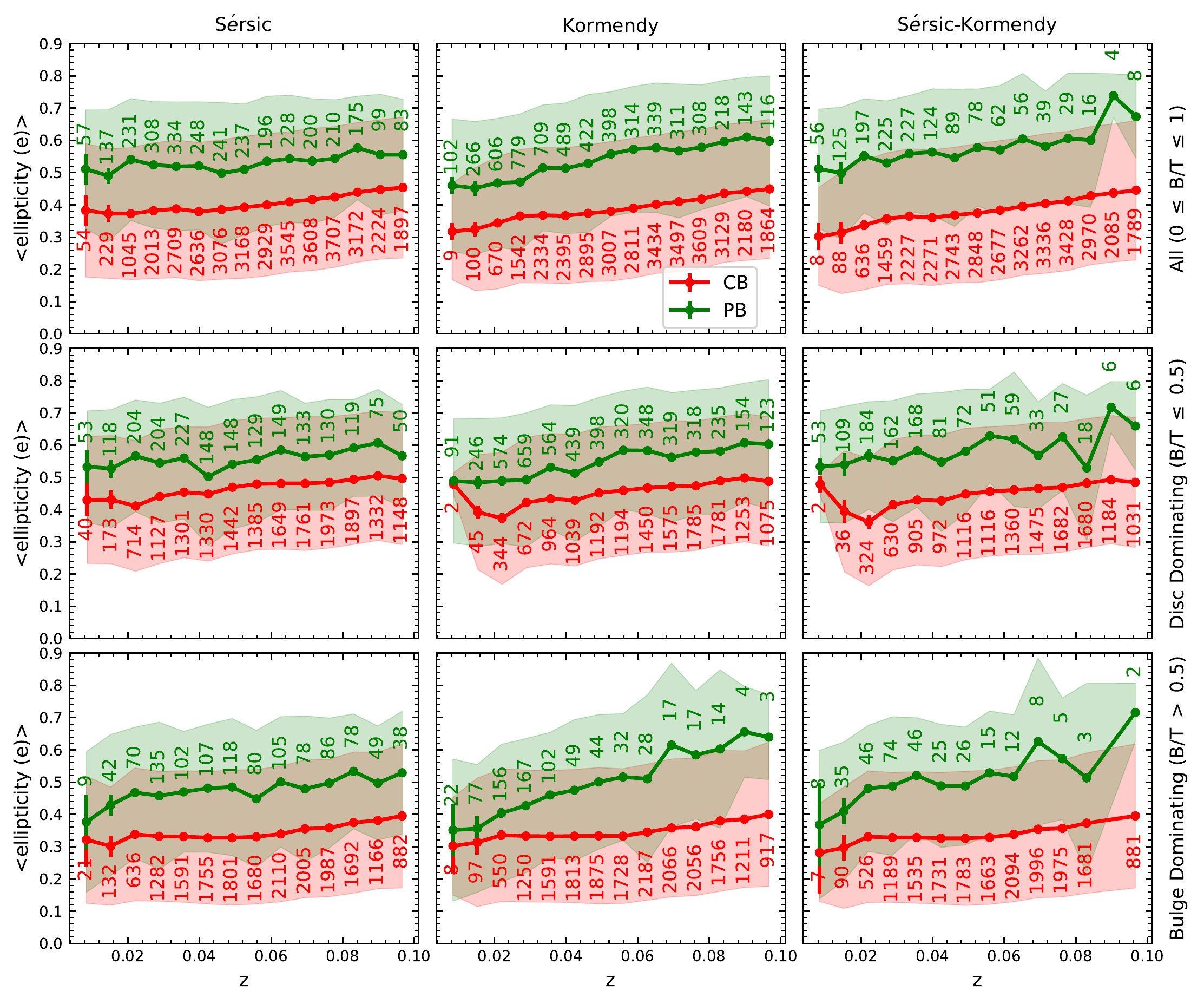}
    \caption{Distribution of mean deprojected bulge ellipticity ($<e=1-b/a>$) with redshift in the SDSS survey. Left, middle, and right columns show S\'ersic, Kormendy, and S\'ersic-Kormendy classifications of bulges respectively. Top, middle, and bottom rows represent all, disc dominating, and bulge dominating galaxies respectively. Red, and green color curves represent the distribution of classical bulges, and pseudo-bulges bulges respectively. Shaded regions show standard deviation from mean and vertical bars represent statistical uncertainty due to number of galaxies. The numbers of classical bulge and pseudo-bulge galaxies at any point are shown with their respective colors. All panels show that bulges are moving close to the round shape with decreasing redshift.}
    \label{fig:ellipticity_with_z}
\end{figure*}

Classical bugles are supported by the random motion of stars that makes them more round in shape as compare to the pseudo-bulges which are supported by the ordered rotational motion of the stars. Rotational motion of the stars in pseudo-bulges provide them flat shape. In its evolution, galaxy goes through several dynamical and morphological changes due to secular evolution and interactions with other galaxies \citep{Kumar.etal.2021}. To trace the morphological changes in the bulges during the evolution of the galaxies, we calculated mean ellipticities ($<e>$) of the classical bulges and pseudo-bulges as a function of redshift. For this purpose, we deprojected all the bulges to minimize the effect of disc inclination on projected shape of the bulges (mainly pseudo-bulges). Fig.~\ref{fig:ellipticity_with_z} shows the evolution of the mean bulge ellipticity of classical bulges and pseudo-bulges with redshift for SDSS data. Red and green curves represent classical bulges and pseudo-bulges respectively. The standard dispersion from mean is displayed using shaded regions around mean values, and statistical uncertainty due the number of galaxies is shown with vertical bars. The numbers of classical bulge and pseudo-bulge galaxies at any point are shown with their respective colors. We have demonstrated S\'ersic, Kormendy, and S\'ersic-Kormendy bulge classification schemes in first, middle, and right columns respectively. Rows, starting from the top, show the ellipticity evolution in all, disc dominating, bulge dominating galaxies respectively.

One thing we can clearly notice from this figure is that the mean ellipticity of the pseudo-bulges is always higher than the classical bulges irrespective to the classification scheme and redshit. It means that the classical bulges are generally rounder than the pseudo-bulges when seen in face-on projection of galaxies. On the other hand, all the panels show decreasing mean ellipticity with decreasing redshift for both types of bulges (when considering points with statistically significant count of galaxies for reasonable mean ellipticity). The reducing mean ellipticity of the bulges implies that the both types of bulges are getting rounder and rounder with the evolution of the galaxies.

The effect of bulge classification scheme on the evolution of mean bulge ellipticity in all, disc dominating, and bulge dominating galaxies can be marked by comparing all the panels of top, middle, and bottom rows respectively. From all the rows, we can see that Kormendy and S\'ersic-Kormendy classification show quite similar declining trends in mean ellipticity for both types of bulges. S\'ersic classification method exhibits shallower decline in mean ellipticity of classical bulges than the other two classification methods. But, at low-redshift, the ellipticity of pseudo-bulges remains nearly unaffected in three classification methods. In the similar manner, one can observe the effect of galaxy type on the distribution of bulge ellipticity in S\'ersic, Kormendy, and S\'ersic-Kormendy classifications by comparing all the panels of left, middle, and right columns respectively. From all the columns, one can clearly see that both types of bulges in bulge domination galaxies show lower mean ellipticity than the disc dominating galaxies. It is true for whole redshift range and for three classifications when considering points with statistically significant number of galaxies. Also, the rate of declining mean ellipticity is steeper in bulge dominating galaxies than in disc dominating galaxies. Now, we can conclude that the shape of massive bulges is more axisymmetric than the low-mass bulges at whole redshift range. Also, the high-mass pseudo-bulges are moving rapidly towards the axisymmetry.

One should note that we have removed the sample of barred galaxies in our analysis. The observed fraction of the barred galaxies ranges from $30\%$ to $60\%$ in the local volume \citep{Aguerri.etal.2009, Masters.etal.2011, Sodi.etal.2015, Diaz-Garcia.etal.2016}. This fraction declines when we move from low-redshift to high-redshift \citep{Sheth.etal.2008, Melvin.etal.2014}. Therefore, it may possible that some galaxies with bulges had formed bar during their evolution. The conversion of bulged galaxies into barred galaxies can affect the mean ellipticity of the bulges. But, how much it will influence evolution of the bulges remains the question for further study.

\subsection{Evolution of the Bulges with Galaxies}
\label{sec:gal_prop_with_z}
\begin{figure*}
    \centering
    \includegraphics[width=\textwidth]{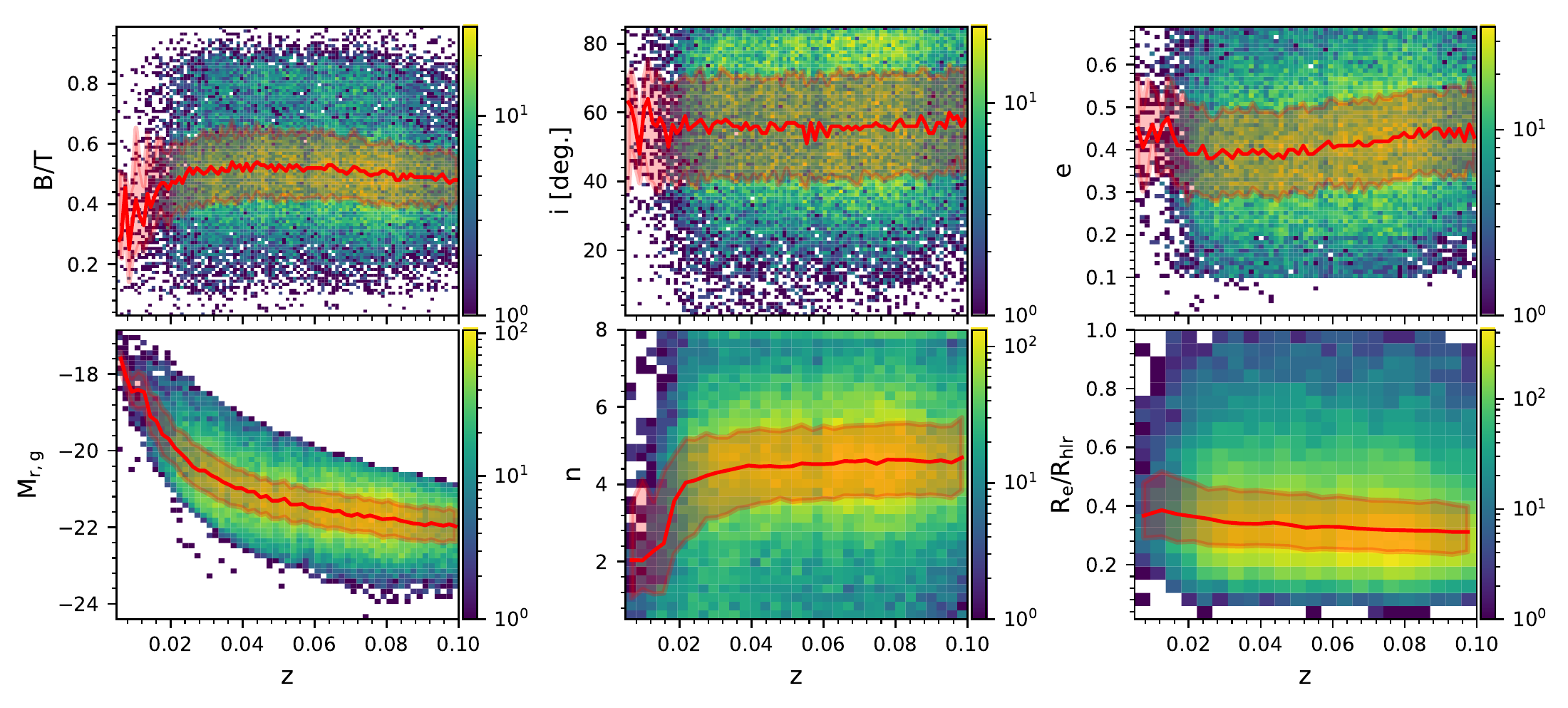}
    \caption{Distribution and evolution of different photometric properties of the galaxies with the redshift ($z$) from SDSS survey. On the y axis, $B/T$ = bulge to total light ratio, $i$ = disc inclination, e = bulge ellipticity, $M_{r,g}$ = r band galaxy absolute magnitude, n = bulge S\'ersic index, and $R_{\rm e}/R_{\rm hlr}$ = bulge to disc half-light radius ratio. The number density of the galaxies is represented by the color on log scale. The red solid curves show the median of the data and shaded region represents the $\pm 25 \%$ dispersion of the data from median.}
    \label{fig:gal_prop_with_z}
\end{figure*}

The different components of the galaxies evolve together with the evolution of the Universe. Therefore, it will be interesting to understand the evolution of the photometric properties of the galaxies with the redshift and their correlations with the evolution of the bulges. In the Fig.~\ref{fig:gal_prop_with_z}, we have shown the distribution and evolution of bulge to total light ratio ($B/T$), disc inclination ($i$), bulge ellipticity ($e$), r band galaxy absolute magnitude (M$_{\rm r,g}$), bulge S\'ersic index ($n$), and bulge to disc half light radius ratio (R$_{\rm e}$/R$_{\rm hlr}$) with redshift ($z$) from SDSS survey. The color of the distribution represents the number density of galaxies on log scale. The red solid curves represents the median of the data and shaded color show the $\pm 25 \%$ dispersion of the data from median. 

From the first column of the first row, one can see that the median of bulge to total light ratio ($B/T$) is decreasing with decreasing redshift. The Universe is becoming disc dominating (dynamically cool) as it is getting older and older. In the local Universe, median + 25$\%$ curve is below the $B/T=0.5$ cut-off i.e. matter fraction in disc component is higher than the bulge component. This suggests that nearly 75$\%$ of the local Universe is dynamically cool. Similar signatures of dynamical cooling can be found from the second and third columns of the first row which show that the medians of inclination ($i$) is nearly constant however the bulge ellipticity ($e$) is increasing with decreasing redshift. Increase in the bulge ellipticity at the constant disc inclination is the indication of the deviation of the bulges from their classical nature. Note that the less number of data points near zero ellipticity in third column of first row is the result of our tight criteria of maximum $10\%$ error in any quantity. Also, the increase in median ellipticity here should not be confused with decrease in mean ellipticity in Fig.~\ref{fig:ellipticity_with_z}. Here, we have considered inclined galaxies.

First column of the second row shows that the median of r band galaxy absolute magnitude (M$_{\rm r,g}$) is increasing with decreasing redshift. These two quantity, magnitude and redshift, have strong and obvious correlation because at a given redshift/distance, we cannot observe an object fainter than the limit of the telescope. Therefore, we are seeing increased number of the fainter galaxies with decreasing redshift. 
The second column of the second row illustrates the decreasing median of S\'ersic index ($n$) with decreasing redshift. This also indicates the dynamical cooling of the galaxies with the evolution of the Universe. The third column of the second row shows the evolution of the bulge to disc semi-major half light radius ratio (R$_{\rm e}$/R$_{\rm hlr}$) with redshift. Though the change is very small, but it is increasing linearly indicating either faster growth of the bulges than the discs or the growing ellipticity of the bulges. 

\begin{figure*}
    \centering
    \includegraphics[width=0.55\textwidth]{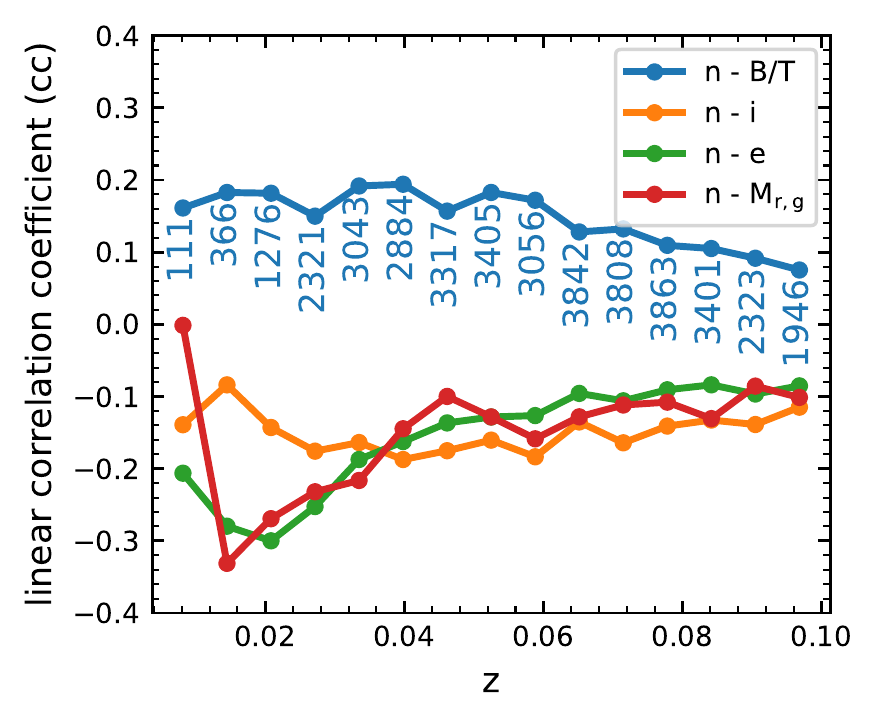}
    \caption{Linear correlation coefficient of the bulge S\'ersic index ($n$) with different photometric properties of the galaxies in different redshift bins of SDSS survey. The blue, orange, green, and red color solid curves represent the linear correlation of bulge S\'ersic index with bulge to total light ratio ($B/T$), disc inclination ($i$), bulge ellipticity ($e$), and r band galaxy absolute magnitude (M$_{\rm r,g}$) respectively. The number of galaxies at any point are also shown at respective redshift.}
    \label{fig:bulge_corr_coef}
\end{figure*}

We have also calculated the Pearson's linear correlation coefficient (cc) of the bulge S\'ersic index ($n$) with other photometric properties of the galaxies in different redshift bins of the SDSS survey. The linear correlation coefficient of two variables tells how the one variable changes with the change in the other variable. Its value lies between -1 (for a perfect negative correlation) and +1 (for a perfect positive correlation). In the Fig.~\ref{fig:bulge_corr_coef}, the blue, orange, green, and red color solid curves represent the linear correlation of bulge S\'ersic index with bulge to total light ratio ($B/T$), disc inclination ($i$), bulge ellipticity ($e$), and r band galaxy absolute magnitude (M$_{\rm r,g}$) respectively. At high-redshift, $B/T$ has very weak positive correlation. It weakly develops positive correlation with decreasing redshift. Disc inclination has very weak and nearly constant anti-correlation at all redshifts. Bulge ellipticity and absolute galaxy magnitude show the weakly increasing negative correlation as we move towards the low-redshift Universe. However, they show sudden decrease in the negative correlation at very low-redshift. Only bulge ellipticity and absolute galaxy magnitude show significant, but small, anti-correlation just before the sudden drop at low-redshift.

\subsection{Comparison with Local Volume Survey}
\label{sec:compare_sdss_s4g}
\begin{figure*}
    \centering
    \includegraphics[width=0.7\textwidth]{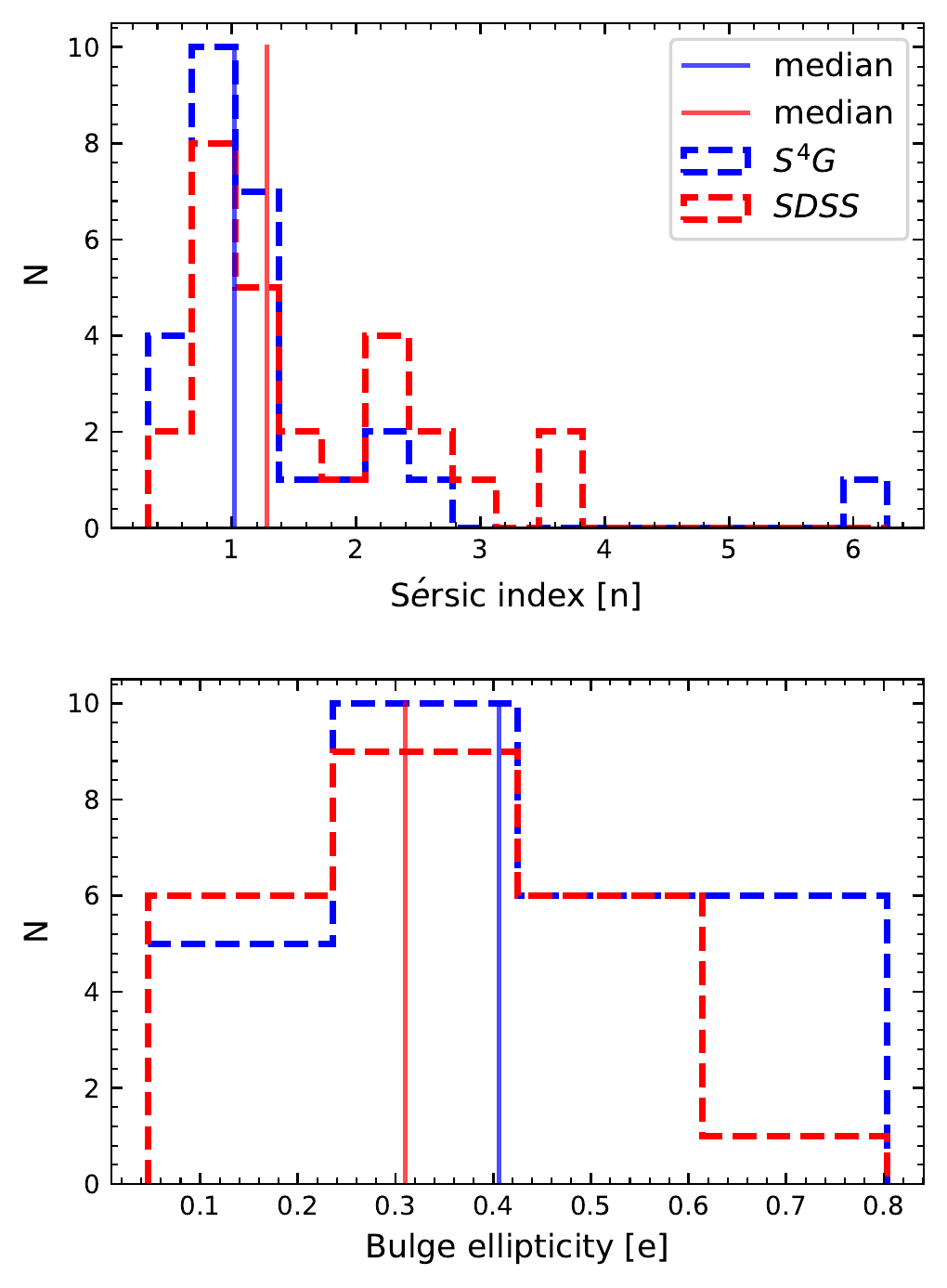}
    \caption{Comparison of SDSS survey with $S^4G$ survey. Top panel shows the S\'ersic index histogram and bottom panel shows the bulge ellipticity histogram of 27 galaxies, common in both survey, which are fitted by S\'ersic component. The red and blue filled histograms show the galaxies from SDSS and $S^4G$ survey respectively and vertical lines represent the medians of respective data. SDSS data over-estimates S\'ersic index and under-estimates bulge ellipticity.}
    \label{fig:compare_sdss_s4g}
\end{figure*}

The SDSS data provides two-component fittings of the galaxies. But, in reality, a galaxy can have more or less than two components. Therefore, two-component fittings can provide biased parameters of the galaxies which do not have exactly two components \citep{Aguerri.etal.2005, Laurikainen.etal.2005, Gadotti.2009, Weinzirl.et.al.2009, Mendez-Abreu.etal.2014, Mendez-Abreu.etal.2017, Gao.etal.2019}. So, it is always good to go for the multi-component fittings of the galaxies. The robustness of the multi-component fittings over the two-component fittings has been discussed in \cite{Salo2015} for barred and unbarred galaxies. The two-component fittings of barred galaxies over estimate the mass, size and S\'ersic index of the bulges but there is no effect on unbarred galaxies. In this subsection, we will use $S^4G$ data to investigate the effect of multi-component fittings on our results. In the Fig.~\ref{fig:compare_sdss_s4g}, we have shown the histograms of S\'ersic index (top panel) and bulge ellipticity (bottom panel) for 27 galaxies, common in both survey, which are fitted by S\'ersic profile. The red and blue filled histograms show the galaxies from SDSS and $S^4G$ survey respectively and vertical lines represent the medians of respective data. We removed all the galaxies having bar and/or point source while making these histograms. The median of S\'ersic indices is 1.02 (standard deviation = 1.16) for $S^4G$ galaxies and 1.28 (standard deviation =0.95) for SDSS galaxies. On the other hand, the median of bulge ellipticity for $S^4G$ is 0.41 (standard deviation = 0.21) and for SDSS is 0.31 (standard deviation =0.19). We can clearly see the clues of over-estimation of S\'ersic index and under estimation of bulge ellipticity in SDSS data. In the estimation of structure parameters of galaxies, GALFIT is better than the GIM2D \citep{Haussler.etal.2007}. Hence, we can say that our result will not be affected by the multi-component fitting rather it will stand in the support of our results with much better confidence that the fraction of pseudo-bulges is increasing as Universe is getting older and older.

\begin{figure*}
    \centering
    \includegraphics[width=0.7\textwidth]{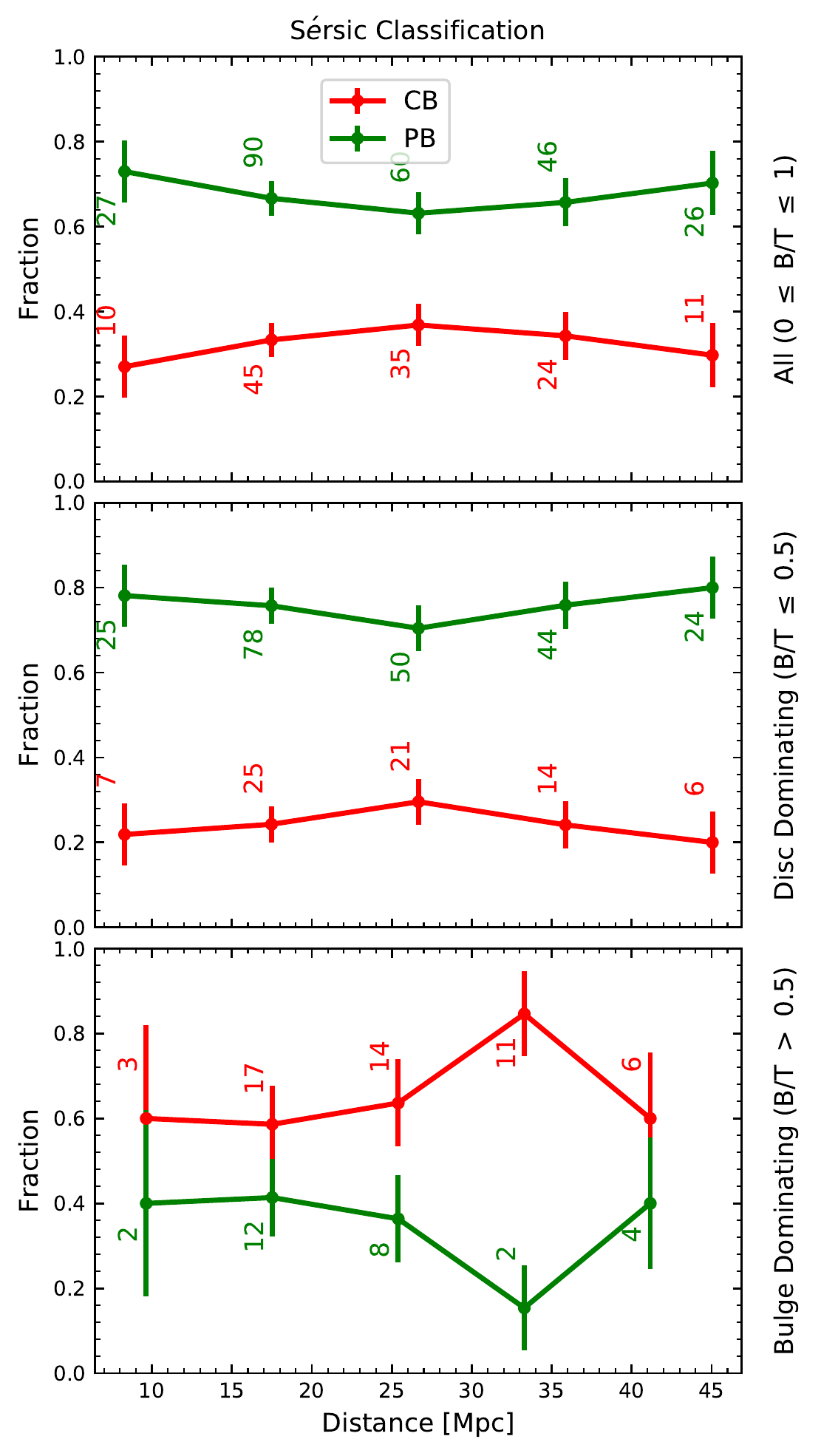}
    \caption{Fractional distribution of classical and pseudo-bulges with distance in the $S^4G$ survey. Top, middle, and bottom rows represent all, disc dominating, and bulge dominating galaxies respectively. Red and green curves represent classical bulges and pseudo-bulges respectively. The numbers of classical bulge and pseudo-bulge galaxies at any point are shown with their respective colors. Average fraction of the bulges here is equivalent to the local volume bulge fraction in SDSS data.}
    \label{fig:dist_bulge_s4g}
\end{figure*}

We have also calculated the distribution of the bulges with distance for $S^4G$ survey after removing galaxies having bar and/or point source. Fig.~\ref{fig:dist_bulge_s4g} shows the distribution of classical and pseudo-bulges in the local Universe for $S^4G$ survey. Top, middle, and bottom rows of this figure represent all, disc dominating, and bulge dominating galaxies respectively. Red and green curves represent classical bulges and pseudo-bulges respectively. Here, we have shown only S\'ersic classification of the bulges because all three classifications give over all similar qualitative trends with small quantitative difference. From top two panels of the figure, one can easily notice that the pseudo-bulges dominates over classical bulges over whole range of distance. This domination is similar to the SDSS data in local volume. On comparison in very local volume, it seems that $S^4G$ data show larger fraction of pseudo-bulges than the SDSS data. This is because all the points in $S^4G$ data are equivalent to one point in SDSS data of the Fig.~\ref{fig:bulge_with_z}. If we take average of all the $S^4G$ points, it will give comparable fraction as that in SDSS data. For example, average fraction of pseudo-bulges in all $S^4G$ data is 0.66, which is very close to the fraction of pseudo-bulges in all SDSS data within statistical uncertainties.

In the bulge dominating galaxies (bottom panel of Fig.~\ref{fig:dist_bulge_s4g}), the fraction of pseudo-bulges is less than the fraction of classical bulges i.e classical bulges are dominating over pseudo-bulges. This is opposite to that in upper two panels but, it is equivalent to the fraction of bulges in bulge dominating galaxies of SDSS data. If we look at top and middle panels of  Fig.~\ref{fig:dist_bulge_s4g}, we can see that both of them are showing more or less similar fractions of bulges. It means that the over all fractional distribution of the bulges in $S^4G$ data is governed by the disc dominating galaxies not by the bulge dominating galaxies. We found that the data which is fitted with S\'ersic profile has more than $78\%$ disc dominating galaxies. Since the $S^4G$ survey is limited by the size and magnitude of the galaxies \citep{Sheth2010}. Therefore the fractional distribution of the bulges in the bottom panel is biased by the survey limitations. We are most probably seeing the merger dominated massive galaxies.

\section{Effect of Disc Inclination and Instrument's Resolution}
\label{sec:effect_of_inc_res}
\begin{figure*}
    \centering
    \includegraphics[width=0.7\textwidth]{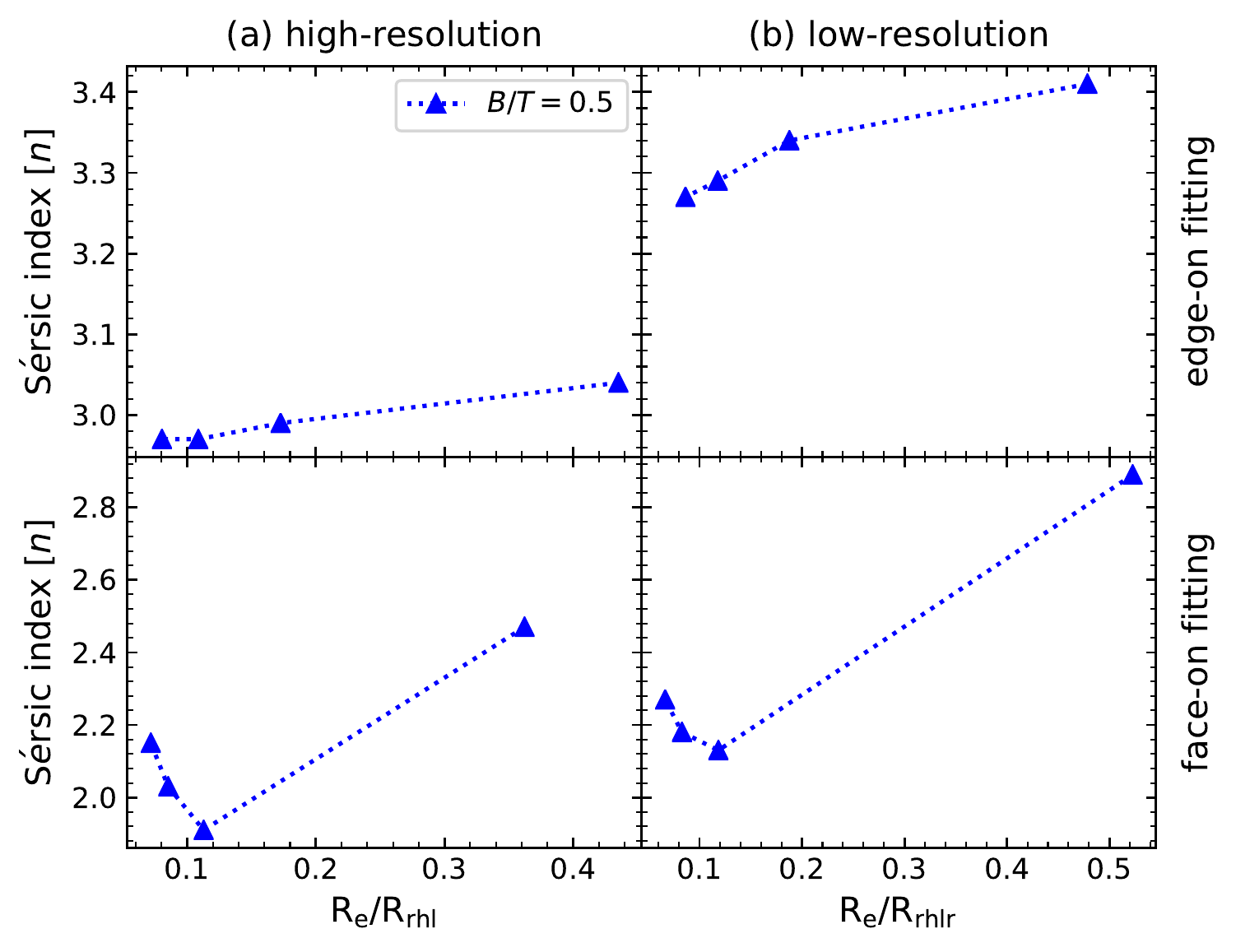}
    \caption{Effect of disc inclination and telescope's resolution on classical bulges of model galaxies. Left and right columns show high-resolution and low-resolution galaxies respectively. Top and bottom rows represent edge-on and face-on fittings of the model galaxies respectively. The blue upper triangle solid curves represent the models $B/T=0.5$ respectively and dots are just connecting points.}
    \label{fig:re_rhlr_ns_model}
\end{figure*}

The state of art cosmological simulations and observational studies suggest that the Universe was dominated by mergers in the early epoch of galaxy formation \citep{White.Rees.1978, Springel.etal.2006, Sinha.etal.2012}. The violent relaxation of merged galaxies usually lead to the formation of classical bulges \citep{Kauffmann.et.al.1993, Hopkins.et.al.2009, Naab.et.al.2014}. Once formed, it is not an easy task to transform a classical bulge into a pseudo-bulge \citep{Kumar.etal.2021} unless there is a central instability (e.g. bar) which can make it dynamically cool \citep{Kanak.etal.2012, Kanak.etal.2016} by transferring significant angular momentum \citep{kataria.das.2019}. But, instead of being dominated by classical bulges, the local Universe is dominated by the pseudo-bulges. Where have classical bulges gone in the evolution of galaxies? Is there any role of the two-component fitting? How do the disc surface density and inclination affect the two-component fitting? In the search of these questions as well as to test the reliability of S\'ersic index method, we have generated $N-$body disc galaxies with classical bulges and performed two-dimensional bulge-disc decomposition using GALFIT. 

Fig.~\ref{fig:re_rhlr_ns_model} shows the effect of disc inclination (rows) and spatial resolution (columns) on the S\'ersic index of classical bulges present in discs with a range of surface densities or in a range of R$_{\rm e}$/R$_{\rm hlr}$. In this figure, blue upper triangle solid curves represent the galaxy model with $B/T=0.5$. We have conducted a similar analysis for several values of $B/T$. However, we show only one for brevity, given our focus on the importance of disc surface density with respect to the central bulge. In this section, we show that only the S\'ersic index is not a reliable quantity to separate two classes of the bulges. We emphasize these points because it is one of the two parameters used in our new S\'ersic-Kormendy classification of bulges. At a given telescope's resolution, the S\'ersic index in edge-on fittings is always greater than the S\'ersic index in face-on fittings. The edge-on decomposition of the model galaxies provides the same S\'ersic index as reported by \cite{Dehnen1993} for pure Hernquist profile. The reasons for this difference between face-on and edge-on fitting are (1) Hernquist density profile which is not truncated at large radius in our models and (2) the real size of bulges can be traced only in edge-on projection of galaxies. Hence, the S\'ersic indices of the classical bulges depend on the disc inclination.

Each curve in Fig.~\ref{fig:re_rhlr_ns_model} represents the dependence of the S\'ersic index of a given classical bulge on the size (or surface density) of the hosting disc. The S\'ersic index of edge-on galaxies increases with increasing R$_{\rm e}$/R$_{\rm hlr}$ or we can say that the S\'ersic index of edge-on galaxies increases with decreasing size of hosting disc because all the bulges have same scale radius in our models. But the S\'ersic index of face-on galaxies first decreases and then increases with increasing R$_{\rm e}$/R$_{\rm hlr}$. In other words, one can think that the classical bulges easily pop-up in the low surface density discs. Similar results have been noticed in Fig.~\ref{fig:bulge_with_ReRd}, where we saw decreasing classical bulge fraction with increasing R$_{\rm e}$/R$_{\rm hlr}$ and then again increasing. The spatial resolution of the telescope also plays a crucial role in the determination of S\'ersic index. The low-resolution image usually takes large number of stars in each pixel which increases central density of the galaxy. This increased central density mainly contributes to the bulge in the term of increasing S\'ersic index. Therefore, the S\'ersic index in low-resolution fittings is always greater than the high-resolution fittings. These results have direct significance on the fraction of the bulges at high-redshift where the spatial resolution of the telescope decreases by a factor of distance. In Fig.~\ref{fig:bulge_with_z}, Some of the classical bulges at high-redshift could be the result of decreasing spatial resolution of the telescope.

\section{Effect of data selection criteria}
\label{sec:effect_of_data_selection}
\begin{figure*}
    \centering
    \includegraphics[width=\textwidth]{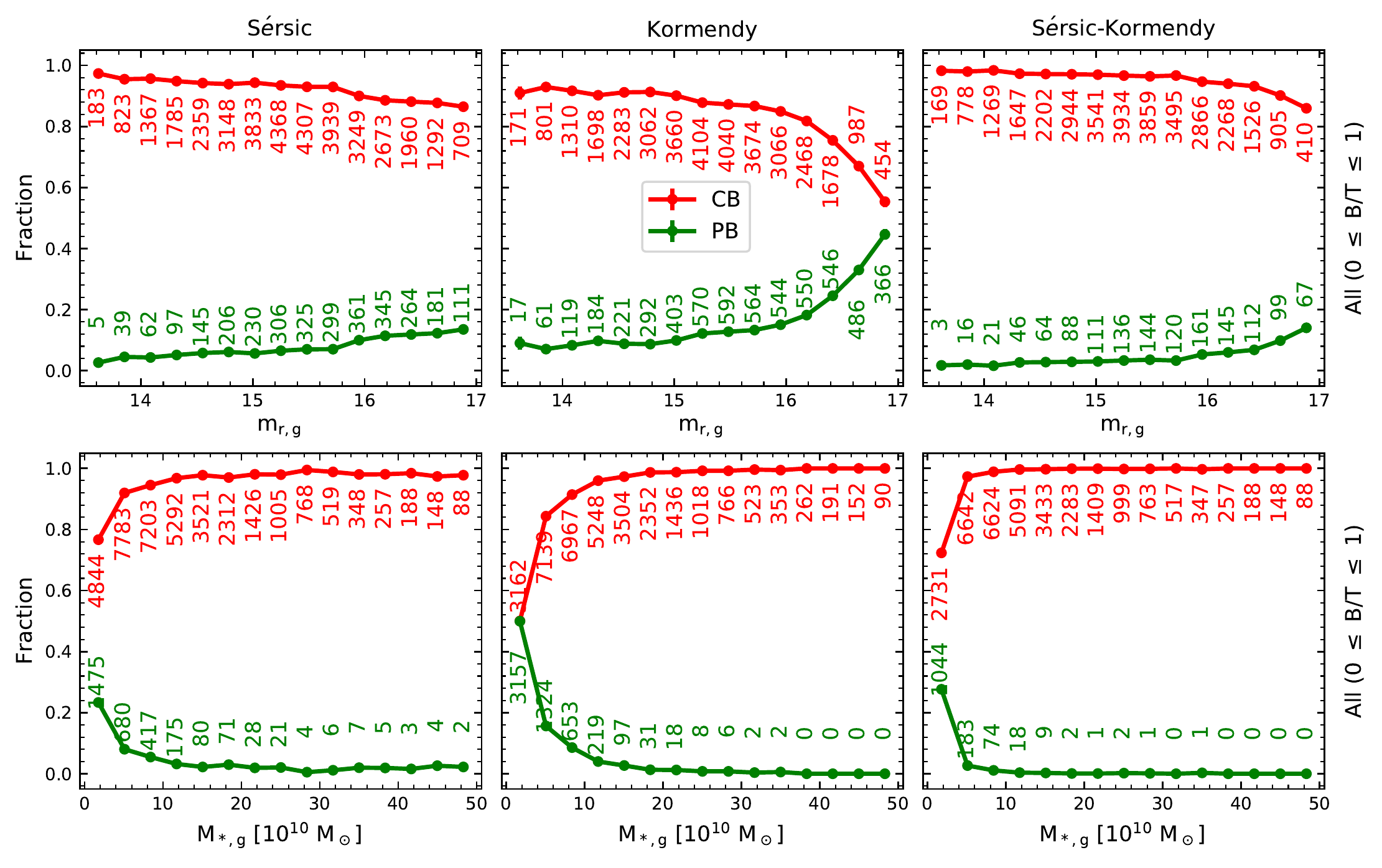}
    \caption{Fractional distribution of classical and pseudo-bulges with r-band apparent magnitude (m$_{\rm r,g}$) and stellar mass (M$_{*,g}$) of all the galaxies in the SDSS survey. Left, middle, and right columns show S\'ersic, Kormendy and S\'ersic-Kormendy classifications respectively. Red and green curves represent classical bulges and pseudo-bulges respectively. The numbers of classical bulge and pseudo-bulge galaxies at any point are shown with their respective colors. The fraction of pseudo-bulges increase in faint and low mass galaxies.}
    \label{fig:bulge_with_mag_mass}
\end{figure*}
The results, we have presented for SDSS data, are based on a sample of galaxies which are best represented by two components with least uncertainty. Imposing this strict constraint on the parent data reduces the galaxies from faint and low-mass end of the distribution, see Fig.~\ref{fig:sample_sdss}. Now, one can ask what will be the effect of this bias on our results? To understand the effect of reducing faint and low-mass galaxies from parent sample, we calculated the distribution of classical bulges and pseudo-bulges as a function of apparent magnitude and stellar mass of galaxies. It is shown in Fig.~\ref{fig:bulge_with_mag_mass}. The left, middle, and right columns of this figure show the distribution of bulges based on S\'ersic, Kormendy, and S\'ersic-Kormendy classifications respectively. Top row shows the fraction of bulges with r-band apparent magnitude of galaxies, whereas bottom row displays fractional distribution of bulges with stellar mass of the galaxies. The number of classical bulge and pseudo-bulge galaxies at any point of magnitude and stellar mass are also shown with respective colors.

From Fig.~\ref{fig:bulge_with_mag_mass}, it is clear that the fraction of the pseudo-bulges increases when we move towards the faint and low-mass end of galaxies in all the classification schemes. On the other hand, the fraction of classical bulges increases when we go from faint and low-mass end to bright and high-mass end of the galaxies. If we had not imposed our strict selection criteria, we would have seen increase in pseudo-bulge fraction as compare to that reported in Section~\ref{sec:bulge_with_z}. We have verified it for parent sample but, for the brevity, is not shown here. Also, the inclusion of faint and low-mass galaxies will not affect our results discussed in Section~\ref{sec:ellipticity_with_z} as those results are based on the mean of the bulge ellipticity of statistically significant number of galaxies. Finally, we would like to emphasize that including large number of galaxies will not affect our results though it will only reduce the statistical uncertainties of our results.

\section{Discussion}
\label{sec:discussion}
In previous section, we have presented our results on the evolution of the bulges since $z=0.1$ redshift. The effect of instrument's spatial resolution and galaxy inclination is shown using Milky Way type model galaxies. The correlation of the bulges with various photometric parameters and effect of multi-component fitting of the galaxy is also discussed. In this section, we interpret our results on the evolution of the bulges in observations.

The classification of bulges in classical bulge and pseudo-bulge categories has always been the topic of debate. Different people use different dividing criteria which suite their needs e.g $n=2$ S\'ersic index \citep{Fisher2006_proce, Fisher.Drory.2008}, Kormendy relation \citep{Gadotti.2009}, Color index \citep{Lackner.Gunn.2012}, velocity dispersion \citep{Fabricius.etal.2012}, central stellar mass density \citep{Cheung.etal.2012}, etc. We defined bulges into classical and pseudo categories using a combination of S\'ersic index and Kormendy relation which we call S\'ersic-Kormendy classification and compared properties of the bulges in three classification schemes. We found that the results based on S\'ersic classification and Kormendy classification deviate more from each other with increasing bulge to disc semi-major axis ratio. Results from combined S\'ersic-Kormendy classification lies in between those from two individual methods.

The fraction of classical bulges and pseudo-bulges varies with the evolution of the Universe. As we go from low-redshift to high-redshift, the fraction of pseudo-bulges decreases while the fraction of classical bulge increases. At redshift $z \approx 0.016$ for S\'erisc-Kormendy bulge classification, both types of bulges contribute equally. Pseudo-bulges are thought to be formed due to secular evolution of galaxies. The ubiquitous distribution of pseudo-bulge hosting disc galaxies in the local volume raise the question at hierarchical nature of the Universe \citep{Weinzirl.et.al.2009, Kormendy.et.al.2010, Fisher.Drory.2011}. Since the existence of zoom-in and hydrodynamic simulations, many studies have shown the importance of gas against the destruction (or heating) of the disc in merger events. These studies show that the pseudo-bulge can also form in mergers in contrary to the secular evolution \citep{Governato.etal.2009, Hopkins.etal.2009b, Moster.etal.2010, Guedes.etal.2013, Okamoto2013}. Other studies suggest to improve the initial condition of galaxy formation \citep{Peebles.2020}. A proper understanding of the physical processes and/or the adjustment to the current cosmological model is still needed for better representation of the observable Universe.

We found that a large fraction of the pseudo-bulges are rarer in density as compare to the classical bulges relative to their hosting discs. \cite{Gadotti.2009} also found that the pseudo bulges less are concentrated than the classical bulges at given bulge to disc mass ratio. High stellar density discs usually go through instabilities e.g. bars. \cite{Lutticke.etal.2000} studied a sample of edge-on galaxies from RC3 catalogue (Third Reference Catalogue of Bright Galaxies) and found that $45\%$ of all bulges are boxy/peanut in shape and explained by the presence of bar. Rare density of pseudo-bulges is in agreement with previous studies where \citep{Erwin.Debattista.2017} claims that the semi-major axes of boxy/peanut shape bulges range from one-quarter to three-quarters of the full bar size. If the classical bulges are small in mass, these instabilities can hide them in (or sometime erode completely) and can result in composite bulges as seen in some observations \citep{Erwin.et.al.2015, BlanaDiaz.etal.2018, Erwin.etal.2021}. The increasing fraction of pseudo-bulges with decreasing redshift could be the result of these hidden low mass classical bulges \citep{Kanak.2015}. Further, our result shows the bulge dominated systems also have higher fraction of pseudo bulges at low-redshift which is opposite to earlier claims \cite{Gadotti.2009}. Though our results supports the composite bulge scenario. A detailed multi-component decomposition of galaxies can reveal what fraction of the pseudo-bulges are composite bulges. Further these techniques can certainly help in improving our understanding of the galaxy evolution. 

The two-dimensional photometric decomposition of the galaxies gives not only the quantitative information of the bulge and disc components but, it also provides the tool to understand the evolution of bulges. From the sub-sample of near face-on galaxies in SDSS data, we found that massive bulges are more round than the low mass bulges at all redshift. In local volume, classical bulges show mean ellipticity in the range 0.2 to 0.3, whereas pseudo-bulges show in the range 0.3 to 0.5 for S\'ersic-Kormendy bulge classification which is consistent with previous studies of spiral galaxies \citep{Fathi.Peletier.2003, Mendez-Abreu.etal.2008}. Also, the mean ellipticity of both types of bulges grows with the increasing redshift. It indicates a clear morphological evolution of the bulges since $z=0.1$. The decreasing mean ellipticity and increasing pseudo-bulge fraction point towards the secular evolution in the later time of Universe. Recent cosmological simulations have also shown that the rate of violent interactions (mergers) between galaxies decreases with the evolution of the universe and slowly, flyby interactions dominate the evolution \citep{Sinha.etal.2012, an.etal.2019}.

The spatial resolution of the observing instrument also play a crucial role in the identification of the bulge type Fig.\ref{fig:re_rhlr_ns_model}. At low spatial resolution, a pseudo-bulge can be mistakenly classified as classical bulge in photometric decomposition of the bulge using S\'ersic profile. Since the spatial resolution of the instrument decreases when we see a high-redshift object. Hence, the fraction of the pseudo-bulge at high-redshift could be higher than what we have calculated from the sample of photometric bulge-disc decomposition. A further investigation is needed to tightly constrain the fraction of the cold and hot fraction of stellar matter in the Universe at different redshift. Calculation of the exact fraction of ordered and random stellar orbits in disc galaxies at different redshift will help us in better understanding of the initial cosmological conditions and/or the underlying baryonic physics.

Our study is based on two-component photometric decomposition. The effect of multi-component decomposition of the galaxies is very crucial to understand the real statistics of the stellar matter distribution in various components of the galaxy. Two-component fitting usually gives higher S\'ersic index and lower ellipticity in case of the barred galaxies (Figure \ref{fig:compare_sdss_s4g}). This can lead to small fraction of the pseudo-bulges and elongated bulges. Sometimes, photometric and kinematic decomposition of cold and hot components show contrary classification of the bulges \citep{Gadotti.2009}. For example, the excess light in the center of galaxy due to nuclear activity gives higher S\'ersic index in photometric decomposition. The detailed relation in photometric and kinematic decomposition is necessary for better photometric classification criterion.

\section{Summary}
\label{sec:summary}
To understand the evolution of the bulges with the evolution of the Universe, we have used the archival data of two-dimensional bulge-disc decomposition of galaxies from SDSS DR7 \citep{Simard2011}. This data is constraint to nearly 40000 galaxies those can be well represented by two components. The classical bulges and pseudo-bulges are separated using $n=2$ S\'ersic index, Kormendy relation, and a combination of two. To explore the effect of multi-component fitting, we have also used the archival data of two-dimensional multi-component decomposition of local volume galaxies from $S^4G$ survey \citep{Salo2015}. Further, we have simulated Milky Way type model galaxies to investigate the effect of spatial resolution and disc inclination. The main findings of our analysis are as follows:-

(i) The fraction of pseudo-bulges dominates over classical bulges in the local Universe. As we go towards the high-redshift, the fraction of pseudo-bulges decreases smoothly. In the history of the Universe, there came a point (z $\approx$ 0.016 for S\'ersic-Kormendy classification) when the fraction of classical bulges and pseudo-bulges was 50-50$\%$. Universe is dominated by classical bulged galaxies before this point during its evolution while it is dominated by pseudo-bulges as soon it crosses this point.

(ii) In the local Universe, disc dominating galaxies show more pseudo-bulges as compare to bulge dominating galaxies. The fractional distributions of pseudo-bulges in Kormendy, and S\'ersic-Kormendy classifications are quite similar. But, S\'ersic classification shows lower pseudo-bulges than other two classification at low-redshift.

(iii) The fraction of the pseudo-bulges increases with increasing bulge to disc semi-major half-light ratio until $R_{\rm e}/R_{\rm hlr} \approx 0.6$. Compact and shorter bulges, as compared to their hosting discs, are usually classical in nature however, pseudo-bulges are diffuse and longer. In other words, concentrated discs harbour pseudo-bulges while rarer discs host classical bulges.

(iv) At large bulge to disc semi-major axis ratio, Kormendy classification shows more pseudo-bulges than other two classifications. This difference is more pronounced in disc dominating galaxies. Bulge dominating galaxies show very similar fraction in all classifications. For better division between classical and pseudo-bulges, we recommend to use a combination of S\'ersic index and Kormendy relation based bulge classification.

(v) The mean ellipticity of pseudo-bulges is greater than the mean ellipticity of classical bulges in whole redshift range and it decreases with decreasing redshift indicating that the bulges are getting more axisymmetric with the evolution of galaxies. High-mass bulges are progressing towards the axisymmetry at more steep rate than the low-mass bulges.

(vi) In local Universe, nearly 75$\%$ of the visible matter is dominated by ordered rotational motion in disc galaxies. The existence of the rotation dominated Universe challenges the hierarchical nature of the Universe which suggests that most of the Universe should be dispersion dominated.

(vii) The evolution of the bulge does not have significantly strong correlation with the evolution of the photometric properties of galaxy e.g. bulge to total light ratio, disc inclination, bulge ellipticity, absolute magnitude of galaxy etc. Only absolute magnitude of galaxies and ellipticity of bulges show a negative correlation of $\approx 0.3$ on the scale of unity in the low-redshift Universe that also drops in very local Universe.

(viii) Our results are consistent with the local volume survey $S^4G$. The multi-component fitting of the galaxies does not fade our conclusions. It stands with strong support of our results discussed in this article. 

The tight constraint on the fraction of the classical bulges and pseudo-bulges with the redshift can be verified with next generation telescopes e.g. JWST, TMT, SKA etc. These telescope will provide unprecedented high-resolution view of the universe. Using the data obtained with these next generation telescope, we can better understand the underlying mechanism which played major role in the stability of discs during the growth and evolution of the structures.

\section{Appendix: Comparison of Bulges and Discs}
\label{app:z_vs_Re_Rd}
\begin{figure*}
    \centering
    \includegraphics[width=\textwidth]{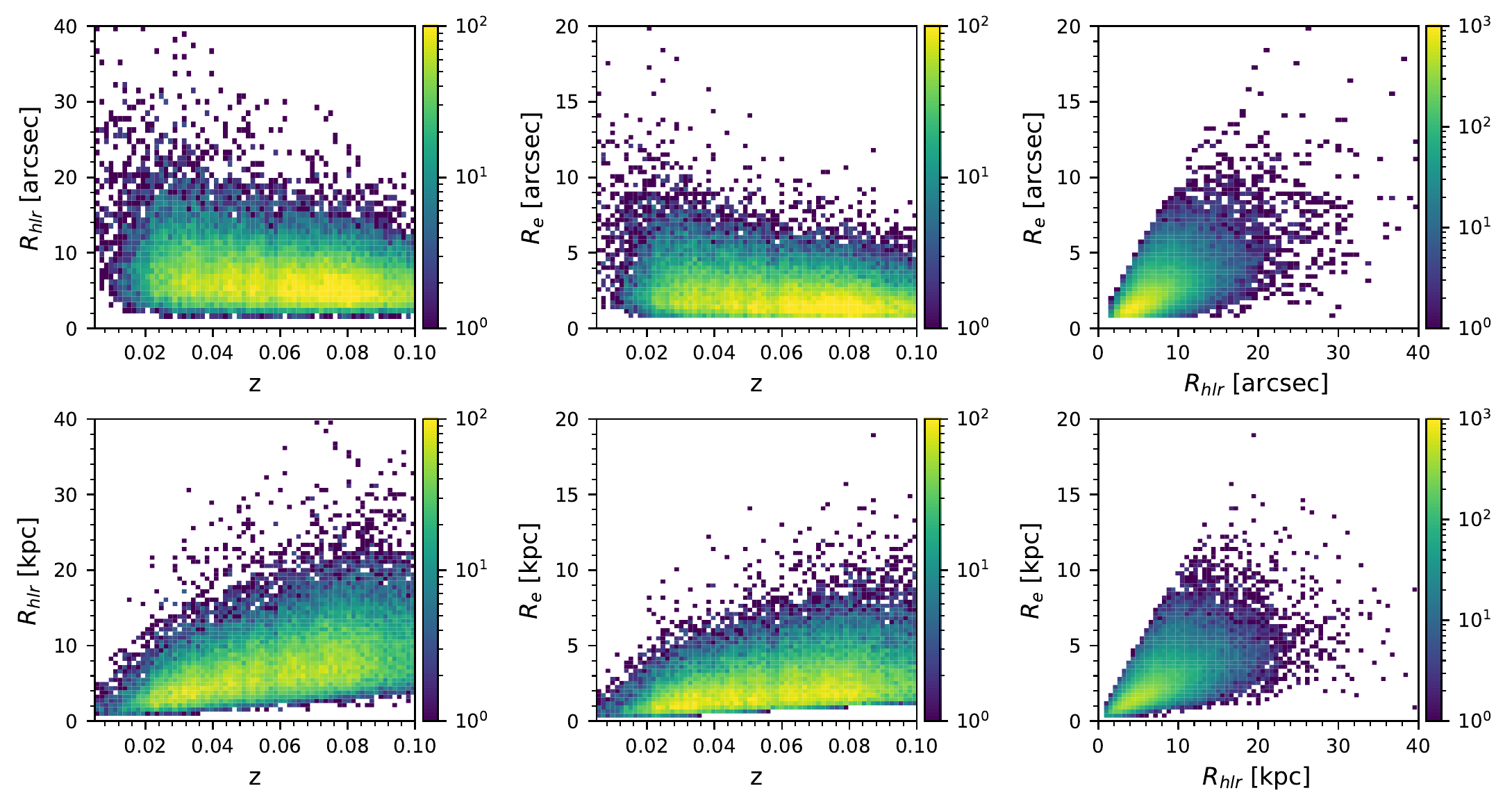}
    \caption{Sizes of the bulges and discs in our sample. Left and middle columns demonstrate absolute sizes of discs and bulges respectively as a function of redshift, whereas right column shows their comparison. Top and bottom rows display apparent sizes and physical sizes respectively.}
    \label{fig:z_vs_Re_Rd}
\end{figure*}

In Fig.\ref{fig:z_vs_Re_Rd}, we have shown the apparent and physical sizes of the bulges and discs as a function of redshift. Top and bottom rows show apparent sizes and physical sizes respectively. Left and middle columns display absolute sizes of discs and bulges respectively, whereas right column shows their comparison. There are some galaxies that fall out of the shown range of the axes. However, for better visualization, range of the axes are chosen arbitrarily. As expected, apparent sizes of the bulges and discs decrease with increasing redshift, while physical sizes increase with increasing redshift. Comparison of bulges and discs show notable positive correlation. On average, larger discs host larger bulges and smaller discs hot smaller bulges. This correlation shows increasing dispersion with increasing disc size.

\section{Appendix: Bulge to Disc Central Brightness}

\label{app:bulge_disc_bright}
\begin{figure*}
    \centering
    \includegraphics[width=0.7\textwidth]{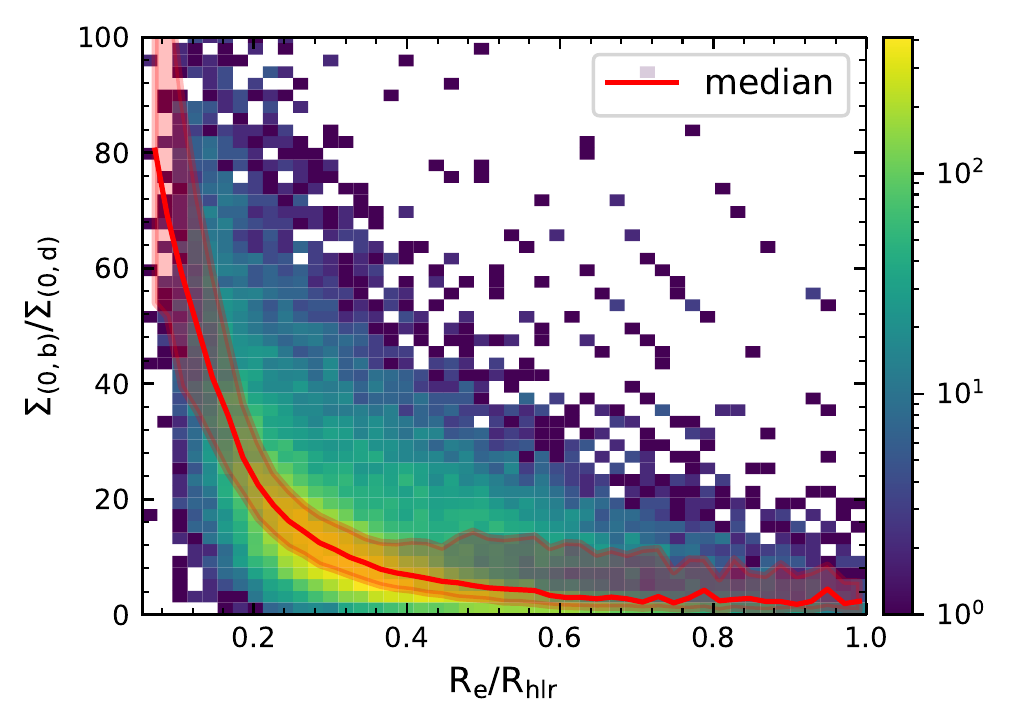}
    \caption{Distribution of the bulge to disc central brightness ratio for the sample selected from SDSS survey with the bulge to disc semi-major half-light radius. The red curve shows the median of the distribution and shaded region represents $\pm 25\%$ deviation from median.}
    \label{fig:bulge_disc_bright}
\end{figure*}

In Fig.~\ref{fig:bulge_disc_bright}, we have shown the ratio of bulge central brightness ($\Sigma_{(0,b)}$) to the disc central brightness ($\Sigma_{(0,d)}$) as a function of the bulge to disc semi-major half-light radius ratio. This ratio is calculated using the following expression,
\begin{equation}
    \frac{\Sigma_{(0,b)}}{\Sigma_{(0,d)}}=\frac{B}{D} \times \frac{R_{hlr}^2}{R_{e}^2}
    \label{eqn:bulge_disc_bright}
\end{equation}
where, $\frac{B}{D}$ is the photometric bulge to disc light ratio obtained from two-component decomposition of the galaxies. $R_{\rm e}$ and $R_{\rm hlr}$ are the semi-major half-light radii of the bulge and disc components respectively. This figure is clearly showing that the bulge concentration is decreasing with increasing bulge to disc semi-major half-light radius ratio. It is evident that the $R_{\rm e}/R_{\rm hlr}$ can be used as an indicator to measure the bulge concentration relative to the hosting disc.

	\begin{savequote}[100mm]
``The key benefit of this Universe is that you have option to choose.''
\qauthor{\textbf{$-$ Frederick Carl Frieseke}}
\end{savequote}

\chapter[Bulgeless Galaxies in the Illustris TNG50 Simulations: A Test for Angular Momentum Problem]{Bulgeless Galaxies in the Illustris TNG50 Simulations: A Test for Angular Momentum Problem}
\label{chapter7}

\section{Introduction}
\label{sec:intro}
The Hubble classification of galaxies \citep{Hubble.1926} separates galaxies into different morphological classes according to the relative importance of the dispersion and rotation components of stars \citep[also see, ][]{de_Vaucouleurs.1959A, de_Vaucouleurs.1959B, van_den_Bergh.2007}. Spiral galaxies are the rotationally dominated systems with varying fractions of dispersion components at their centers, commonly known as bulges. In the photometric decomposition of the galaxies, the excess light above the exponential disk profile is defined as the bulge \citep[see the book,][]{Laurikainen.2016}. Bulges show a variety of surface brightness profiles displaying steep to flat variation with radius \citep{Fisher.2006Oct, Fisher.etal.2008Aug, Fisher.etal.2008Oct, Kumar.etal.2021, Kumar.etal.2022Aug}. Bulges with steep and flat surface brightness profiles are known as classical bulges and pseudo-bulges respectively \citep{Kormendy.1993, Andredakis.Sanders.1994, Kormendy.2006, Drory.Fisher.2007}.

Several theoretical and numerical studies have shown that the merger of two galaxies results in the formation of classical bulges \citep{Kauffmann.et.al.1993, Baugh.et.al.1996, Aguerri.etal.2001, Bournaud.etal.2005, Hopkins.etal.2009Aug, Naab.et.al.2014}. The kinematic and morphological similarities of classical bulges and elliptical galaxies further support the fact that classical bulges are formed in violent relaxation processes \citep{Bournaud.etal.2007, Naab.et.al.2014}. On the other hand, pseudo-bulges are thought to form during the secular evolution of galaxies \citep{Kormendy.1993, Heller.Shlosman.1994, Kormendy.Kennicutt.2004, laurikainen.etal.2009}. Classical bulges are usually redder than the pseudo-bulges in color. This is because their star formation histories differ significantly. Using observations of nearby galaxies, \cite{Fisher.2006May} concluded that the pseudo-bulges show long lasting star formation, whereas classical bulges show episodic star formation. Galaxies with pseudo-bulges also exhibit the presence of rings, bars and spirals in their nuclear regions \citep{Andredakis.Sanders.1994, Carollo.etal.1997, Erwin.Sparke.2002, Fisher.2006May, Fisher.etal.2008Aug}.

A significant fraction of spiral galaxies have been observed to have no bulges or very small bulges. This class of spiral galaxies are referred to as pure disk or flat galaxies or bulgeless galaxies \citep{Karachentsev.etal.1993, Karachentsev.etal.1999}. Nearly $15-30\%$ of disk galaxies in the local volume are bulgeless \citep{Kautsch.etal.2006, Barazza.etal.2008, Cameron.etal.2009, Karachentsev.Karachentsev.2019}. These galaxies show large amount of neutral and atomic hydrogen gas throughout their disks \citep{Matthews.van_Driel.2000, Kaisin.etal.2020}. However, they exhibit low global star formation rates because of low stellar surface densities and high stellar velocity dispersion \citep{van_der_Hulst.etal.1993, Karachentseva.etal.2020, Banerjee.etal.2010}. 

It is well known that there is a tight correlation between the mass of supermassive black-holes (SMBHs) and the velocity dispersion (or luminosity) of bulges, which is in accordance to theories of hierarchical structure formation \citep{Graham.etal.2001, Peng.2007, Wang.Kauffmann.2008}. In the context of such studies, the presence of SMBHs and nuclear activities in bulgeless galaxies \citep{Filippenko.etal.2003, Satyapal.etal.2016, Subramanian.etal.2016, Bohn.etal.2022} cannot be explained.

Bulges galaxies cannot be explained in the context of a merger dominated Universe. The existence of bulgeless galaxies is thus a challenge to the standard $\Lambda$CDM model of galaxy formation \citep{Kormendy.etal.2010, D'Onghia.Burkert.2004}. In the $\Lambda$CDM cosmological framework, galaxies form in the hierarchical manner. Several small structures merge over the course of evolution to form the large structures we see today \citep{White.Rees.1978, Aguerri.etal.2001, Bournaud.etal.2005, Baugh.2006}. Numerically, it has been shown that the major mergers (collision of nearly equal mass galaxies) disrupt the disks and lead to the formation of elliptical galaxies \citep{Barnes.1992, Cox.Loeb.2008}. Though the minor mergers (collision of significantly unequal mass galaxies) do not disrupt the disks, they can heat the disks dynamically to form the massive bulges \citep{Hopkins.etal.2010, Das.etal.2012, Kumar.etal.2021}. Therefore, mergers are expected to form spheroid like structures and not pure disks.

Recent advancements in numerical simulations have shown that the feedback-driven outflows of gas in dwarf galaxies can prevent the formation of classical bulges \citep{D'Onghia.Burkert.2004, Governato.etal.2010, Robertson.etal.2004}. These simulations involve exceptionally intense feedback-driven gas removal from the center of the low-mass galaxies to produce bulgeless systems. Furthermore, simulated bulgeless galaxies are centrally concentrated, small in size, and have lower angular momentum than their observational counterparts \citep{D'Onghia.Burkert.2004, Piontek.Steinmetz.2011}. In cosmology, this controversy is known as the angular momentum problem or angular momentum catastrophe \citep{van_den_Bosch.etal.2001, D'Onghia.Burkert.2004}. The rotation curves of bulgeless galaxies reveal that most of them are dark matter dominated systems. Dark matter is thought to stabilize the disk against external perturbations, like the classical bulge does in galaxy disks \citep{Samland.Gerhard.2003}.

There is plenty of literature available on the formation and stability of bulgeless galaxies in isolation and in the non-cosmological context. But their study in the cosmological environment is still not understood. In this chapter, we investigate the formation of bulgeless galaxies in the cosmological setting using the state-of-art cosmological simulations, Illustis TNG. We aim to test how successful the TNG galaxy formation model is in reproducing bulgeless galaxies and whether they are comparable to those observed. Using the Illustris TNG data products we have  estimated the fraction of bulgeless galaxies, their dark matter content, angular momentum, and merger histories.

We have arranged the chapter as follows: In Section~\ref{sec:tng_sims}, we have briefly described the Illustris TNG suite of cosmological simulations. We have talked about the automated pipeline for photometric decomposition of simulated galaxies in Section~\ref{sec:morp_decomp}. Identification of the bulgeless galaxies is described in Section~\ref{sec:bulgeless_and_bulged_gal} along with the control sample for comparison.

\section{The Illustris TNG simulations}
\label{sec:tng_sims}
To study the formation of the bulgeless galaxies in cosmological context, we have used the sample of galaxies from the IllustrisTNG project\footnote{https://www.tng-project.org/}. The IllustrisTNG project, TNG in short, is `the next generation' follow-up of the Illustris project \citep{Genel.etal.2014, Vogelsberger.etal.May2014, Vogelsberger.etal.Oct2014, Nelson.etal.2015, Sijacki.etal.2015}. TNG is a suite of gravo-magnetohydrodynamical, large scale, cosmological simulations \citep{Nelson.etal.May2019}. It comprises three cosmological volumes, namely TNG50, TNG100, and TNG300 having box sizes approximately 50, 100, and 300 Mpc respectively. For each cosmological volume, there have been performed low-resolution and dark matter only runs as well. The three boxes enable the study of different aspects of structure formation. TNG300 is useful in the investigation of large scale structures (e.g., clustering of the galaxies), and TNG50 is capable in the understanding of small scale properties (e.g., structure of the galaxies). On the other hand, TNG100 acts as a bridge between TNG300 and TNG50 for detailed study of structure formation. TNG100 and TNG300 have been introduced in \cite{Marinacci.etal.2018, Naiman.etal.2018, Nelson.etal.2018, Pillepich.etal.Mar2018, Springel.etal.2018}. TNG50 is discussed in \cite{Nelson.etal.May2019, Nelson.etal.Dec2019, Pillepich.etal.2019}. Now, all three runs are publicly available in their entirety \citep{Nelson.etal.May2019}. The data release includes snapshots, halo and subhalo catalogs, and merger trees for each run.

All the TNG simulations are performed with the moving-mesh and massively parallel code AREPO \citep{Springel.2010} starting from the redshit z=127 to z=0. The initial conditions for the simulations are $\Omega_{\rm \Lambda,0} = 0.6911$, $\Omega_{\rm m,0} = 0.3089$, $\Omega_{\rm b,0} = 0.0486$, $n_{\rm s} = 0.9667$, $\sigma_{\rm 8} = 0.8159$, and $h = 0.6774$ that are consistent with \cite{planck.collaboration.2016}. TNG galaxy formation model includes physics of gas cooling, star formation, stellar evolution, metal enrichment, stellar feedback, supermassive black-hole growth, supermassive black-hole feedback, magnetic field, etc \citep{Vogelsberger.etal.2013, Weinberger.etal.2017, Pillepich.etal.Jan2018}. TNG has successfully reproduced several observational results in different regimes. For example, $g-r$ color bimodality in SDSS galaxies \citep{Nelson.etal.2018}, galaxy correlation function in SDSS \citep{Springel.etal.2018}, [Eu/Fe] spread at low metallicities in the Milky Way \citep{Naiman.etal.2018}, mass-size relation of late- and early-type galaxies \citep{Genel.etal.2018}, mass-metallicity relation of gas \citep{Torrey.etal.2019}, dark matter mass and circular velocity of the Milky Way mass galaxies \citep{Lovell.etal.2018}, morphology of Pan-STARRS galaxies \citep{Rodriguez-Gomez.etal.2019}, etc.

\section{Pipeline for morphological decomposition}
\label{sec:morp_decomp}
In this study, we have selected sample galaxies from TNG50-1 (hereafter TNG50) run because this is the highest resolution run with full physics available in IllustrisTNG suite of simulations. In IllustrisTNG, galaxies are the subhalos identified by the SUBFIND alogrithm \citep{Springel.etal.2001}. We have used the available catalog of the subhalos and restricted ourselves to the galaxies having stellar masses greater than 10$^9$M$_\odot$. This choice removes the dwarf galaxies and provides the sample galaxies having significant number of stellar particles ($>$ 2.2$\times$10$^{4}$) that is necessary for the reliable study of the galactic components. The mass constraint on the catalog returns a total of 2512 sample galaxies.

For the morphological decomposition of these galaxies, we have used the up-to-date version of the two-dimensional photometric decomposition tool, GALFIT \citep{Peng.etal.2002, Peng.etal.2010}. GALFIT is a C language based least-square fitting code that uses Levenberg–Marquardt algorithm for searching the optimal parameters to the fitting. It is capable of fitting multiple components simultaneous. GALFIT takes the image of the galaxy in fits format and initial guess parameters as input for the fitting. To perform the multi-component fitting on such a large sample of the galaxies, we have developed an automated python language based pipeline. It acts as a wrapper for the GALFIT code. This pipeline reads the TNG galaxies for given list of subhalo IDs, estimates the center of mass for stellar component, calculates the total stellar angular momentum within the radius containing 90$\%$ of the stellar mass, rotates all the particles so that the total angular momentum within radius containing 90$\%$ of stellar mass aligns in $z-$direction of the Cartesian co-ordinates, makes fits format image of galaxy in face-on view, finds initial guess for the fitting profile, generates input file for GALFIT, executes GALFIT, and stores the output in a txt file.

Our complete pipeline consists of 3 separate scripts. The first script of the pipeline calculates the center of mass (CM) for the galaxy. Then it estimates the radius containing 90$\%$ of the stellar mass (r$_{90}$) and measures total angular momentum within this radius ($\vec{L}_{90}$). Both the quantities along with the IDs of the galaxies are stored in a text file. The second script of the pipeline uses the parameters estimated in the first script to rotate the galaxy in such a way that $\vec{L}_{90}$ aligns in $z-$direction of the Cartesian co-ordinates. Then it measures the complex amplitude of m=2 Fourier mode in x-y plane using the following equation,
\begin{align}
    A_{m} (<R) = \sum_j m_j e^{-im\phi_j},
    \label{eqn:fourier_mode}
\end{align}
where A$_{m}$ is the complex amplitude of $m$th Fourier mode within radius R, $m_{\rm j}$ is the mass of $j$th
particle, $\phi_{\rm j}$ is the azimuth angle of $j$th
particle, and $i$ is the iota \citep[see e.g.,][]{Kumar.etal.2021, Kumar.etal.2022Jan}. The normalized amplitude of m=2 Fourier mode ($|$A$_{2}$$|$/$|$A$_{0}$$|$), phase angle ($\tan^{-1}[\frac{Im(A_{2})}{Re(A_{2})}]$), and radius along with IDs of galaxies are saved in a text file. The third script of the pipeline reads the output from first script to save a fits image in face-on projection. It also makes use of the outputs from second script to guess the initial parameters of the components to be fitted. For example, if one wants to fit the central component of the galaxy, it will take the phase angle of m=2 Fourier mode as a guess for the position angle.

GALFIT provides several in-built profiles for the fitting. Currently, we have included only S\'ersic, Exponential disk, Edge-on disk, and Ferrer profiles in our pipeline. The S\'ersic profile for the two-dimensional decomposition is expressed as,
\begin{equation}
    \Sigma (R) = \Sigma_{\rm e} \exp\left[-b_{\rm n} \left\{\left(\frac{R}{R_{\rm e}}\right)^{\frac{1}{n}} - 1\right\}\right]
    \label{eqn:galfit_sersic}
\end{equation}
where R$_{\rm e}$ is the effective radius, $\Sigma_{\rm e}$ is the surface density at the effective radius, and b$_{\rm n}$ is a function of S\'ersic index $n$. The exponential disk profile for face-on decomposition is given by,
\begin{equation}
    \Sigma_{\rm d}(R) = \Sigma_{\rm d0} \exp\left(-{\frac{R}{R_{\rm s}}}\right)
    \label{eqn:galfit_face_disk}
\end{equation}
where $\Sigma_{\rm d0}$ is the surface density at center of the disk and $R_{\rm s}$ is scale radius of the disk. The edge-on disk profile for edge-on decomposition is modeled by
\begin{equation}
    \Sigma_{\rm d}(R,z) = \Sigma_{\rm d0} \left(\frac{R}{R_{\rm s}}\right) K_{1}\left(\frac{R}{R_{\rm s}}\right) \sech^{2}\left(\frac{z}{z_{\rm 0}}\right)
    \label{eqn:galfit_edge_disk}
\end{equation}
where $K_{1}$ is the Bessel function and $z_{\rm 0}$ is the scale height of the disk. 

While performing 2D photometric decomposition, we have centred face-on galaxy in a box of size 3r$_{90}$ with grid (or pixel) size equal to the gravitational softening ($\epsilon$) of the snapshot. The choice of box size (i.e. 3r$_{90}$) is arbitrary. The only motivation for this choice is to consider most of the stellar mass during the fitting. 



Previous studies of galaxy bulges suggest that the S\'ersic index $n=2$ is a good proxy for distinguishing between classical bulges and pseudobulges \citep{Fisher.2006Oct, Fisher.etal.2008Aug}. Generally, classical bulges have a S\'ersic index $n>2$ while pseudobulges have a S\'ersic index $n<2$. We have applied this classification scheme for bulges in our study.

\section{Sample of bulgeless galaxies and galaxies with bulge}
\label{sec:bulgeless_and_bulged_gal}
The sample of  bulgeless galaxies is selected using the single S\'ersic profile fitting as done in \cite{Grossi.etal.2018} for observational galaxies. First, we performed single S\'ersic component fitting on the whole sample of TNG50 galaxies with our pipeline. Then, we selected all those galaxies which are best fitted with single S\'ersic component, and have S\'ersic index less than 1.5. We removed all the unresolved galaxies that have effective radius smaller than three softening length of the simulation. All these constraints return a sample of 16 unbarred bulgeless galaxies in TNG50 run.

Next, we execute our pipeline with two S\'ersic components and selected those galaxies that are best fitted with two S\'ersic profiles. Then we removed all the galaxies that were best fitted with single S\'ersic profile. From this sample, we separated disk and bulge/bar components using S\'ersic index, effective radius, and normalized amplitude of m=2 Fourier mode. The component with S\'ersic index less than 1.5 and the largest effective radius is considered disk and the other component is considered bulge/bar. If the normalized amplitude of m=2 Fourier mode is less than 0.1, we call the smaller component bulge; otherwise, we call it bar. Bulge with S\'ersic index less than 2 are pseudo-bulge and bulge with S\'ersic index greater than 2 are classical bulge. Further, we removed galaxies that have bulge/bar effective radius smaller than two softening length of the simulation. Similar analogy is used for three component fitting.

\begin{figure*}
    \centering
    \includegraphics[width=0.8\textwidth]{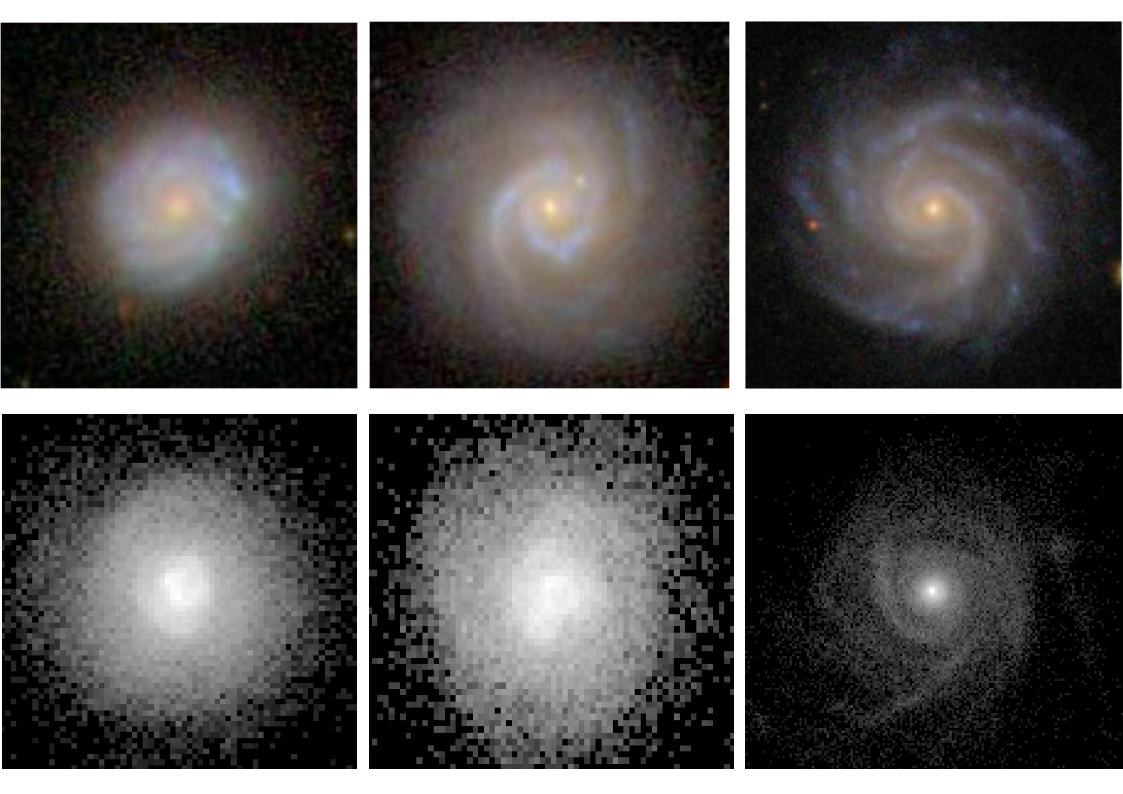}
    \caption{Visual comparison of observed bulgeless galaxies with those in TNG50 simulation. Top row show three observed bulgeless galaxies and bottom row display bulgeless galaxies formed in TNG50 simulation. Note that in TNG50 images, we have only shown stellar component.}
    \label{fig:observed_simulated_bulgeless}
\end{figure*}

In Fig.~\ref{fig:observed_simulated_bulgeless}, we have shown example of bulgeless galaxies from observation and TNG50 simulation. Top row panels of the figure display observed examples of bulgeless galaxies drawn from Sloan Digital Sky Survey (SDSS) and bottom row panels show the simulated examples of bulgeless galaxies taken from TNG50 data. These are pure disk galaxies. One can barely notice the central bulge component in these galaxies. Note that in TNG50 images, we have only shown stellar component.

\section{Results}
\label{sec:results}

\subsection{fraction of bulgeless galaxies}
\label{subsec:fraction_of_bulgeless}
\begin{figure*}
    \centering
    \includegraphics[width=0.9\textwidth]{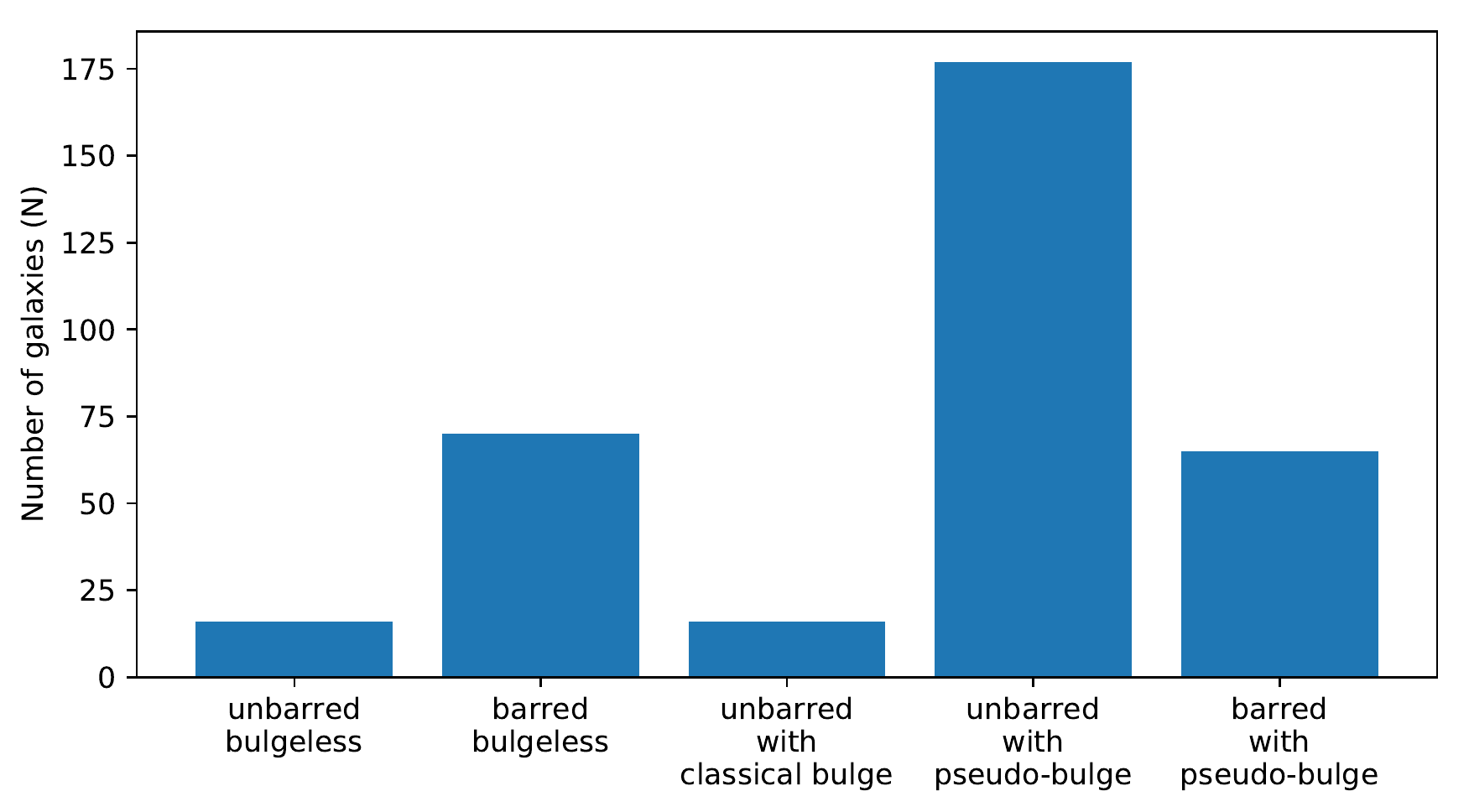}
    \caption{Number distribution of different classes of galaxies in TNG50 simulation. These galaxies are identified using our two dimensional multi-component fitting pipeline.}
    \label{fig:bulgeless_and_bulged_gal}
\end{figure*}
Local volume observation of spiral galaxies shows that approximately $15-30\%$ of disk galaxies are bulgeless \citep{Kautsch.etal.2006, Barazza.etal.2008, Cameron.etal.2009, Karachentsev.Karachentsev.2019}. The fraction of barred bulgeless galaxies is around $87\%$ of all bulgeless galaxies \citep{Barazza.etal.2008}. To compare the statistics of bulgeless galaxies in observations and TNG50 simulation, we took all the galaxies which are best fitted by disk and bulge or/and bar components. We found that $25\%$ of TNG50 disk galaxies are bulgeless. The fraction of barred bulgeless galaxies is $81\%$ of bulgeless galaxies. The fraction of bulgeless galaxies in TNG50 is reasonably good match with the observed fraction in local volume. In Fig.~\ref{fig:bulgeless_and_bulged_gal}, we have shown different classes of galaxies and their numbers found in the TNG50 simulation.

\subsection{Specific angular momentum}
\label{subsec:specific_angmom}
\begin{figure*}
    \centering
    \includegraphics[width=0.65\textwidth]{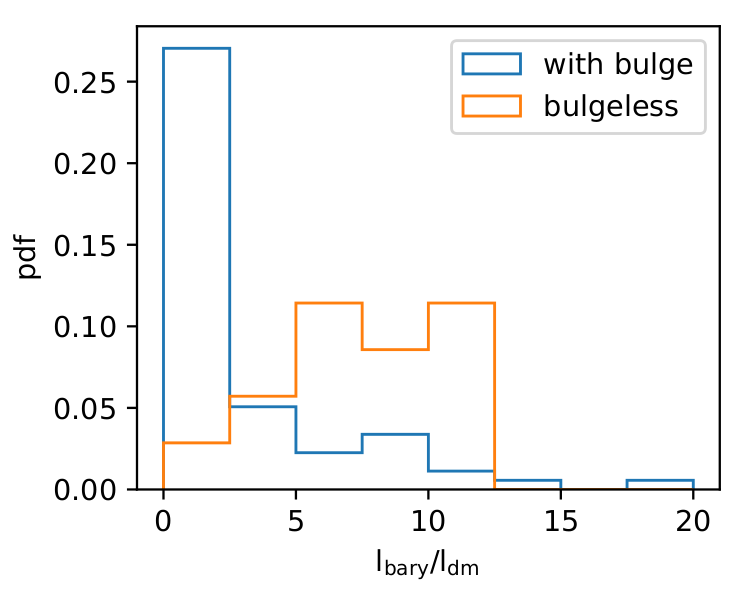}
    \caption{baryon to dark matter specific angular momentum ratio distribution of bulgeless galaxies and galaxies with bulge. A significant fraction of bulgeless galaxies shows higher specific angular momentum as compare to the galaxies with bulge.}
    \label{fig:baryon_to_dark_angmom}
\end{figure*}
Observations have shown that bulgeless galaxies are rich in specific angular momentum \citep{van_den_Bosch.etal.2001, D'Onghia.Burkert.2004}. To test whether bulgeless galaxies in TNG50 are also rich in angular momentum or not, we have calculated their specific angular momenta within two half stellar mass radius \footnote{Here, we define specific angular momentum as $l = L/M$, where $L = \sqrt{(\sum_{i} L_{x,i})^{2} + (\sum_{i} L_{y,i})^{2} + (\sum_{i} L_{z,i})^{2}}$, $M = \sum_{i} m_{i}$, and $i$ runs over all the particles within a fixed radius.}. In Fig.~\ref{fig:baryon_to_dark_angmom}, we have shown the ratio of baryonic and dark matter angular momenta for bulgeless galaxies and galaxies with a bulge. The distribution for bulgeless galaxies and galaxies with bulges are respectively coded with orange and blue colours. It is evident that the bulgeless galaxies are rich in specific angular momentum. The majority of galaxies with bulge are distributed in the low specific angular momentum region.

We have also estimated specific angular momenta of stellar and gas components separately and computed their ratios with specific angular momentum of dark matter. Similar to the baryon to dark matter specific angular momentum ratio, both, gas to dark matter and stellar to dark matter specific angular momentum ratios exhibit distinct distribution for bulgeless galaxies and galaxies with bulges. For brevity, we have not included these results in this chapter.

\subsection{Baryonic to dark matter mass ratio}
\label{subsec:baryon_to_dark_matter}
\begin{figure*}
    \centering
    \includegraphics[width=0.65\textwidth]{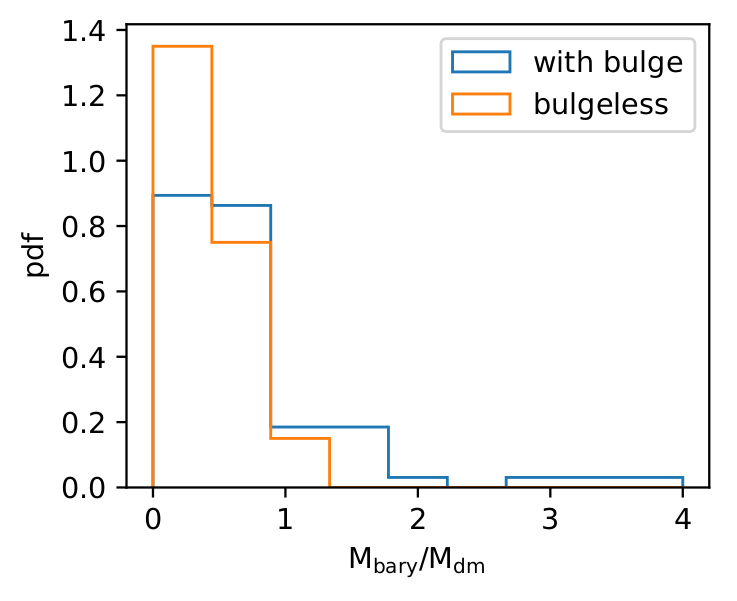}
    \caption{baryon to dark matter mass ratio distribution of bulgeless galaxies and galaxies with bulge. Most of the bulgeless galaxies lies in $0<\rm M_{bary} / \rm M_{dm}<1$ range and galaxies with falls in wider range. It means bulgeless galaxies are dark matter dominating as compare to galaxies with bulge.}
    \label{fig:baryon_to_dark_mass}
\end{figure*}
Bulgeless galaxies are generally dark matter dominated systems \citep{Samland.Gerhard.2003}. The presence of dark matter stabilizes the disk against external disturbances. Here, we have calculated baryonic to dark matter mass ratio with two half stellar mass radius for bulgeless galaxies and for the control sample of galaxies with bulge. Fig.~\ref{fig:baryon_to_dark_mass} represents baryonic to dark matter mass ratio distribution of bulgeless galaxies and galaxies with bulge in TNG50 simulation. Unlike stellar metallcity and specific angular momentum, baryonic to dark matter mass ratio of bulgeless galaxies is not very different from galaxies with bulges. Both the distributions peaks in $0<\rm M_{bary} / \rm M_{dm}<1$ range. However, concentrated distribution of bulgeless galaxies and wider distribution of galaxies with bulges indicate that bulgeless galaxies are dark matter dominated.

\subsection{Stellar metallcity}
\label{subsec:stellar_metallicity}
\begin{figure*}
    \centering
    \includegraphics[width=0.65\textwidth]{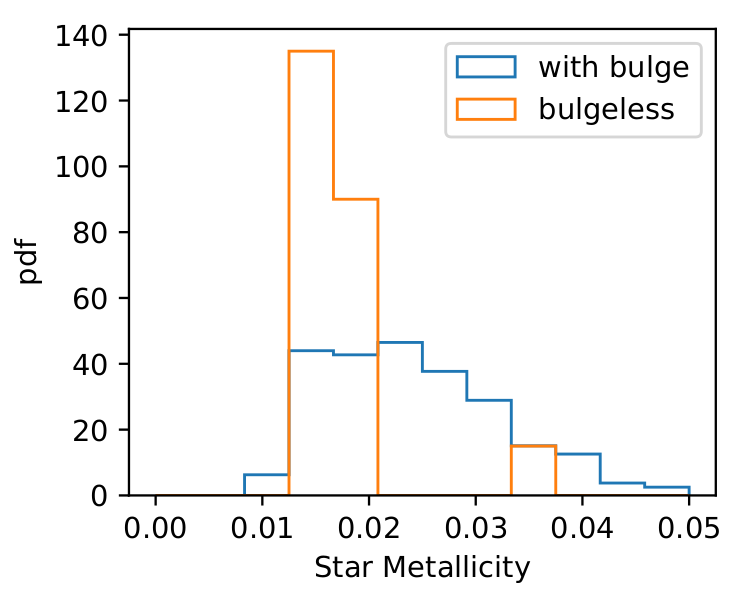}
    \caption{Stellar metallicity distribution of bulgeless galaxies and galaxies with bulge. Bulgeless galaxies are metal-deficient and show narrow metallicity range. However, galaxies with bulge exhibit large metallicity range.}
    \label{fig:stellar_metallicity}
\end{figure*}
In Fig.~\ref{fig:stellar_metallicity}, we have shown the stellar metallicity distribution of bulgeless galaxies and compared it with the stellar metallcity of galaxies with bulge. Bulgeless galaxies and galaxies with bulges are respectively represented with orange and blue colours. Here, we have considered all the elements above helium (He) as metal and have shown mass weighted average stellar metallicity within two half stellar mass radius. It is clear that most of the bulgeless galaxies are metal-deficient as compared to galaxies with bulges. Galaxies with bulges show a wide range of metallicities, whereas bulgeless galaxies exhibit a small metallicity range. For the majority of bulgeless galaxies, stellar metallicity is less than the solar metallicity (solar metallicity $\sim$ 0.02). Observed bulgeless galaxies also typically exhibit low metallicity \citep{Matthews.Gallagher.1997}.

\section{Conclusions}
\label{sec:conclusions}
We have investigated properties of bulgeless galaxies using Illustris TNG50 cosmological data products at redshift zero. To find the bulgeless galaxies, we have developed an automated python language based pipeline for the two dimensional photometric decomposition of galaxies. Our pipeline uses the GALFIT code at its core to decompose the simulated galaxies. The main findings of our study are the following.

\begin{itemize}
    \item In TNG50, approximately $25\%$ of disk galaxies are bulgeless. Out of the bulgeless galaxies around $81\%$ are barred. These fractions of bulgeless and barred bulgeless galaxies are in agreement with the observed fraction of bulgeless galaxies in the local volume.
    
    \item Specific angular momentum of bulgeless galaxies in TNG50 is higher than the galaxies with bulges. The majority of bulgeless galaxies have baryon to dark matter specific angular momentum ratio from 2 to 13. The gas component of bulgeless galaxies shows a higher specific angular momentum as compared to galaxies with bulges.
    
    \item We found that the bulgeless galaxies in TNG50 are dark matter dominated. However, galaxies with bulges have a wide range of baryon to dark matter fraction.
    
    \item Bulgeless galaxies in TNG50 show lower stellar metallicity values compared to control sample of galaxies with bulges. Most of the bulgeless galaxies show mass weighted average metallicities less than the solar metallicity value.
\end{itemize}

So to summarise we find that the bulgeless galaxies in TNG50 reasonably represent the characteristic properties of observed bulgeless galaxies. Thus, we conclude that the galaxy formation model of TNG50 is capable of producing bulgeless galaxies analogous to their observed counterpart.

	\begin{savequote}[100mm]
``Everything dies, from the smallest blade of grass to the biggest galaxy.''
\qauthor{\textbf{$-$ Stephen R. Donaldson}}
\end{savequote}

\chapter[Concluding Summary and Future Directions]{Concluding Summary and Future Directions}
\label{chapter8}

\section{Summary}
\label{sec:summary}
Previous chapters have  discussed details of the individual work completed during the PhD period, including one on-going project. Here, we provide a brief summary of the work presented in this thesis. 

In the Chapter~\ref{chapter3}, we have investigated the effect of minor flybys on the bulges, disks, and spiral arms of Milky Way mass galaxies for two types of bulges -- classical bulges and boxy/peanut pseudo-bulges. Our N-body simulations comprise of two disk galaxies of mass ratios 10:1 and 5:1. We varied the distance of closest approach (pericenter distance) keeping the disks of the galaxies in their orbital plane. We performed photometric and kinematic bulge-disk decomposition at regular time steps and traced the evolution of the disk size, spiral structure, bulge S\'ersic index, bulge mass, and bulge angular momentum. Our results show that the main effect on the disks is disk thickening, which is seen as the increase in the ratio of disk scale height to scale radius. The strength of the spiral structure $A_{2}/A_{0}$ shows small oscillations about the mean time-varying amplitude in the pseudo-bulge host galaxies. The flyby has no significant effect on the non-rotating classical bulges, which shows that these bulges are extremely stable in galaxy interactions. However, the pseudo-bulges become dynamically hotter in flybys indicating that flybys may play an important role in accelerating the rate of secular evolution in disk galaxies. This effect on pseudo-bulges is a result of their rotational nature as part of the bar. Also, flybys do not affect the time and strength of bar buckling.

In the Chapter~\ref{chapter4}, we have tested whether the excitation of vertical breathing motion in a galaxy disk are directly linked to tidal interactions by constructing a set of $N$-body models (with mass ratio 5:1) of unbound, single fly-by interactions with varying orbital configurations. We first reproduce the well-known result that such fly-by interactions can excite strong transient spirals (lasting for $\sim 2.9-4.2~Gyr$) in the outer disc of the host galaxy. The generation and strength of the spirals are shown to vary with the orbital parameters (the angle of interaction, and the orbital spin vector). Furthermore, we demonstrate that our fly-by models exhibit coherent breathing motion whose amplitude increases with height. The amplitudes of the breathing motion show characteristic modulation along the azimuthal direction, with compression breathing motions coinciding with the peaks of the spirals and expanding breathing motions falling in the inter-arm regions -- a signature of a spiral-driven breathing motion. These breathing motions in our models end when the strong tidally-induced spiral arms fade away. Thus, it is the tidally-induced spirals which drives the large-scale breathing motions in our fly-by models, and the dynamical role of the tidal interaction in this context is indirect.

In the Chapter~\ref{chapter5}, we have shown that the dark matter halo shape affects bar formation and bar buckling. We have performed N-body simulations of bar buckling in non-spherical dark matter halos and traced bar evolution for 8~Gyr. We find that bar formation is delayed in oblate halos, resulting in delayed buckling whereas bars form earlier in prolate halos leading to earlier buckling. However, the duration of first buckling remains almost comparable. All the models show two buckling events but the most extreme prolate halo exhibits three distinct buckling features. Bars in prolate halos also show buckling signatures for the longest duration compared to spherical and oblate halos. Since ongoing buckling events are rarely observed, our study suggests that most barred galaxies may have more oblate or spherical halos rather than prolate halos. Our measurement of BPX structures also shows that prolate halos promote bar thickening and disc heating more than oblate and spherical halos.

In the Chapter~\ref{chapter6}, we have quantified the evolution of bulges since $z=0.1$ using photometric parameters of nearly 39,000 unbarred disc galaxies from SDSS DR7 which are well represented by two components. We adopted a combination of the S\'ersic index and Kormendy relation to separate classical bulges and disc-like pseudo-bulges. We found that the fraction of pseudo-bulges smoothly increases, and the fraction of classical bulges smoothly decreases as the Universe gets older. In the history of the Universe, there comes a point ($z \approx 0.016$) when classical bulges and pseudo-bulges become equal in number. The fraction of pseudo-bulges rises with increasing bulge to disc half-light radius ratio until R$_{\rm e}$/R$_{\rm hlr} \approx 0.6$ suggesting concentrated disc is the most favourable place for pseudo-bulge formation. The mean ellipticity of pseudo-bulges is always greater than that of classical bulges and it decreases with decreasing redshift indicating that the bulges tend to be more axisymmetric with evolution. Also, the massive bulges are progressing towards axisymmetry at a steeper rate than the low-mass bulges. There is no tight correlation of bulge S\'ersic index evolution with other photometric properties of the galaxies. Using the sample of multi-component fitting of $S^4G$ data and $N-$body galaxy models, we have verified that our results are consistent and even more pronounced with multi-component fitting and high-resolution photometry.

In the Chapter~\ref{chapter7}, we have studied bulgeless galaxies in the Illustris TNG50 simulations to test its galaxy formation model. We selected all the redshift zero galaxies with stellar mass greater than 10$^9$M$_{\odot}$ and separated sample of bulgeless galaxies using two dimensional photometric decomposition. We calculated distributions of parameters for bulgeless galaxies and compared with a control sample of galaxies with bulge. We found that bulgeless galaxies in TNG50 exhibit low stellar metallicity, high specific angular momentum, and large dark matter fraction as compared to the galaxies with bulge. Nearly $25\%$ of disk galaxies in TNG50 are bulgeless and $81\%$ of these bulgeless galaxies are barred. Fraction of bulgeless galaxies in TNG50 matches well with observed fraction. We conclude that the TNG50 physics of galaxy formation is capable of producing observed characteristics of bulgeless galaxies in local volume.

\section{Future Directions}
\label{sec:future_directions}
The work presented in this thesis opens up various directions to explore in the future. Here, we list some possible projects to carry forward based on our individual work. 

The work we presented in chapter~\ref{chapter3} showed the effect of fly-by interactions on the evolution of bulge, disk, and spiral arms in a Milky Way mass galaxy. We performed purely $N-$body simulations of single minor fly-by interactions. For simplicity, we considered fly-by interactions on prograde orbits and varied the paricenter distances. Galaxies also contain a significant amount of gas, the fly-by interactions can occur between similar mass galaxies, the orbit of the interaction can be inclined, and a galaxy may experience more than one fly-by interaction simultaneously. These possibilities open the door for future projects where one can explore the consequences of these varying parameters on disk thickening, spiral arm formation, and growth of bulges. These possible projects will also include the evolution of galaxies in cosmological environments, such as  groups and clusters of galaxies.

Using purely $N-$body simulation of fly-by interactions with various orbital configuration, we showed that the excitation of wave-like breathing motion in the Milky Way is related to its spiral arms. The effect of tidal interactions is indirect. Our simulations do not account for the presence of gas and bar. The Milky Way hosts a strong bar at its center. One can study the contribution of bar on the wave-like breathing motion. The role of breathing motion on disk thickening and gas flaring will also be interesting to investigate. Studying breathing motion can be useful to understand the interaction history of the Milky Way and its siblings.

One of our studies showed that the dark matter distribution around the disk galaxies affects the formation of bars and the resultant boxy/peanut/x-shape (BPX) bulges. The shape of dark matter halo leaves its imprint on baryonic matter. A detailed study of BPX bulges in simulated galaxies may give us hints of dark matter distribution in observed disk galaxies. In this study, we did not include central bulge, super massive black hole (SMBH), and gas physics. It will be interesting to study the role of a central potential and gas physics on bar buckling and resultant BPX bulge formation in non-spherical dark matter distributions.

In our work on the cosmic evolution of disk-like pseudo-bulges, we have shown that the fraction of pseudo-bulges is increasing as the Universe becomes older. A large fraction of the local volume is rotation dominated but in the $\Lambda$CDM model of hierarchical structure formation, we expect a dispersion dominated Universe. Our results are thus useful for constraining the formation and evolution of bulges. These results may also be important for setting the initial conditions in cosmological simulations.

	
	

	\bibliographystyle{mnras}
	\bibliography{galaxies_bib}
	\printthesisindex
\end{document}